\documentclass[prd,showpacs,amsmath,amssymb,floatfix]{revtex4}
\usepackage{graphicx,color,dcolumn,booktabs}
\def\feynslash#1{#1\!\!\!\slash}
\begin{document}

\title{The molecular systems composed of the charmed mesons in the $H\bar{S}+h.c.$ doublet}

\author{Lei-Lei Shen}
\author{Xiao-Lin Chen}
\email{chenxl@pku.edu.cn}
\author{Zhi-Gang Luo}
\author{Peng-Zhi Huang}
\author{Shi-Lin Zhu}\email{zhusl@pku.edu.cn}
\affiliation{Department of Physics
and State Key Laboratory of Nuclear Physics and Technology\\
and Center of High Energy Physics, Peking University, Beijing
100871, China }

\author{Peng-Fei Yu}
\author{Xiang Liu\footnote{Corresponding author}}\email{xiangliu@lzu.edu.cn}
 \affiliation{School of Physical
Science and Technology, Lanzhou University, Lanzhou 730000, China}

\date{\today}

\begin{abstract}

We study the possible heavy molecular states composed of a pair of
charm mesons in the H and S doublets. Since the P-wave
charm-strange mesons $D_{s0}(2317)$ and $D_{s1}(2460)$ are
extremely narrow, the future experimental observation of the
possible heavy molecular states composed of $D_s/D_s^\ast$ and
$D_{s0}(2317)/D_{s1}(2460)$ may be feasible if they really exist.
Especially the possible $J^{PC}=1^{--}$ states may be searched for
via the initial state radiation technique.

\end{abstract}

 \maketitle

\section{Introduction}\label{sec1}

The new family of the charmonium or charmonium-like states include
$X(3872)$, $Y(3940)$, $Y(4260)$, $Z(3930)$, $X(3940)$, $Y(4325)$,
$Y(4360)$, $Y(4660)$, $Z^+(4430)$, $Z^+(4050), Z^+(4250)$ and
$Y(4140)$ etc
\cite{2003-Choi-p262001-262001,2005-Choi-p182002-182002,
2005-Aubert-p142001-142001, 2006-Uehara-p82003-82003,
2007-Abe-p82001-82001, 2007-Aubert-p212001-212001,
2007-Yuan-p182004-182004, 2007-Wang-p142002-142002,
2008-Choi-p142001-142001, 2008-Mizuk-p72004-72004,
2009-Aaltonen-p242002-242002}. Many states sit on the the
threshold of two charmed mesons, which inspired some of them
(especially those charged ones) to be candidates of heavy
moleculues \cite{2004-Swanson-p197-202,close1,
2008-Liu-p94015-94015,2009-Liu-p17502-17502,liu1,liu2,liu3,liu4,liu5}.

In the heavy quark limit, the S-wave and P-wave heavy mesons can
be categorized into three doublets: $H=(0^-,1^-)$, $S=(0^+,1^+)$,
$T=(1^+,2^+)$. We collect their masses from PDG in Table \ref{HS}.
The bottom mesons in the $S$ doublet are still missing
experimentally. Thus, we will adopt the theoretical predictions of
the bottom meson masses in the $S$ doublet when we study the heavy
flavor molecular system composed of the bottom and anti-bottom
mesons.

In the framework of the meson exchange model, we have investigated
the possible loosely bound molecular states composed of a pair of
heavy mesons in Refs. \cite{liu1,liu2,liu3,liu4,liu5,liu6}. In
this work, we will investigate the possible heavy molecular system
constructed by the charmed and anti-charmed mesons, where one
meson is in the $H$ doublet and the other one is in the $S$
doublet. In the following, we denote the heavy flavor molecular
system as the $H\bar S+h.c.$ system for the convenience.

The $H\bar{S}+h.c.$ system can be categorized into four
subsystems: $[PP_0^*]$, $[PP_1^\prime]$, $[P^* P_0^*]$ and $[P^*
P_1^\prime]$. They correspond to different quantum number
combinations $0^-+0^+$, $0^-+1^+$, $1^-+0^+$ and $1^-+1^+$,
respectively. Since charmed mesons belong to the fundamental
representation of flavor $SU(3)$, the system constructed by the
charmed meson and anti-charmed meson forms an octet and a singlet:
$3\otimes \bar 3=8\oplus1$ as illustrated in Table
\ref{wavefunction}. The parameter $c=\mp1$ in the flavor wave
functions corresponds to the charge parity $C=\pm1$ respectively
for the neutral systems as pointed out in Refs.
\cite{liu1,liu2,liu3,liu4}.

This paper is organized as follows. We review the formalism in
Section \ref{sec2} and present the results in Section \ref{sec3}.
The last section is a short summary.

\section{Theoretical framework}
\label{sec2}

\subsection{The potential model}

The potential model is an effective approach to study the two-body
bound state problem. For the $H\bar{S}+h.c.$ system, the
scattering between the charmed and anti-charmed meson occurs via
exchanging the light pseudoscalar, scalar and vector mesons, which
play the role of providing long-distant, intermediate-distance and
short-distance forces. At the hadron level, there exist two types
of diagrams in the scattering of the charmed and anti-charmed
mesons, i.e., the cross and direct diagrams as shown in Table
\ref{diagrams}. The exchanged light mesons relevant to the four
subsystems $[PP_0^*]$, $[PP_1^\prime]$, $[P^*P_0^*]$ and
$[P^*P_1^\prime]$ are also presented in Table \ref{diagrams}. When
writing out the scattering amplitude, the monopole form factor is
introduced at every interaction vertex to compensate the off-shell
effect of the exchanged light meson
\begin{eqnarray}
\mathcal{F}(q)=\frac{\Lambda^2-m^2}{\Lambda^2-q^2},
\end{eqnarray}
where the phenomenological cutoff parameter $\Lambda$ is about 1
GeV. $q$ and $m$ denote the four-momentum and the mass of the
exchanged meson.

\renewcommand{\arraystretch}{1.6}
\begin{center}
\begin{table}[htb]
\begin{tabular}{cccccccccccc}
\toprule[1pt]
$H$ doublet\\
\begin{tabular}{c||cccc|cccccc }\toprule[1pt]
&\multicolumn{4}{c}{charmed-up(down) meson  }&&&&\multicolumn{2}{c}{charmed-strange meson  }&\\
\cline{2-10}

\raisebox{1.0ex}{$J^{P}$}&charged&mass (MeV)&neutral&mass (MeV)&
&&&charged&mass (MeV)\\\cline{1-10}

$0^-$&$D^{\pm}$&1869.3& $D^{0}$&1864.5&&&&$D_s^{\pm}$&1968.2
\\
$1^-$&$D^{*\pm}$&2010.0&$D^{*0}$&2006.7&&&&$D_s^{*\pm}$&2112.0
\\\midrule[1pt]
&\multicolumn{4}{c}{bottom-up(down) meson  }&&&&\multicolumn{2}{c}{bottom-strange meson  }&\\
\cline{2-10}

\raisebox{1.0ex}{$J^{P}$}&charged&mass (MeV)&neutral&mass (MeV)&
&&&neutral&mass (MeV)\\\cline{1-10}

$0^-$&$B^{\pm}$&5279.1& $B^{0}$&5279.5&&&&$B_s^{0}$&5365.1
\\
$1^-$&$B^{*\pm}$&5325.1&$B^{*0}$&5325.1&&&&$B_s^{*0}$&5412.0
\\\bottomrule[1pt]
\end{tabular}\\

$S$ doublet\\
\begin{tabular}{c||cccc|cccccc }\toprule[1pt]
&\multicolumn{4}{c}{charmed-up(down) meson  }&&&&\multicolumn{2}{c}{charmed-strange meson  }&\\
\cline{2-10}

\raisebox{1.0ex}{$J^{P}$}&charged&mass (MeV)&neutral&mass (MeV)&
&&&charged&mass (MeV)\\\cline{1-10}

$0^+$&$D_0^{\pm}$&2403& $D_0^{0}$&2352&&&&$D_{s0}^{\pm}$&2317.3
\\
$1^+$&$D_1^{\prime\pm}$&-&$D_1^{\prime0}$&2427&&&&$D_{s1}^{\prime\pm}$&2458
\\\midrule[1pt]
&\multicolumn{4}{c}{bottom-up(down) meson  }&&&&\multicolumn{2}{c}{bottom-strange meson  }&\\
\cline{2-10}

\raisebox{1.0ex}{$J^{P}$}&charged&mass (MeV)&neutral&mass (MeV)&
&&&neutral&mass (MeV)\\\cline{1-10}

$0^+$&$B_0^{\pm}$&-& $B_0^{0}$&-&&&&$B_{s0}^{0}$&-
\\
$1^+$&$B_1^{\prime\pm}$&-&$B_1^{\prime0}$&-&&&&$B_{s1}^{\prime0}$&-
\\\bottomrule[1pt]
\end{tabular}
\end{tabular}
\caption{The masses of the heavy mesons in the $H$ and $S$ doublets
\cite{pdg}. \label{HS}}
\end{table}
\end{center}

\renewcommand{\arraystretch}{1.8}
\begin{center}
\begin{table}[htb]

\begin{tabular}{cccccccccccc}
&(a) $[PP_0^*]$ system&(b) $[PP_1^\prime]$ system&\\
\raisebox{-10em}{\scalebox{0.7}{\includegraphics{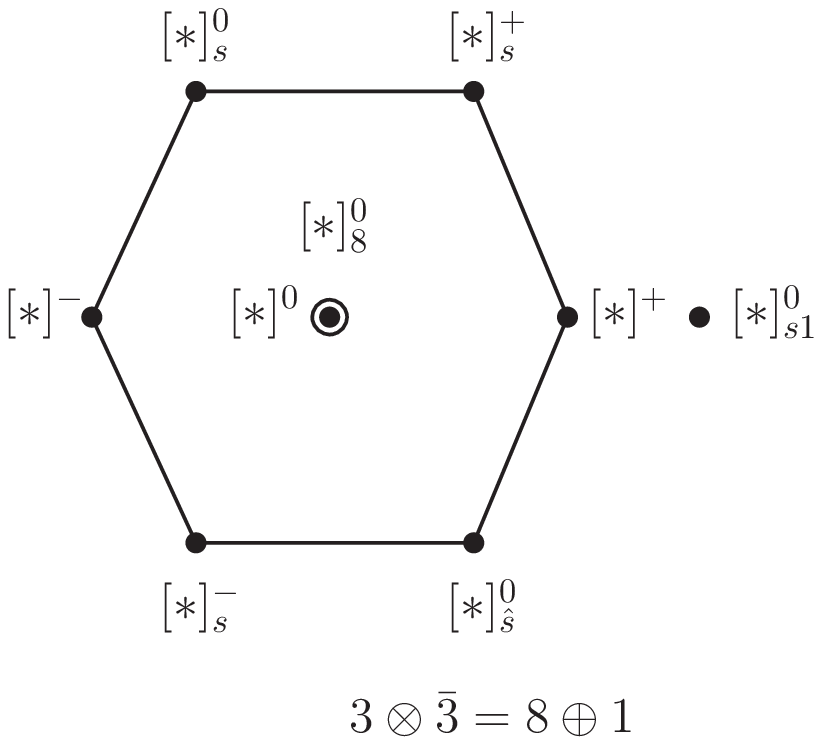}}}&

\begin{tabular}{c|c}\toprule[1pt]
States&Wave function\\\midrule[1pt]

$[PP_0^*]^+$&$\frac{1}{\sqrt{2}}[D^+\bar{D}_0^0 +c\bar{D}^0D_0^+]$\\
\raisebox{0em}{$[PP_0^*]^0$}&$\frac{1}{2}[(D^0\bar{D}_0^0+c\bar{D}^0 D_0^0)$\\
&$-(D^+D_0^-+c{D}^- D_0^+)]$\\
$[PP_0^*]^-$&$\frac{1}{\sqrt{2}}[D^-\bar{D}_0^0 +c\bar{D}^0D_0^-]$\\

$[PP_0^*]_s^+$&$\frac{1}{\sqrt{2}}[D_s^+\bar{D}_0^0 +c\bar{D}^0D_{s0}^+]$\\
{$[PP_0^*]_s^0$}&$\frac{1}{\sqrt{2}}[D_s^+D_0^-+c{D}^- D_{s0}^+]$\\
$[PP_0^*]_s^-$&$\frac{1}{\sqrt{2}}[D_s^-\bar{D}_0^0 +c\bar{D}^0D_{s0}^-]$\\

$[PP_0^*]_{\hat s}^0$&$\frac{1}{\sqrt{2}}[D^+D_{s0}^-+c{D}_s^- D_{0}^+]$\\

\raisebox{0em}{$[PP_0^*]_8^0$}&$\frac{1}{2}[(D^0\bar{D}_0^0+c\bar{D}^0 D_0^0)$\\
&$+(D^+D_0^-+c{D}^- D_0^+)]$\\
$[PP_0^*]_{s1}^0$&$\frac{1}{\sqrt{2}}[D_s^+D_{s0}^{-}+cD_s^-
D_{s0}^+]$

\\\bottomrule[1pt]
\end{tabular}
&
\begin{tabular}{c|c}\toprule[1pt]
States&Wave function\\\midrule[1pt]

$[PP_1^\prime]^+$&$\frac{1}{\sqrt{2}}[D^+\bar{D}_1^{\prime0} +c\bar{D}^0D_1^{\prime+}]$\\
\raisebox{0em}{$[PP_1^\prime]^0$}&$\frac{1}{2}[(D^0\bar{D}_1^{\prime0}+c\bar{D}^0 D_1^{\prime0})$\\
&$-(D^+D_0^-+c{D}^- D_0^+)]$\\
$[PP_1^\prime]^-$&$\frac{1}{\sqrt{2}}[D^-\bar{D}_1^{\prime 0} +c\bar{D}^0D_1^{\prime-}]$\\

$[PP_1^\prime]_s^+$&$\frac{1}{\sqrt{2}}[D_s^+\bar{D}_1^{\prime0} +c\bar{D}^0D_{s1}^{\prime+}]$\\
{$[PP_1^\prime]_s^0$}&$\frac{1}{\sqrt{2}}[D_s^+D_1^{\prime-}+c{D}^- D_{s1}^{\prime+}]$\\
$[PP_1^\prime]_s^-$&$\frac{1}{\sqrt{2}}[D_s^-\bar{D}_1^{\prime0} +c\bar{D}^0D_{s1}^{\prime-}]$\\

$[PP_1^\prime]_{\hat s}^0$&$\frac{1}{\sqrt{2}}[D^+D_{s1}^{\prime-}+c{D}_s^- D_{1}^{\prime+}]$\\

\raisebox{0em}{$[PP_1^\prime]_8^0$}&$\frac{1}{2}[(D^0\bar{D}_1^{\prime0}+c\bar{D}^0 D_1^{\prime0})$\\
&$+(D^+D_1^{\prime-}+c{D}^- D_1^{\prime+})]$\\
$[PP_1^\prime]_{s1}^0$&$\frac{1}{\sqrt{2}}[D_s^+D_{s1}^{\prime-}+cD_s^-
D_{s1}^{\prime+}]$

\\\bottomrule[1pt]
\end{tabular}

\\

&

(c) $[P^*P_0^*]$ system&(d) $[P^*P_1^\prime]$ system
\\
&
\begin{tabular}{c|c}\toprule[1pt]
States&Wave function\\\midrule[1pt]

$[P^*P_0^*]^+$&$\frac{1}{\sqrt{2}}[D^{*+}\bar{D}_0^0 +c\bar{D}^{*0}D_0^+]$\\

\raisebox{0em}{$[P^*P_0^*]^0$}&$\frac{1}{2}[(D^{*0}\bar{D}_0^0+c\bar{D}^{*0} D_0^0)$\\
&$-(D^{*+}D_0^-+c{D}^{*-} D_0^+)]$\\
$[P^*P_0^*]^-$&$\frac{1}{\sqrt{2}}[D^{*-}\bar{D}_0^0 +c\bar{D}^{*0}D_0^-]$\\

$[P^*P_0^*]_s^+$&$\frac{1}{\sqrt{2}}[D_s^{*+}\bar{D}_0^0 +c\bar{D}^{*0}D_{s0}^+]$\\
{$[P^*P_0^*]_s^0$}&$\frac{1}{\sqrt{2}}[D_s^{*+}D_0^-+c{D}^{*-} D_{s0}^+]$\\
$[P^*P_0^*]_s^-$&$\frac{1}{\sqrt{2}}[D_s^{*-}\bar{D}_0^0 +c\bar{D}^{*0}D_{s0}^-]$\\

$[P^*P_0^*]_{\hat s}^0$&$\frac{1}{\sqrt{2}}[D^{*+}D_{s0}^-+c{D}_s^{*-} D_{0}^+]$\\

\raisebox{0em}{$[P^*P_0^*]_8^0$}&$\frac{1}{2}[(D^{*0}\bar{D}_0^0+c\bar{D}^{*0} D_0^0)$\\
&$+(D^{*+}D_0^-+c{D}^{*-} D_0^+)]$\\
$[P^*P_0^*]_{s1}^0$&$\frac{1}{\sqrt{2}}[D_s^{*+}D_{s0}^{-}+cD_s^{*-}
D_{s0}^+]$

\\\bottomrule[1pt]
\end{tabular}
&
\begin{tabular}{c|c}\toprule[1pt]
States&Wave function\\\midrule[1pt]

$[P^*P_1^\prime]^+$&$\frac{1}{\sqrt{2}}[D^{*+}\bar{D}_1^{\prime0} +c\bar{D}^{*0}D_1^{\prime+}]$\\
\raisebox{0em}{$[P^*P_1^\prime]^0$}&$\frac{1}{2}[(D^{*0}\bar{D}_1^{\prime0}+c\bar{D}^{*0} D_1^{\prime0})$\\
&$-(D^{*+}D_0^-+c{D}^{*-} D_0^+)]$\\
$[P^*P_1^\prime]^-$&$\frac{1}{\sqrt{2}}[D^{*-}\bar{D}_1^{\prime0} +c\bar{D}^{*0}D_1^{\prime-}]$\\

$[P^*P_1^\prime]_s^+$&$\frac{1}{\sqrt{2}}[D_s^{*+}\bar{D}_1^{\prime0} +c\bar{D}^{*0}D_{s1}^{\prime+}]$\\
{$[P^*P_1^\prime]_s^0$}&$\frac{1}{\sqrt{2}}[D_s^{*+}D_1^{\prime-}+c{D}^{*-} D_{s1}^{\prime+}]$\\
$[P^*P_1^\prime]_s^-$&$\frac{1}{\sqrt{2}}[D_s^{*-}\bar{D}_1^{\prime0} +c\bar{D}^{*0}D_{s1}^{\prime-}]$\\

$[P^*P_1^\prime]_{\hat s}^0$&$\frac{1}{\sqrt{2}}[D^{*+}D_{s1}^{\prime-}+c{D}_s^{*-} D_{1}^{\prime+}]$\\

\raisebox{0em}{$[P^*P_1^\prime]_8^0$}&$\frac{1}{2}[(D^{*0}\bar{D}_1^{\prime0}+c\bar{D}^{*0} D_1^{\prime0})$\\
&$+(D^{*+}D_1^{\prime-}+c{D}^{*-} D_1^{\prime+})]$\\
$[P^*P_1^\prime]_{s1}^0$&$\frac{1}{\sqrt{2}}[D_s^{*+}D_{s1}^{\prime-}+cD_s^{*-}
D_{s1}^{\prime+}]$

\\\bottomrule[1pt]
\end{tabular}&&&&

\end{tabular}
\caption{The flavor wave functions of the $[PP_0^*]$,
$[PP_1^\prime]$, $[P^*P_0^*]$ and $[P^*P_1^\prime]$ systems. We
use $[\cdots]^{a}_{b}$ to name the different states, where the
superscript $a$ denotes the charge of the state and subscript $b$
is introduced to distinguish the different states in a subsystem.
\label{wavefunction}}
\end{table}
\end{center}

According to the effective Lagrangian describing the interaction
between the light and charmed mesons, one can write down the
scattering amplitude between the charmed and anti-charmed mesons.
Such a system can be described as
\begin{eqnarray}
|J,J_Z\rangle=\sum_{\lambda_1,\lambda_2}\langle
J_1,\lambda_1;J_2,\lambda_2|J,J_Z\rangle |p_1,p_2\rangle,
\end{eqnarray}
where $J$ and $J_i$ ($i=1,2$) denote the angular momentum of the
system and the componentents. The scattering amplitude
$i\mathcal{M}(J,J_Z)$ is related to the interaction potential in
the momentum space in terms of the Breit approximation
\begin{eqnarray}
\mathcal{V}(\mathbf{q})=-\frac{1}{\sqrt{\prod_i 2m_i \prod_f
2m_f}}\mathcal{M}(J,J_Z) \; .
\end{eqnarray}
Here, $m_{i}$ and $m_j$ denote the masses of the initial and final
statrs respectively. The potential in the coordinate space
$\mathcal{V}(\mathbf{r})$ is obtained after Fourier
transformation.

\subsection{The effective Lagrangian}

The effective Lagrangian describing the interaction between the
light and heavy flavor mesons is constructed with the help of the
chiral symmetry and heavy quark symmetry \cite{lag}
\begin{eqnarray}
  \label{eq:lag}
  \mathcal{L} &=&
  igTr[H_b\gamma_\mu\mathcal{A}^\mu_{ba}\gamma_5\bar{H}_a] +
  ig'Tr[S_b\gamma_\mu\mathcal{A}^\mu_{ba}\gamma_5\bar{S}_a] +\{
  ihTr[S_b\gamma_\mu\mathcal{A}^\mu_{ba}\gamma_5\bar{H}_a] + h.c.\} \nonumber\\ 
  && + i\beta{}Tr[H_bv^\mu(\mathcal{V}_\mu-\rho_\mu)_{ba}\bar{H}_a] +
  i\lambda{}Tr[H_b\sigma^{\mu\nu}F_{\mu\nu}(\rho)_{ba}\bar{H}_a] \nonumber\\
  && + i\beta{'}Tr[S_bv^\mu(\mathcal{V}_\mu-\rho_\mu)_{ba}\bar{S}_a] +
  i\lambda{'}Tr[S_b\sigma^{\mu\nu}F_{\mu\nu}(\rho)_{ba}\bar{S}_a] \nonumber\\
  && + \{i\zeta{}Tr[H_b\gamma^\mu(\mathcal{V}_\mu-\rho_\mu)_{ba}\bar{S}_a]
  +  i\mu{}Tr[H_b\sigma^{\mu\nu}F_{\mu\nu}(\rho)_{ba}\bar{S}_a] + h.c.\} \nonumber\\ 
  && + g_\sigma{}Tr[H_a\sigma\bar{H}_a] +
  g'_{\sigma}Tr[S_a\sigma\bar{S}_a] + \Big\{\frac{h_\sigma}{f_\pi}Tr[S_a(\partial_\mu\sigma)\gamma^\mu\bar{H}_a] +
  h.c.\Big\}. \label{lagrangian} 
\end{eqnarray}
The $H$ and $S$ fields, which correspond to the $(0^-,1^-)$ and
$(0^+,1^+)$ doublets are defined respectively
\begin{eqnarray}
  \label{eq:doublets}
  H_a &=& \frac{1+\feynslash{v}}{2}[P_{a\mu}^{*}\gamma^\mu + iP_a\gamma_5],\nonumber\\
  \bar{H}_b &=&\gamma^0 H^\dag \gamma^0= [P^{*\dag}_{a\mu}\gamma^\mu + iP_a^\dag\gamma_5]\frac{1+\slash \!\!\! v}{2},\nonumber\\
    S_a &=& \frac{1+\slash \!\!\! v}{2}[P_{1a}^{\prime\mu}\gamma_\mu\gamma_5-P^*_{0a}],\nonumber\\
  \bar{S}_a &=& \gamma^0 S^\dag \gamma^0= [P^{\prime\dag}_{1a\mu}\gamma^\mu\gamma_5 - P^{*\dag}_{0a}]
  \frac{1+\slash \!\!\! v}{2},
\end{eqnarray}
where $v=(1,0,0,0)$. Here, the annihilation operations
$P,\,P^*_\mu,\,P_0^*,\,P_{1\mu}^\prime$ are of dimension $3/2$ and
satisfy the normalization relations
\begin{eqnarray*}
\langle 0|P|Q\bar{q}(0^-)\rangle=\sqrt{M_H},\qquad
\langle 0|P^*_\mu|Q\bar{q}(0^-)\rangle=\epsilon_\mu\sqrt{M_H},\\
\langle 0|P_0^*|Q\bar{q}(0^+)\rangle=\sqrt{M_S},\qquad \langle
0|P_{1\mu}^\prime|Q\bar{q}(1^+)\rangle=\epsilon_\mu\sqrt{M_S}.
\end{eqnarray*}
In Eq. (\ref{lagrangian}), the expansion of the axial vector gives
\begin{eqnarray}
\mathcal{A}_{\mu}=\frac{1}{2}\big(\xi^{\dag}\partial_{\mu}\xi-\xi\partial_{\mu}\xi^{\dag}\big)=
\frac{i}{f_{\pi}}\partial_{\mu}\mathcal{M}+\cdot\cdot\cdot
\end{eqnarray}
with $\xi=\exp(i\mathbb{P}/f_{\pi})$ and $f_\pi=132$ MeV.
$\rho^\mu_{ba}=ig_{_V}\mathbb{V}^\mu_{ba}/\sqrt{2}$ and
$F_{\mu\nu}(\rho)=\partial_\mu\rho_\nu - \partial_\nu\rho_\mu +
[\rho_\mu,{\ } \rho_\nu]$. The octet pseudoscalar and nonet vector
meson matrices read as
\begin{eqnarray}
\mathbb{P}&=&\left(\begin{array}{ccc}
\frac{\pi^{0}}{\sqrt{2}}+\frac{\eta}{\sqrt{6}}&\pi^{+}&K^{+}\\
\pi^{-}&-\frac{\pi^{0}}{\sqrt{2}}+\frac{\eta}{\sqrt{6}}&
K^{0}\\
K^- &\bar{K}^{0}&-\frac{2\eta}{\sqrt{6}}
\end{array}\right),\quad
\mathbb{V}=\left(\begin{array}{ccc}
\frac{\rho^{0}}{\sqrt{2}}+\frac{\omega}{\sqrt{2}}&\rho^{+}&K^{*+}\\
\rho^{-}&-\frac{\rho^{0}}{\sqrt{2}}+\frac{\omega}{\sqrt{2}}&
K^{*0}\\
K^{*-} &\bar{K}^{*0}&\phi
\end{array}\right).\label{vector}
\end{eqnarray}

\begin{center}
\begin{table}[htb]
\begin{tabular}{c|ccc|ccc}\toprule[1pt]
&\multicolumn{3}{c|}{\it Direct diagram} &\multicolumn{3}{c}{\it
Crossed diagram}
\\
&\multicolumn{3}{c|}{\scalebox{0.5}{\includegraphics{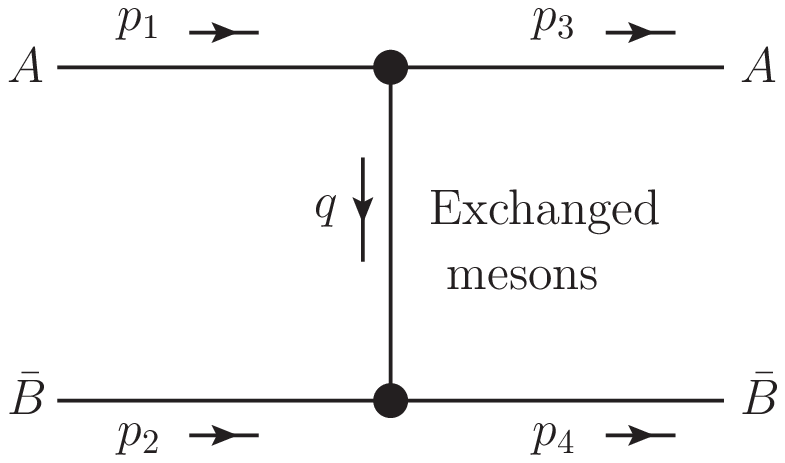}}}&\multicolumn{3}{c}{\scalebox{0.5}{\includegraphics{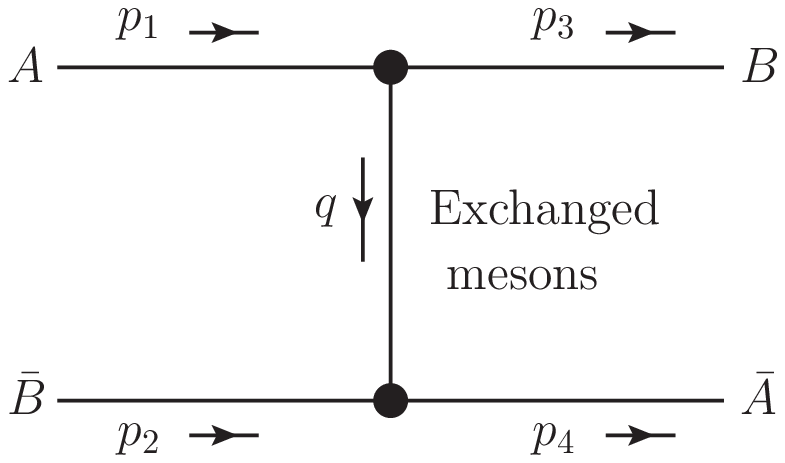}}}
\\\cline{1-7}
\raisebox{-0.5em}{{\it {\bf \underline{Subsystems}}}} & {\it
Pseudoscalar} & {\it Vector} & {\it Scalar} & {\it Pseudoscalar} &
{\it Vector} & {\it Scalar}
\\
&
$(\mathbb{P})$&$(\mathbb{V})$&($\mathbb{S})$&$(\mathbb{P})$&$(\mathbb{V})$&$(\mathbb{S})$\\\cline{2-7}
$[PP_0^*]$&&$\checkmark$&$\checkmark$&$\checkmark$&&\\
$[PP_1^\prime]$& &$\checkmark$&$\checkmark$&&$\checkmark$&$\checkmark$\\
$[P^*P_0^*]$&&$\checkmark$&$\checkmark$&&$\checkmark$&$\checkmark$\\
$[P^*P_1^\prime]$&$\checkmark$&$\checkmark$&$\checkmark$&$\checkmark$&$\checkmark$&$\checkmark$
\\\toprule[1pt]
\end{tabular}
\caption{The direct and crossed scattering diagrams with the
exchanged light mesons for the $H\bar{S}+h.c.$ system
\label{diagrams}.}
\end{table}
\end{center}

After expanding Eq. (\ref{eq:lag}), the effective Lagrangian of
the pseudoscalar mesons with heavy flavor mesons reads
\begin{eqnarray}\label{eq:lag-p-exch}
  \mathcal{L}_{\mathcal{D}^*\mathcal{D}^*\mathbb{P}} &=& \frac{g}{f_\pi}\varepsilon_{\alpha\mu\beta\nu}(2iv^\alpha{}P^{*\mu}_{b}P^{*\nu\dag}_{a})\partial^\beta{}\mathbb{P}_{ba},\\
  \mathcal{L}_{\mathcal{D}^*\mathcal{D}\mathbb{P}} &=& i\frac{2g}{f_\pi}(P_bP^{*\dag}_{a\lambda}-P^*_{b\lambda}P^\dag_{a})\partial^\lambda{}\mathbb{P}_{ba},\\
  \mathcal{L}_{\mathcal{D}_1'\mathcal{D}_1'\mathbb{P}} &=& \frac{g'}{f_\pi}\varepsilon_{\alpha\mu\beta\nu}(2iv^\alpha)P_{1b}^{\prime\mu}P_{1a}^{\prime\nu\dag}
  \partial^\beta{}\mathbb{P}_{ba},\\
  \mathcal{L}_{\mathcal{DD}^*_0\mathbb{P}} &=& \frac{h}{f_\pi}(2iv_{\mu})(P^*_{0b}P^\dag_a - P_bP^{*\dag}_{0a})\partial^\mu{}\mathbb{P}_{ba},\\
  \mathcal{L}_{\mathcal{D}^*\mathcal{D}_1'\mathbb{P}} &=& i\frac{h}{f_\pi}(P'_{1b}\cdot{}P^{*\dag}_a + P^*_b\cdot{}P^{\prime\dag}_{1a})(2iv)\cdot\partial{}\mathbb{P}_{ba}\;
  .
\end{eqnarray}
The effective Lagrangian of the vector mesons with heavy flavor
mesons reads
\begin{eqnarray}\label{eq:lag-v-exch}
  \mathcal{L}_{\mathcal{DD}\mathbb{V}} &=& i\frac{\beta{}g_V}{\sqrt{2}}P_bP_a^\dag(2iv_\mu)\mathbb{V}^\mu_{ba},\\
  \mathcal{L}_{\mathcal{D}^*\mathcal{D}\mathbb{V}} &=& -2\frac{\lambda{}g_V}{\sqrt{2}}(2iv^\lambda)\varepsilon_{\lambda\alpha\beta\mu}(P_bP^{*\mu\dag}_a - P_b^{*\mu}P^\dag_a)(\partial^\alpha{}\mathbb{V}^\beta)_{ba},\\
  \mathcal{L}_{\mathcal{D}^*\mathcal{D}^*\mathbb{V}} &=& -i\frac{\beta{}g_V}{\sqrt{2}}P_b^*\cdot{}P^{*\dag}_a(2iv\cdot{}\mathbb{V}_{ba})-i4\frac{\lambda{}g_V}{\sqrt{2}}P^{*\mu}_bP^{*\nu\dag}_a(\partial_\mu{}\mathbb{V}_\nu - \partial_\nu{}\mathbb{V}_\mu)_{ba},\\
  \mathcal{L}_{\mathcal{D}_0^*\mathcal{D}_0^*\mathbb{V}} &=& -i\frac{\beta'g_V}{\sqrt{2}}P^*_{0b}P^{*\dag}_{0a}(2iv\cdot{}\mathbb{V}_{ba}),\\
  \mathcal{L}_{\mathcal{D}_0^*\mathcal{D}^*\mathbb{V}} &=&
  -2\frac{\zeta{}g_V}{\sqrt{2}}(P_b^{*\mu}P^{*\dag}_{0a} +
  P^*_{0b}P^{*\mu\dag}_{a})\mathbb{V}_{\mu{}ba}
  -2\frac{\varpi g_V}{\sqrt{2}}(2iv^\alpha)(P^{*\beta}_bP^{*\dag}_{0a} -
  P^*_{0b}P^{*\beta\dag}_{a})(\partial_\alpha{}\mathbb{V}_\beta - \partial_\beta{}\mathbb{V}_\alpha)_{ba},\\
  \mathcal{L}_{\mathcal{D}_1'\mathcal{D}_1'\mathbb{V}} &=& i\frac{\beta'g_V}{\sqrt{2}}P'_{1b}\cdot{}P^{\prime\dag}_{1a}(2iv\cdot{}\mathbb{V}_{ba}) + i4\frac{\lambda'g_V}{\sqrt{2}}P'_{1b\mu}P^{\prime\dag}_{1a\nu}(\partial^\mu{}\mathbb{V}^\nu - \partial^\nu{}\mathbb{V}^\mu)_{ba},\\
  \mathcal{L}_{\mathcal{D}_1'\mathcal{D}\mathbb{V}} &=&
  i2\frac{\zeta{}g_V}{\sqrt{2}}(P_bP_{1a}^{\prime\mu\dag} -
  P^{\prime\mu}_{1b}P^\dag_a)\mathbb{V}_{\mu{}ba} +
  i2\frac{\varpi{}g_V}{\sqrt{2}}(2iv^\alpha)(P_bP_{1a}^{\prime\beta\dag}
  + P^{\prime\beta}_{1b}P^\dag_a)(\partial_\alpha{}\mathbb{V}_\beta - \partial_\beta{}\mathbb{V}_\alpha)_{ba},\\
  \mathcal{L}_{\mathcal{D}_1'\mathcal{D}^*\mathbb{V}} &=&
  \frac{\zeta{}g_V}{\sqrt{2}}(2iv^\lambda)\varepsilon_{\lambda\nu\mu\delta}(P_b^{*\nu}P_{1a}^{\prime\delta\dag}-P_{1b}^{\prime\delta}P_a^{*\nu\dag})\mathbb{V}^\mu_{ba} +
  2\frac{\varpi{}g_V}{\sqrt{2}}\varepsilon_{\nu\alpha\beta\mu}(P^{*\nu}_bP^{\prime\mu\dag}_{1a}
  + P^{\prime\mu}_{1b}P^{*\nu\dag}_{a})(\partial^\alpha{}\mathbb{V}^\beta - \partial^\beta{}\mathbb{V}^\alpha)_{ba},\\
  \mathcal{L}_{\mathcal{DD}_1^\prime{}\mathbb{V}} &=&
  i2\frac{\zeta{}g_V}{\sqrt{2}}(P_bP^{\prime\mu\dag}_{1a} - P^\dag_aP^{\prime\mu}_{1b})\mathbb{V}_{\mu{}ba}
  -i2\frac{\varpi{}g_V}{\sqrt{2}}(2iv^{\beta})(P_bP^{\prime\alpha\dag}_{1a}
  + P^{\prime\alpha}_{1b}P^+_a)(\partial_\alpha{}\mathbb{V}_\beta - \partial_\beta{}\mathbb{V}_\alpha)_{ba},\\
  \mathcal{L}_{\mathcal{D}_1'\mathcal{D}_0^*\mathbb{V}} &=& i2\frac{\lambda'g_V}{\sqrt{2}}(2iv^\lambda)\varepsilon_{\lambda\alpha\beta\mu}(P^*_{0b}P^{\prime\mu\dag}_{1a} +
  P^{\prime\mu}_{1b}P^{*\dag}_{0a})(\partial^\alpha{}\mathbb{V}^\beta)_{ba}\;
  .
\end{eqnarray}
The effective Lagrangian of the $\sigma$ meson interacting with
heavy flavor mesons reads
\begin{eqnarray}\label{eq:lag-s-exch}
  \mathcal{L}_{\mathcal{DD}\sigma} &=& -2g_\sigma\sigma{}P_bP^\dag_b,
\end{eqnarray}
\begin{eqnarray}
  \mathcal{L}_{\mathcal{D}^*\mathcal{D}^*\sigma} &=& 2g_\sigma\sigma{}P^*_b\cdot{}P^{*\dag}_b,
\end{eqnarray}
\begin{eqnarray}
  \mathcal{L}_{\mathcal{D}_0^*\mathcal{D}_0^*\sigma} &=& 2g'_\sigma\sigma{}P^*_{0b}P^{*\dag}_{0b},
\end{eqnarray}
\begin{eqnarray}
  \mathcal{L}_{\mathcal{D}_1'\mathcal{D}_1'\sigma} &=& -2g'_\sigma\sigma{}P'_{1b}\cdot{}P^{\prime\dag}_{1b},
\end{eqnarray}
\begin{eqnarray}
  \mathcal{L}_{\mathcal{DD}_1'\sigma} &=& -i\frac{2h_\sigma}{f_\pi}(P'_{1b\mu}P^\dag_b - P_bP^{\prime\dag}_{1b\mu})\partial^\mu\sigma,
\end{eqnarray}
\begin{eqnarray}
  \mathcal{L}_{\mathcal{D}^*\mathcal{D}^*_0\sigma} &=& -\frac{2h_\sigma}{f_\pi}(P^*_{0b}P^{*\lambda\dag}_b + P^{*\lambda}_bP^{*\dag}_{0b})\partial_\lambda\sigma,
\end{eqnarray}
\begin{eqnarray}
  \mathcal{L}_{\mathcal{D}^*\mathcal{D}_1^{\prime}\sigma}&=&i\frac{2h_\sigma}{f_\pi}v^\alpha\varepsilon_{\alpha\mu\lambda\nu}(P_{1a}^{\prime\mu}P_a^{*\nu\dag}-P_a^{*\nu}P_{1a}^{\prime\mu\dag})\partial^\lambda\sigma.
\end{eqnarray}
The values of the parameters $g^{(\prime)}$, $\beta^{(\prime)}$,
$\lambda^{(\prime)}$, $h$, $\varpi$, $\zeta$,
$g_{\sigma}^{(\prime)}$ and $h_\sigma$ are discussed in Refs.
\cite{lag,hpz}.

\subsection{The effective potential of the $H\bar{S}+h.c.$ system}

We list the effective potentials of the states in the
$H\bar{S}+h.c.$ system in Table \ref{potential}. Here $V_{i}^j$,
$Q_{i}^j$, $W_{i}^j$ and $O_{i}^j$ are the sub-potentials
corresponding to the subsystems $[PP_0^*]$, $[PP_1^\prime]$,
$[P^*P_0^*]$ and $[P^*P_1^\prime]$ respectively. The parameters in
front of the sub-potentials in Table \ref{potential} are from the
coefficients in the octet pseudoscalar and nonet vector meson
matrices.

\renewcommand{\arraystretch}{2}

\begin{table}[htb]
\begin{center}
\begin{tabular}{c||c}\toprule[1pt]
States&Effective potential\\\midrule[1pt]

$[PP_0^*]^{\pm,0}$&$[-\frac{1}{2}{V}_{\rho}^{Direct}(r)+\frac{1}{2}V_{\omega}^{Direct}(r)+
V_{\sigma}^{Direct}(r)]+c[-\frac{1}{2}V_{\pi}^{Cross}(r)
+\frac{1}{6}V_{\eta}^{Cross}(r)]$\\
$[PP_1^\prime]^{\pm,0}$&$[-\frac{1}{2}Q_{\rho}^{Direct}(r)
+\frac{1}{2}Q_{\omega}^{Direct}(r)+
Q_{\sigma}^{Direct}(r)]+c[-\frac{1}{2}Q_{\rho}^{Cross}(r)
+\frac{1}{2}Q_{\omega}^{Cross}(r)+
Q_{\sigma}^{Cross}(r)]$\\
$[P^*P_0^*]^{\pm,0}$&$[-\frac{1}{2}W_{\rho}^{Direct}(r)+\frac{1}{2}W_{\omega}^{Direct}(r)+
W_{\sigma}^{Direct}(\sigma)]+c[-\frac{1}{2}W_{\rho}^{Cross}(r)
+\frac{1}{2}W_{\omega}^{Cross}(r)+
W_{\sigma}^{Cross}(r)]$\\
\raisebox{-0.5em}{$[P^*P_1^\prime]^{\pm,0}$}&$[-\frac{1}{2}O_{\rho}^{Direct}(r)
+\frac{1}{2}O_{\omega}^{Direct}(r) -\frac{1}{2}O_{\pi}^{Direct}(r)
+\frac{1}{6}O_{\eta}^{Direct}(r) +
O_{\sigma}^{Direct}(r)]$\\
&$+c[-\frac{1}{2}O_{\pi}^{Cross}(r)
+\frac{1}{6}O_{\eta}^{Cross}(r)-\frac{1}{2}O_{\rho}^{Cross}(r)
+\frac{1}{2}O_{\omega}^{Cross}(r)+O_{\sigma}^{Cross}(r)]$
\\\midrule[1pt]


$[PP_0^*]_s^{\pm,0},[PP_0^*]_{\hat{s}}^{0}$&$c[-\frac{1}{3}V_{\eta}^{Cross}(r)]$\\
$[PP_1^\prime]_s^{\pm,0},[PP_1^\prime]_{\hat s}^{0}$&0\\
$[P^*P_0^*]_s^{\pm,0},[P^*P_0^*]_{\hat s}^{0}$&0\\
$[P^*P_1^\prime]_s^{\pm,0},[P^*P_1^\prime]_{\hat
s}^{0}$&$[-\frac{1}{3}O_{\eta}^{Direct}(r)]
+c[-\frac{1}{3}O_{\eta}^{Cross}(r)]$
\\\midrule[1pt]


$[PP_0^*]_{s1}^{0}$&$[V_{\phi}^{Direct}(r)]+c[\frac{2}{3}V_{\eta}^{Cross}(r)]$\\
$[PP_1^\prime]_{s1}^{0}$&$[Q_{\phi}^{Direct}(r)]+c[Q_{\phi}^{Cross}(r)]$\\
$[P^*P_0^*]_{s1}^{0}$&$[W_{\phi}^{Direct}(r)]+c[W_{\phi}^{Cross}(r)]$\\
$[P^*P_1^\prime]_{s1}^{0}$&$[O_{\phi}^{Direct}(r)+\frac{2}{3}O_{\eta}^{Direct}(r)]
+c[O_{\phi}^{Cross}(r) +\frac{2}{3}O_{\eta}^{Cross}(r)]$
\\\midrule[1pt]


$[PP_0^*]_8^{0}$&$[\frac{3}{2}V_{\rho}^{Direct}(r)+\frac{1}{2}V_{\omega}^{Direct}(r)+
V_{\sigma}^{Direct}(r)]+c[\frac{3}{2}V_{\pi}^{Cross}(r)
+\frac{1}{6}V_{\eta}^{Cross}(r)]$\\
$[PP_1^\prime]_8^{0}$&$[\frac{3}{2}Q_{\rho}^{Direct}(r)
+\frac{1}{2}Q_{\omega}^{Direct}(r)+
Q_{\sigma}^{Direct}(r)]+c[\frac{3}{2}Q_{\rho}^{Cross}(r)
+\frac{1}{2}Q_{\omega}^{Cross}(r)+
Q_{\sigma}^{Cross}(r)]$\\
$[P^*P_0^*]_8^{0}$&$[\frac{3}{2}W_{\rho}^{Direct}(r)+\frac{1}{2}W_{\omega}^{Direct}(r)+
W_{\sigma}^{Direct}(r)]+c[\frac{3}{2}W_{\rho}^{Cross}(r)
+\frac{1}{2}W_{\omega}^{Cross}(r)+
W_{\sigma}^{Cross}(r)]$\\
\raisebox{-0.5em}{$[P^*P_1^\prime]_8^{0}$}&$[\frac{3}{2}O_{\rho}^{Direct}(r)
+\frac{1}{2}O_{\omega}^{Direct}(r) +\frac{3}{2}O_{\pi}^{Direct}(r)
+\frac{1}{6}O_{\eta}^{Direct}(r) +
O_{\sigma}^{Direct}(r)]$\\
&$+c[\frac{3}{2}O_{\pi}^{Cross}(r)
+\frac{1}{6}O_{\eta}^{Cross}(r)+\frac{3}{2}O_{\rho}^{Cross}(r)
+\frac{1}{2}O_{\omega}^{Cross}(r)+O_{\sigma}^{Cross}(r)]$
\\\bottomrule[1pt]
\end{tabular}
\caption{The total effective potentials of the states in the
subsystems $[PP_0^*]$, $[PP_1^\prime]$, $[P^*P_0^*]$ and
$[P^*P_1^\prime]$.\label{potential}}
\end{center}
\end{table}

For the $[P{P}_0^*]$ subsystem, the sub-potentials are
\begin{eqnarray}
V^{Direct}_\mathbb{V}(r) &=& -\frac{1}{2}\beta\beta^\prime g_V^2 Y(\Lambda,q_0,m_\mathbb{V},r),\\
V^{Direct}_\sigma (r)&=& -g_\sigma g_{\sigma}^\prime
Y(\Lambda,q_0,m_\sigma,r),\\
V^{Cross}_\mathbb{P}(r) &=& \frac{h^2{q_0^\prime}^2}{f_\pi^2}
Y(\Lambda,q_0^\prime,m_\mathbb{P},r).
\end{eqnarray}
For the $[P{P}_1^\prime]$ subsystem, the sub-potentials are
\begin{eqnarray}
Q^{Direct}_\mathbb{V}(r) &=& -\frac{1}{2}\beta\beta^\prime g_V^2 Y(\Lambda,q_0,m_\mathbb{V},r),\\
Q^{Direct}_\sigma (r)&=& -g_\sigma g_{\sigma}^\prime
Y(\Lambda,q_0,m_\sigma,r),
\end{eqnarray}
\begin{eqnarray}
Q^{Cross}_\mathbb{V} (r)&=& -\frac{g_V^2}{4}(2\zeta^2 -
8\zeta\varpi q_0^\prime + 8\varpi^2{q_0^\prime}^2)
Y(\Lambda,q_0^\prime,m_\mathbb{V},r)\nonumber\\&&+\frac{g_V^2}{12}(8\mu^2
- \frac{2\zeta^2}{m_V^2}) Z(\Lambda,q_0^\prime,m_\mathbb{V},r),
\end{eqnarray}
\begin{eqnarray}
Q^{Cross}_\sigma (r)&=& -\frac{h_\sigma^2}{3f_\pi^2}
Z(\Lambda,q_0^\prime,m_\sigma,r).
\end{eqnarray}
The sub-potentials relevant to the $[P^* {P}_0^*]$ subsystem are
\begin{eqnarray}
W^{Direct}_\mathbb{V}(r) &=& -\frac{1}{2}\beta\beta^\prime g_V^2 Y(\Lambda,q_0,m_\mathbb{V},r),\\
 W^{Direct}_\sigma (r)&=& -g_\sigma g_\sigma^\prime Y(\Lambda, q_0,
m_\sigma, r),\\
W^{Cross}_\mathbb{V} (r) &=& \frac{1}{4}g_V^2(2\zeta^2 -
8\zeta\varpi q_0^\prime + 8\varpi^2{q_0^\prime}^2) Y(\Lambda, q_0^\prime, m_\mathbb{V}, r)\nonumber\\&& -\frac{1}{12}g_V^2(8\mu^2-\frac{2\zeta^2}{m_\mathbb{V}^2}) Z(\Lambda,q_0^\prime, m_\mathbb{V}, r),\\
W^{Cross}_\sigma (r)&=& \frac{h_\sigma^2}{f_\pi^2}\frac{1}{3}
Z(\Lambda, q_0^\prime, m_\sigma, r).
\end{eqnarray}
For the $[P^*\bar{P}_1^\prime]$ subsystem, the relevant potentials
are
\begin{eqnarray}
O^{Direct}_\mathbb{V} (r)&=& -\frac{1}{2}g_V^2\beta\beta^\prime
C[J] Y(\Lambda, q_0,
m_\mathbb{V}, r) - 2g_V^2\lambda\lambda^\prime B[J] Z(\Lambda, q_0, m_V, r),\\
O^{Direct}_\mathbb{P}(r) &=& -\frac{g g^\prime}{f_\pi^2} A[J] Z(\Lambda, q_0, m_\mathbb{V}, r),\\
O^{Direct}_\sigma (r) &=& - g_\sigma g_\sigma^\prime C[J] Y(\Lambda, q_0, m_\mathbb{V}, r),\\
O^{Cross}_\mathbb{V} (r)&=& \frac{1}{4}g_V^2(2\zeta^2 -
8\zeta\varpi q_0^\prime + 8\varpi^2{q_0^\prime}^2)
C[J] Y(\Lambda, q_0^\prime, m_\mathbb{V}, r)  \nonumber\\
&&-\frac{1}{4}g_V^2(\frac{2\zeta^2}{m_\mathbb{V}^2}-8\mu^2)A[J]
Z(\Lambda, q_0,
m_\mathbb{V}, r),\\
O^{cross}_\mathbb{P} (r)&=& -\frac{h^2}{f_\pi^2}{q_0^\prime}^2 C[J] Y(\Lambda, q_0^\prime, m_\mathbb{P}, r),\\
O^{Cross}_\sigma (r) &=& -\frac{h_\sigma^2}{f_\pi^2} A[J]
Z(\Lambda, q_0^\prime, m_\sigma, r)
\end{eqnarray}
with the coefficients $A[J]$, $B[J]$ and $C[J]$
\begin{eqnarray}
\begin{array}{|c|c|c|c|}\toprule[1pt]
&$J=0$&$J=1$&$J=2$\\\hline
$A[J]$&2/3&1/3&-1/3\\
$B[J]$&4/3&2/3&-2/3\\
$C[J]$&1&1&1\\\bottomrule[1pt]
\end{array}\,.
\end{eqnarray}
In the above expressions, the functions $Y(\Lambda, \kappa, m, r)$
and $Z(\Lambda, \kappa, m, r)$ are defined as
\begin{eqnarray}
Y(\Lambda, \kappa, m, r)
&=&\int\frac{d^3q}{(2\pi)^3}e^{i\mathbf{q}\cdot \mathbf{r}}
  \frac{1}{\mathbf{q}^2-\kappa^2-m^2}\bigg(\frac{\Lambda^2-m^2}{\Lambda^2-\mathbf{q}^2+\kappa^2}\bigg)^2 \\ \nonumber
&=& \left\{ \begin{array}{ll}
            \mbox{if $|\kappa|\le m$},\quad -\frac{1}{4\pi
              r}\left( e^{-\zeta_1r} -
            e^{-\zeta_2r}\right) +
            \frac{1}{8\pi}\left(\zeta_2-\frac{\zeta_1^2}{\zeta_2}\right)
            e^{-\zeta_2r} &\\
            \mbox{otherwise},\quad -\frac{1}{4\pi r}\left(
            \cos(\zeta_1^\prime r)- e^{-\zeta_2r}\right) +
            \frac{1}{8\pi}\left(\zeta_2+\frac{\zeta_1^{\prime2}}{\zeta_2}
              \right) e^{-\zeta_2r} &
            \end{array}
     \right.
\end{eqnarray}
and \begin{eqnarray}
Z(\Lambda, \kappa, m, r) &=&
-\frac{1}{r^2}\frac{\partial}{\partial
  r}\left(r^2\frac{\partial}{\partial r}Y(\Lambda, \kappa, m, r)\right),
\end{eqnarray}
where $\zeta_1=\sqrt{m^2-\kappa^2}$,
$\zeta_1^\prime=\sqrt{\kappa^2-m^2}$, and
$\zeta_2=\sqrt{\Lambda^2-\kappa^2}$. We take $q_0=0$ and the mass
differences $q_0^\prime$ as $(m_{P_0}-m_P)$,
$(m_{P^\prime_1}-m_P)$, $(m_{P_0}-m_{P^*})$ and
$(m_{P_1^\prime}-m_{P^*})$ corresponding to the $[PP_0^*]$,
$[PP_1^\prime]$, $[P^*P_0^*]$ and $[P^*P_1^\prime]$ subsystems
respectively.

\section{Numerical results}\label{sec3}

In the following, we present the numerical results for the
different systems. Throughout this work, $\mu$ denotes the reduced
mass of the corresponding system in all figures of the numerical
results. 

\subsection{The $[PP_0^*]$ system}

\subsubsection{$[PP_0^*]^{\pm,0}$}

\begin{center}
\begin{figure}[htb]
\begin{tabular}{cccc}
\scalebox{0.55}{\includegraphics{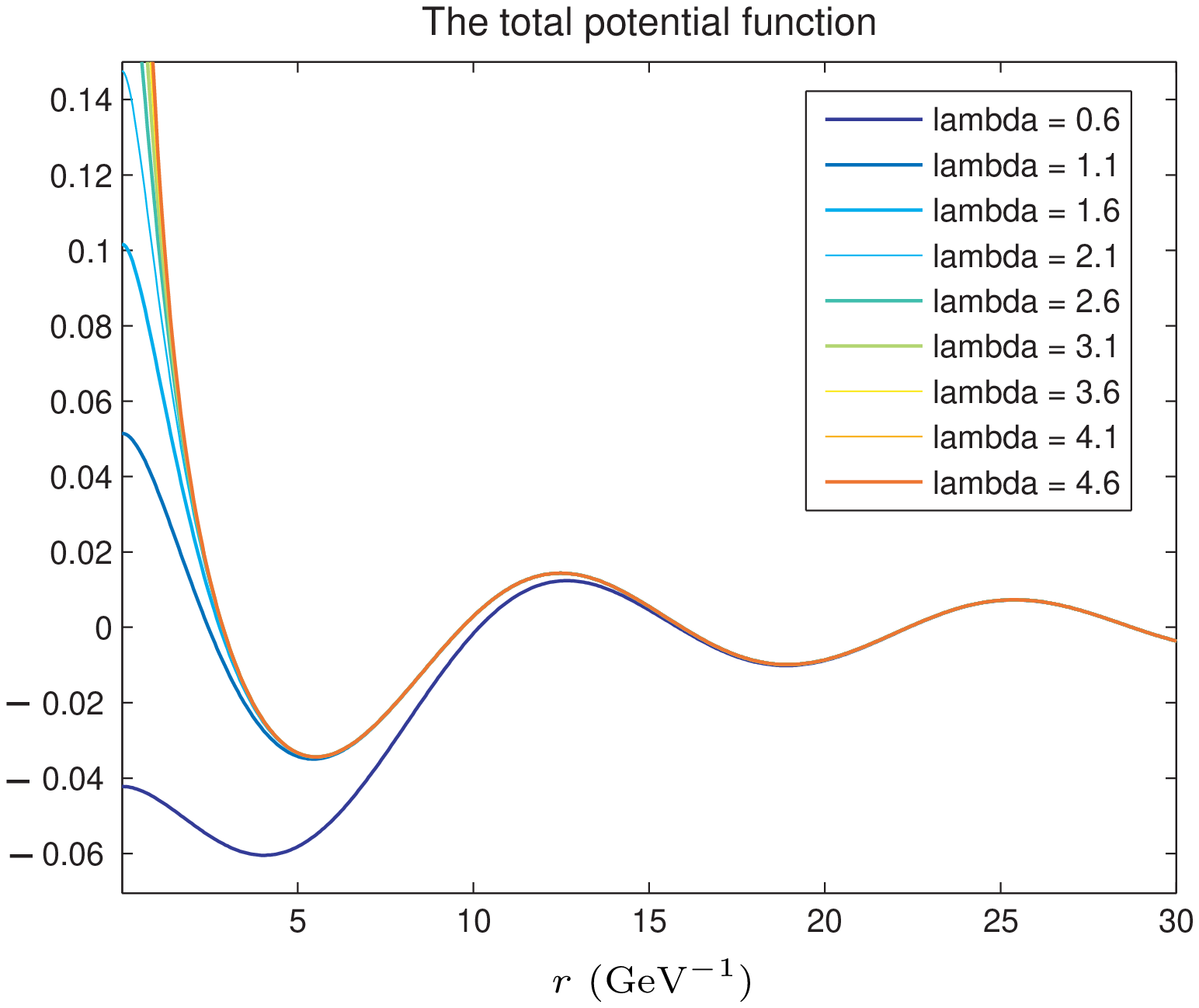}}&\scalebox{0.55}{\includegraphics{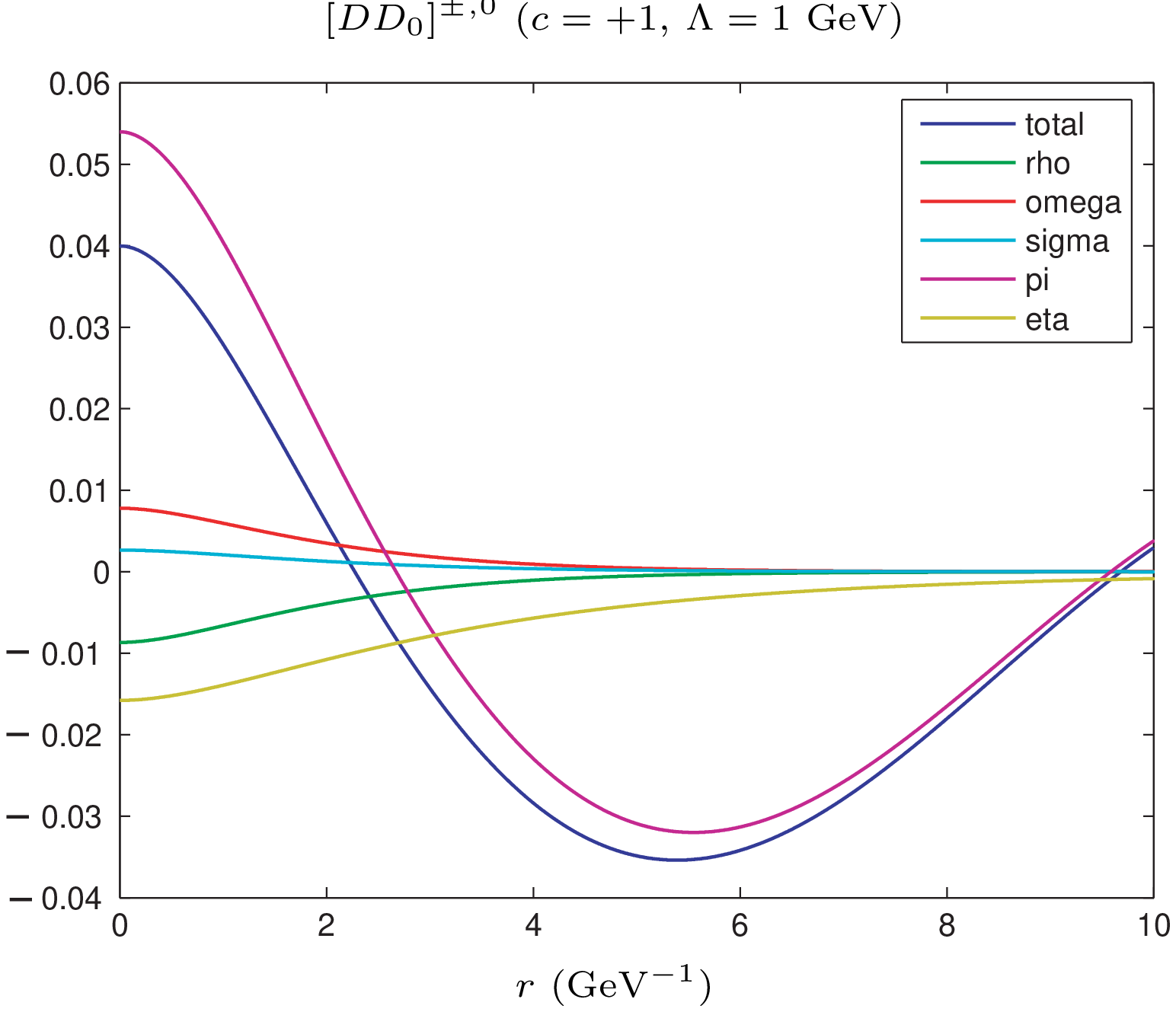}}\\
(a)&(b)\\
\scalebox{0.55}{\includegraphics{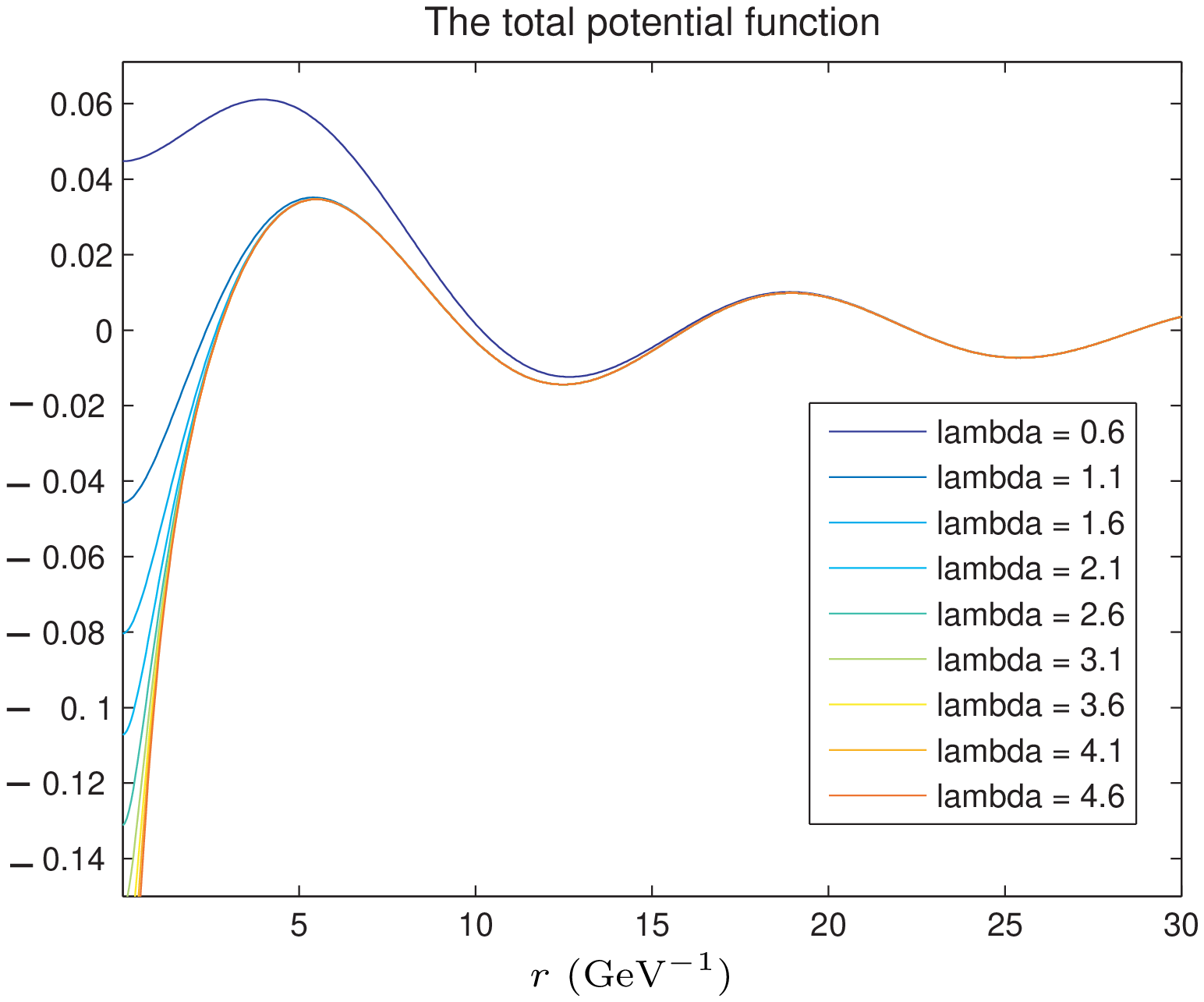}}&\scalebox{0.55}{\includegraphics{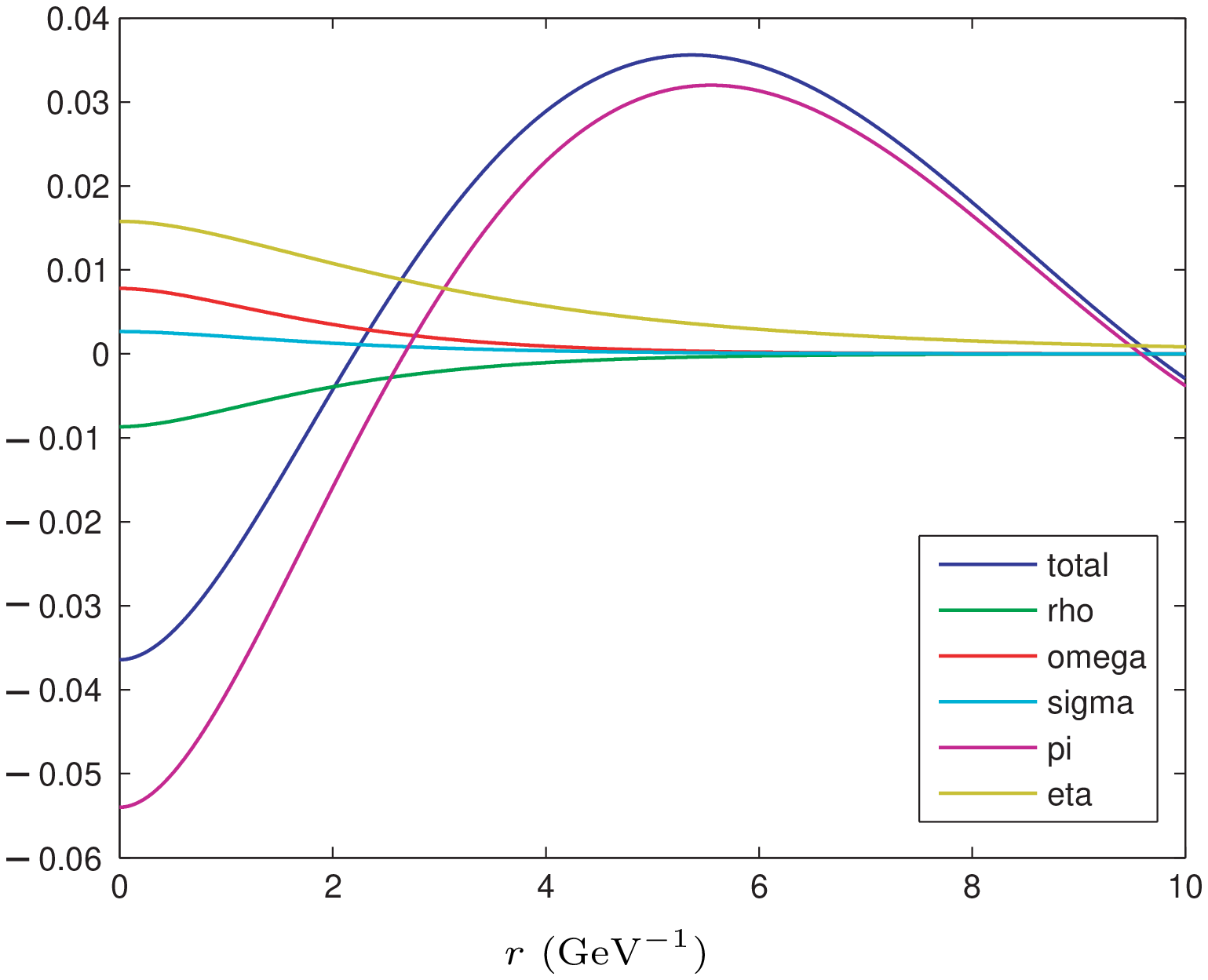}}\\
(c)&(d)
\end{tabular}
\caption{(a) The variation of the total potential of the
$[DD_0]^{\pm,0}$ system $(c=+1)$ with $r$ and $\Lambda$; (b) the
exchange potentials of the $\pi,\,\eta,\,\sigma,\,\rho,\,\omega$
mesons of the $[DD_0]^{\pm,0}$ system $(c=+1)$ with $\Lambda=1$
GeV; (c) the variation of the total potential of the
$[DD_0]^{\pm,0}$ system $(c=-1)$ with $r$ and $\Lambda$; (d) the
exchange potentials of the $\pi,\,\eta,\,\sigma,\,\rho,\,\omega$
mesons of the $[DD_0]^{\pm,0}$ system $(c=-1)$ with $\Lambda=1$
GeV. The above potentials are obtained when $g_\sigma
g_\sigma^\prime>0$ and $\beta\beta^\prime>0$. Here, $|h|=0.56$,
$|g_\sigma|=0.76$, $|g_\sigma^\prime|=0.76$, $|\beta|=0.909$ and
$|\beta^\prime|=0.533$.\label{dd0s1potential}}
\end{figure}
\end{center}

The total potentials of the $[PP_0^*]^{\pm,0}$ system with $c=\pm
1$ and the typical values of $\Lambda$ are presented in Fig.
\ref{dd0s1potential}. The comparison of the total potential with
the partial potentials indicates that the $\pi$ exchange plays the
dominant role in the total effective potential of the
$[PP_0^*]^{\pm,0}$ system with $c=\pm 1$. Here we only illustrate
the potential with $\Lambda=1$ GeV in Fig. \ref{dd0s1potential}
(b) and (d).

Since the one $\pi$ exchange potential is dominant in the total
potential of $[PP_0^*]^{\pm,0}$, it's enough to consider the one
$\pi$ exchange force only when we study the bound state solution
of the $[PP_0^*]^{\pm,0}$ system. The one $\pi$ exchange potential
is proportional to the coupling constant $h$. Thus, we try to
solve the schr\"{o}dinger equation with the obtained one pion
exchange potential under some typical values of $h$.

\begin{center}
\begin{figure}[htb]
\begin{tabular}{cccc}
\scalebox{0.63}{\includegraphics{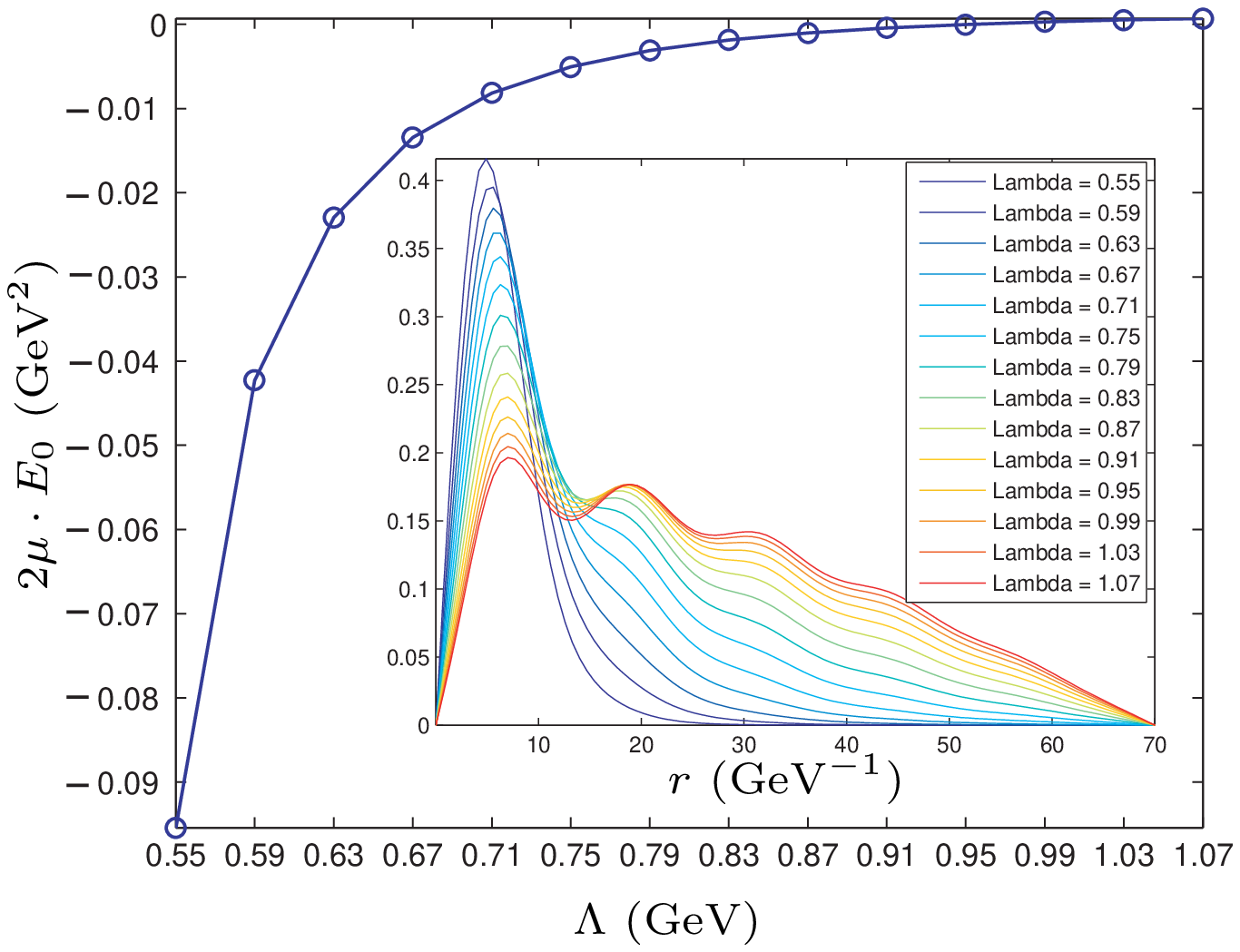}}&\scalebox{0.63}{\includegraphics{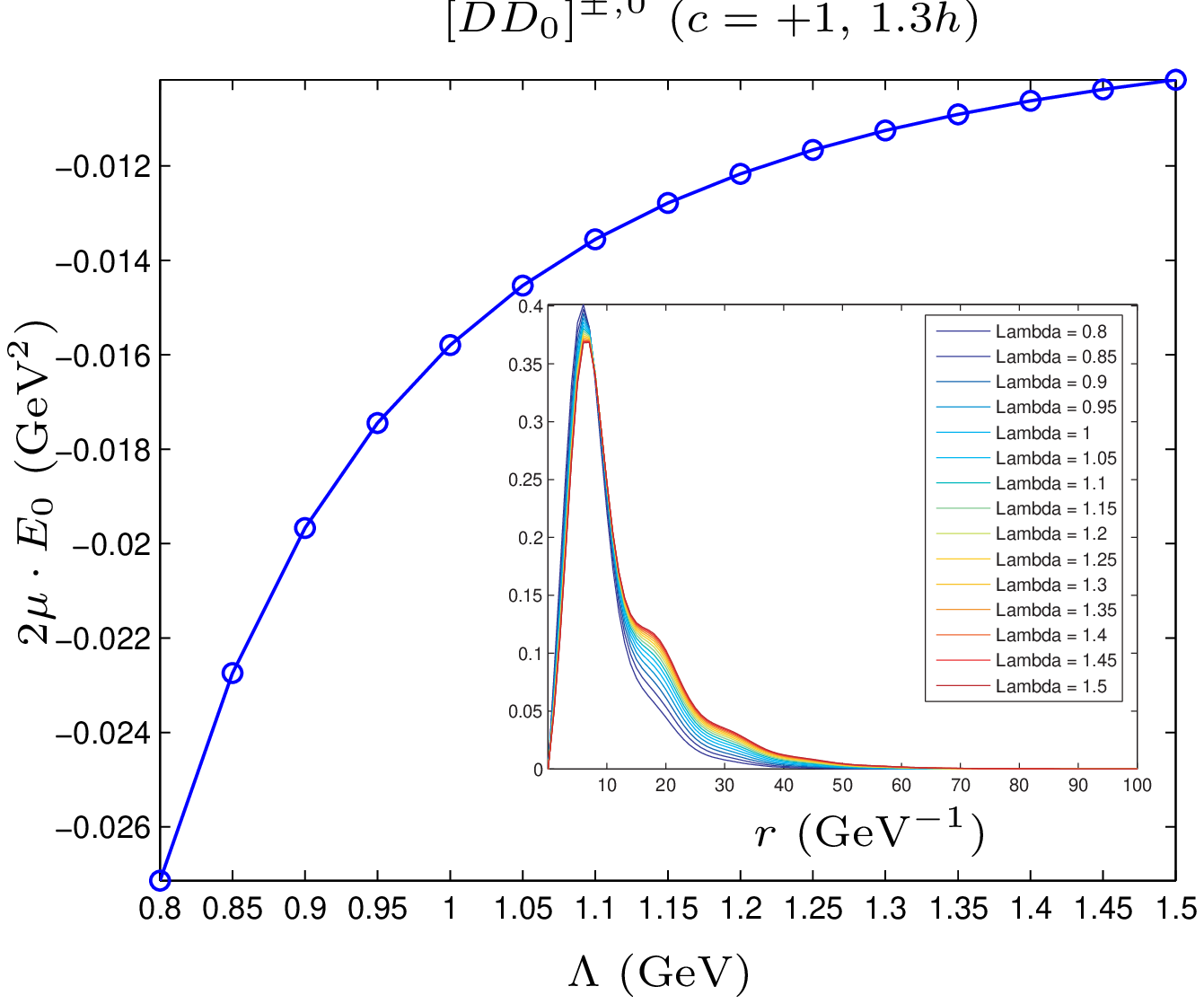}}\\
(a)&(b)\\ \\
\scalebox{0.63}{\includegraphics{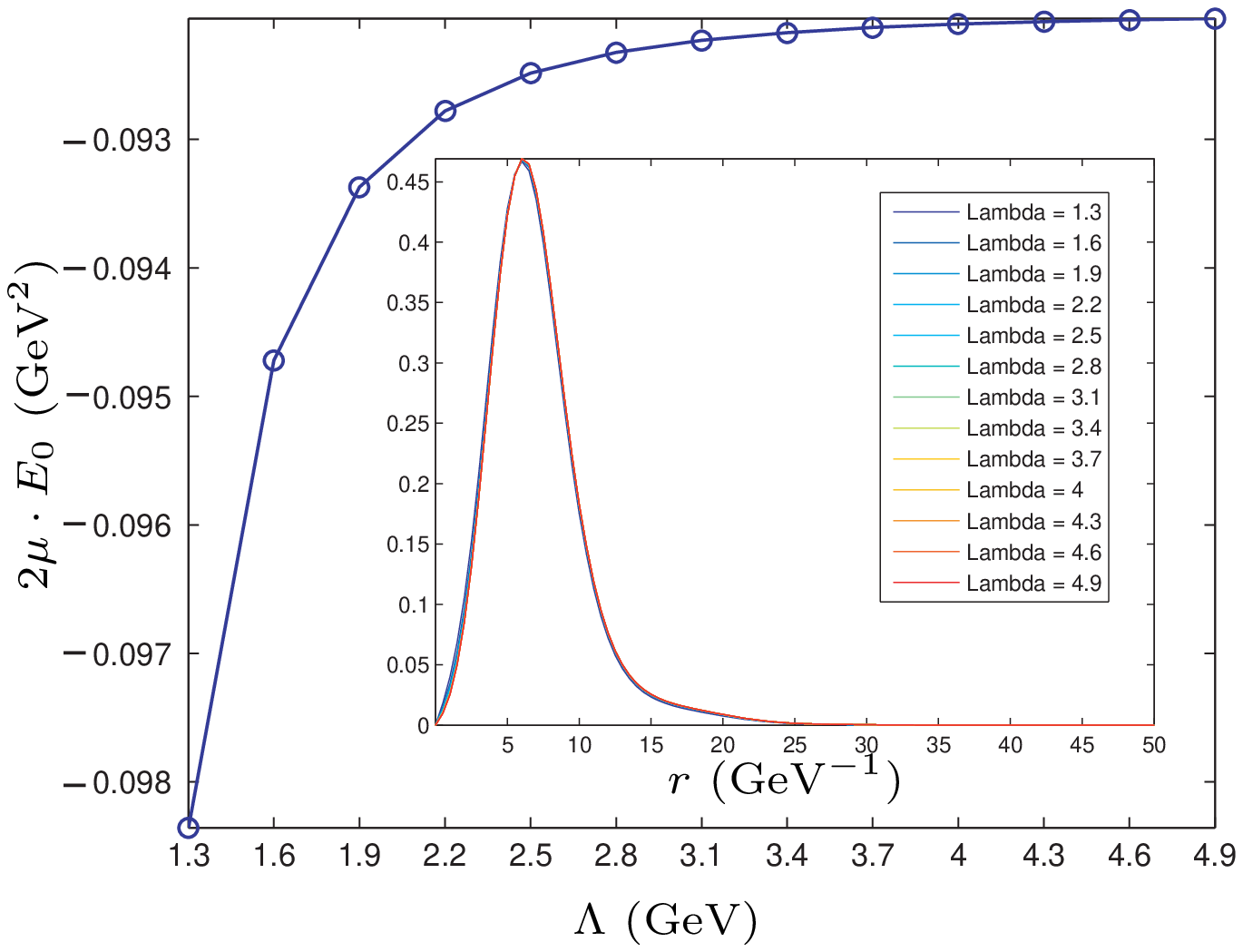}}&\scalebox{0.63}{\includegraphics{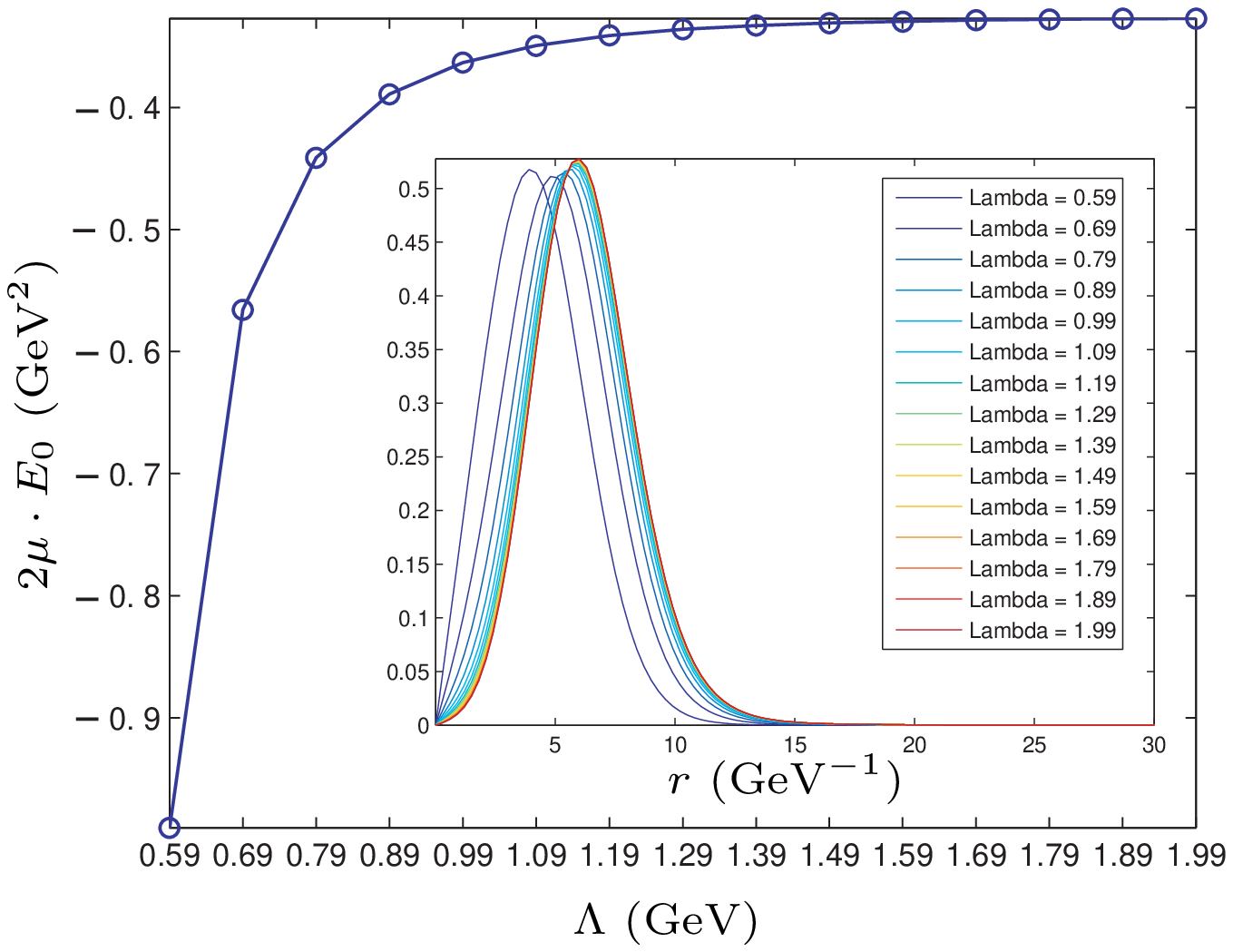}}\\
(c)&(d)
\end{tabular}
\caption{The dependence of the binding energy of the
$[DD_0]^{\pm,0}$ system $(c=+1)$ on $\Lambda$ and the wave
function of this system. Here, we only consider the single $\pi$
exchange potential. (a), (b), (c) and (d) correspond to the case
of $1h$, $1.3h$, $2h$ and $3h$ with $|h|=0.56$
respectively.\label{dds1e}}
\end{figure}
\end{center}

For the $[PP_0^*]^{\pm,0}$ system with $c=+1$, we can find the
bound state solutions numerically. As shown in Fig. \ref{dds1e},
the binding energy $E_0$ decreases when $\Lambda$ becomes larger.
One may further check the corresponding wave function to make sure
whether the bound state solution is reasonable. An example is
shown in Fig. \ref{dds1e} (a) with $h\equiv -0.56$. As $\Lambda$
increases, the wave function oscillates at large distance. In
other words, smaller $\Lambda$ values such as $\Lambda<0.75$ GeV
lead to the bound state wave function with less oscillating
behavior. On the other hand, the reasonable range of $\Lambda$ is
sometimes assumed to be around $1$ GeV.

Enhancing the coupling constant helps to form the bound state of
$[PP_0^*]^{\pm,0}$ system with $c=+1$. For example, there exist a
very reasonable bound state solution with $1.3h$ corresponding to
a shallow binding energy, a reasonable $\Lambda$ and wave function
with good behavior as shown in Fig. \ref{dds1e} (b). If the
coupling constant becomes even larger, the binding energy of the
$[PP_0^*]^{\pm,0}$ system with $c=+1$ becomes deeper. As shown in
Fig \ref{dds1e} (c) and (d), there only exist deeply bound states
for the $[PP_0^*]^{\pm,0}$ system ($c=+1$) with $2h$ and $3h$.

\begin{center}
\begin{figure}[htb]
\begin{tabular}{cccc}
\scalebox{0.63}{\includegraphics{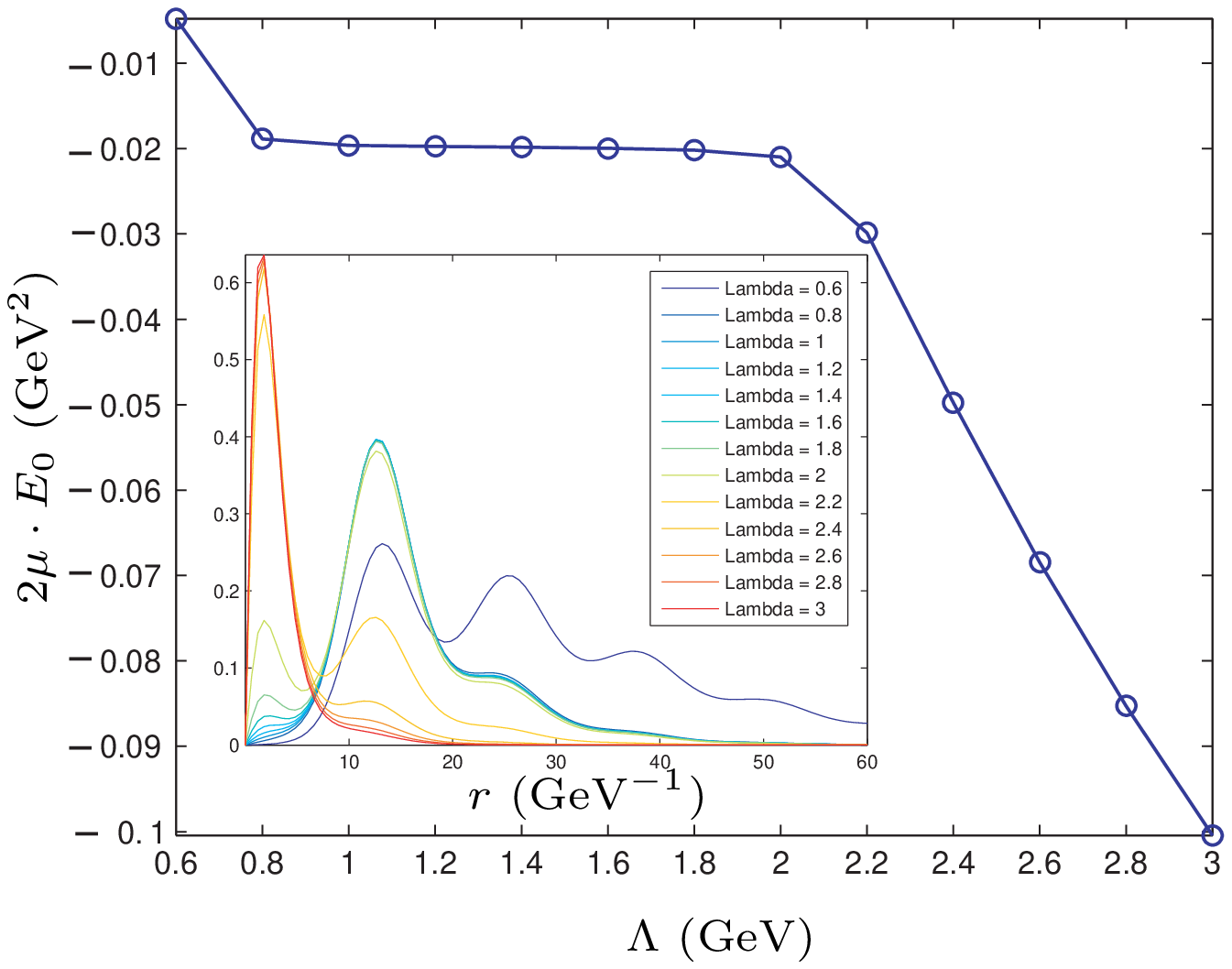}}&\raisebox{0.0em}{\scalebox{0.63}{\includegraphics{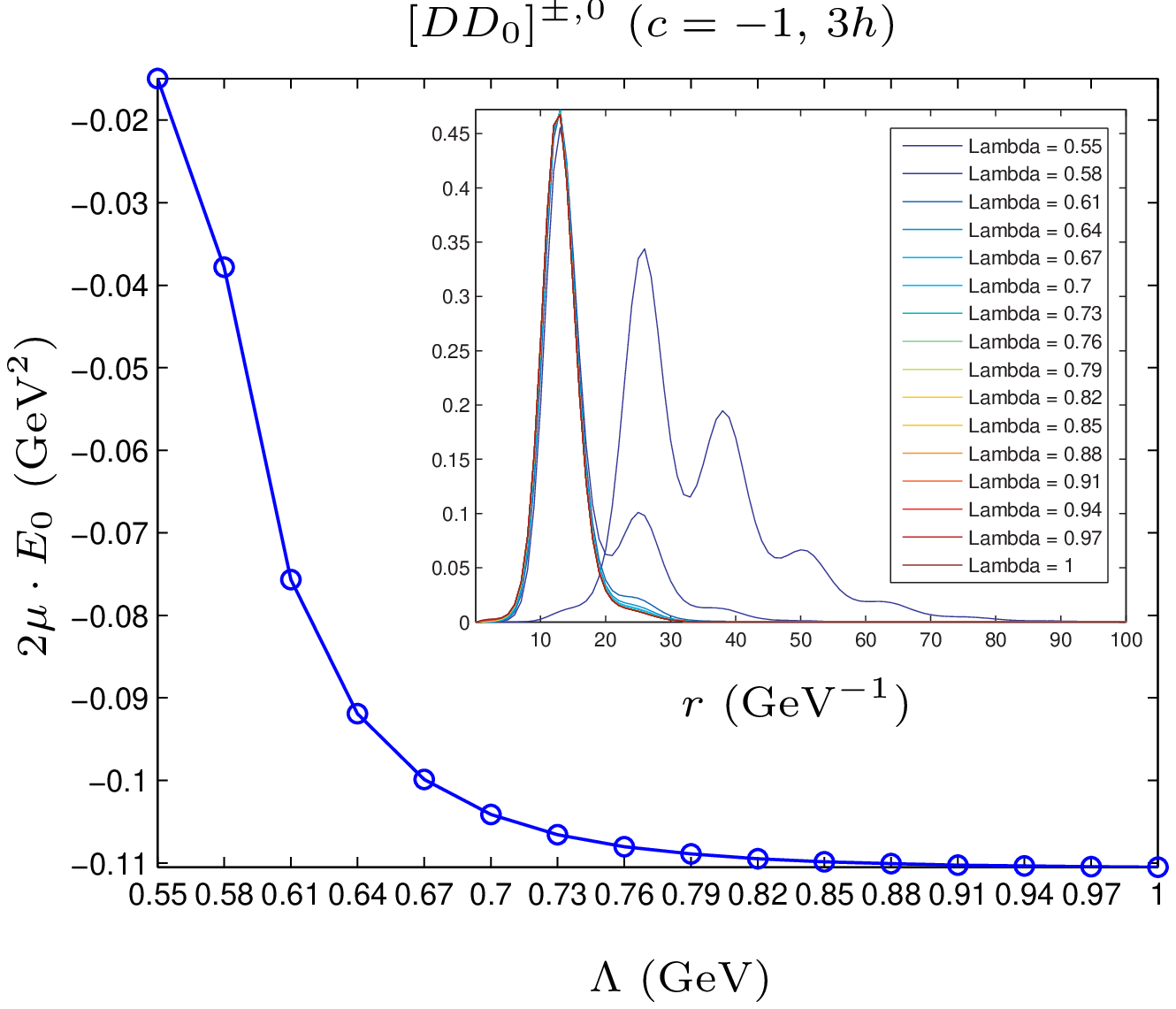}}}\\
(a)&(b)
\end{tabular}
\caption{The dependence of the binding energy of $[DD_0]^{\pm,0}$
system $(c=-1)$ on $\Lambda$ and the wave function of this system.
Here, we only consider the single $\pi$ exchange potential. (a)
and (b) correspond to the case of $2h$ and $3h$ with $|h|=0.56$
respectively. When taking $1h$, we can not find the bound state
solution.\label{dds1-1e}}
\end{figure}
\end{center}

For the $[PP_0^*]^{\pm,0}$ system with $c=-1$, we can not find the
bound state solution with $1h$. With $2h$ and $\Lambda=2.6\sim3$
GeV, the wave function of the $[PP_0^*]^{\pm,0}$ system with
$c=-1$ becomes non-oscillating. When further increasing the
coupling constant to $3h$, there exists a $[PP_0^*]^{\pm,0}$ bound
state with $c=-1$ and $\Lambda$ near 1 GeV. The numerical results
of the $[PP_0^*]^{\pm,0}$ system with $c=-1$ are presented in Fig.
\ref{dds1-1e}.

\subsubsection{$[PP_0^*]_s^{\pm,0},\,[PP_0^*]_{\hat{s}}^{0}$}

\begin{center}
\begin{figure}[htb]
\begin{tabular}{cccc}
\scalebox{0.55}{\includegraphics{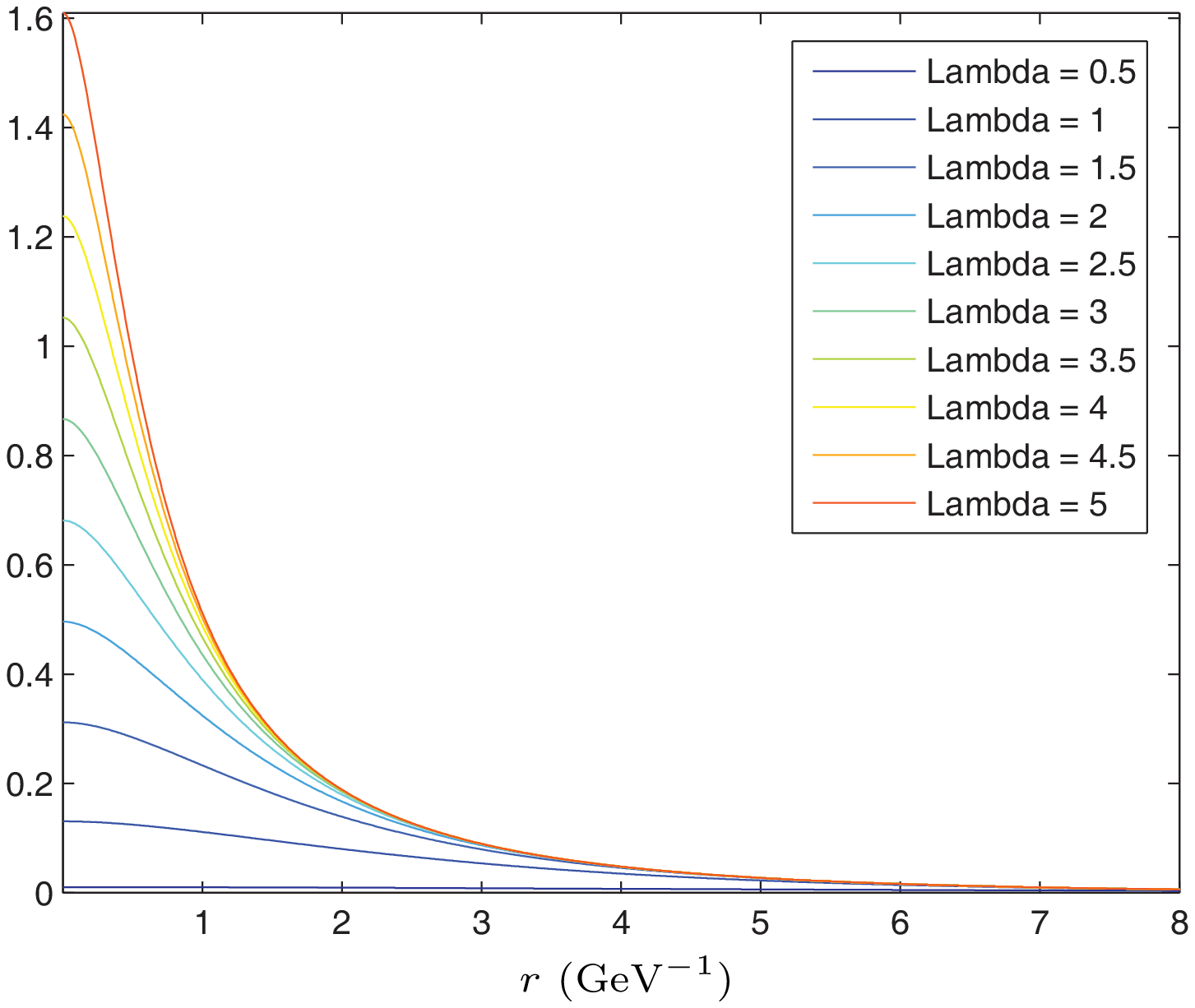}}&\scalebox{0.55}{\includegraphics{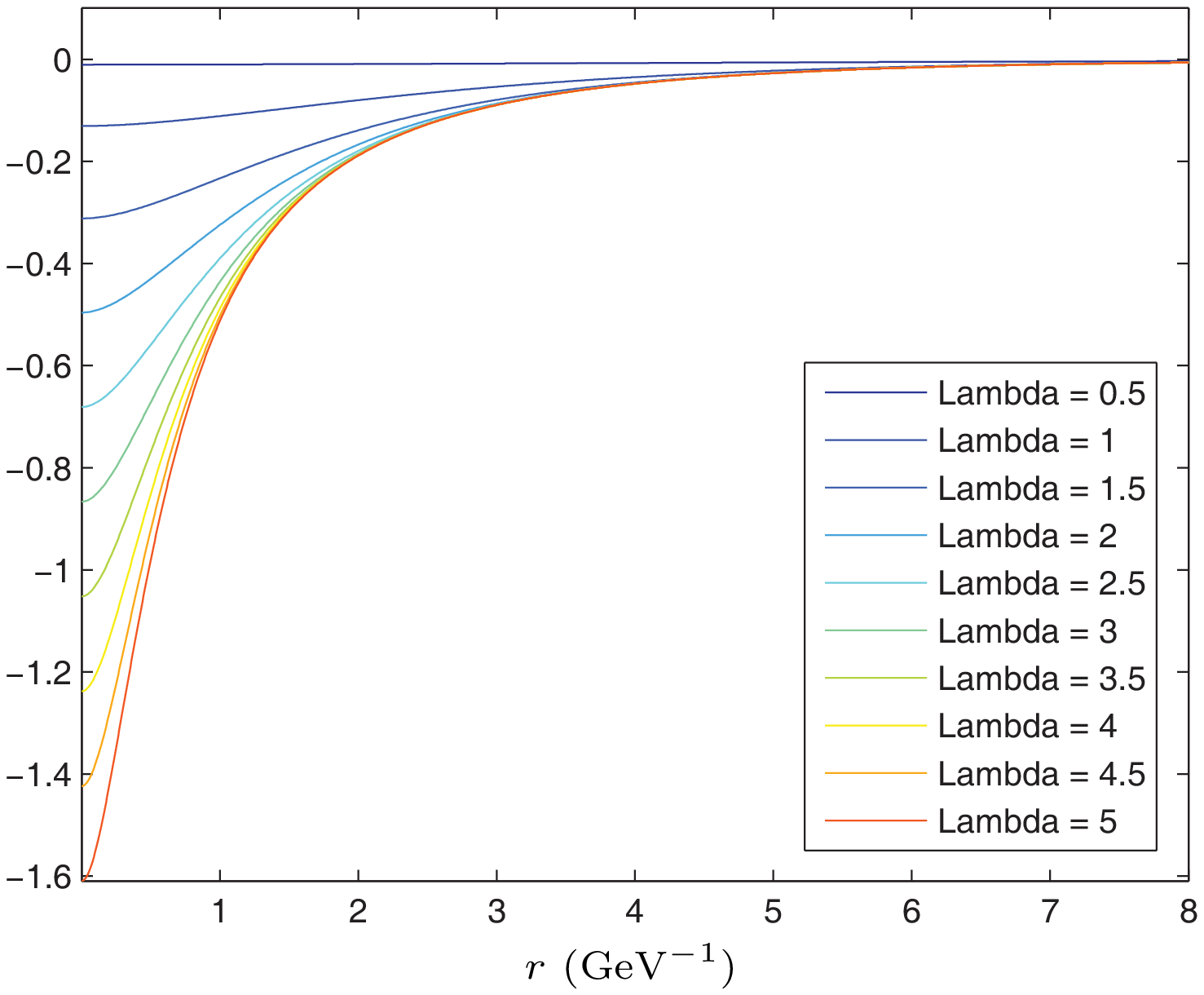}}\\
(a)&(b)
\end{tabular}
\caption{(a) The variation of the total potential of the
$[DD_0]_s^{\pm,0},[DD_0]_{\hat{s}}^{0}$ system $(c=+1)$ with $r$
and $\Lambda$; (b) The variation of the total potential of the
$[DD_0]_s^{\pm,0},[DD_0]_{\hat{s}}^{0}$ system $(c=-1)$ with $r$
and $\Lambda$. Here, we take $|h|=0.56$.\label{dds2-potential}}
\end{figure}
\end{center}

The potential of $[PP_0^*]_s^{\pm,0},\,[PP_0^*]_{\hat{s}}^{0}$ is
shown in Fig. \ref{dds2-potential}, where only the $\eta$ meson
exchange contributes. The effective potential is repulsive and
attractive for $c=+1$ and $c=-1$ respectively. There does not
exist any bound states with $c=+1$.

\begin{center}
\begin{figure}[htb]
\begin{tabular}{cccc}
\scalebox{0.63}{\includegraphics{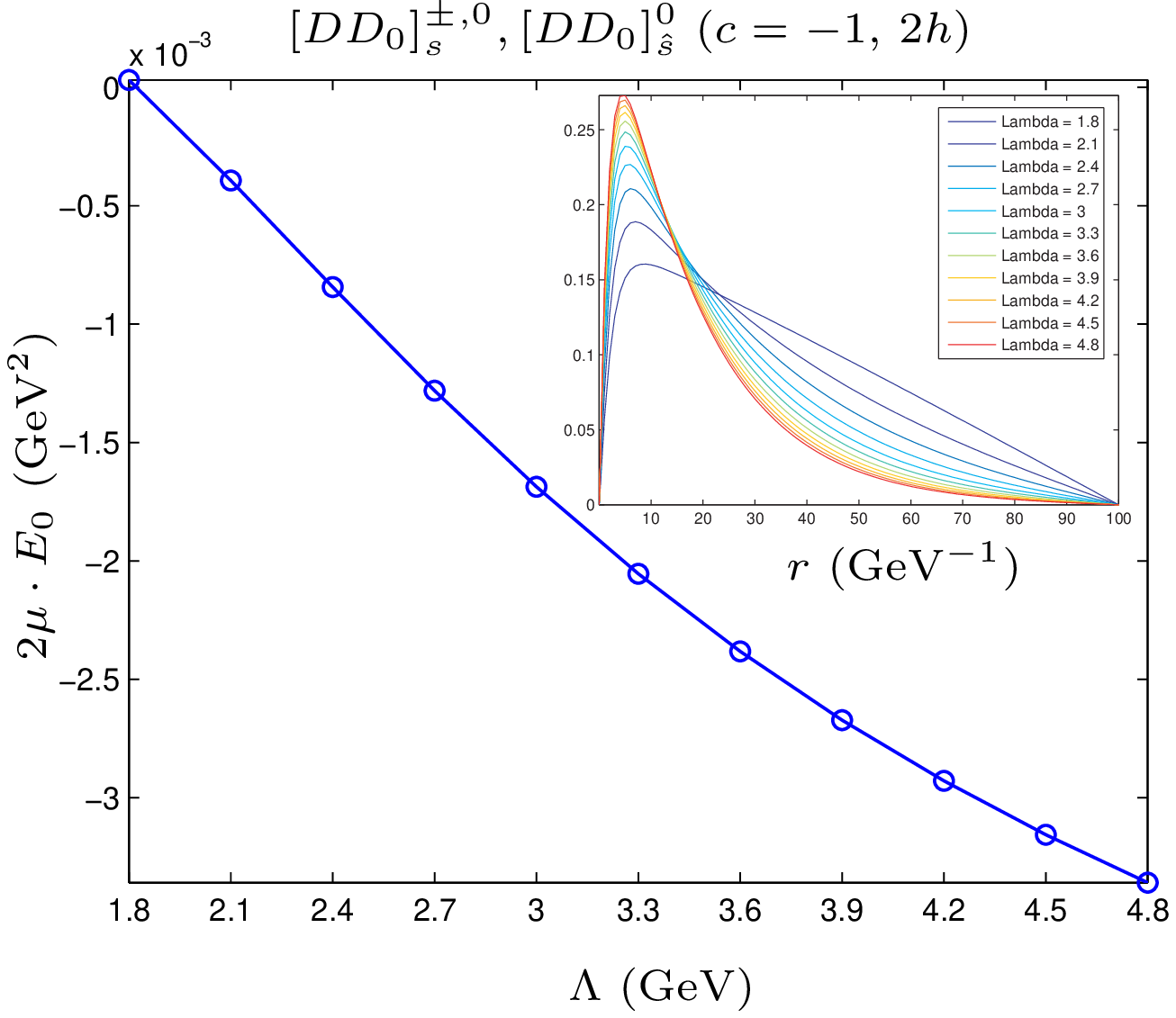}}&\raisebox{0.0em}{\scalebox{0.63}{\includegraphics{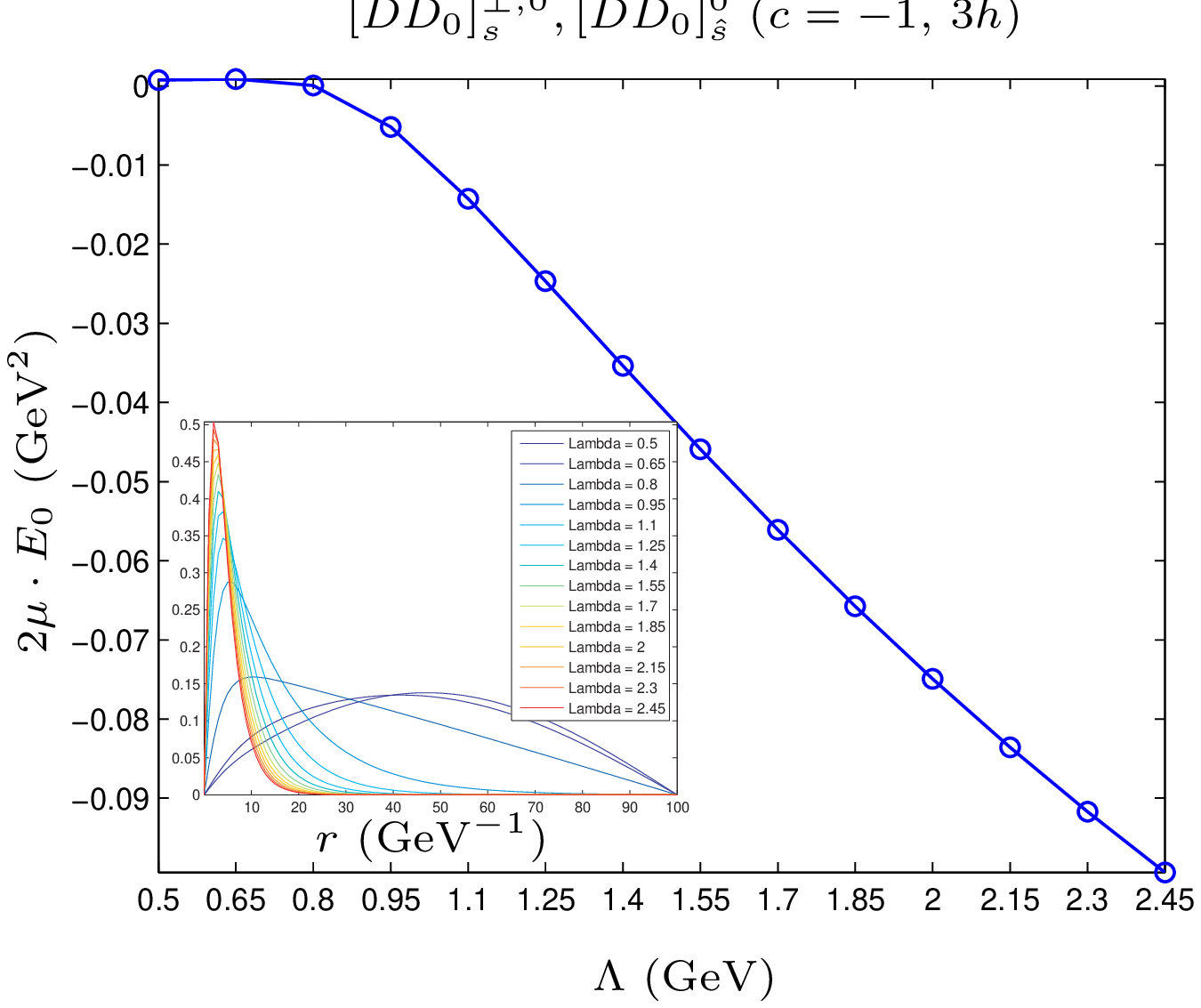}}}\\
(a)&(b)
\end{tabular}
\caption{The dependence of the binding energy of the
$[DD_0]_s^{\pm,0},[DD_0]_{\hat{s}}^{0}$ system $(c=-1)$ on
$\Lambda$ and the wave function of this system. Here, we only
consider the single $\eta$ exchange potential. (a) and (b)
correspond to the case of $2h$ and $3h$ with $|h|=0.56$
respectively. When taking $1h$, we can not find the bound state
solution.\label{dds2-E}}
\end{figure}
\end{center}

For the $[PP_0^*]_s^{\pm,0},\,[PP_0^*]_{\hat{s}}^{0}$ system with
$c=-1$, we can not find the bound state solution with $h\equiv
-0.56$. We illustrate the numerical results with $2h$ and $3h$ in
Fig. \ref{dds2-E}. There may exist a
$[PP_0^*]_s^{\pm,0},\,[PP_0^*]_{\hat{s}}^{0}$ bound state with
$c=-1$ if the coupling constant is around several $h$.

\subsubsection{$[PP_0^*]_{s1}^{0}$}

\begin{center}
\begin{figure}[htb]
\begin{tabular}{cccc}
\scalebox{0.55}{\includegraphics{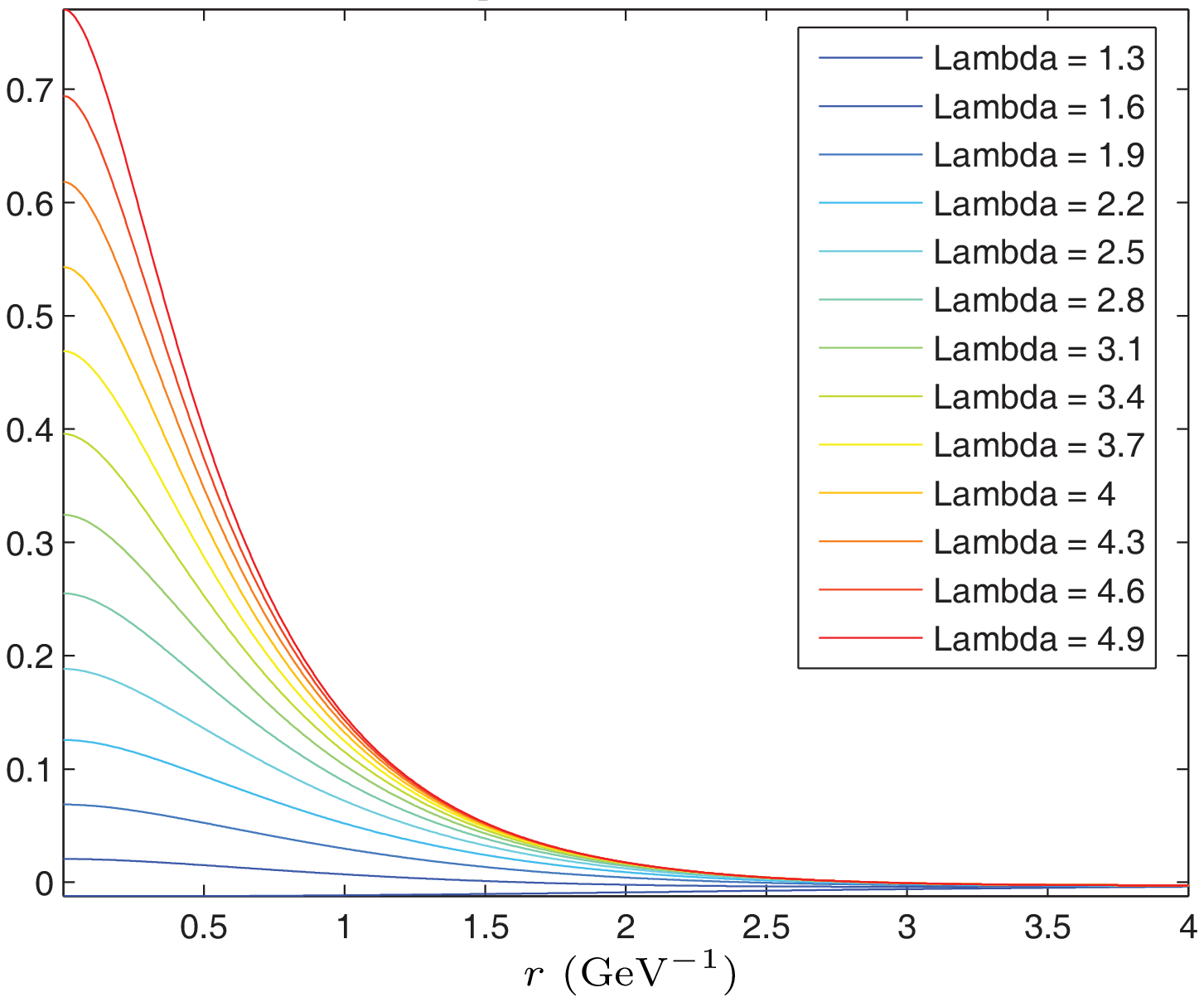}}&\scalebox{0.55}{\includegraphics{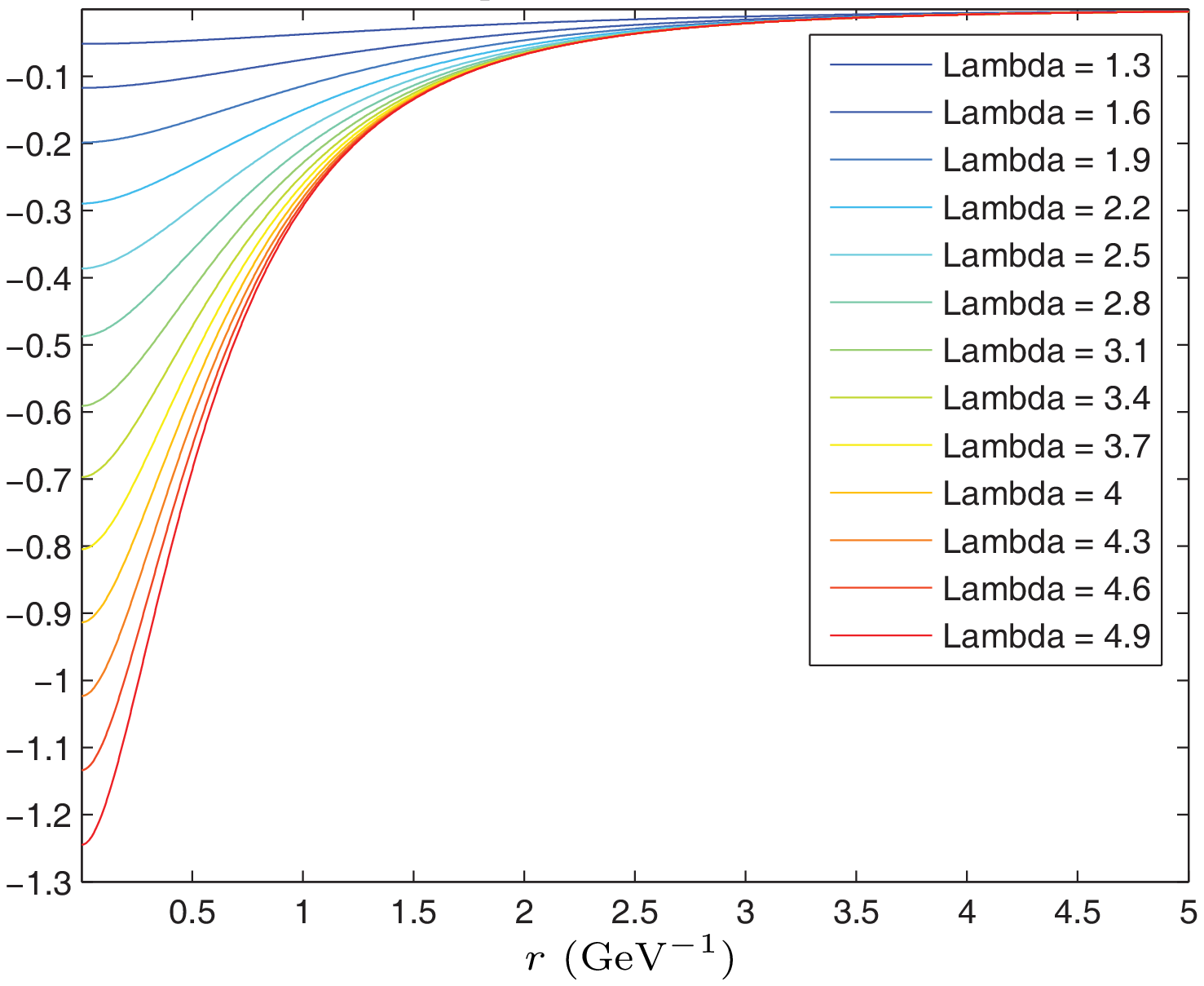}}\\
(a)&(b)\\
\scalebox{0.55}{\includegraphics{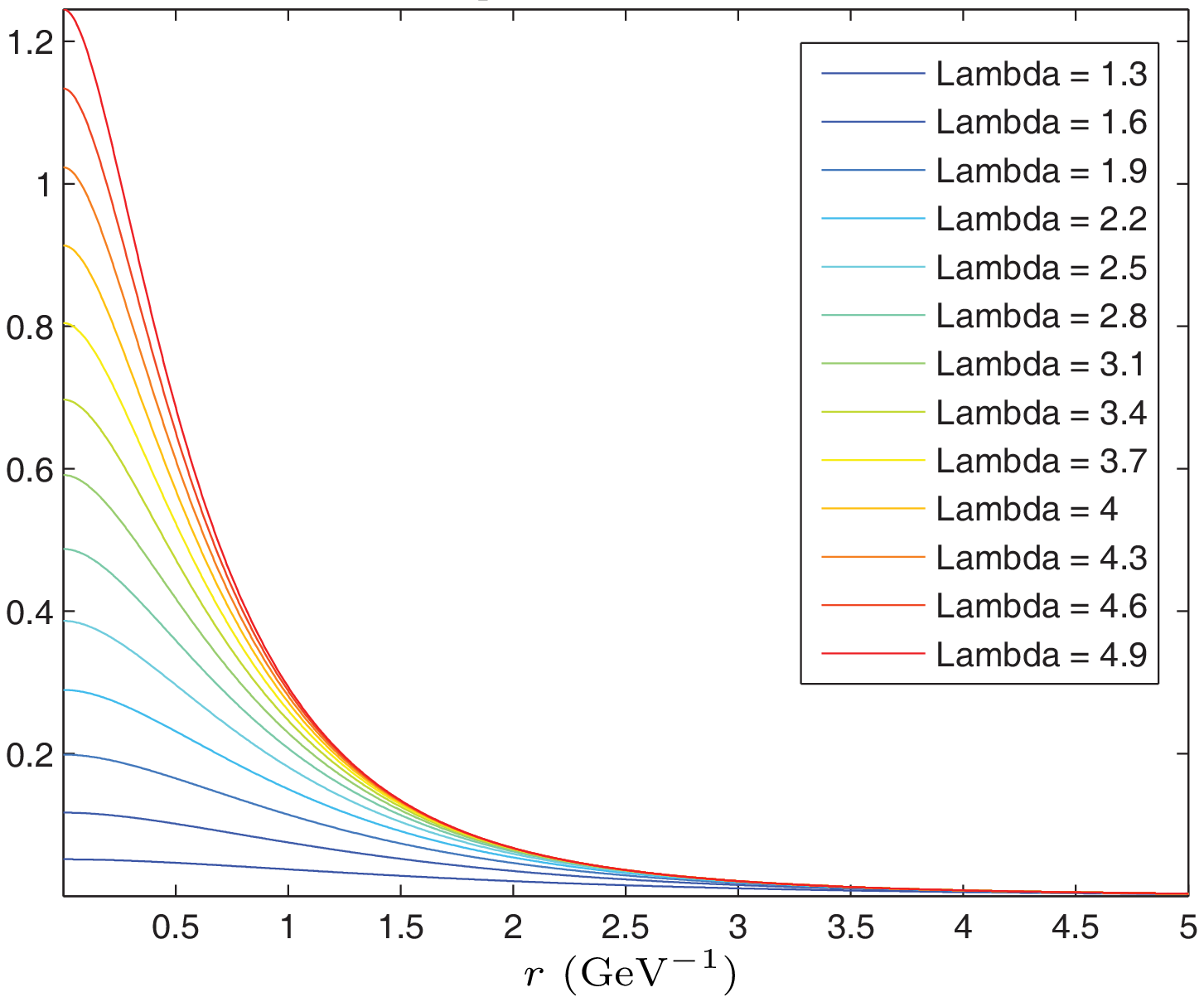}}&\scalebox{0.55}{\includegraphics{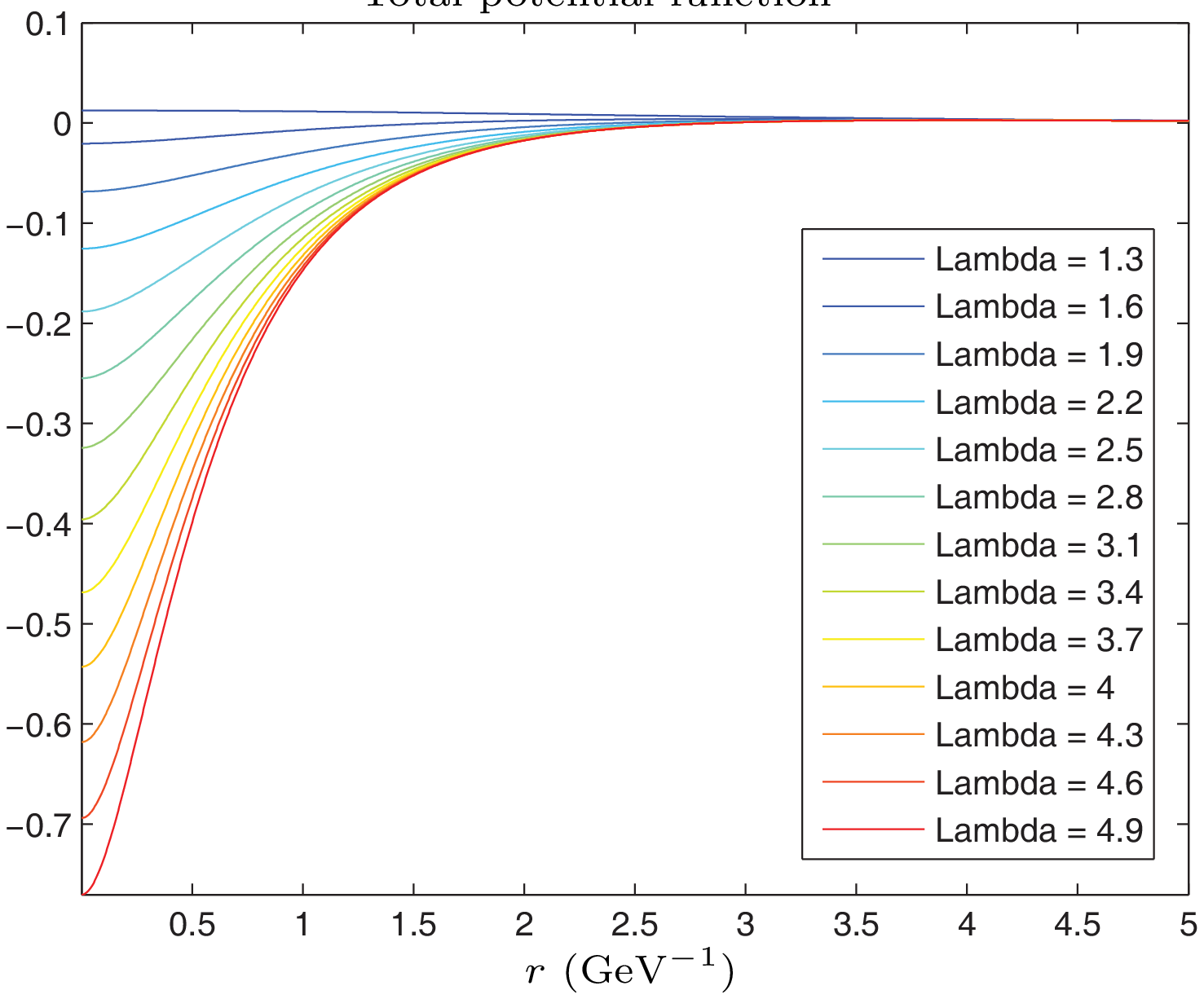}}\\
(c)&(d)\\
\scalebox{0.55}{\includegraphics{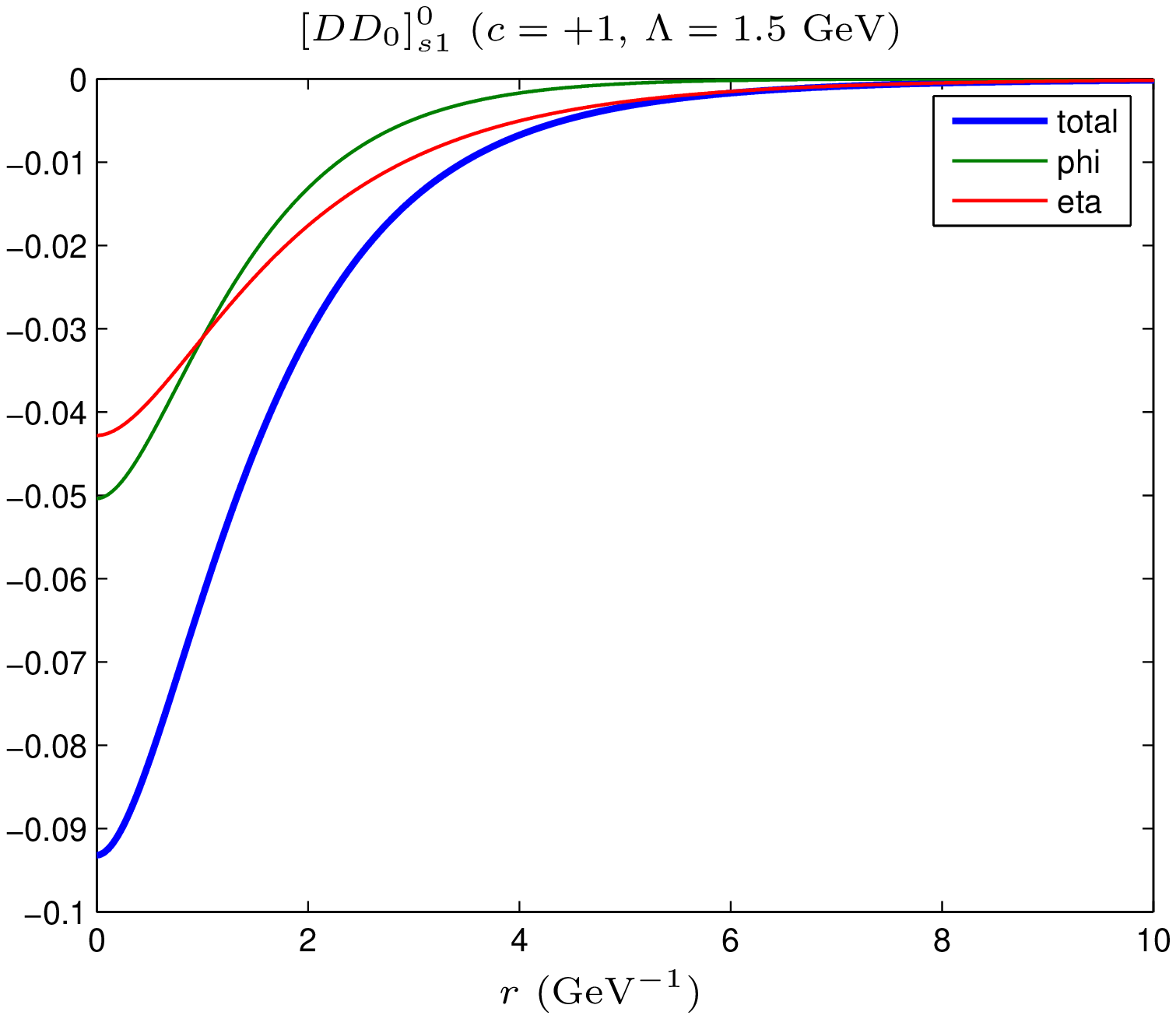}}&\scalebox{0.55}{\includegraphics{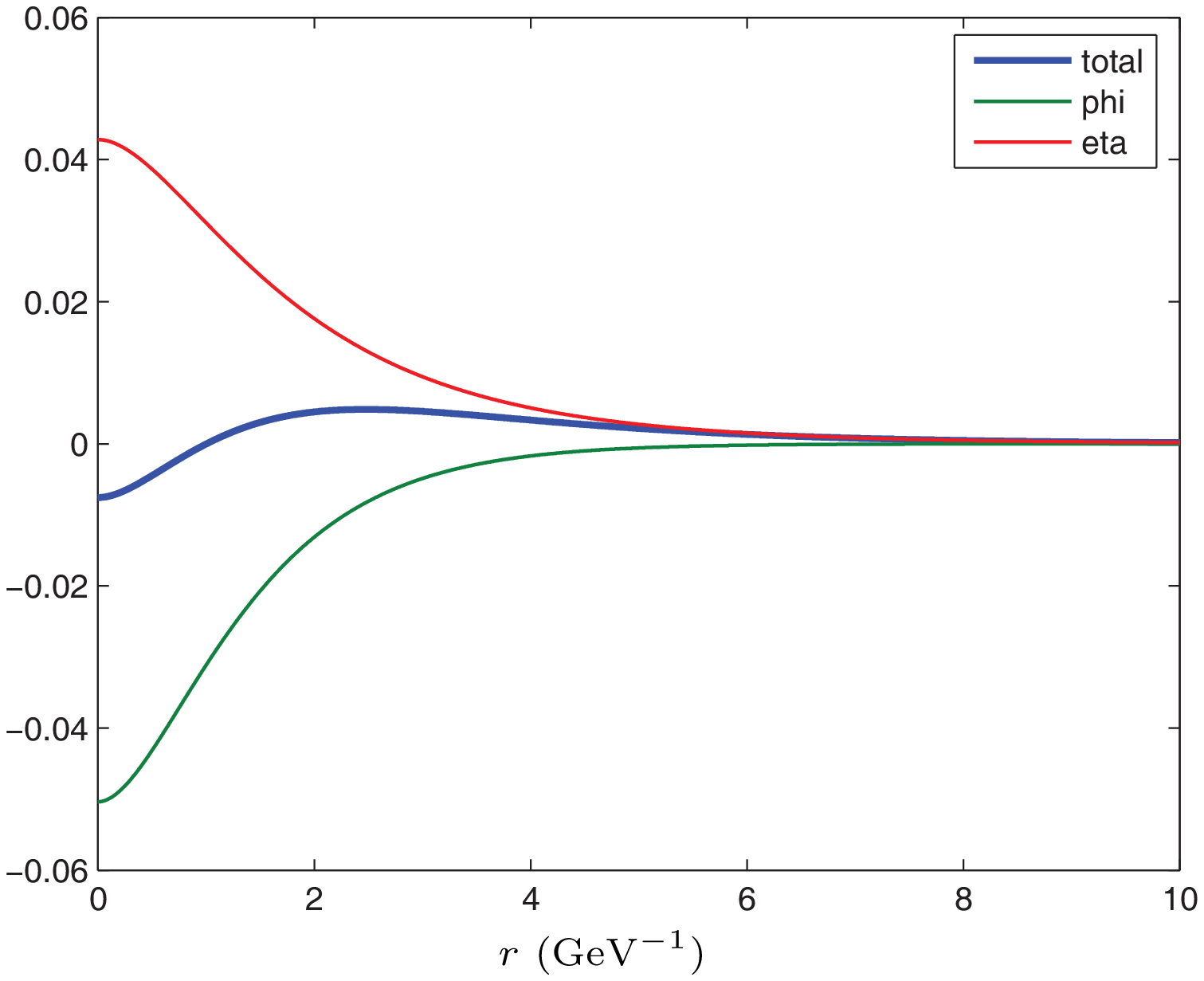}}\\
(e)&(f)
\end{tabular}
\caption{(a), (b), (c) and (d) are the total potentials of the
$[DD_0]_{s1}^{0}$ systems with $(c=+1,\, \beta\beta^\prime>0)$,
$(c=+1,\, \beta\beta^\prime<0)$, $(c=-1,\, \beta\beta^\prime>0)$
and $(c=-1,\, \beta\beta^\prime<0)$ respectively. Under taking
$\Lambda=1.5$ GeV, (e) and (f) are the partial potentials of the
$[DD_0]_{s1}^{0}$ system with $(c=+1,\, \beta\beta^\prime<0)$ and
$(c=-1,\, \beta\beta^\prime<0)$ respectively. Here, we take
$|h|=0.56$, $|\beta|=0.909$ and
$|\beta^\prime|=0.533$.\label{dds3-potential}}
\end{figure}
\end{center}

In Fig. \ref{dds3-potential}, we plot the variation of the
effective potential of the $[PP_0^*]_{s1}^{0}$ system with $r$ and
different coupling constants and the cutoff parameter. The $\eta$
and $\phi$ meson exchange contributes to the total effective
potential. The cutoff $\Lambda$ should be larger than the mass of
$\phi$ meson. The total effective potentials with $(c=+1,
\beta\beta^\prime<0)$ and $(c=-1, \beta\beta^\prime<0)$ are
attractive while the effective potentials with $(c=+1,
\beta\beta^\prime>0)$ and $(c=-1, \beta\beta^\prime>0)$ are
repulsive. The comparison of the total potential with the $\eta$
and $\phi$ meson exchange potential is given in Fig.
\ref{dds3-potential} (e) and (f) with $\Lambda=1.5$ GeV.

\begin{center}
\begin{figure}[htb]
\begin{tabular}{cccc}
\scalebox{0.63}{\includegraphics{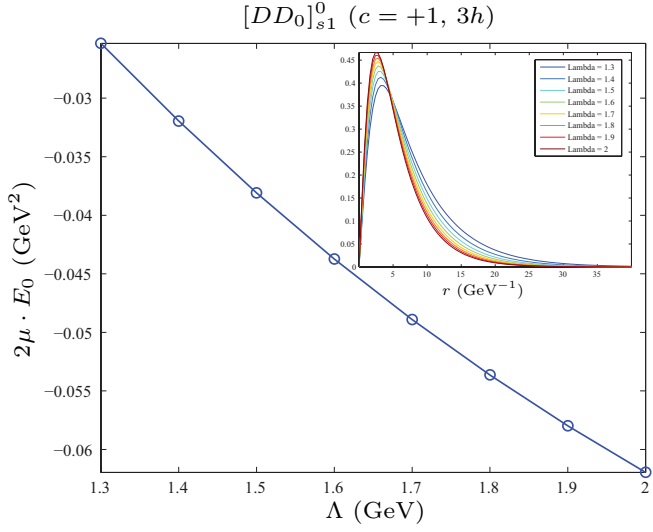}}&\raisebox{0.5em}{\scalebox{0.61}{\includegraphics{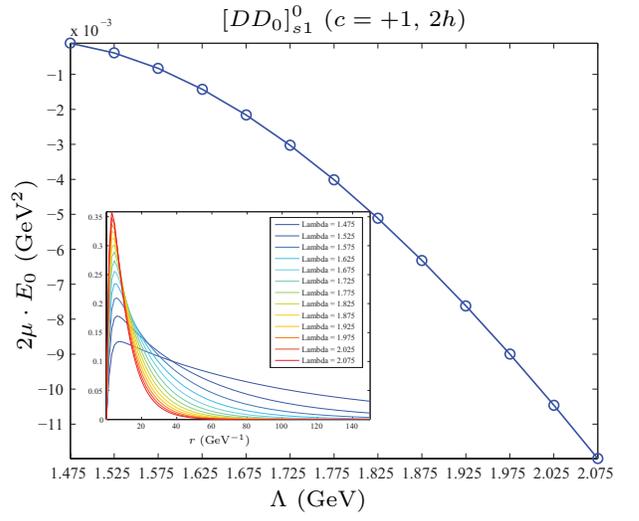}}}\\
(a)&(b)\\
\raisebox{0.0em}{\scalebox{0.63}{\includegraphics{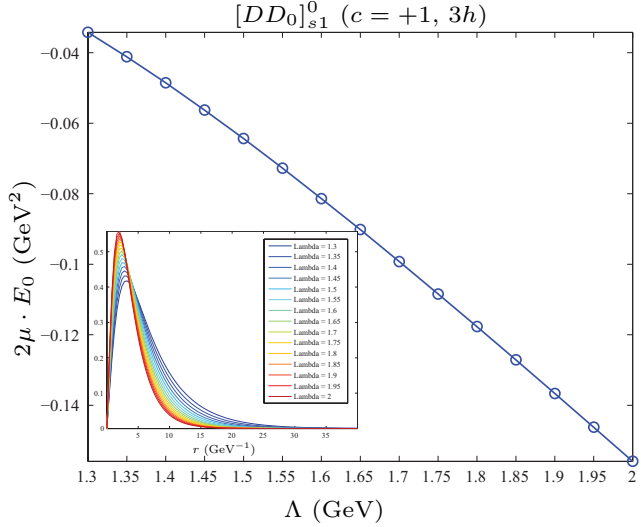}}}& \\
(c)&\\
\end{tabular}
\caption{The dependence of the binding energy of the
$[DD_0]_{s1}^{0}$ system $(c=+1)$ on $\Lambda$ and the wave
function of this system. Here, we consider the contribution of the
$\eta$ and $\phi$ exchange potentials. (a) is for $(c=+1,\,
\beta\beta^\prime>0)$; (b) and (c) are for $(c=+1,\,
\beta\beta^\prime<0)$. When taking $(c=-1,\,
\beta\beta^\prime<0)$, we can not find the bound state solution.
Here, we take $|h|=0.56$, $|\beta|=0.909$ and
$|\beta^\prime|=0.533$.\label{dds3-E}}
\end{figure}
\end{center}

For the $[PP_0^*]_{s1}^{0}$ system with $c=-1$, there does not
exist a bound state not only for the case $\beta\beta^\prime>0$
but also for $\beta\beta^\prime<0$. For the $[PP_0^*]_{s1}^{0}$
system with $c=+1$ and $\beta\beta^\prime>0$, a reasonable
solution of bound state appears with $3h$. For the
$[PP_0^*]_{s1}^{0}$ system with $c=+1$ and $\beta\beta^\prime<0$,
there does not exist a bound state with $1h$. The bound state
solution appears with $2h$ and $3h$. The detailed numerical
results of the $[PP_0^*]_{s1}^{0}$ system with $c=+1$ are shown in
Fig. \ref{dds3-E}.

\subsubsection{$[PP_0^*]_{8}^{0}$}

\begin{center}
\begin{figure}[htb]
\begin{tabular}{cccc}
\scalebox{0.55}{\includegraphics{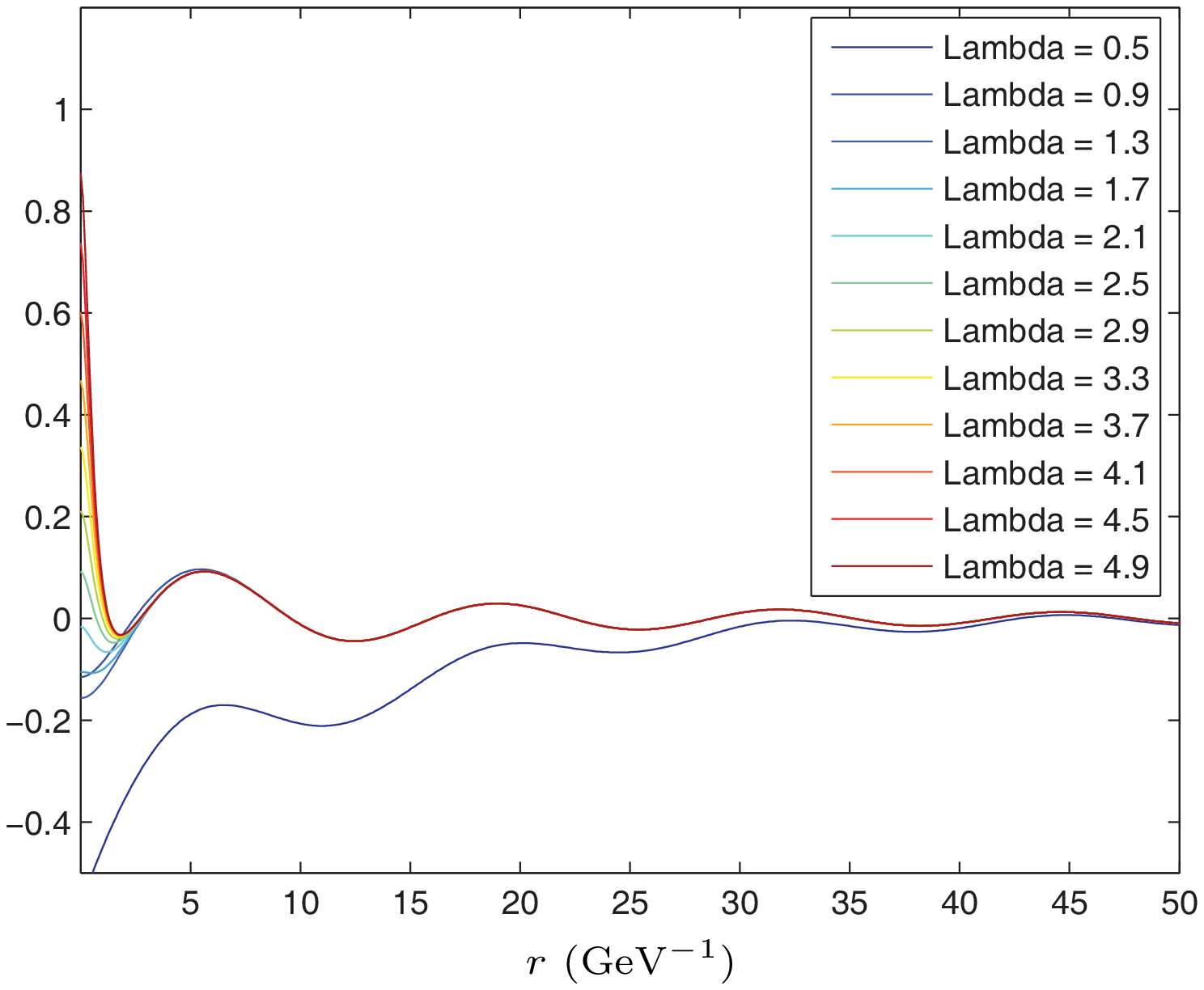}}&\scalebox{0.55}{\includegraphics{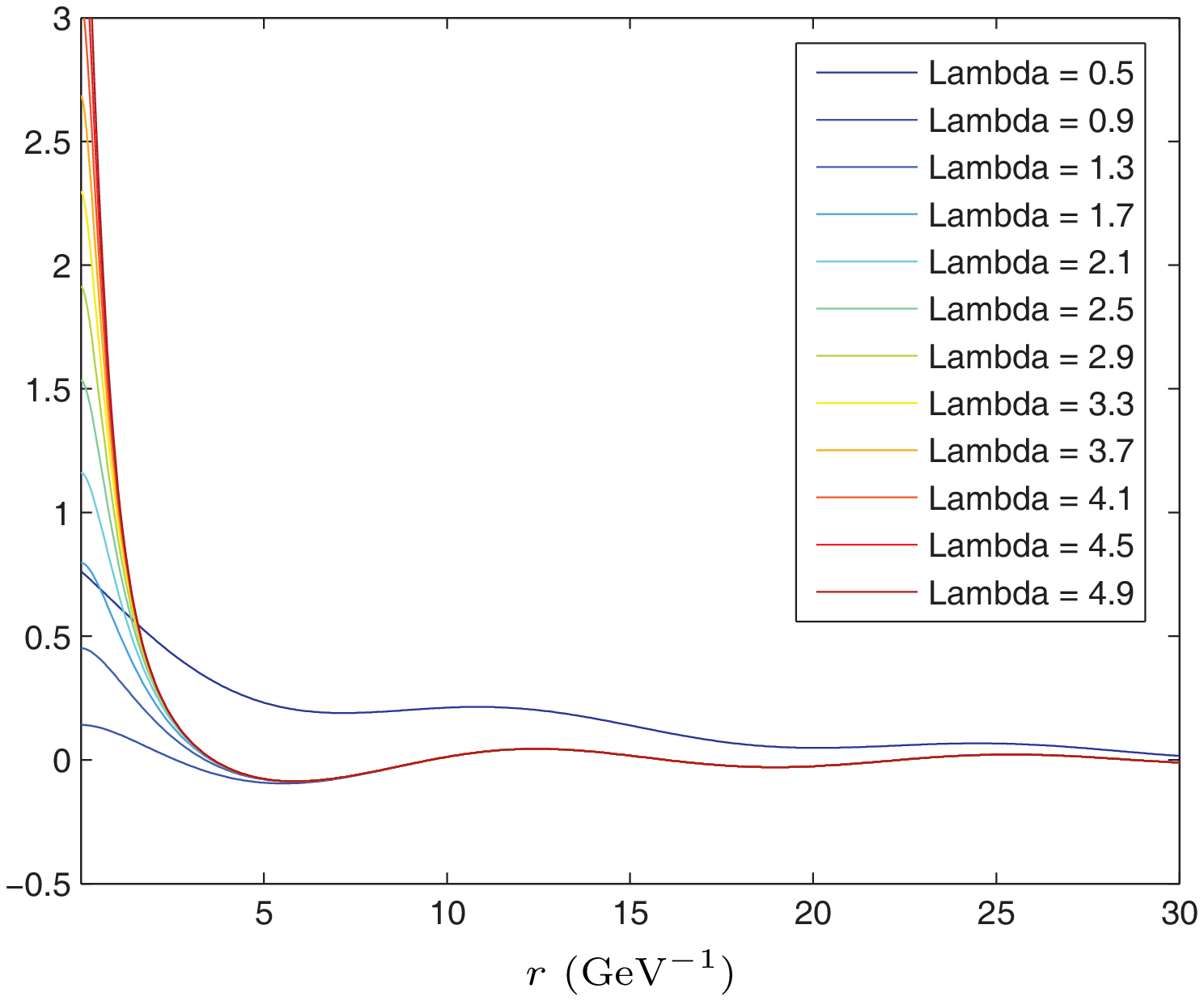}}\\
(a)&(b)\\
\scalebox{0.55}{\includegraphics{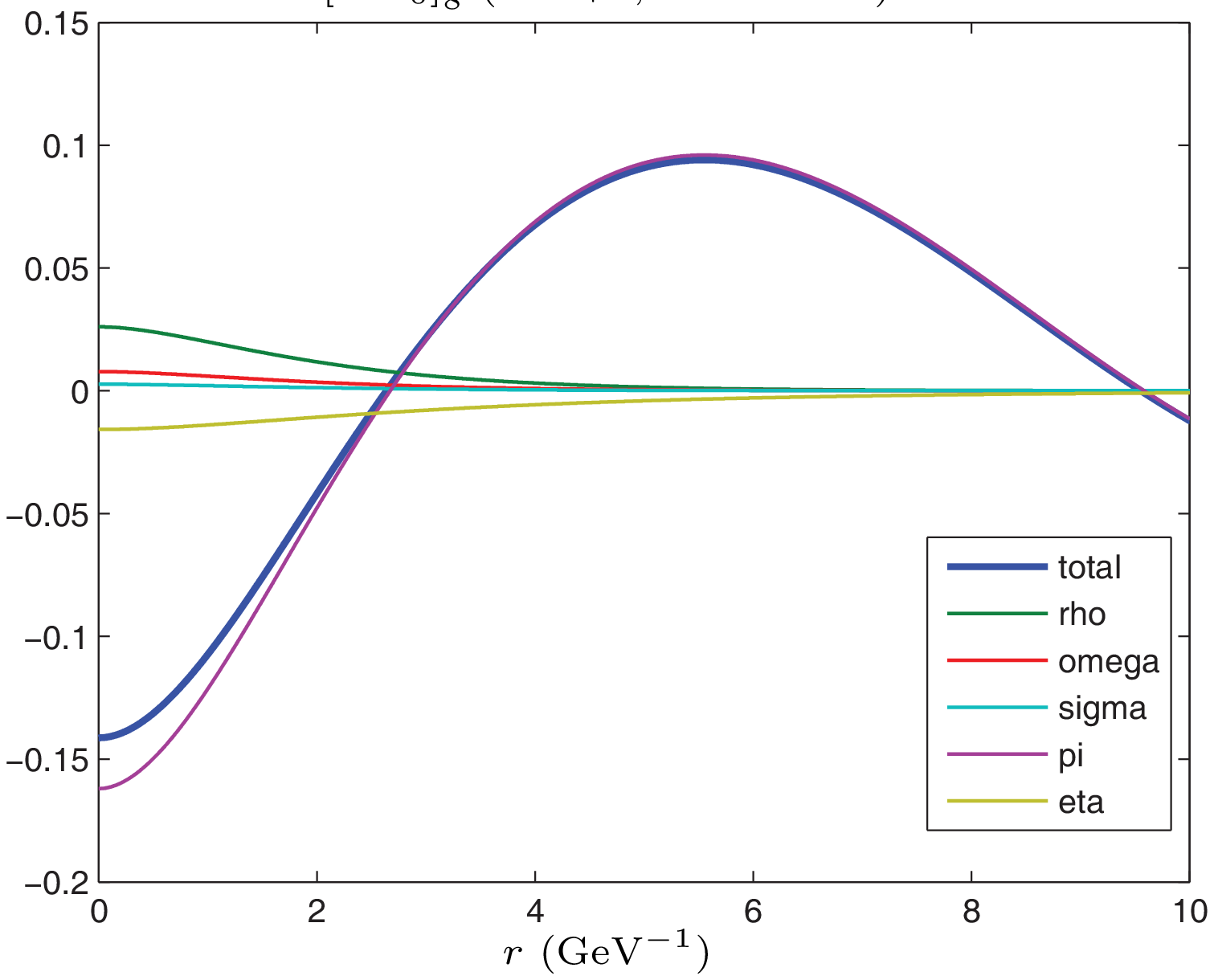}}&\scalebox{0.55}{\includegraphics{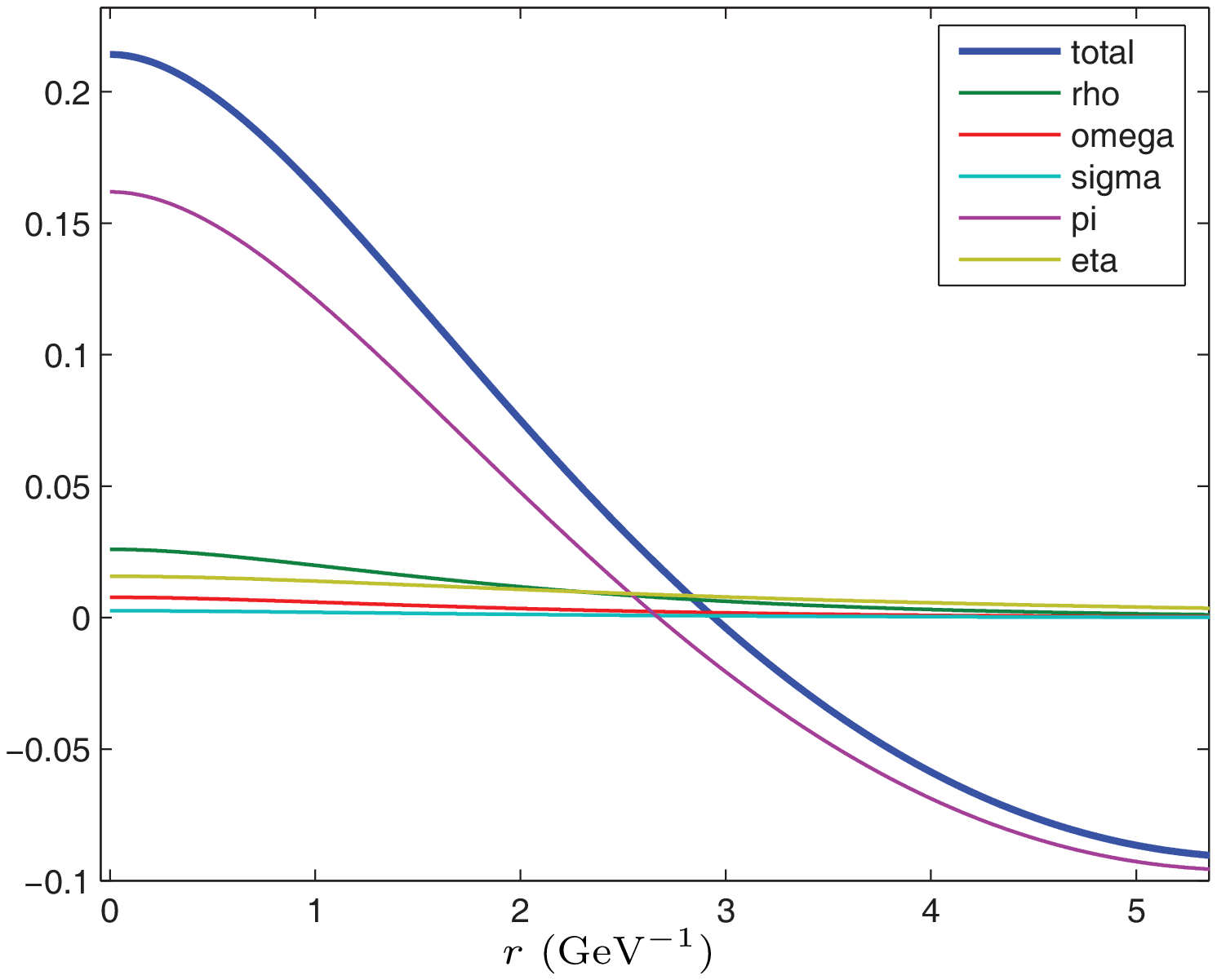}}\\
(c)&(d)
\end{tabular}
\caption{(a) and (b) are the total potentials of the
$[DD_0]_{8}^{0}$ systems with $c=+1$ and $c=-1$ respectively. With
$\Lambda=1$ GeV, diagrams (c) and (d) illustrate the partial
potentials of the $[DD_0]_{8}^{0}$ system with $c=+1$ and $c=-1$
respectively. The above potentials are obtained with $g_\sigma
g_\sigma^\prime>0$ and $\beta\beta^\prime>0$. Here, $|h|=0.56$,
$|g_\sigma|=0.76$, $|g_\sigma^\prime|=0.76$, $|\beta|=0.909$ and
$|\beta^\prime|=0.533$.\label{dd4-potential}}
\end{figure}
\end{center}

The dependence of the total potential of the $[PP_0^*]_{8}^{0}$
system on $\Lambda$ is presented in Fig. \ref{dd4-potential}. The
total effective potential oscillates with $r$. The $\pi$, $\eta$,
$\sigma$, $\rho$ and $\omega$ exchange contributes to the total
exchange potential. With $\Lambda=1$ GeV, the corresponding $\pi$,
$\eta$, $\sigma$, $\rho$ and $\omega$ exchange potentials for the
$[PP_0^*]_{8}^{0}$ systems with $c=\pm 1$ are given in
\ref{dd4-potential} (c) and (d). The $\pi$ exchange potential is
dominant. Therefore, we only consider the single pion exchange
potential when we explore whether there exists the bound state
solution of the $[PP_0^*]_{8}^{0}$ system.

\begin{center}
\begin{figure}[htb]
\begin{tabular}{cccc}
\scalebox{0.63}{\includegraphics{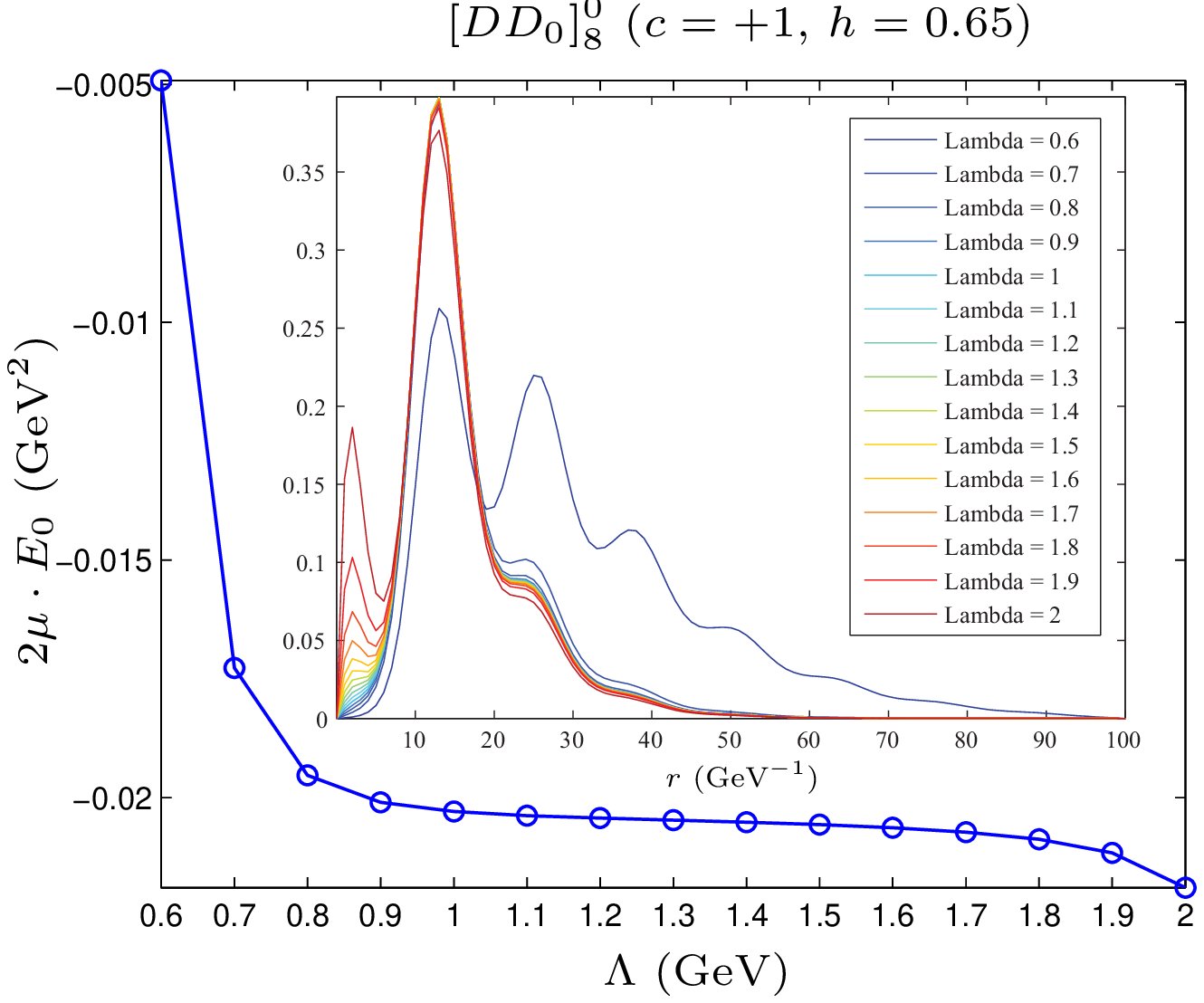}}&\raisebox{0em}{\scalebox{0.63}{\includegraphics{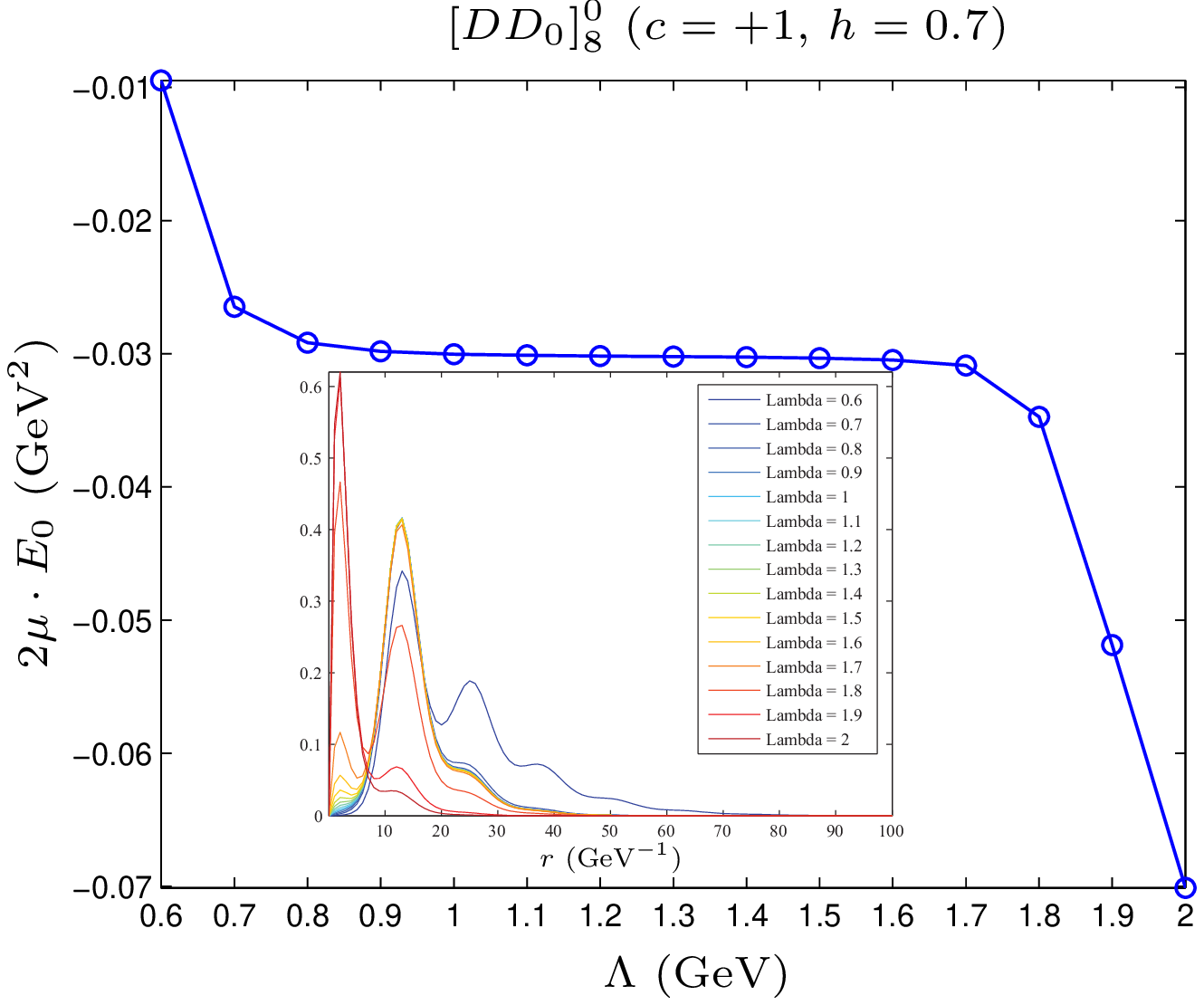}}}\\
(a)&(b)\\
\raisebox{0.0em}{\scalebox{0.63}{\includegraphics{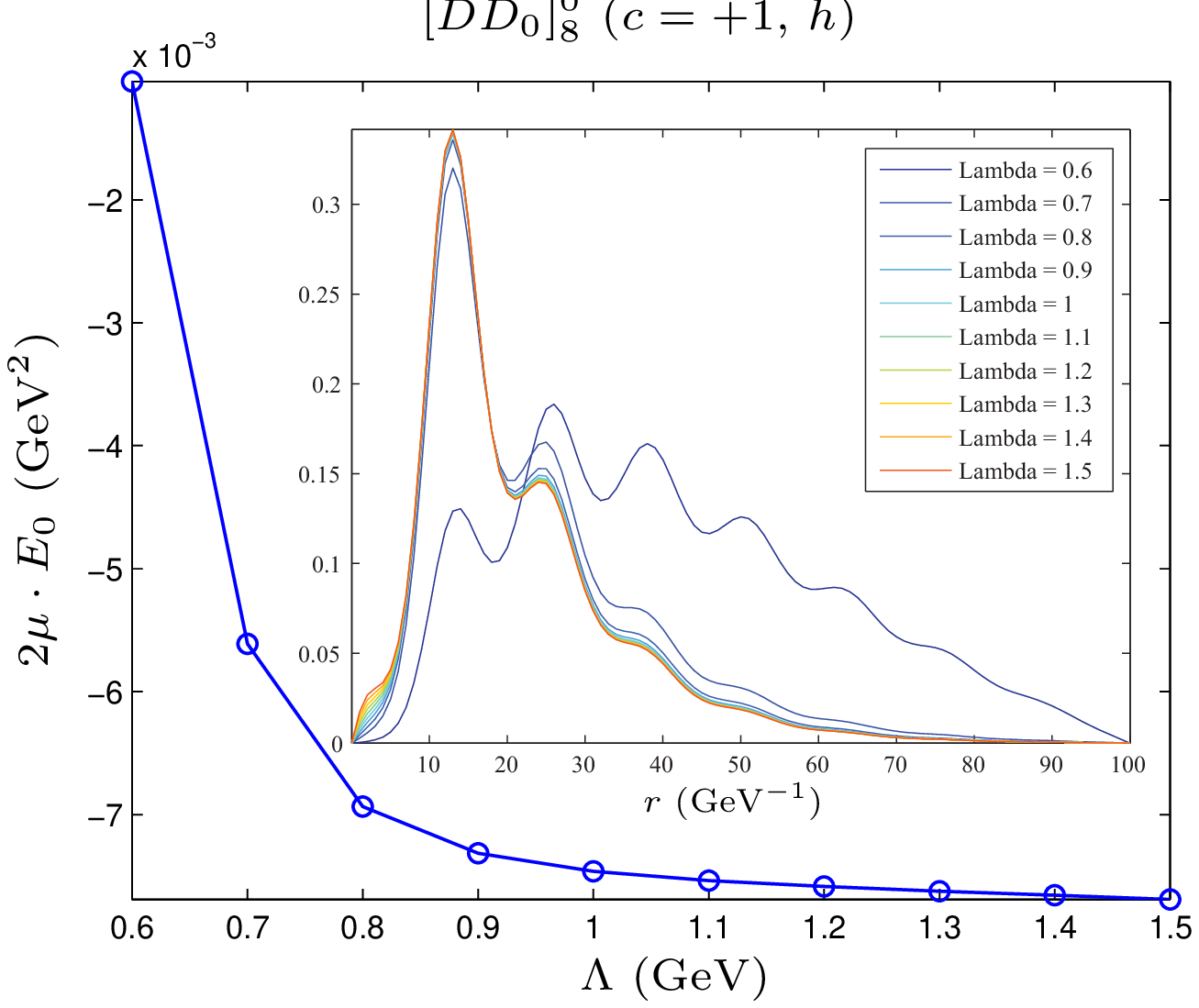}}}& \raisebox{0.0em}{\scalebox{0.63}{\includegraphics{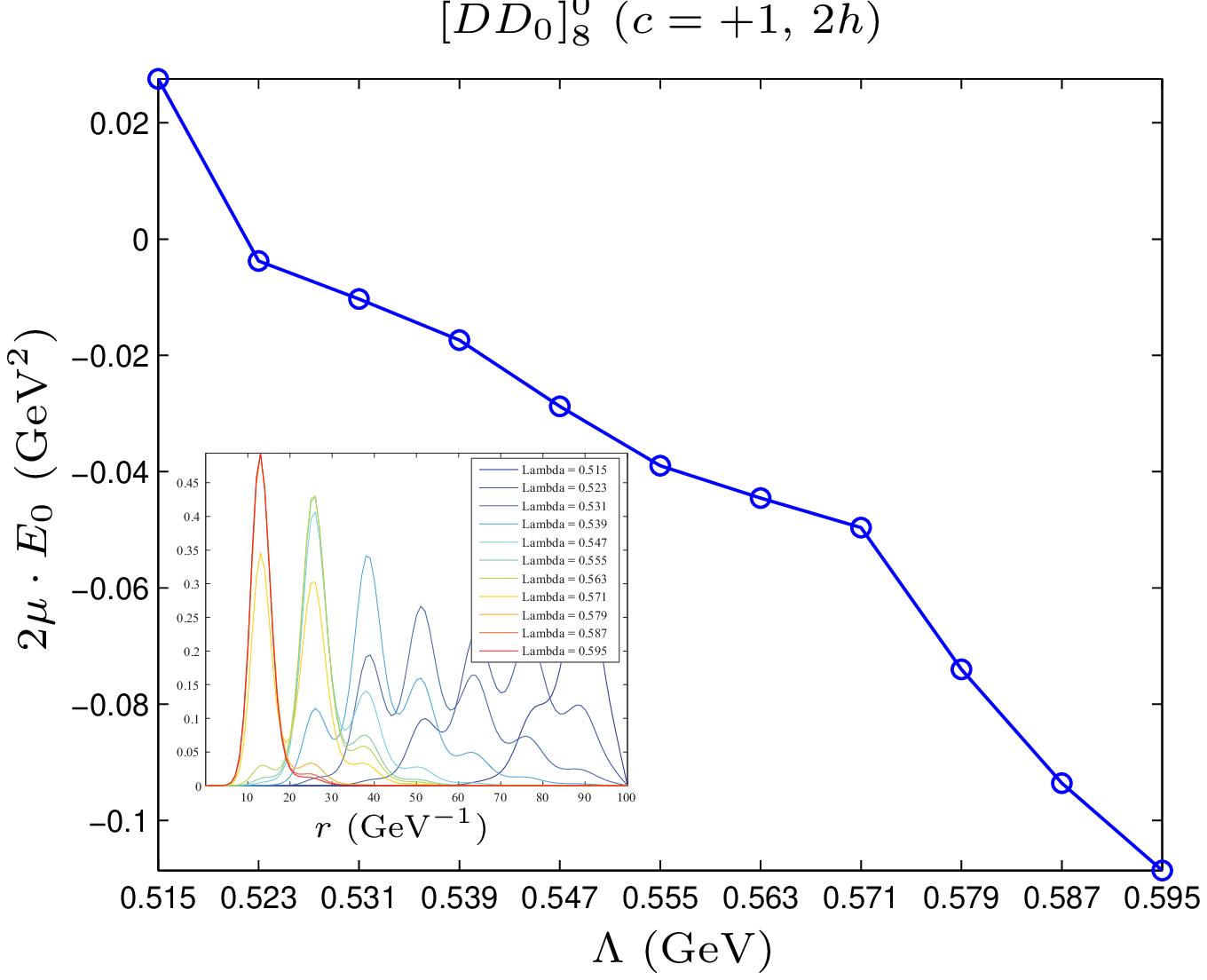}}}\\
(c)&(d)\\
\raisebox{0.0em}{\scalebox{0.63}{\includegraphics{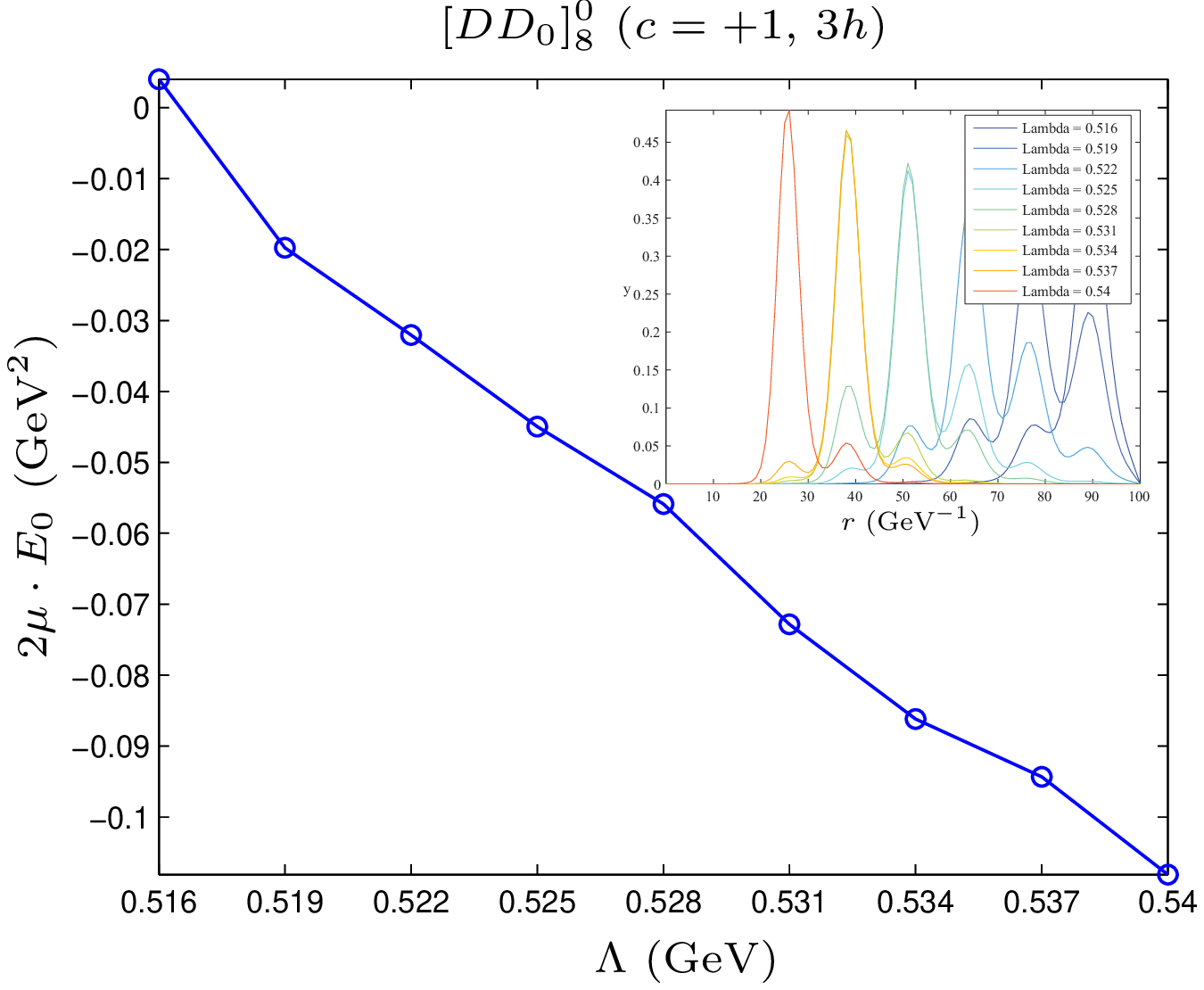}}}&\\
(e)&
\end{tabular}
\caption{The dependence of the binding energy of the
$[DD_0]_{8}^{0}$ system $(c=+1)$ on $\Lambda$ and the wave
function of this system. Here, we only consider the contribution
of the $\pi$ exchange potentials. We take different values of $h$,
in diagrams (a)-(e). The absolute value of $h$ in $1h,\,2h,\,3h$
is 0.56. \label{dd4-c1-E}}
\end{figure}
\end{center}

For the $[PP_0^*]_{8}^{0}$ system with $c=+1$, there exists a
bound state solution. The variation of the binding energy of
$[PP_0^*]_{8}^{0}$ with $c=+1$ is shown in Fig. \ref{dd4-c1-E}
with different values of the coupling constant. The wave functions
is not reasonable when taking $|h|=0.56,\,0.65,\,0.7$, which
corresponds to the shallow binding energy of the
$[PP_0^*]_{8}^{0}$ system with $c=+1$. Increasing the coupling
constant to $2h$ leads to a bound state solution with a reasonable
wave function.

\begin{center}
\begin{figure}[htb]
\begin{tabular}{cccc}
\scalebox{0.63}{\includegraphics{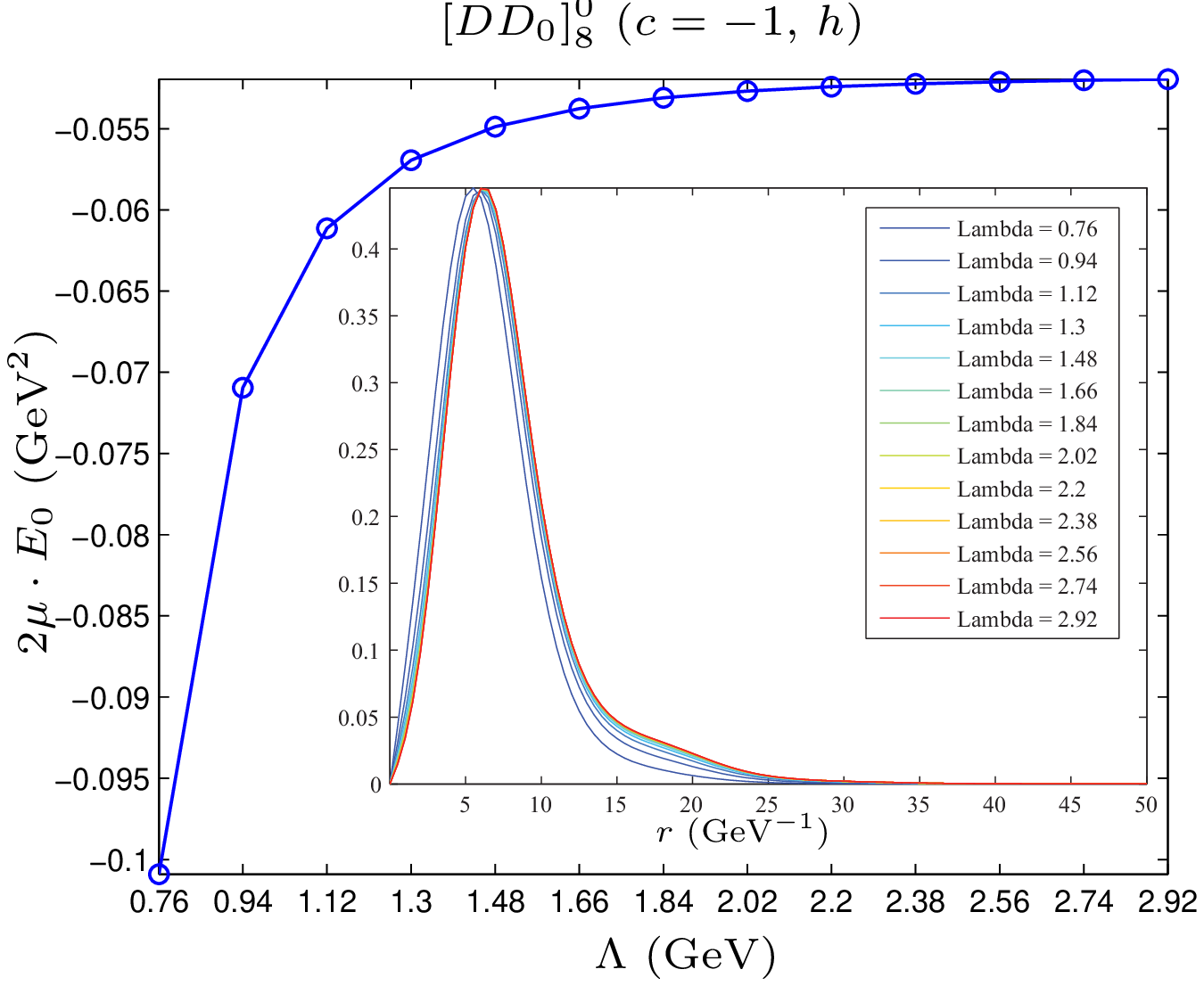}}&\raisebox{0em}{\scalebox{0.63}{\includegraphics{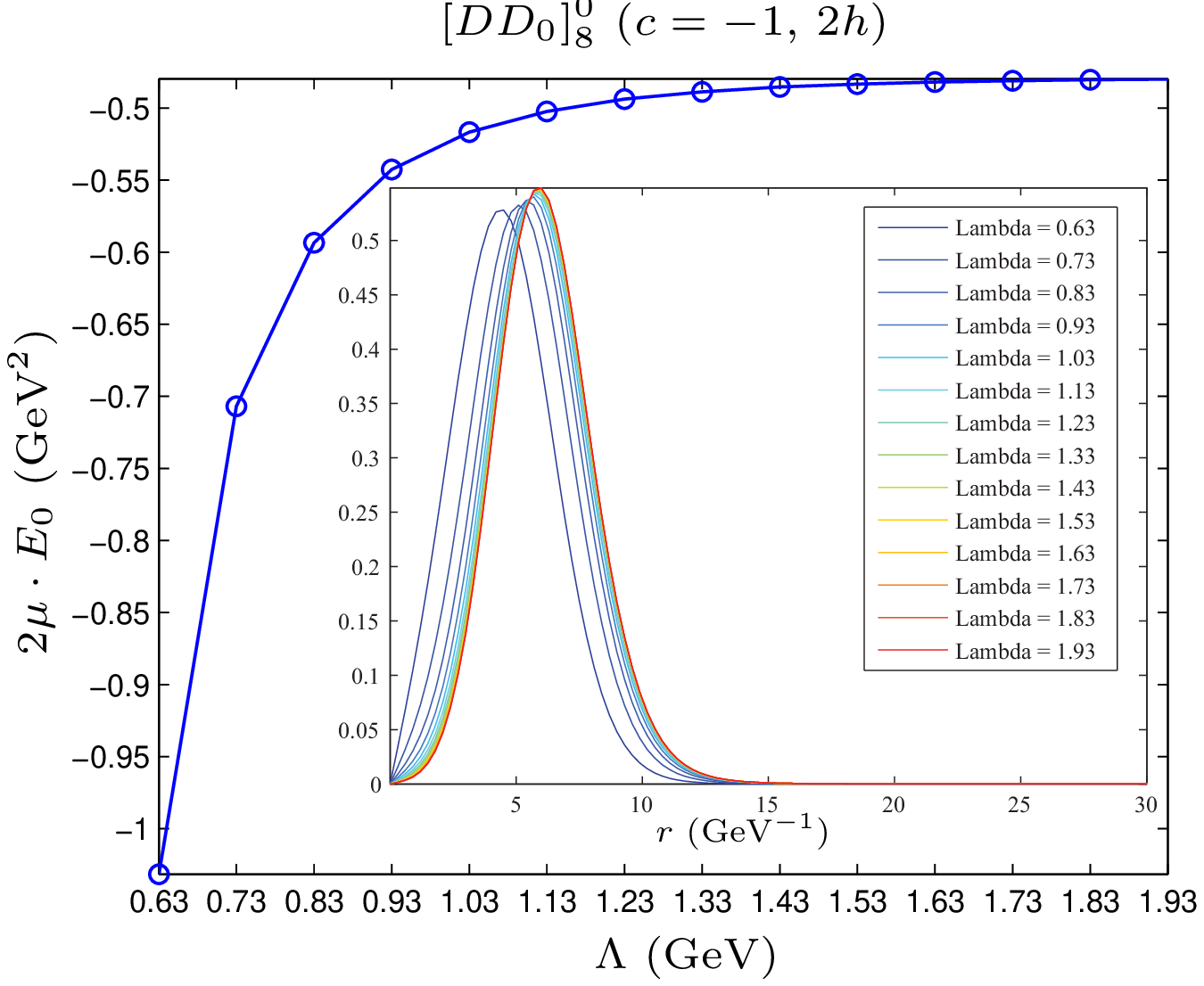}}}\\
(a)&(b)
\end{tabular}
\caption{The binding energy of the $[DD_0]_{8}^{0}$ system
$(c=-1)$ and the wave function of this system. Here, we only
consider the $\pi$ exchange potential with $1h,\,2h$ in
diagrams (a)-(c) respectively. Here, $|h|=0.56$.
\label{dd4-c-1-E}}
\end{figure}
\end{center}

The numerical result presented in Fig. \ref{dd4-c-1-E} shows that
there exists a $[PP_0^*]_{8}^{0}$ bound state with $c=-1$.

\subsection{The $[PP_1^\prime]$ system}

\subsubsection{$[PP_1^\prime]^{\pm,0}$}

\begin{center}
\begin{figure}[htb]
\begin{tabular}{cccc}
\scalebox{0.55}{\includegraphics{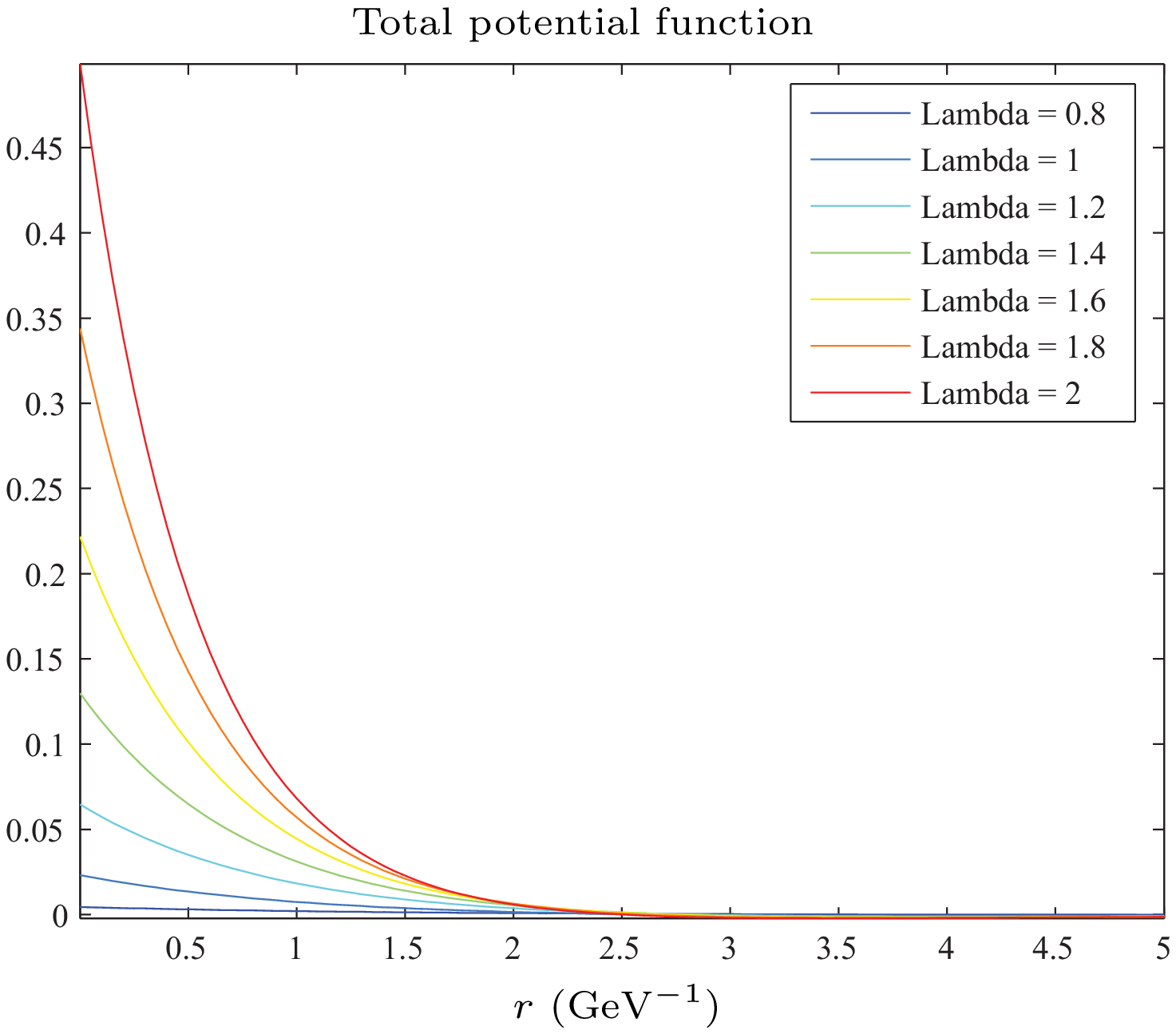}}&\scalebox{0.55}{\includegraphics{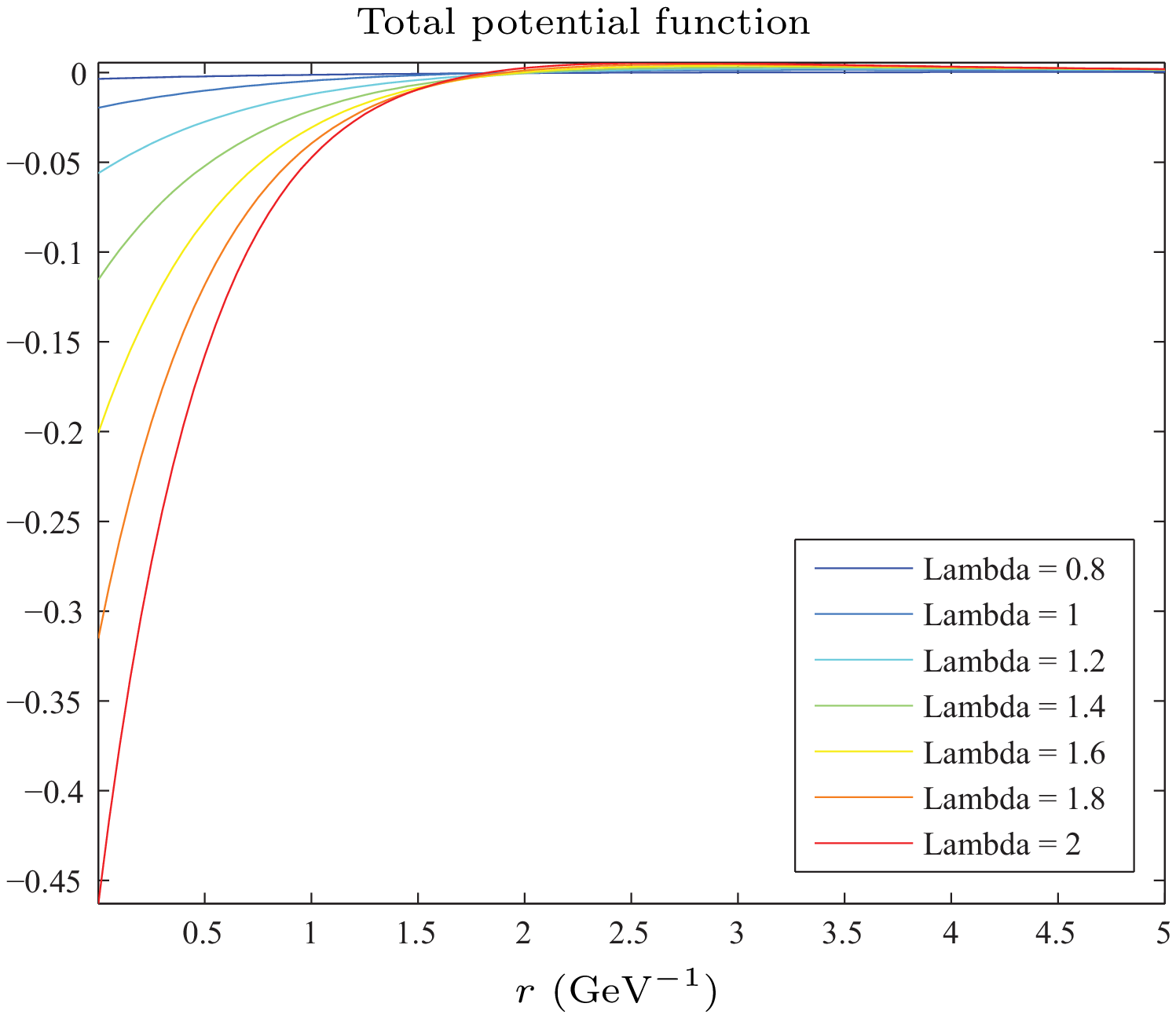}}\\
(a)&(b)\\
\scalebox{0.55}{\includegraphics{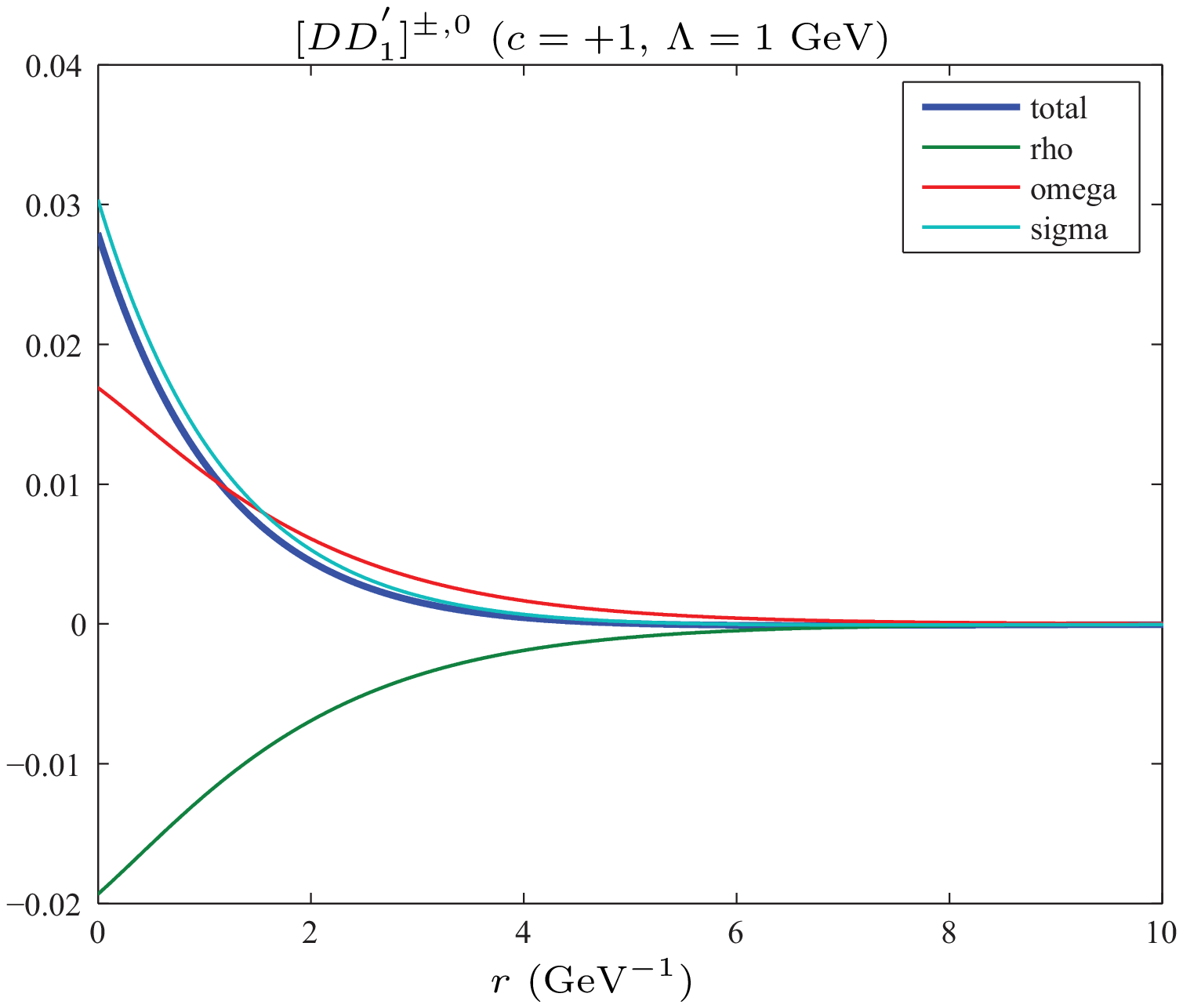}}&\scalebox{0.55}{\includegraphics{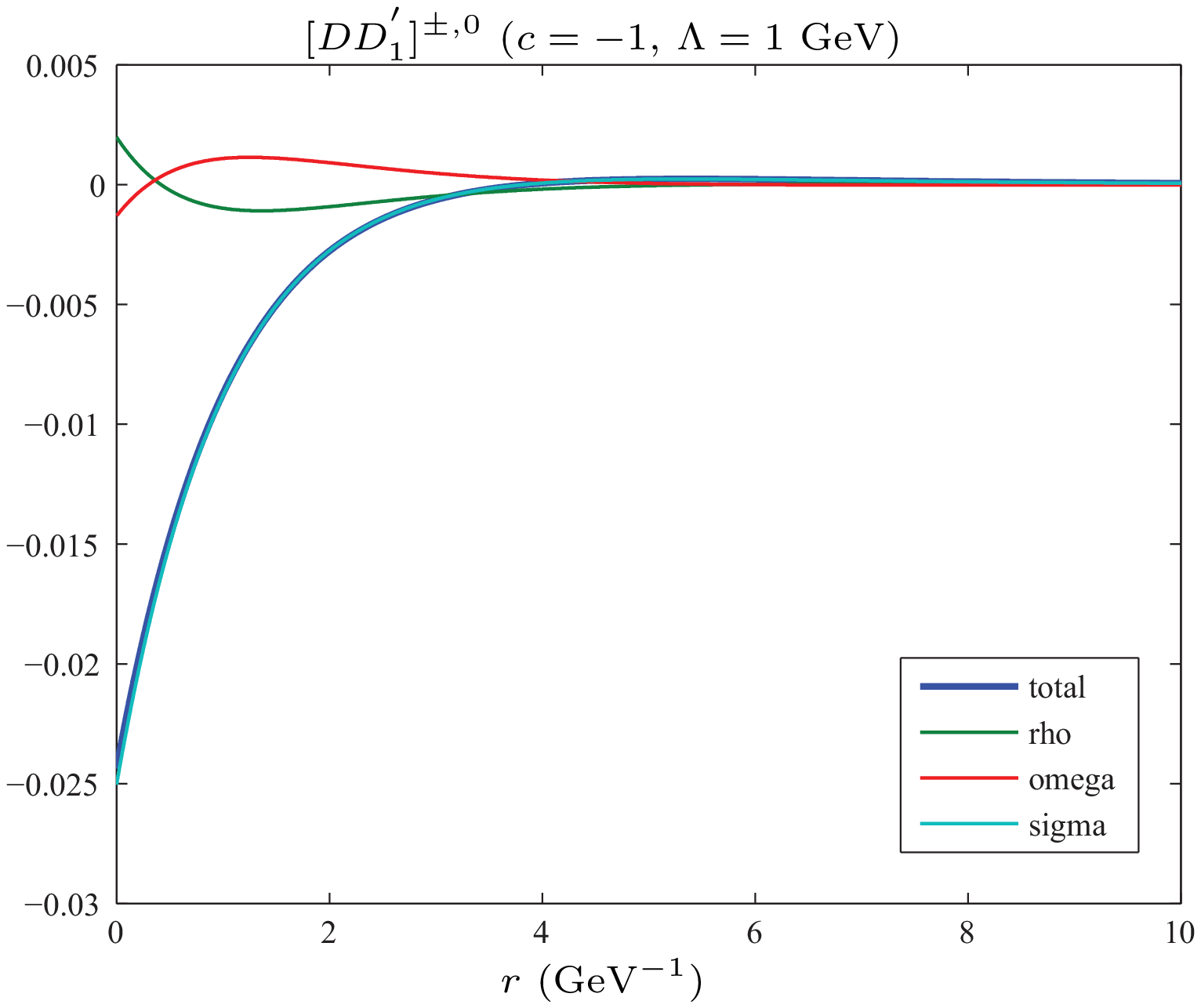}}\\
(c)&(d)
\end{tabular}
\caption{(a) The variation of the total potential of the
$[DD_1^\prime]^{\pm,0}$ system $(c=+1)$ with $r$ and $\Lambda$;
(b) the variation of the total potential of the
$[DD_1^\prime]^{\pm,0}$ system $(c=-1)$ with $r$ and $\Lambda$;
(c) the exchange potentials of the $\sigma,\,\rho,\,\omega$ mesons
of the $[DD_1^\prime]^{\pm,0}$ system $(c=+1)$ with $\Lambda=1$
GeV; (d) the exchange potentials of the $\sigma,\,\rho,\,\omega$
mesons of the $[DD_1^\prime]^{\pm,0}$ system $(c=-1)$ with
$\Lambda=1$ GeV. The above potentials are obtained with $g_\sigma
g_\sigma^\prime>0$, $\beta\beta^\prime>0$ and $\zeta \varpi>0$.
Here, $|g_\sigma|=0.76$, $|g_\sigma^\prime|=0.76$,
$|h_\sigma|=0.323$, $|\beta|=0.909$ and $|\beta^\prime|=0.533$,
$\zeta=0.727$ and $\varpi=0.364$.\label{dd1s1potential}}
\end{figure}
\end{center}

The $\sigma$, $\rho$ and $\omega$ meson exchange contributes to
the total effective potential of the $[PP_1^\prime]^{\pm,0}$
system. Since the $\rho$ and $\omega$ exchange potentials cancel
each other, the $\sigma$ meson exchange potential is dominant. We
only need to consider the $\sigma$ exchange potential in order to
explore whether there exists a $[PP_1^\prime]^{\pm,0}$ bound
state. In Fig. \ref{dd1s1potential}, the effective potential of
the $[PP_1^\prime]^{\pm,0}$ system is presented. The $\sigma$
exchange potential of the $[PP_1^\prime]^{\pm,0}$ system with
$c=+1$ is repulsive. There does not exist a
$[PP_1^\prime]^{\pm,0}$ bound state with $c=+1$.

For the $[PP_1^\prime]^{\pm,0}$ system with $c=-1$, the $\sigma$
exchange potential is attractive. Meanwhile, the cross diagram
plays the dominant role. However, in the cases $g_\sigma
g_\sigma^\prime>0$ and $g_\sigma g_\sigma^\prime<0$, we can not
find a bound state solution of the $[PP_1^\prime]^{\pm,0}$ system
with $c=-1$ and the coupling constants listed in the caption of
Fig. \ref{dd1s1potential} when scanning the range of cutoff
$\Lambda$.

\begin{center}
\begin{figure}[htb]
\begin{tabular}{cccc}
\scalebox{0.63}{\includegraphics{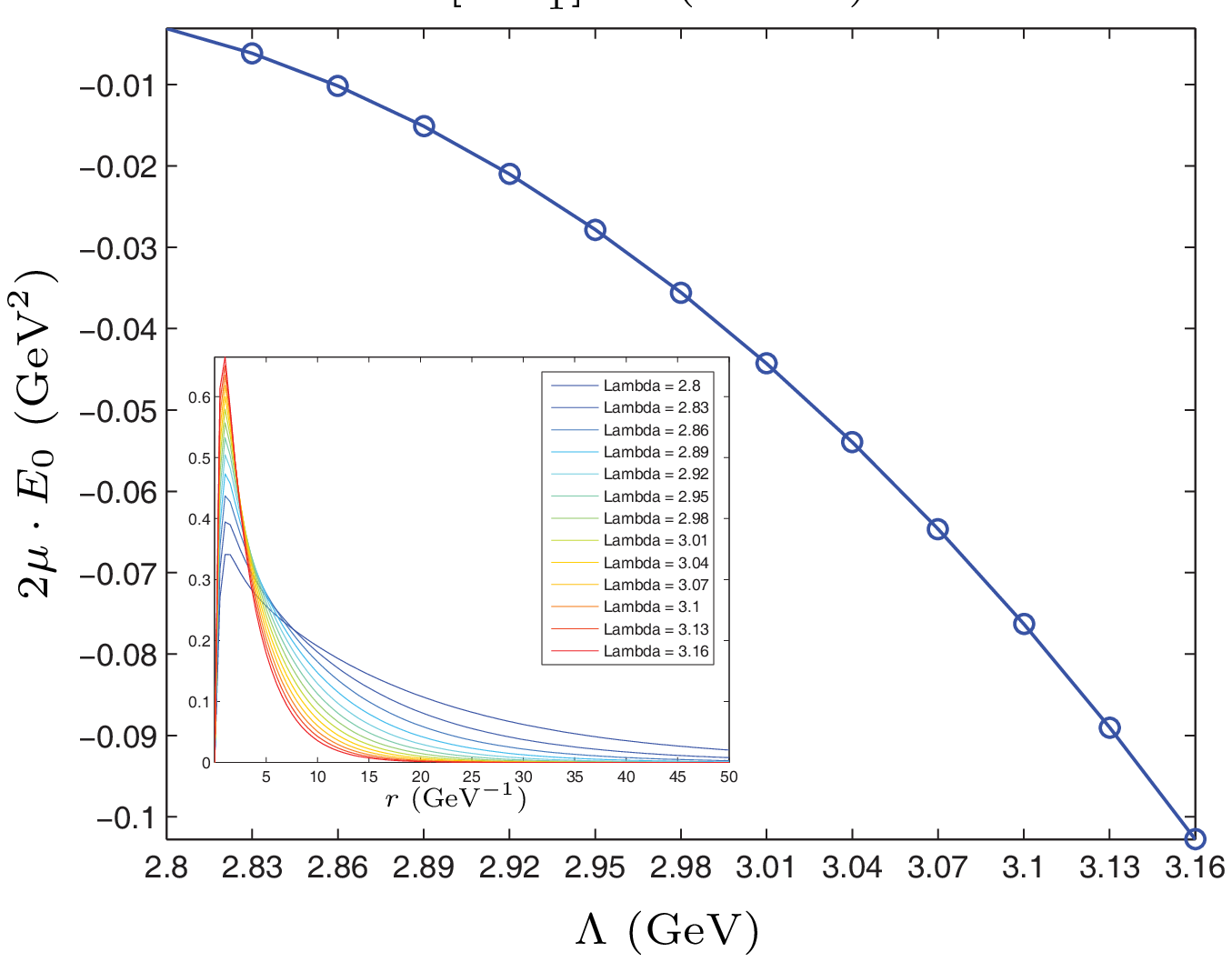}}&\scalebox{0.63}{\includegraphics{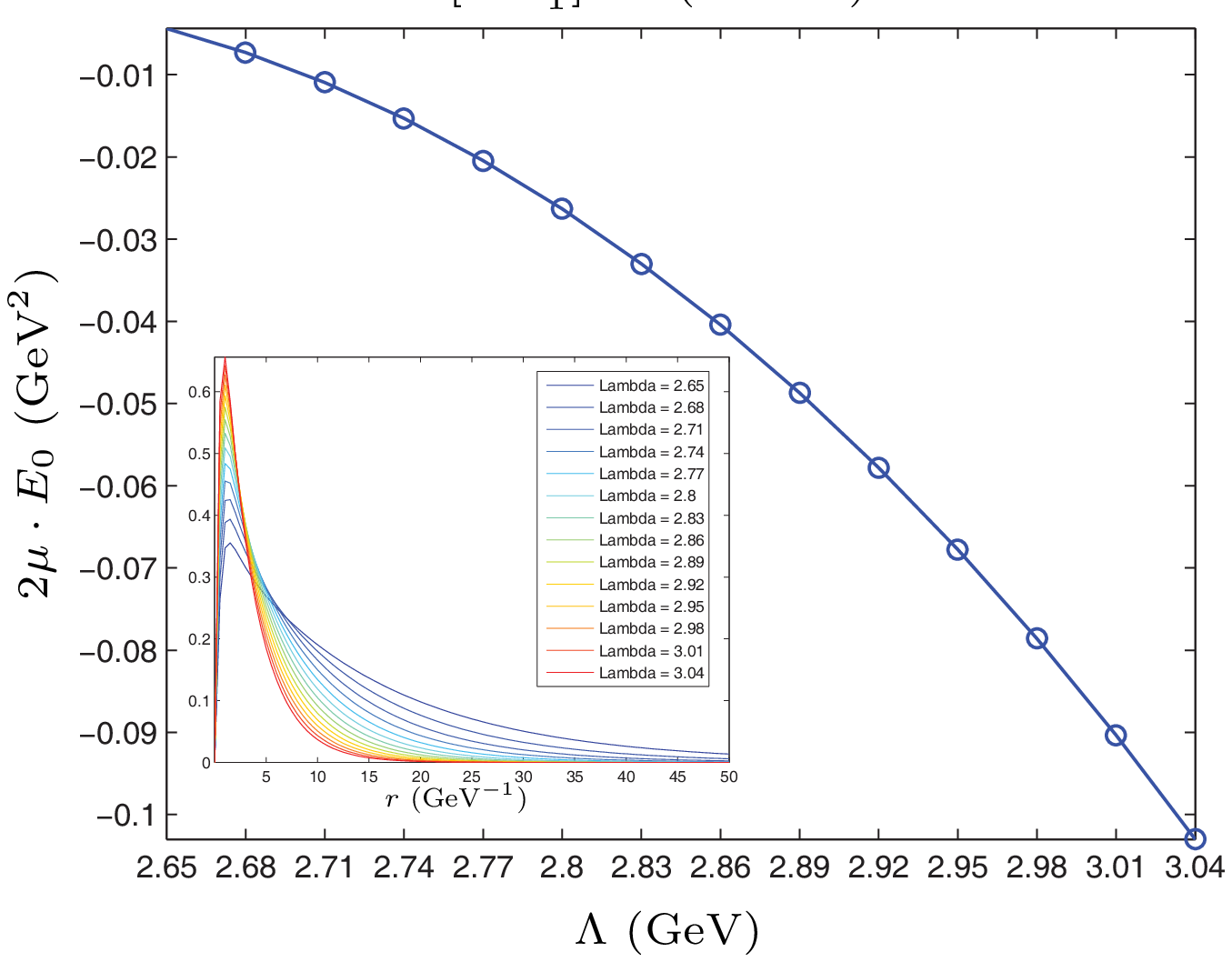}}\\
(a)&(b)
\end{tabular}
\caption{The binding energy of the $[DD_1^\prime]^{\pm,0}$ system
$(c=-1)$ and the wave function of this system with $2h_\sigma$.
Here, we only consider the $\sigma$ exchange potential. (a) and
(b) correspond to the case of $g_\sigma g_\sigma^\prime>0$ and
$g_\sigma g_\sigma^\prime<0$ with $|h_\sigma|=0.323$,
$|g_\sigma|=0.76$ and $|g_\sigma^\prime|=0.76$ respectively.\label{dd1s1e}}
\end{figure}
\end{center}

When taking $2h_\sigma$, we present the variation of the binding
energy of the $[PP_1^\prime]^{\pm,0}$ system with $c=-1$ and the
cutoff $\Lambda$ in Fig. \ref{dd1s1e} in the cases $g_\sigma
g_\sigma^\prime>0$ and $g_\sigma g_\sigma^\prime<0$. However, the
corresponding cutoff $\Lambda$ is far larger than $1$ GeV. Thus,
we tend to conclude that there does not exist a
$[PP_1^\prime]^{\pm,0}$ bound state with $c=-1$.

\subsubsection{$[PP_1^\prime]_s^{\pm,0},\,[PP_1^\prime]_{\hat{s}}^{0}$}

There does not exist the
$[PP_1^\prime]_s^{\pm,0},\,[PP_1^\prime]_{\hat{s}}^{0}$ bound
state system since none of the meson exchange forces is allowed as
indicated in Table \ref{potential}.

\subsubsection{$[PP_1^\prime]_{s1}^{0}$}

\begin{center}
\begin{figure}[htb]
\begin{tabular}{cccc}
\scalebox{0.55}{\includegraphics{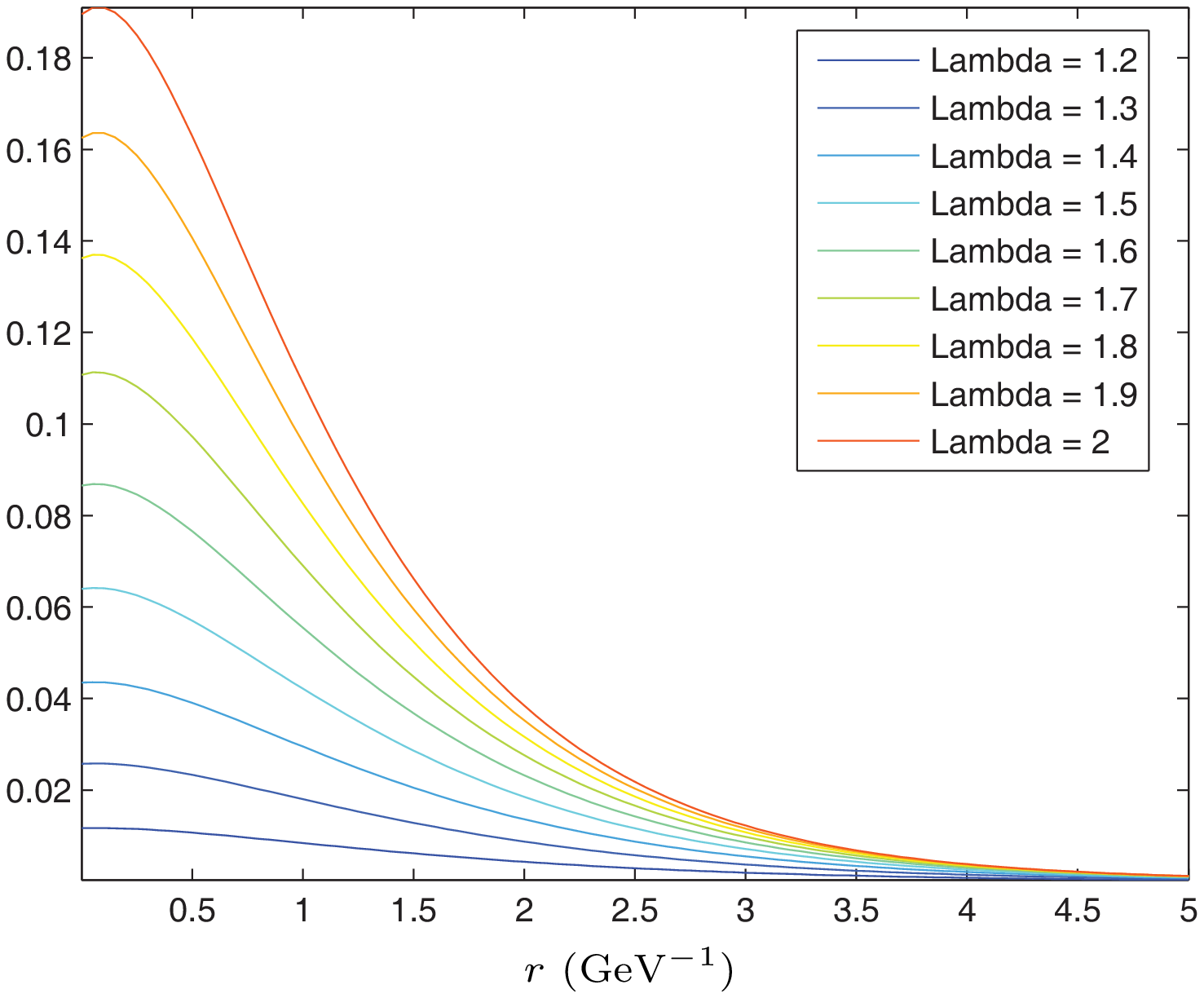}}&\scalebox{0.55}{\includegraphics{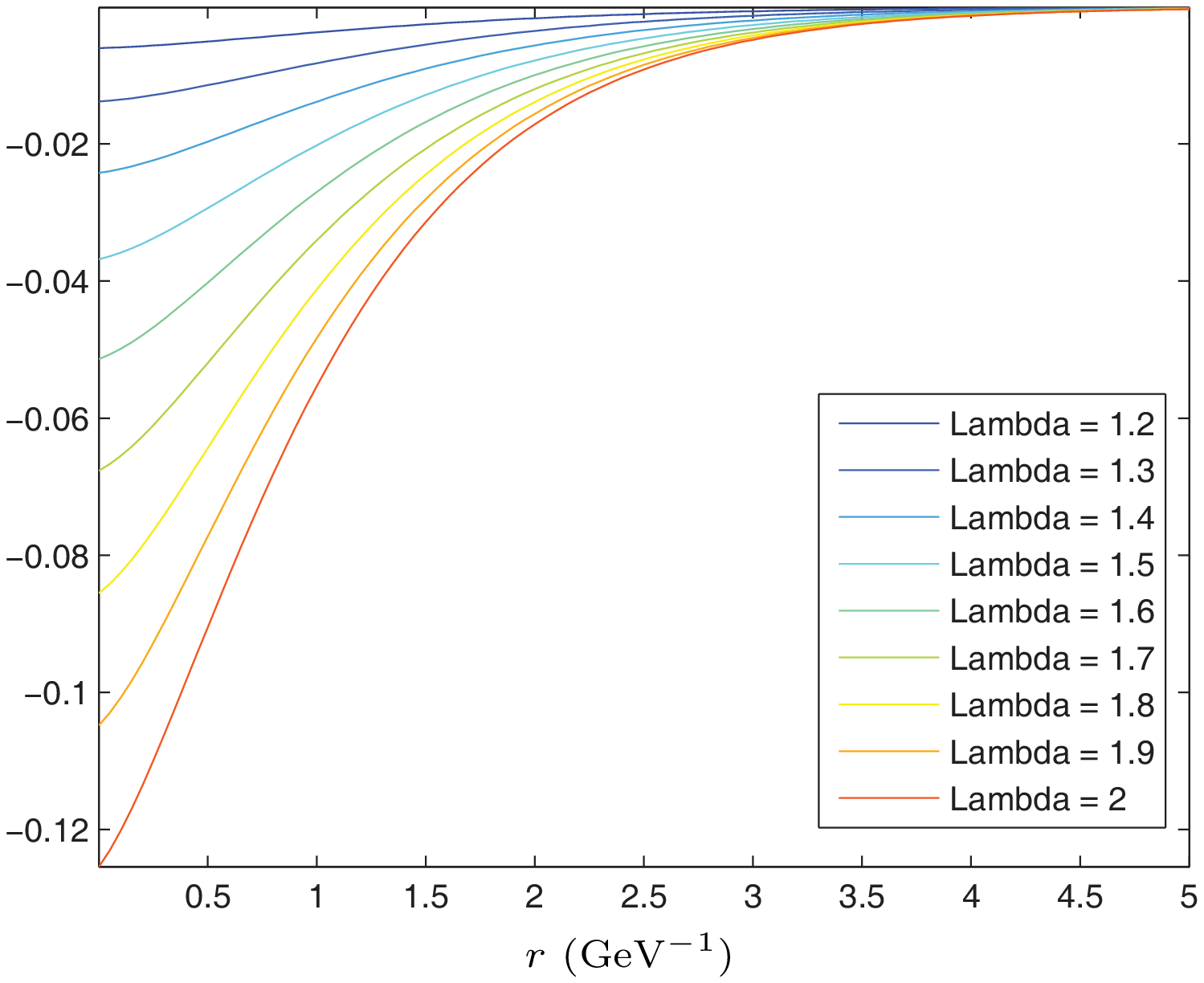}}\\
(a)&(b)\\
\scalebox{0.55}{\includegraphics{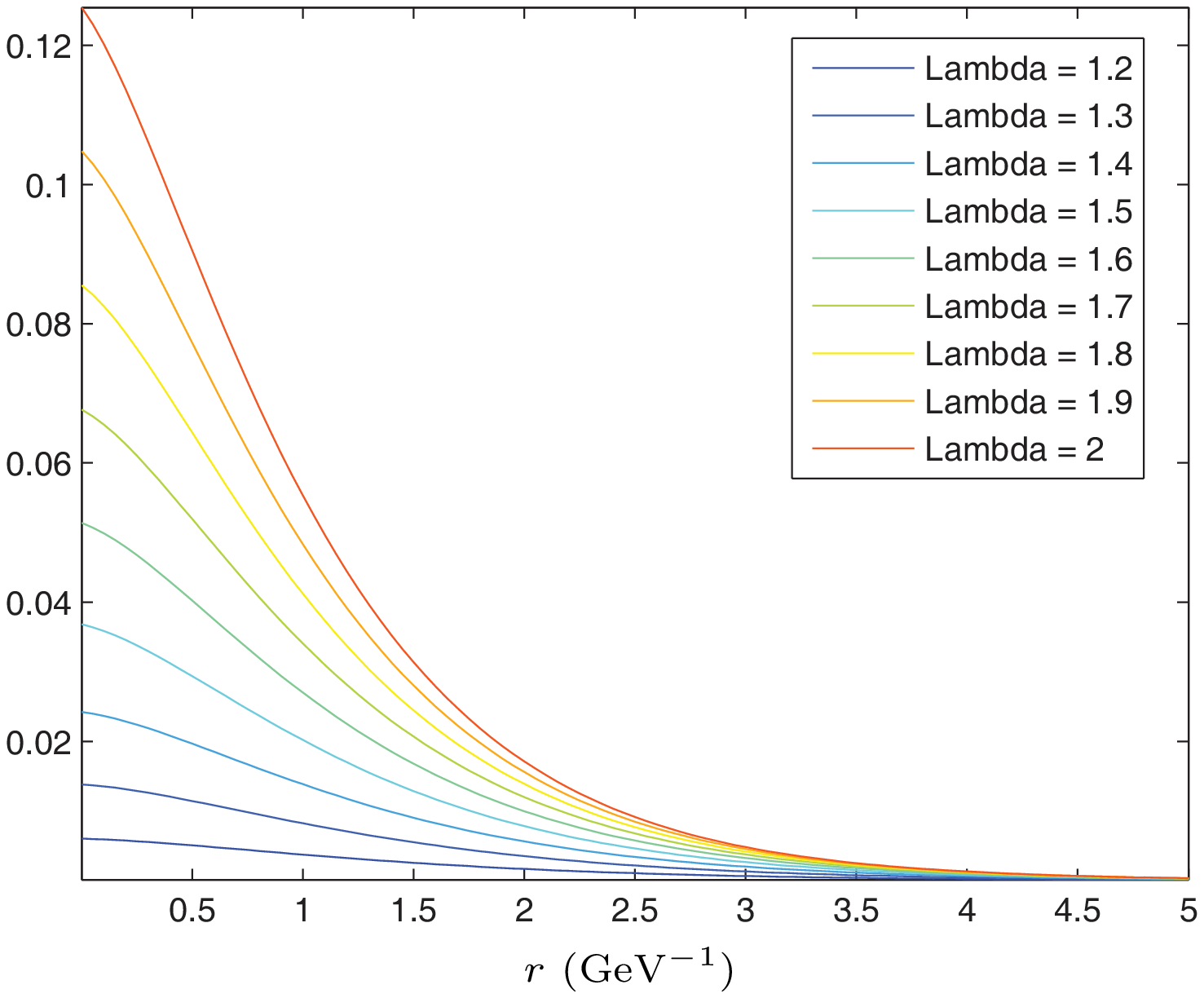}}&\scalebox{0.55}{\includegraphics{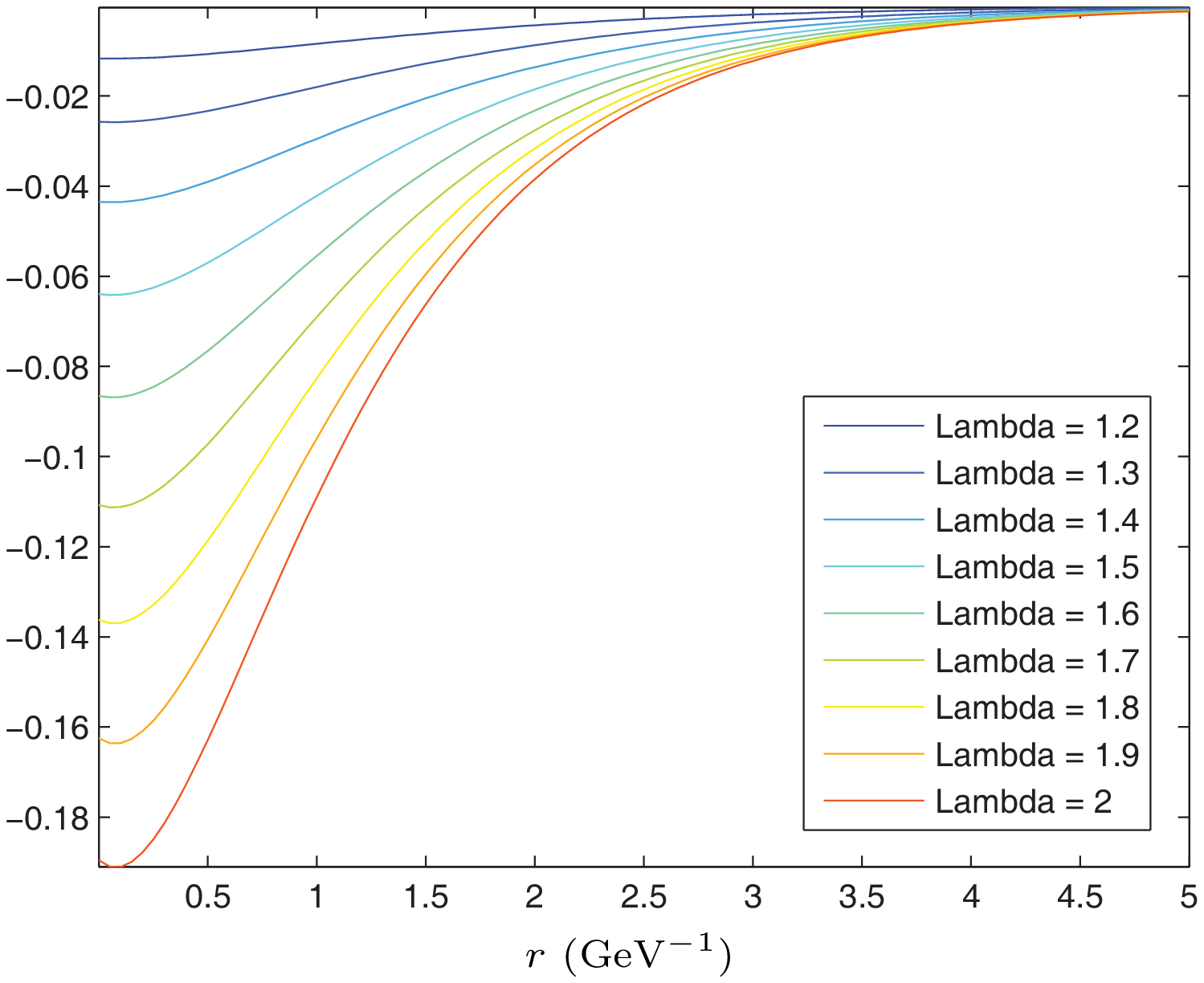}}\\
(c)&(d)
\end{tabular}
\caption{(a), (b), (c) and (d) are the potentials of the
$[PP_1^\prime]_{s1}^{0}$ with $(c=+1,\,\beta\beta^\prime>0)$,
$(c=+1,\,\beta\beta^\prime<0)$, $(c=-1,\,\beta\beta^\prime>0)$ and
$(c=-1,\,\beta\beta^\prime<0)$, respectively. Here, we take
$\zeta\varpi>0$ with $|\beta|=0.909$, $|\beta^\prime|=0.533$,
$|\zeta|=0.727$ and $|\varpi|=0.364$. \label{dd1s3potential}}
\end{figure}
\end{center}

\begin{center}
\begin{figure}[htb]
\begin{tabular}{cccc}
\scalebox{0.55}{\includegraphics{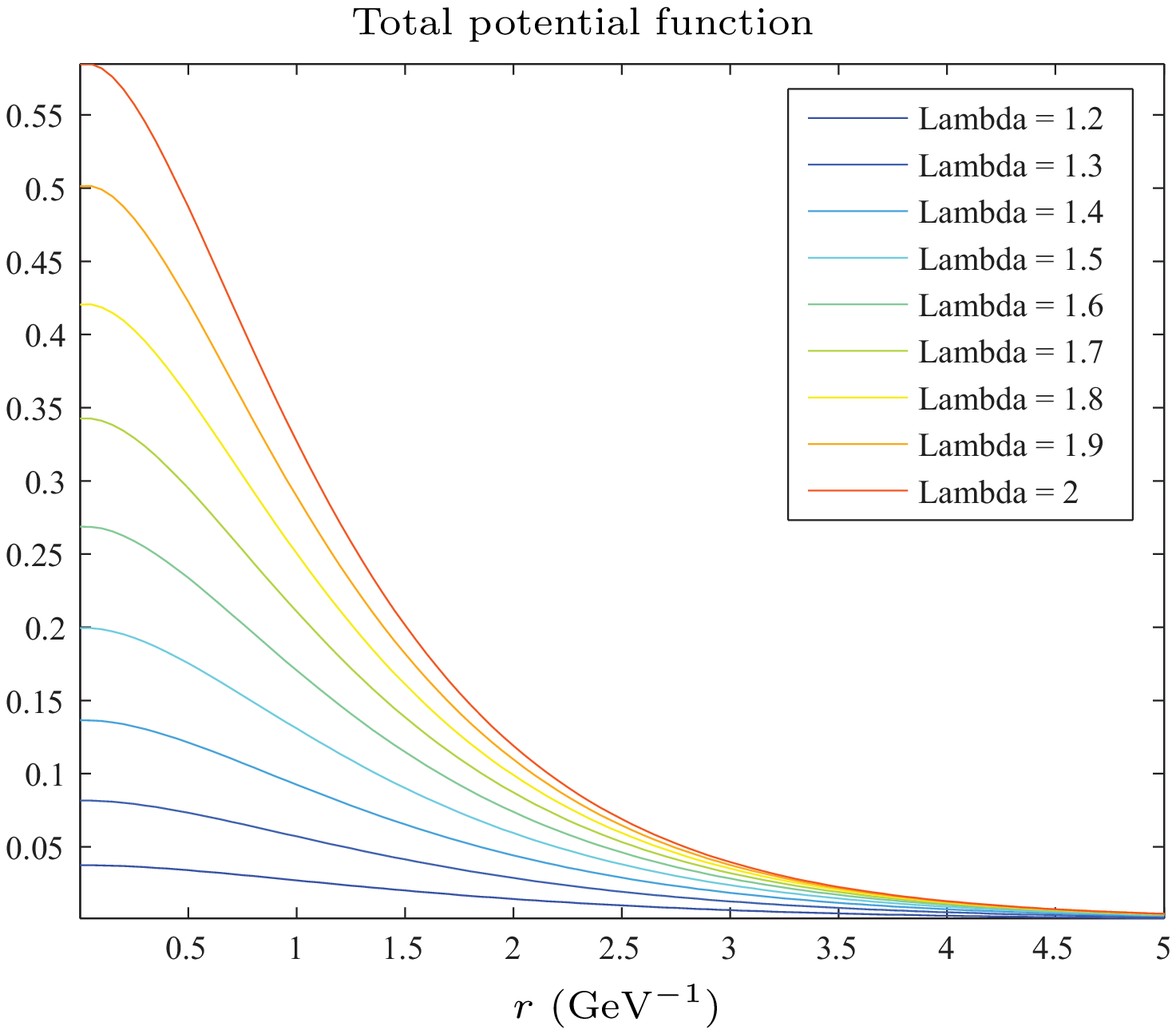}}&\scalebox{0.55}{\includegraphics{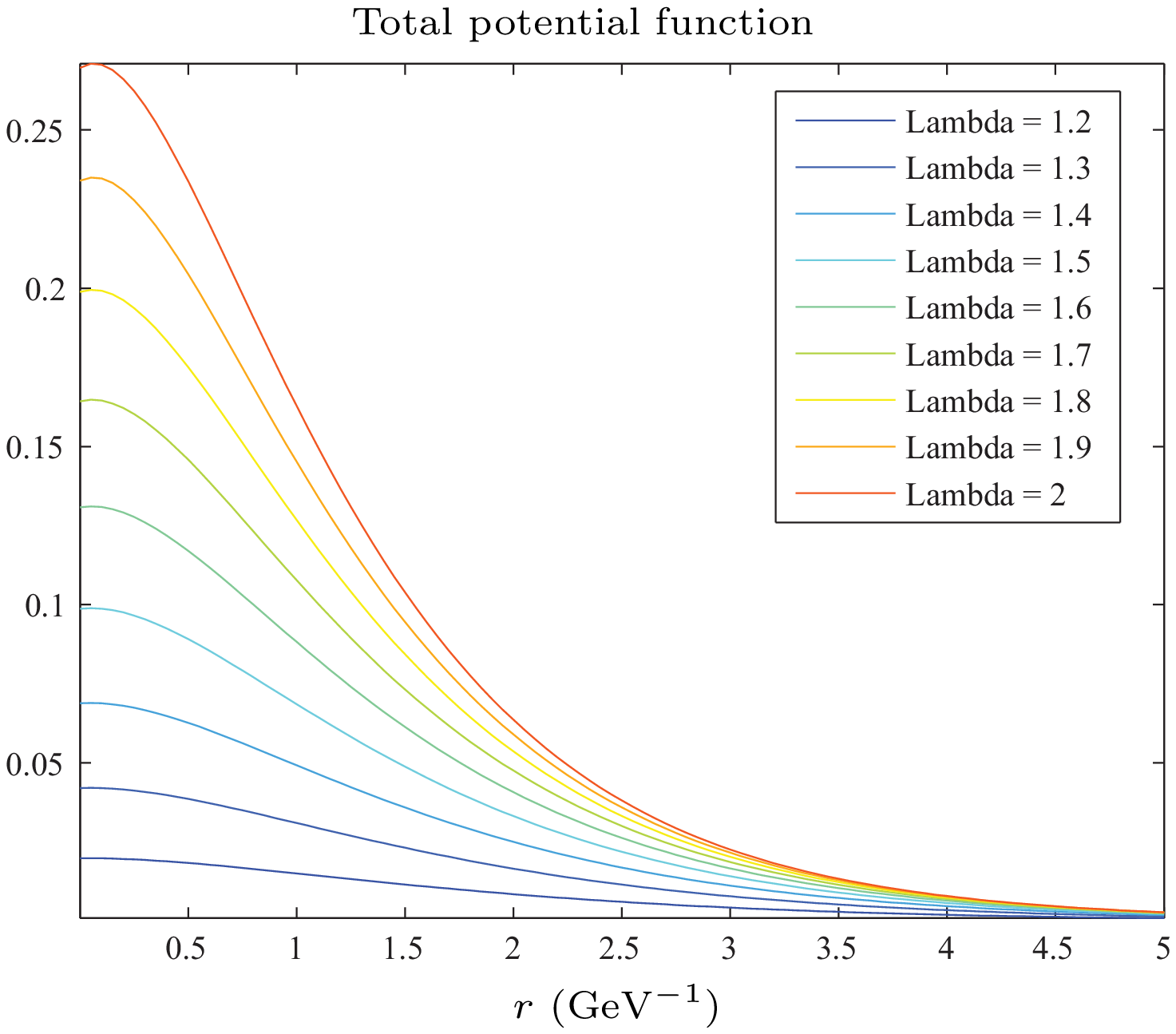}}\\
(a)&(b)\\
\scalebox{0.55}{\includegraphics{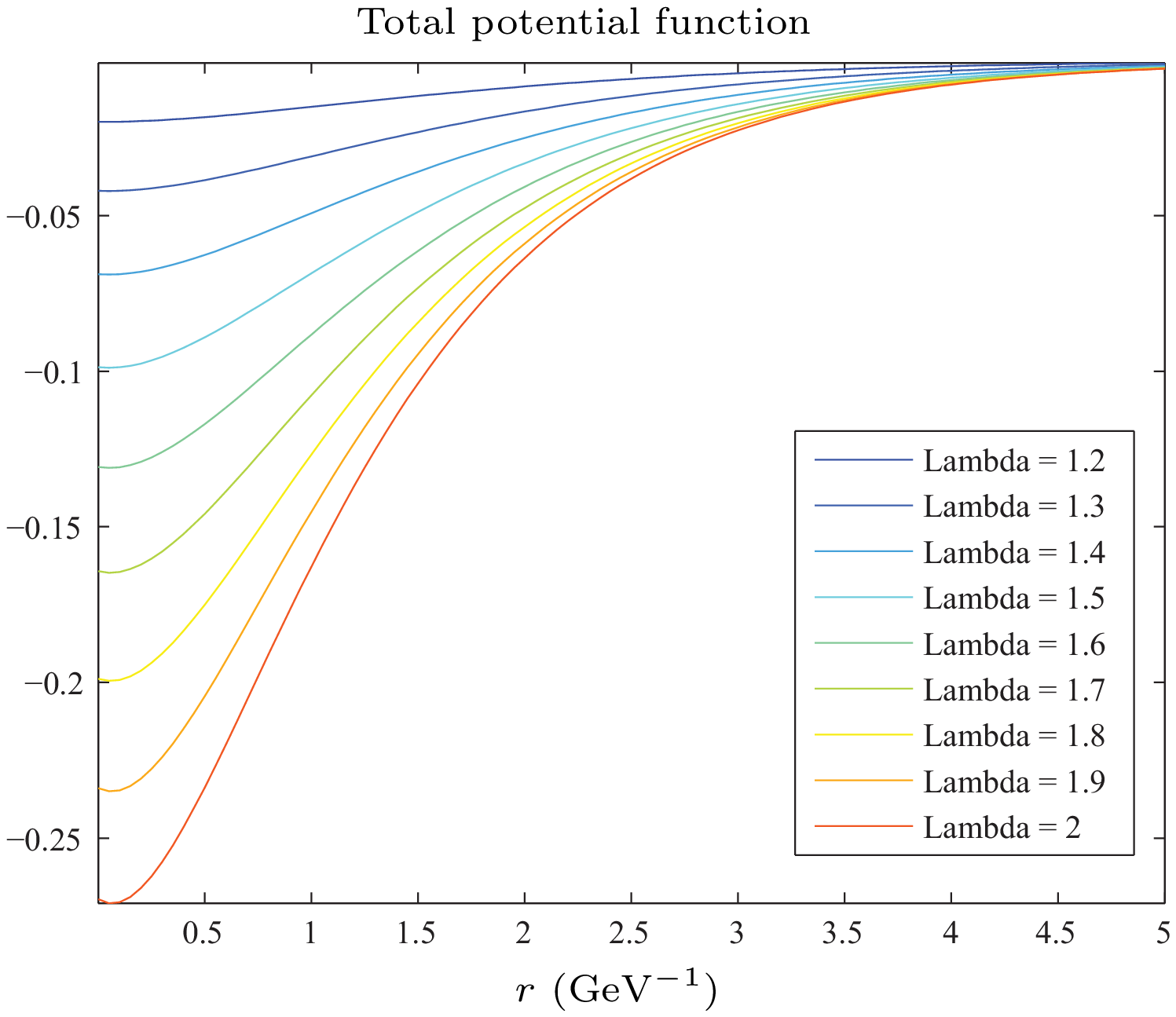}}&\scalebox{0.55}{\includegraphics{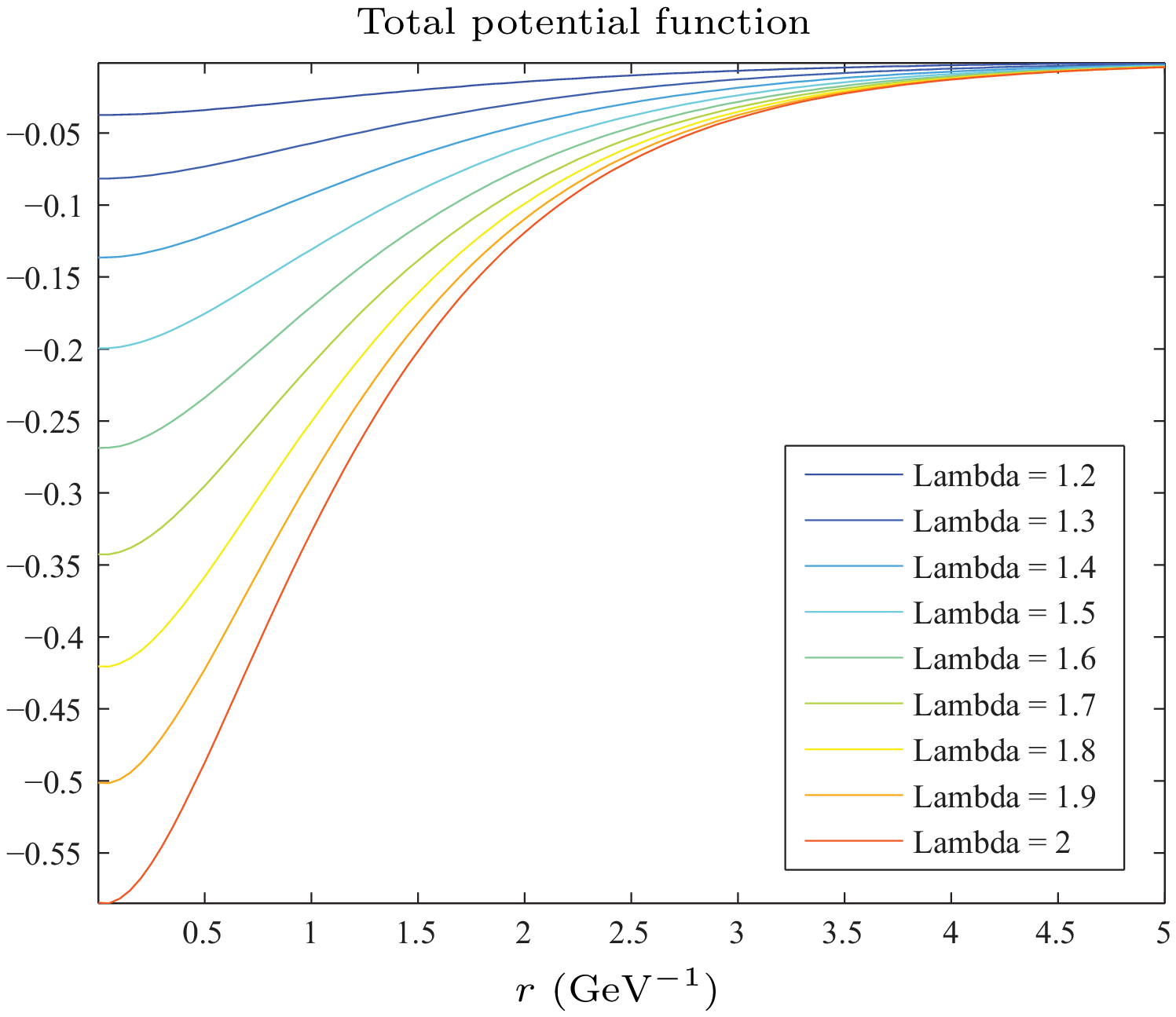}}\\
(c)&(d)
\end{tabular}
\caption{(a), (b), (c) and (d) are the potentials of the
$[PP_1^\prime]_{s1}^{0}$ with $(c=+1,\,\beta\beta^\prime>0)$,
$(c=+1,\,\beta\beta^\prime<0)$, $(c=-1,\,\beta\beta^\prime>0)$ and
$(c=-1,\,\beta\beta^\prime<0)$, respectively. Here, we take
$\zeta\varpi<0$ with $|\beta|=0.909$, $|\beta^\prime|=0.533$,
$|\zeta|=0.727$ and $|\varpi|=0.364$. \label{dd1s3potential-1}}
\end{figure}
\end{center}

Here the effective potential of the $[PP_1^\prime]_{s1}^{0}$
arises from the $\phi$ meson exchange. There exist eight
combinations of the parameters $c$, $\beta\beta^\prime$ and
$\zeta\varpi$. With different parameters, the dependence of the
effective potential of the $[PP_1^\prime]_{s1}^{0}$ system with
$c=\pm 1$ on $r$ is shown in Figs.
\ref{dd1s3potential}-\ref{dd1s3potential-1}.

\begin{center}
\begin{figure}[htb]
\begin{tabular}{cccc}
\scalebox{0.8}{\includegraphics{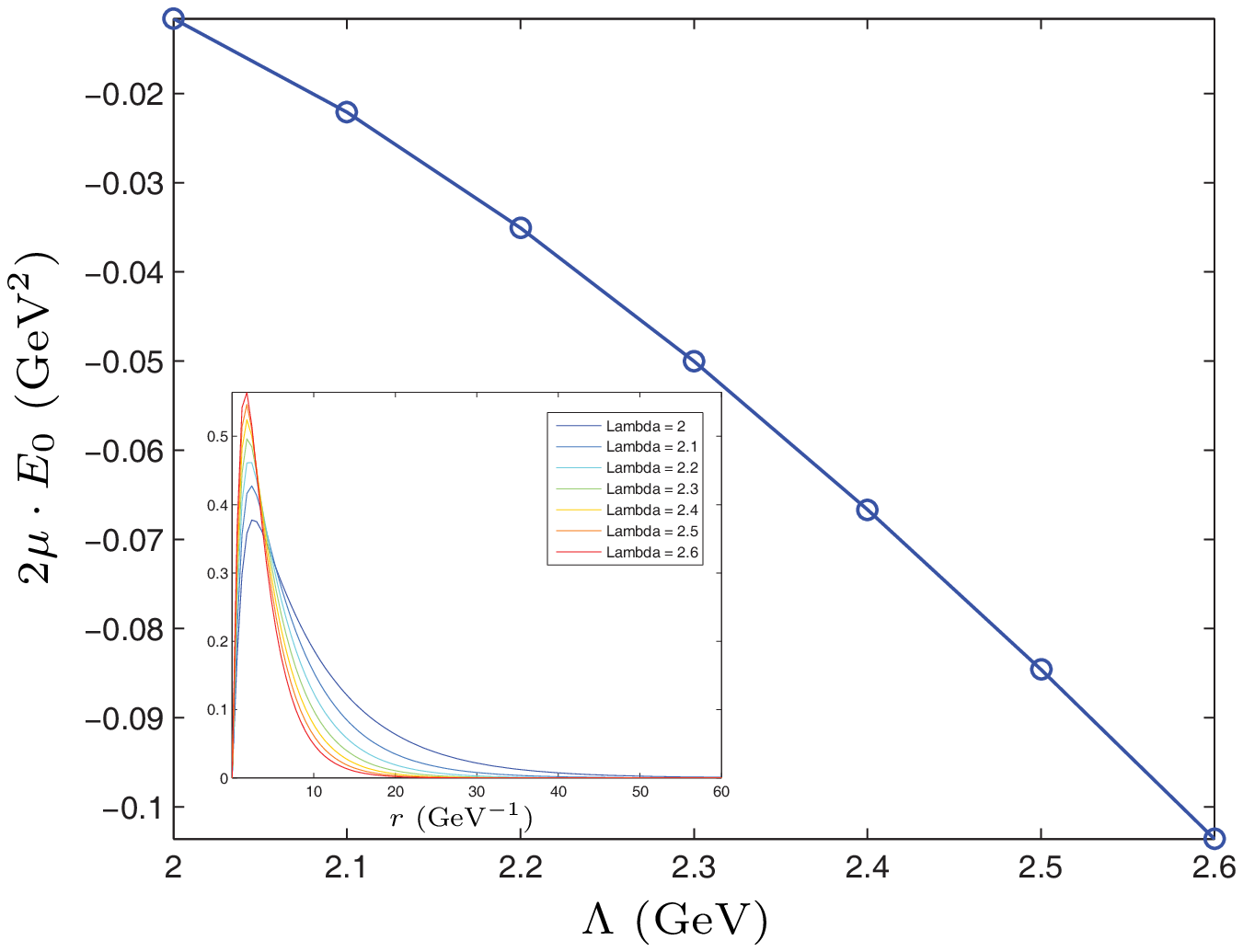}}
\end{tabular}
\caption{The dependence of the binding energy of the
$[PP_1^\prime]_{s1}^{0}$ system on $\Lambda$ with $c=-1$,
$\beta\beta^\prime<0$ and $\zeta\varpi<0$ with $|\beta|=0.909$,
$|\beta^\prime|=0.533$, $|\zeta|=0.727$ and $|\varpi|=0.364$.
\label{dd1s3-E}}
\end{figure}
\end{center}

When solving the Schr\"{o}dinger equation, the cutoff $\Lambda$ is
expected to be larger than the mass of the $\phi$ meson. We only
find bound state solutions of $[PP_1^\prime]_{s1}^{0}$ with
$c=-1$, $\beta\beta^\prime <0$ and $\zeta\varpi <0$, which is
shown in Fig. \ref{dd1s3-E}.

\subsubsection{$[PP_1^\prime]_{8}^0$}

\begin{center}
\begin{figure}[htb]
\begin{tabular}{cccc}
\scalebox{0.55}{\includegraphics{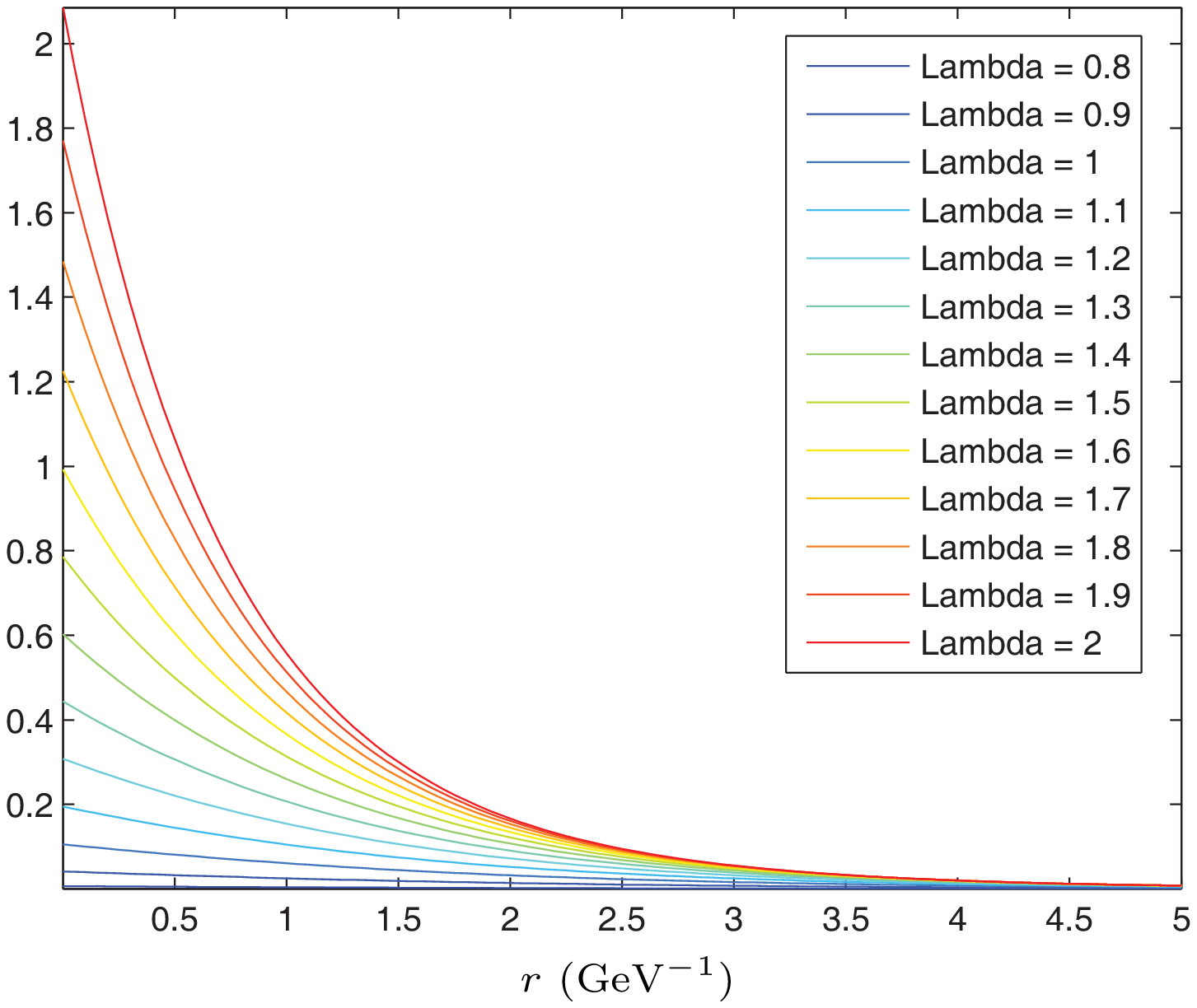}}&\scalebox{0.55}{\includegraphics{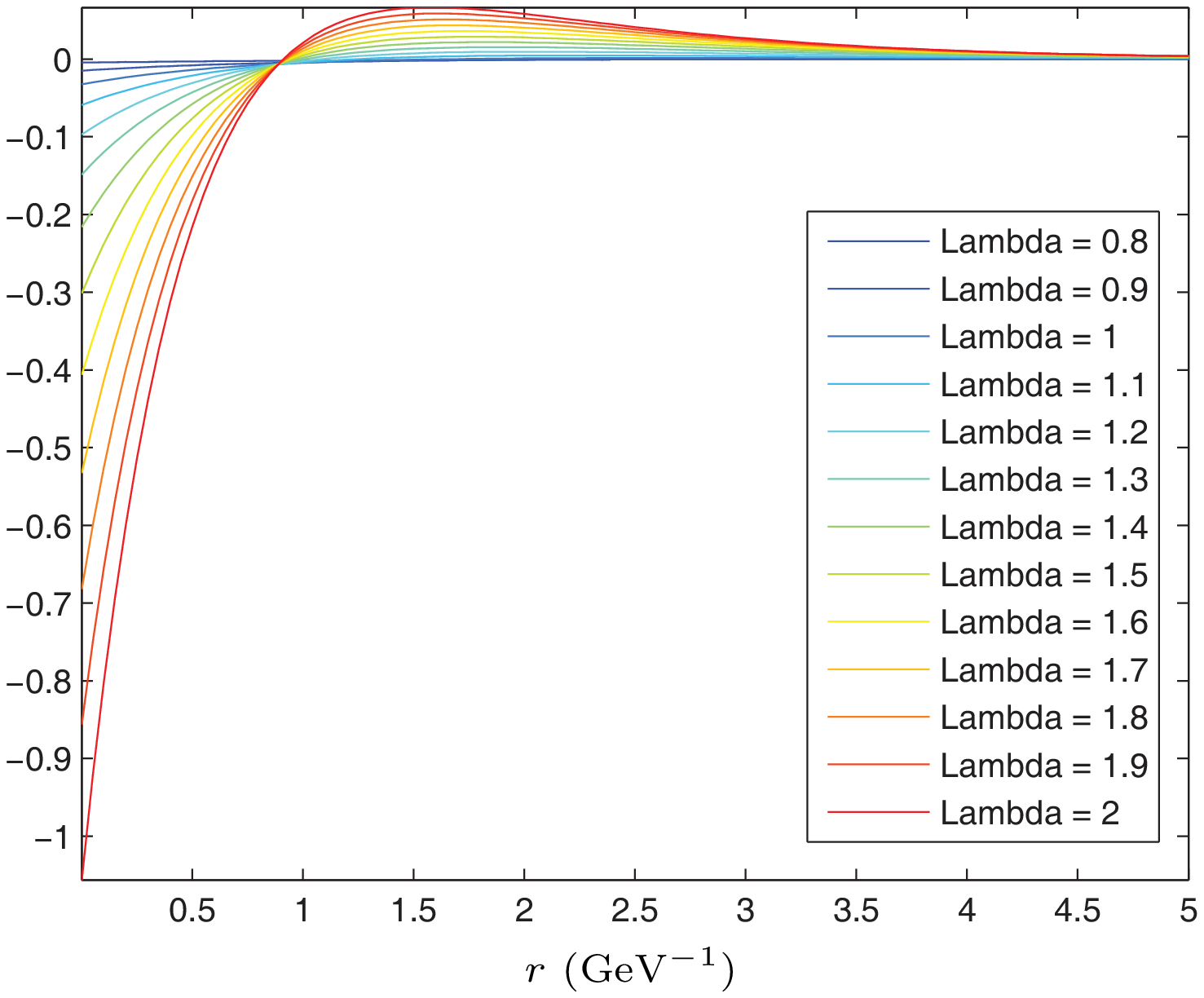}}\\
(a)&(b)\\
\scalebox{0.55}{\includegraphics{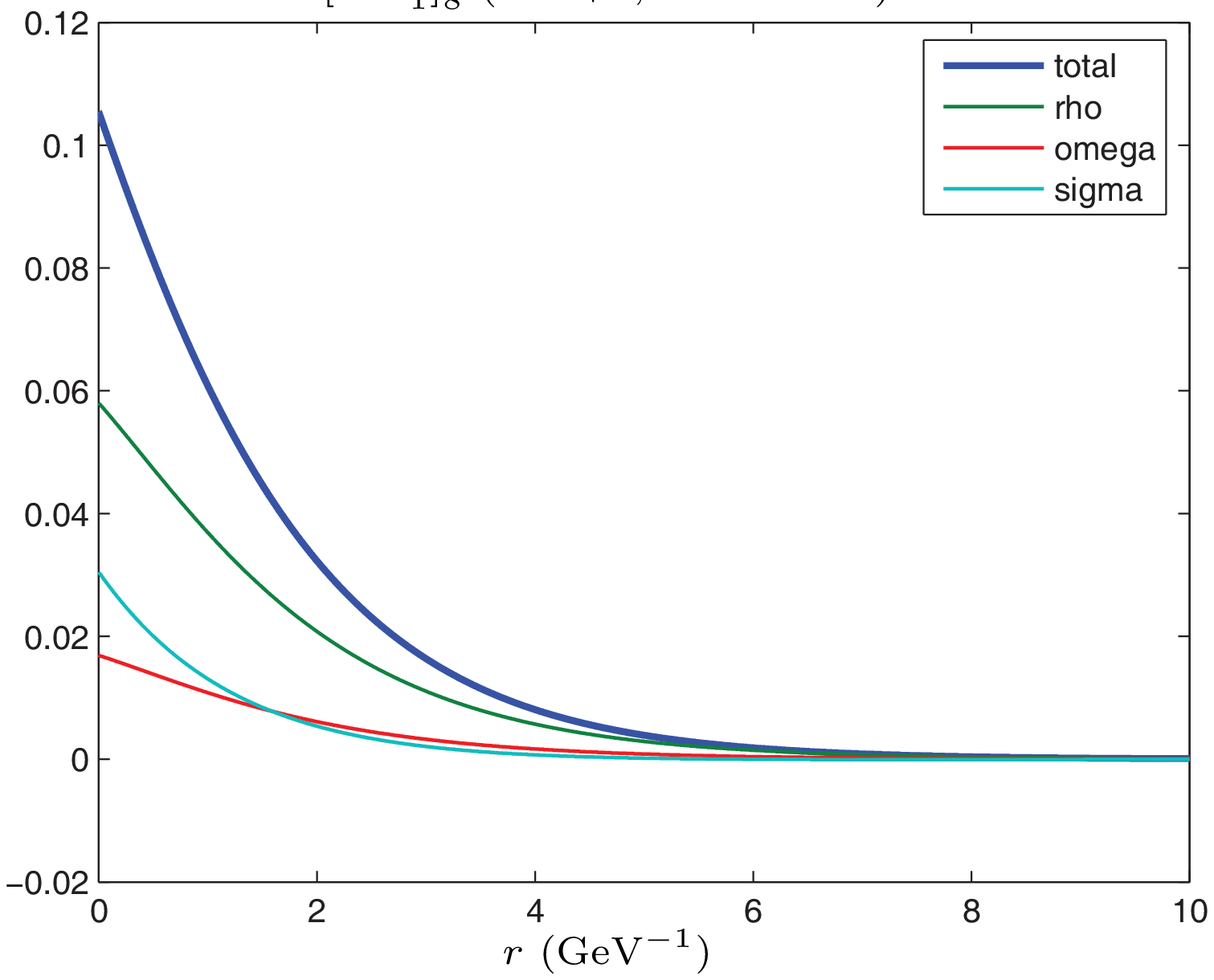}}&\scalebox{0.55}{\includegraphics{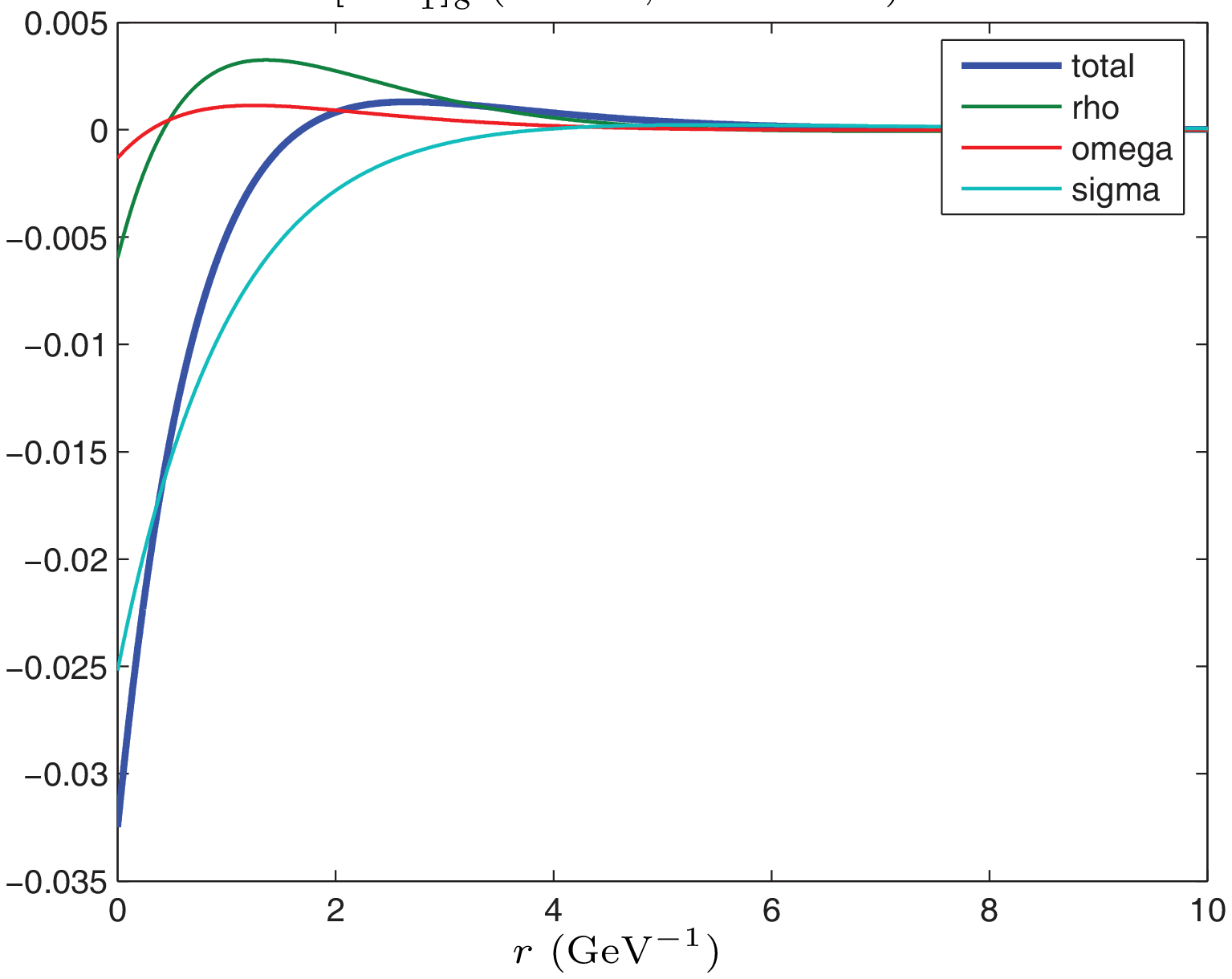}}\\
(c)&(d)
\end{tabular}
\caption{(a) and (b) are the total potentials of the
$[PP_1^\prime]_8^{0}$ system with $c=+1$ and $c=-1$ respectively
when $g_\sigma g_\sigma^\prime>0$, $\beta\beta^\prime>0$ and
$\zeta\mu>0$ with $|\beta|=0.909$, $|\beta^\prime|=0.533$,
$|\zeta|=0.727$ and $|\varpi|=0.364$. With $\Lambda=1$ GeV, we
show the comparison of the total potential and the
$\sigma,\,\rho,\,\omega$ exchange potentials, where (c) and (d)
illustrate the result of the $[PP_1^\prime]_8^{0}$ system with
$c=+1$ and $c=-1$ respectively.\label{dd1s4potential}}
\end{figure}
\end{center}

The $\sigma$, $\rho$, $\omega$ exchange contributes to the total effective potential of $[PP_1^\prime]_{8}^0$ system.
In Fig. \ref{dd1s4potential}, the variation of the total effective potential of $[PP_1^\prime]_{8}^0$ to $r$ and different
typical $\Lambda$ values is given. The comparison of the total potential with the $\sigma$, $\rho$ and $\omega$
exchange potential is presented in Fig. \ref{dd1s4potential} (c) and (d) with $\Lambda=1$ GeV.

\begin{center}
\begin{figure}[htb]
\begin{tabular}{cccc}
\scalebox{0.6}{\includegraphics{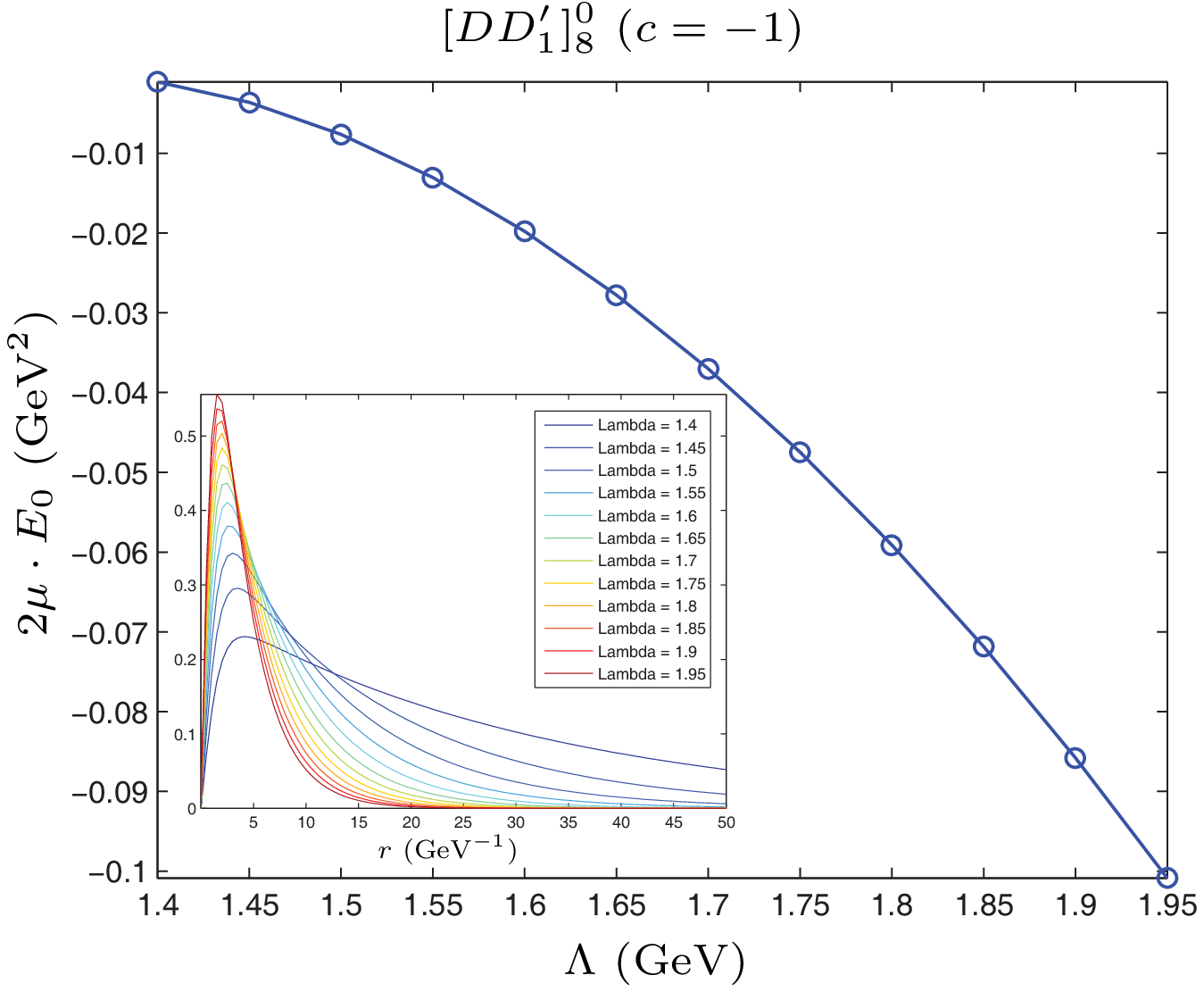}}&\scalebox{0.6}{\includegraphics{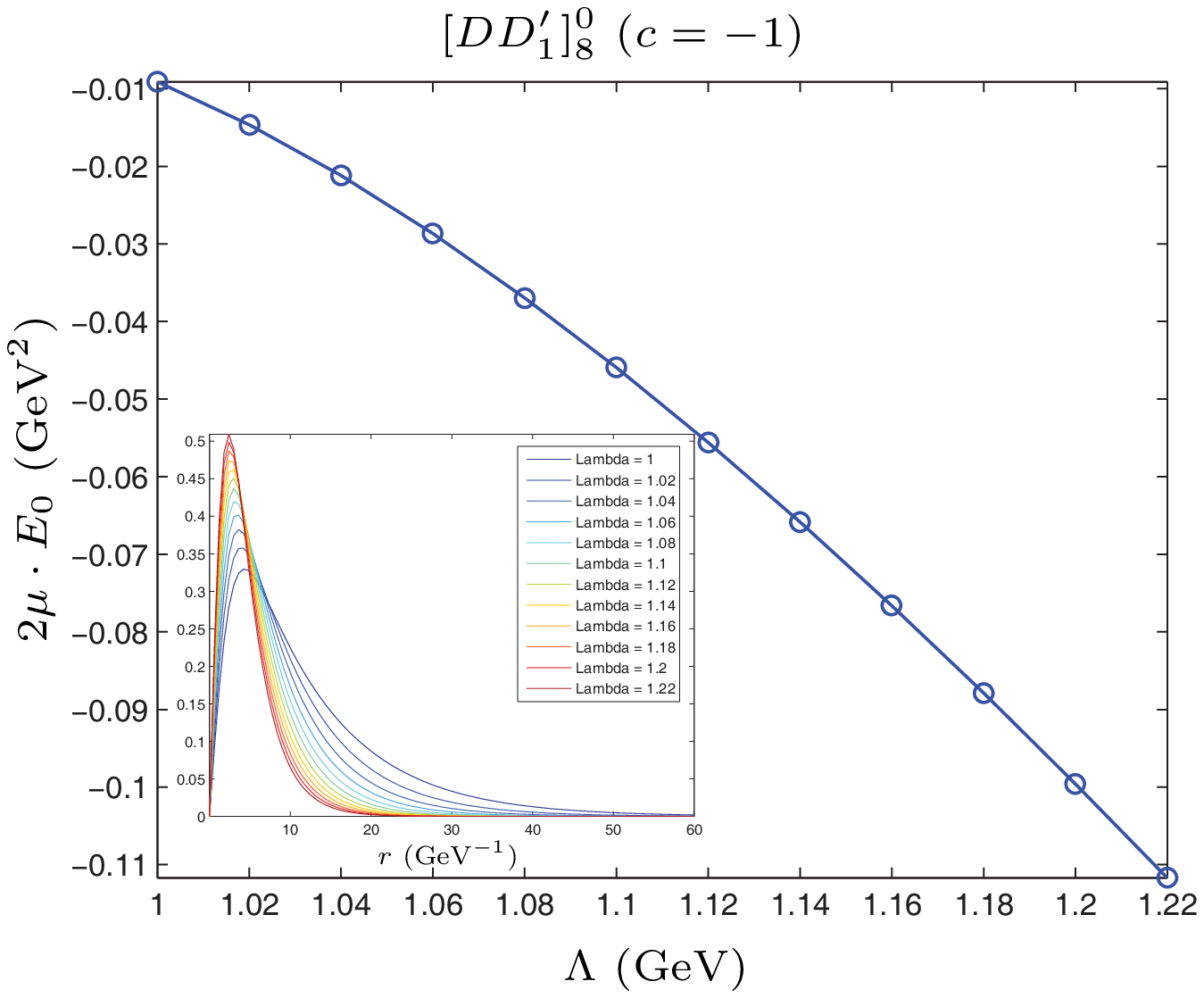}}\\
(a)&(b)\\
\scalebox{0.6}{\includegraphics{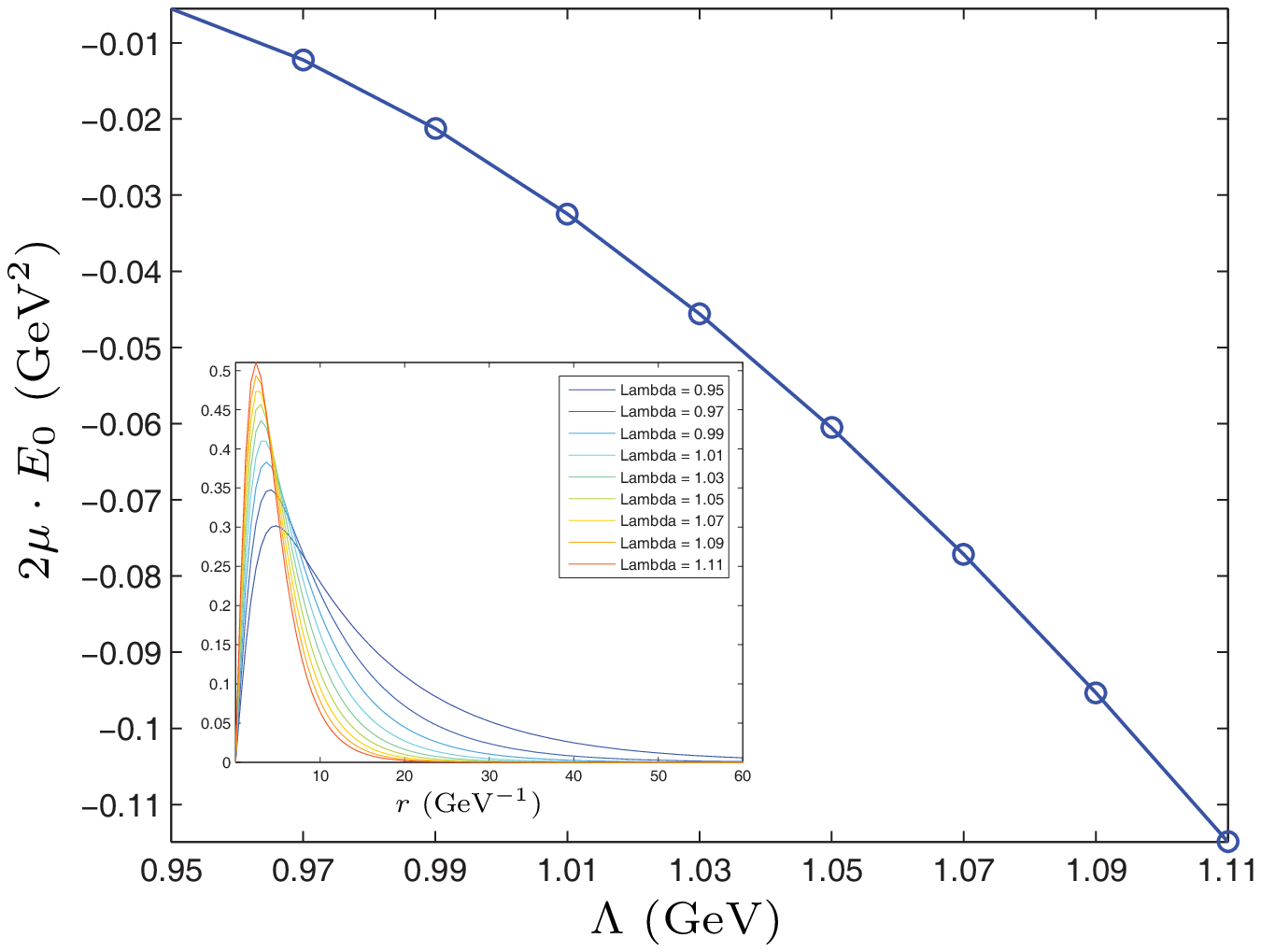}}&\scalebox{0.6}{\includegraphics{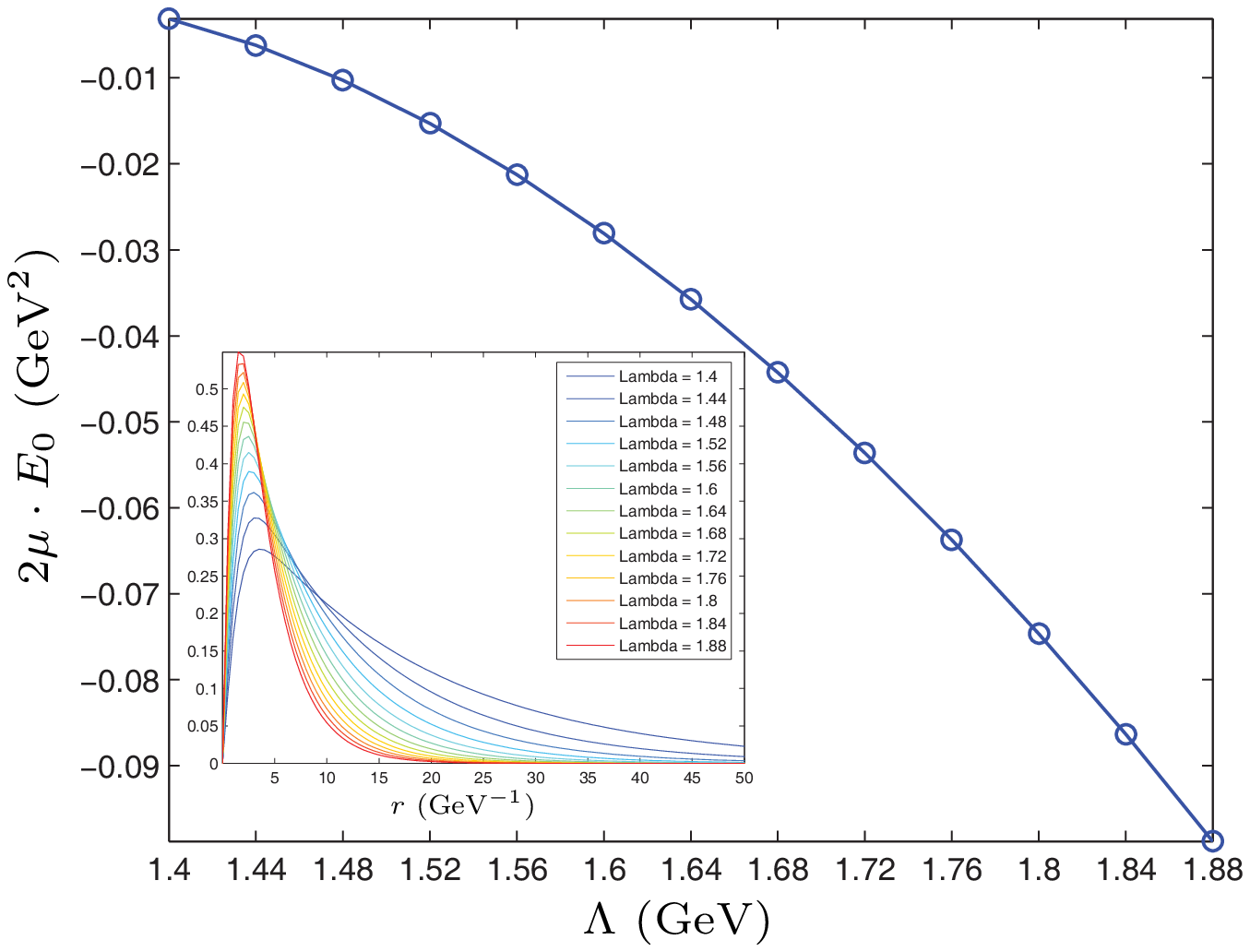}}\\
(c)&(d)\\
\scalebox{0.6}{\includegraphics{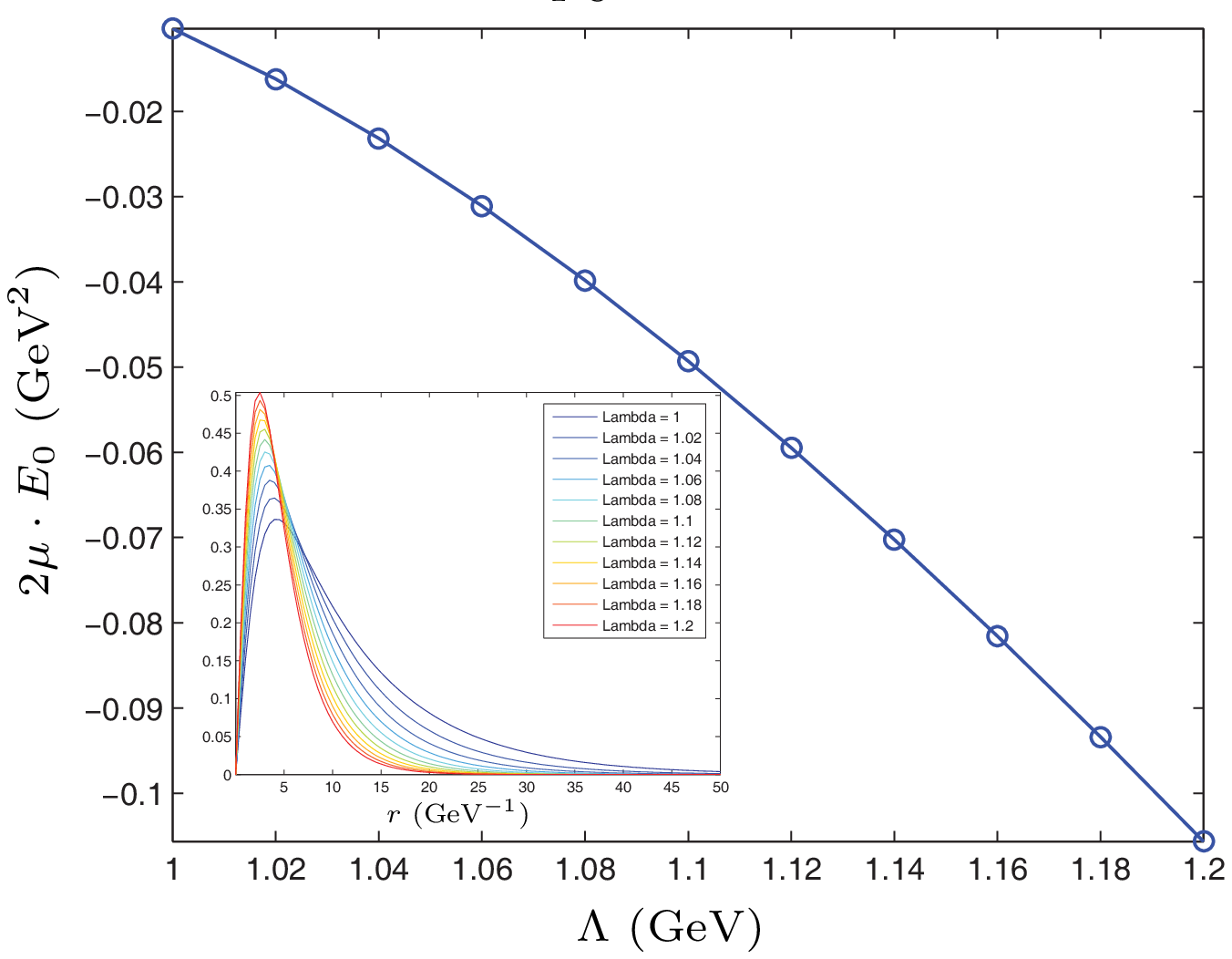}}&\scalebox{0.6}{\includegraphics{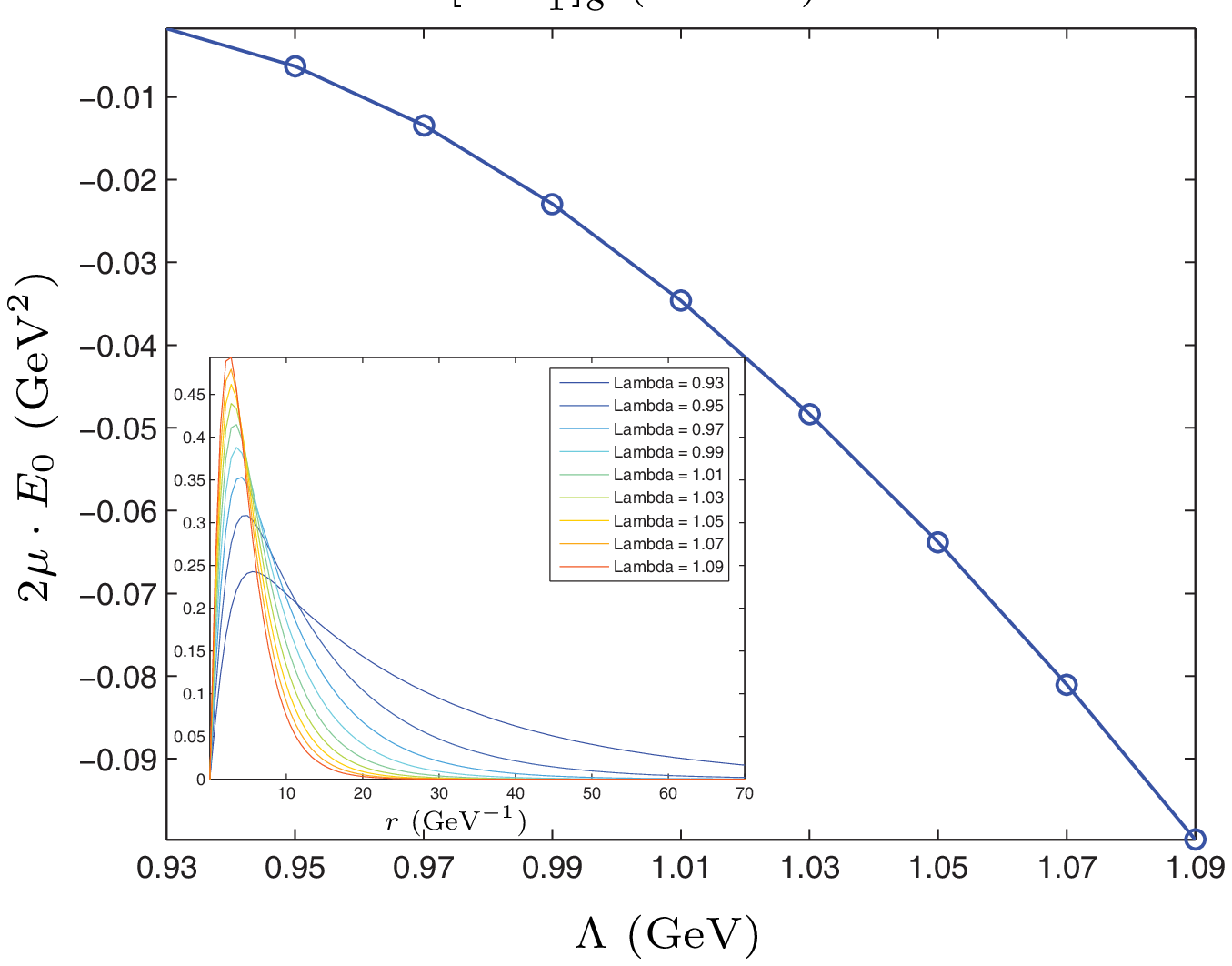}}\\
(e)&(f)\\
\end{tabular}
\caption{The bound state solution of the $[PP_1^\prime]_8^{0}$
with $c=-1$. (a), (b), (c), (d), (e), and (f) correspond to the
results with $(g_\sigma g_\sigma^\prime>0,\,
\beta\beta^\prime<0,\, \zeta \varpi>0)$, $(g_\sigma
g_\sigma^\prime>0,\, \beta\beta^\prime>0,\, \zeta \varpi<0)$,
$(g_\sigma g_\sigma^\prime>0,\, \beta\beta^\prime<0,\, \zeta
\varpi<0)$, $(g_\sigma g_\sigma^\prime<0,\, \beta\beta^\prime<0,\,
\zeta \varpi>0)$, $(g_\sigma g_\sigma^\prime<0,\,
\beta\beta^\prime>0,\, \zeta \varpi<0)$ and $(g_\sigma
g_\sigma^\prime<0,\, \beta\beta^\prime<0,\, \zeta \varpi<0)$
respectively. When taking $(g_\sigma g_\sigma^\prime>0,\,
\beta\beta^\prime>0,\, \zeta \varpi>0)$ or $(g_\sigma
g_\sigma^\prime<0,\, \beta\beta^\prime>0,\, \zeta \varpi>0)$, we
can not find the bound state solution. Here, $|g_\sigma|=0.76$,
$|g_\sigma^\prime|=0.76$, $|h_\sigma|=0.323$, $|\beta|=0.909$ and
$|\beta^\prime|=0.533$, $\zeta=0.727$ and $\varpi=0.364$.
\label{dd1s4-E}}
\end{figure}
\end{center}

Since the total potential is repulsive for the case of $c=+1$, we only explore wether there exists the bound state $[PP_1^\prime]_{8}^0$ system
with $c=-1$. The combinations of coupling constants $g_\sigma g_\sigma^\prime$, $\beta\beta^\prime$ and $\zeta\varpi$
are eight. However, we find the bound state solution of $[PP_1^\prime]_{8}^0$ system
with $c=-1$ with six parameter combinations just listed in the caption of Fig. \ref{dd1s4-E}.

\subsection{The $[P^*P_0^*]$ system}

\subsubsection{$[P^*P_0^*]^{\pm,0}$}

\begin{center}
\begin{figure}[htb]
\begin{tabular}{cccc}
\scalebox{0.55}{\includegraphics{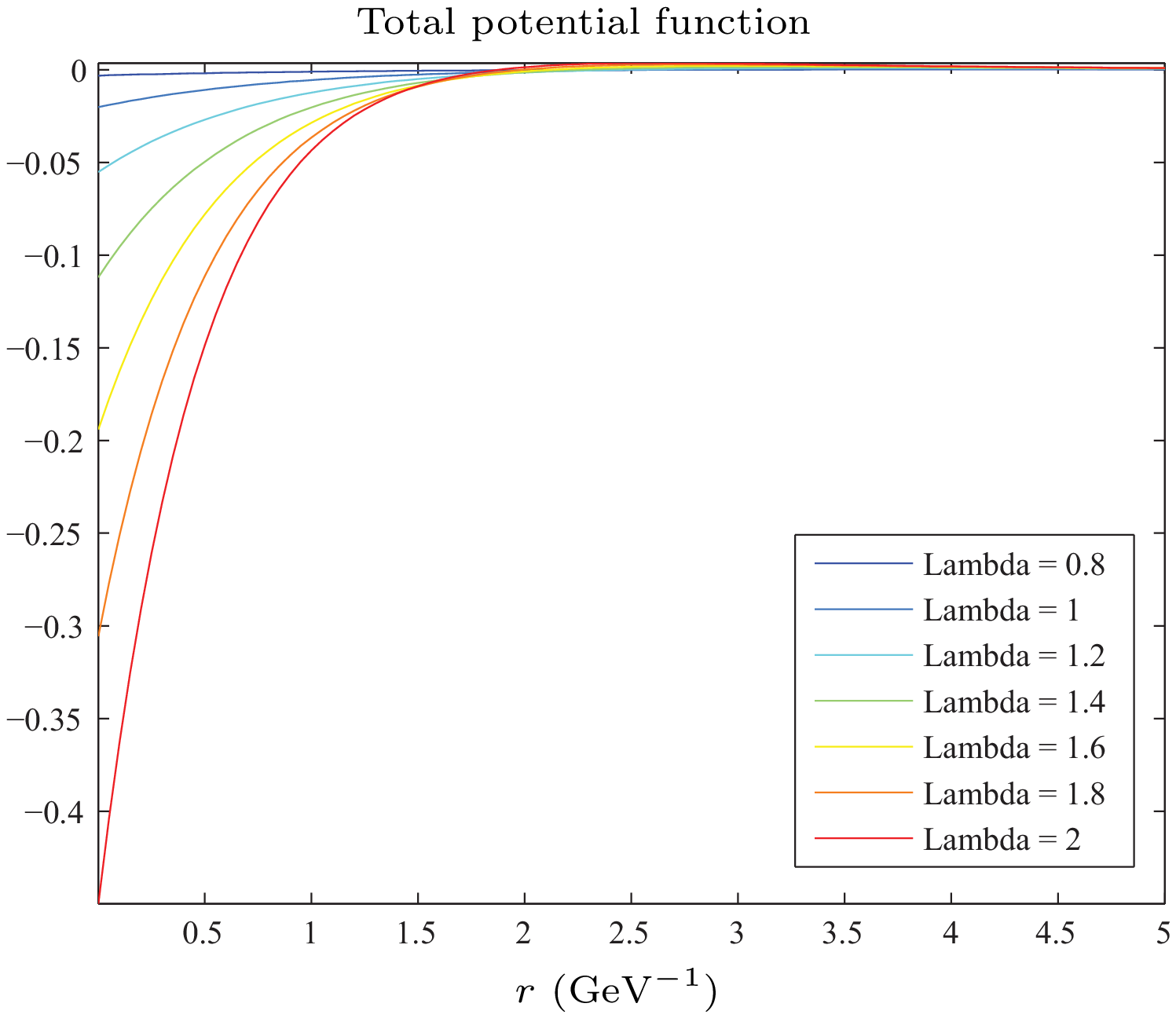}}&\scalebox{0.55}{\includegraphics{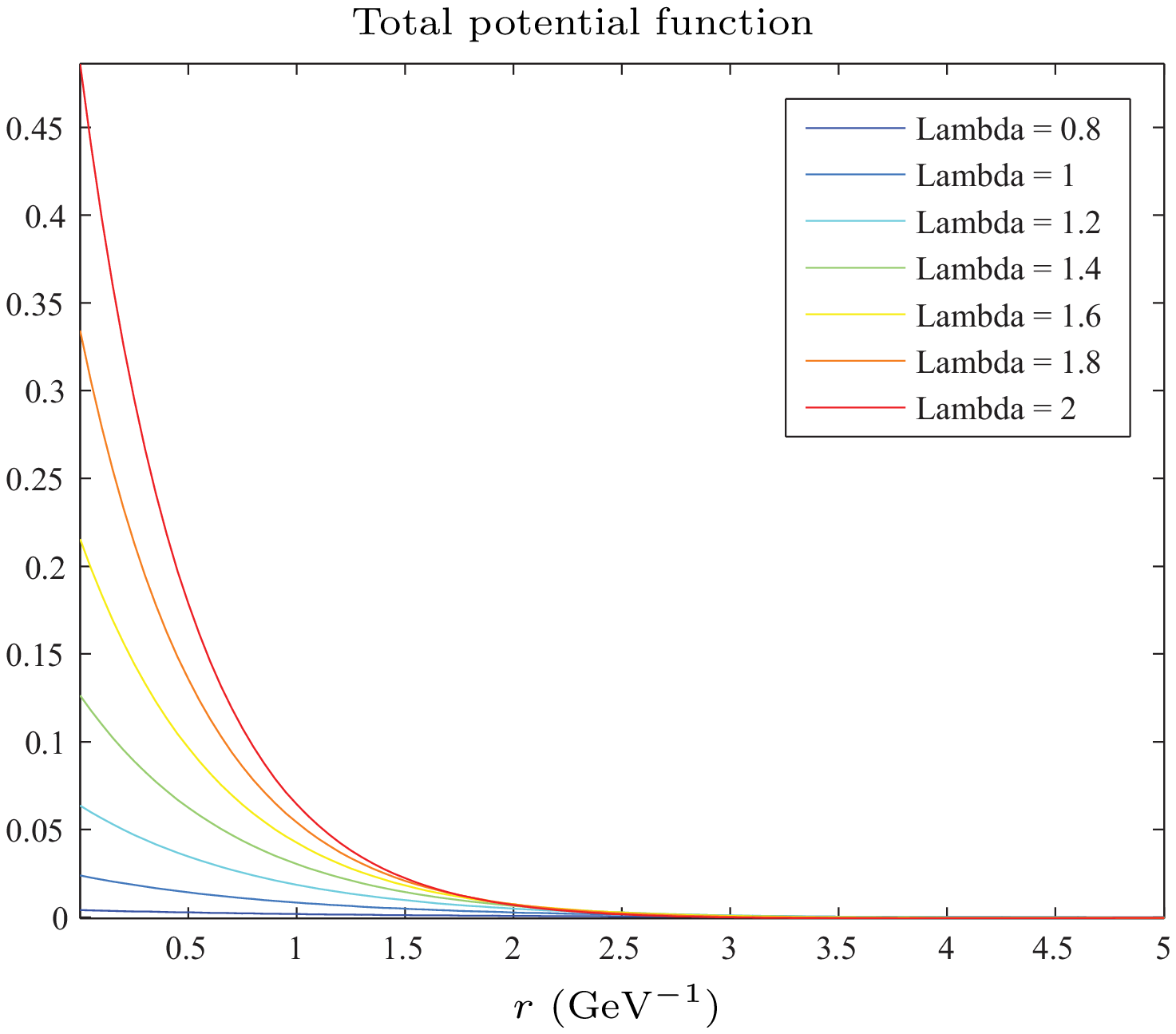}}\\
(a)&(b)\\
\scalebox{0.55}{\includegraphics{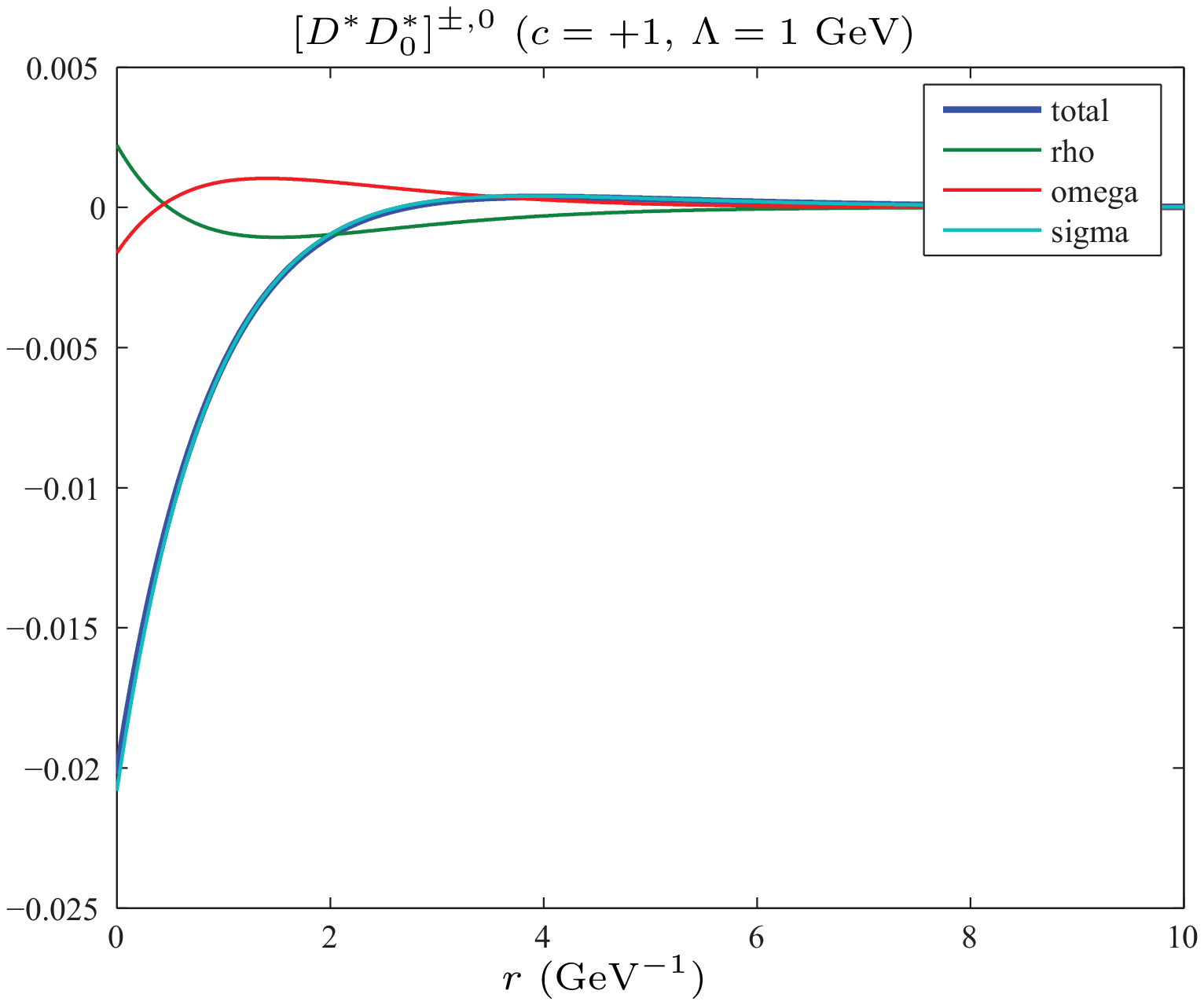}}&\scalebox{0.55}{\includegraphics{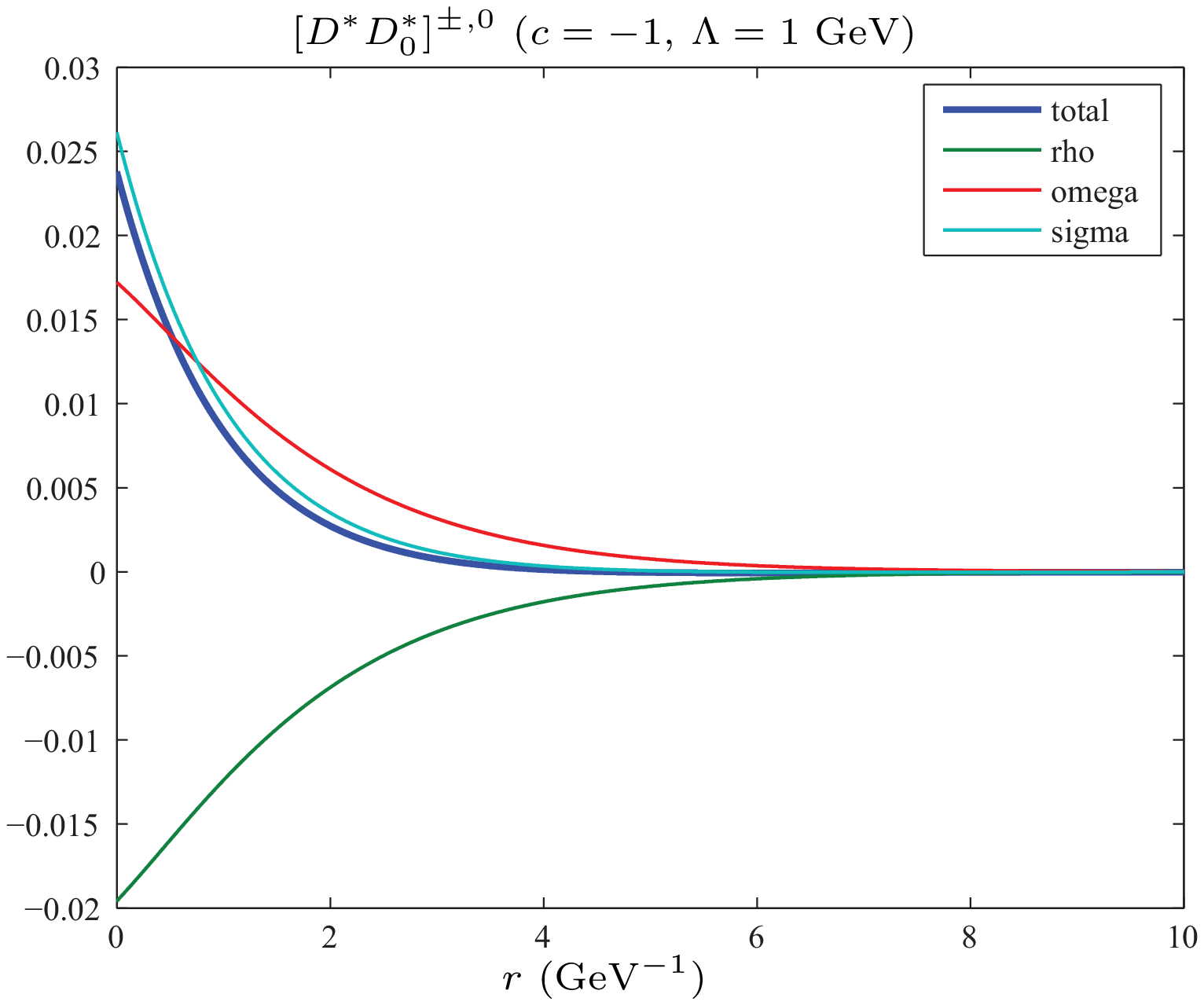}}\\
(c)&(d)
\end{tabular}
\caption{(a) The variation of the total potential of the
$[D^*D_0^*]^{\pm,0}$ system $(c=+1)$ with $r$ and $\Lambda$; (b)
the variation of the total potential of the $[D^*D_0^*]^{\pm,0}$
system $(c=-1)$ with $r$ and $\Lambda$; (c) the exchange
potentials of the $\sigma,\,\rho,\,\omega$ mesons of the
$[DD_1^\prime]^{\pm,0}$ system $(c=+1)$ with $\Lambda=1$ GeV; (d)
the exchange potentials of the $\sigma,\,\rho,\,\omega$ mesons of
the $[D^*D_0^*]^{\pm,0}$ system $(c=-1)$ with $\Lambda=1$ GeV. The
above potentials are obtained with $g_\sigma g_\sigma^\prime>0$,
$\beta\beta^\prime>0$ and $\zeta \varpi>0$. Here,
$|g_\sigma|=0.76$, $|g_\sigma^\prime|=0.76$, $|h_\sigma|=0.32$,
$|\beta|=0.909$ and $|\beta^\prime|=0.533$, $\zeta=0.727$ and
$\varpi=0.364$.\label{dsds0s1potential}}
\end{figure}
\end{center}

The total effective potential of the $[P^*P_0^*]^{\pm,0}$ system
arises from the $\sigma$, $\rho$ and $\omega$ meson exchange.
Since the $\rho$ and $\omega$ exchange potentials cancel each
other, the $\sigma$ exchange potential is dominant in the total
effective potential of $[P^*P_0^*]^{\pm,0}$. We only consider the
$\omega$ exchange potential contribution to explore whether there
exists a $[P^*P_0^*]^{\pm,0}$ bound state . In Fig.
\ref{dsds0s1potential}, the effective potential of the
$[P^*P_0^*]^{\pm,0}$ system is presented.

The $\sigma$ exchange potential of $[PP_1^\prime]^{\pm,0}$ system
with $c=-1$ is repulsive. Thus there does not exist a
$[PP_1^\prime]^{\pm,0}$ bound state with $c=-1$.

For the $[P^*P_0^*]^{\pm,0}$ system with $c=+1$, the $\sigma$
exchange potential is attractive. Meanwhile, the cross diagram
plays the dominant role to the $\sigma$ exchange potential.
However, in the $g_\sigma g_\sigma^\prime>0$ and $g_\sigma
g_\sigma^\prime<0$ two cases, we can not find bound state
solutions of the $[P^*P_0^*]^{\pm,0}$ system with $c=-1$ and the
coupling constants listed in the caption of Fig.
\ref{dsds0s1potential} when scanning the range of the cutoff
$\Lambda$.

\begin{center}
\begin{figure}[htb]
\begin{tabular}{cccc}
\scalebox{0.8}{\includegraphics{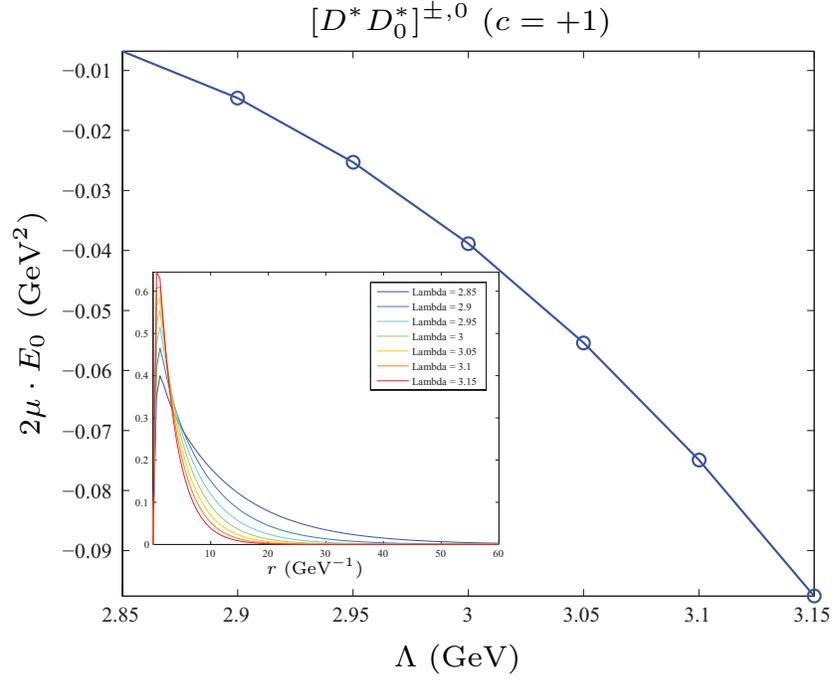}}\\
\end{tabular}
\caption{The binding energy of the $[D^*D_0^*]^{\pm,0}$ system
$(c=+1)$ and its wave function with $2h_\sigma$. Here, we only
consider the $\sigma$ exchange potential and take $g_\sigma
g_\sigma^\prime<0$ with $|h_\sigma|=0.323$, $|g_\sigma|=0.76$ and
$|g_\sigma^\prime|=0.76$.\label{dsds0s1-E}}
\end{figure}
\end{center}

When taking $2h_\sigma$, we obtain the binding energy of the
$[P^*P_0^*]^{\pm,0}$ system with $c=+1$ dependent on the cutoff
$\Lambda$ in Fig. \ref{dsds0s1-E} only in the $g_\sigma g_\sigma^\prime<0$ case.
However, the corresponding cutoff $\Lambda$ is far larger than
usual $1$ GeV. Thus, the $[P^*P_0^*]^{\pm,0}$ system with $c=+1$
seems impossible to form the bound state.

\subsubsection{$[P^*P_0^*]_s^{\pm,0},\,[P^*P_0^*]_{\hat{s}}^{0}$}

There does not exist the
$[P^*P_0^*]_s^{\pm,0},\,[P^*P_0^*]_{\hat{s}}^{0}$ bound state
since no suitable meson exchange force is allowed as indicated in
Table \ref{potential}.

\subsubsection{$[P^*P_0^*]_{s1}^{0}$}

\begin{center}
\begin{figure}[htb]
\begin{tabular}{cccc}
\scalebox{0.55}{\includegraphics{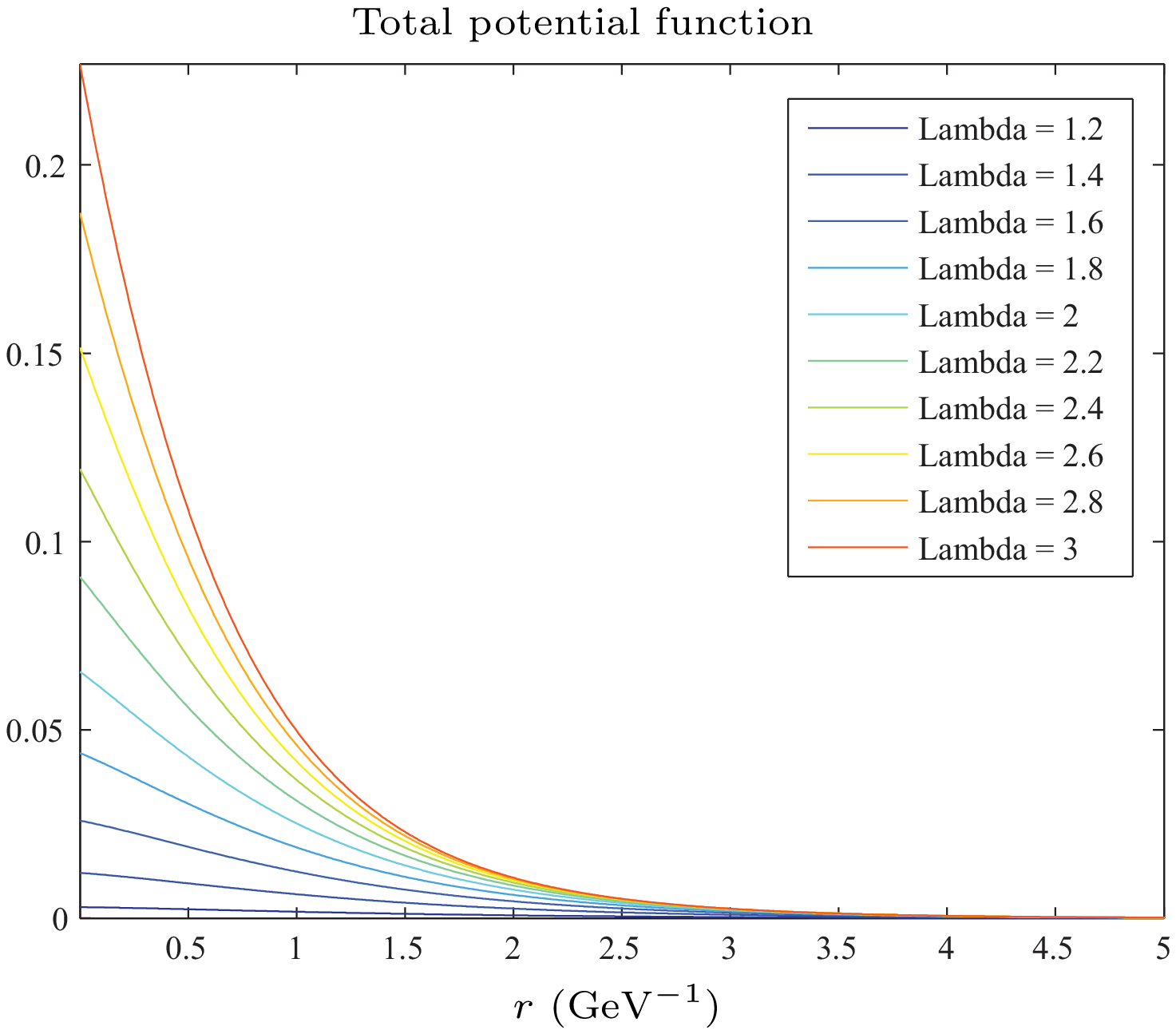}}&\scalebox{0.55}{\includegraphics{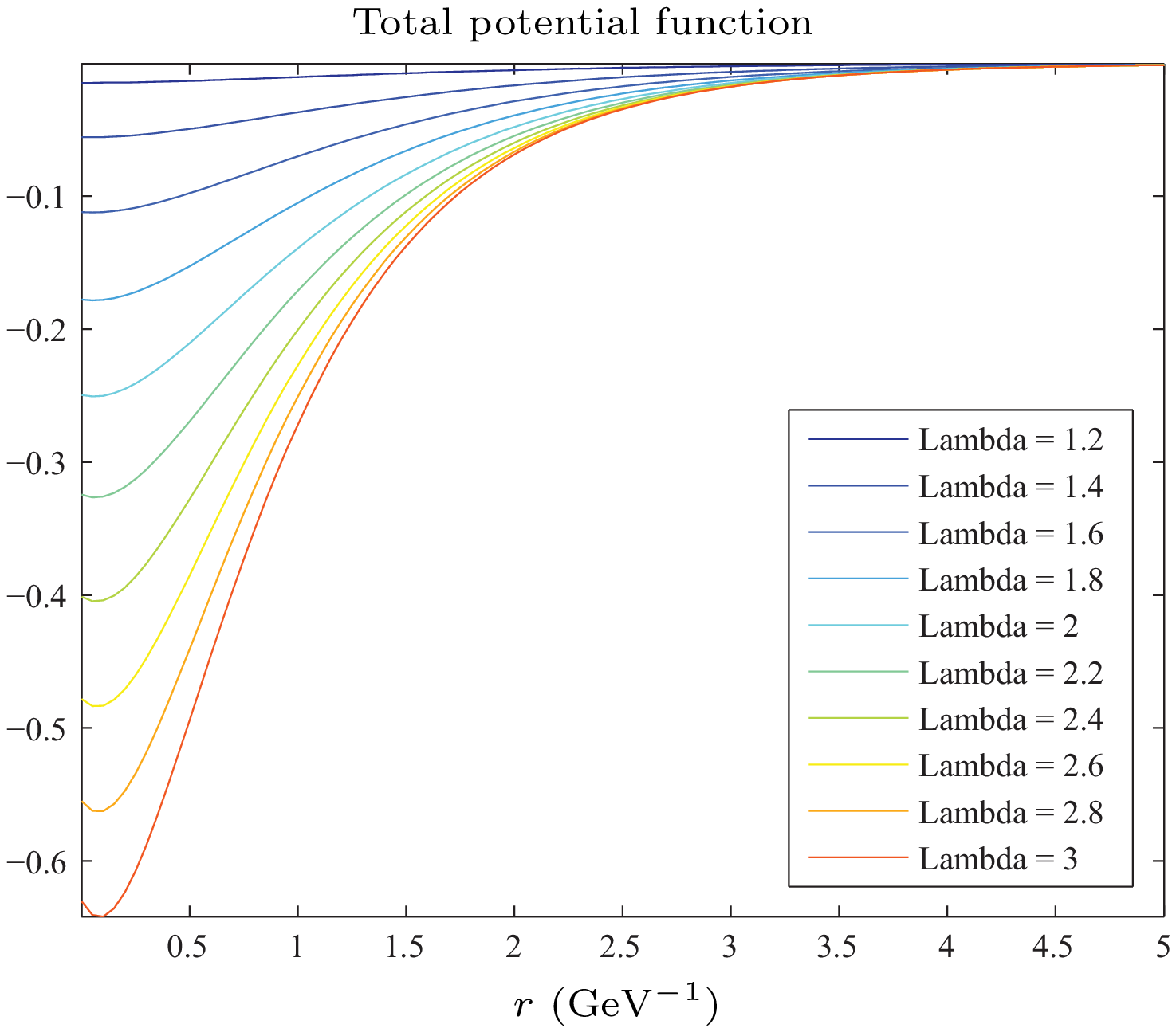}}\\
(a)&(b)\\
\scalebox{0.55}{\includegraphics{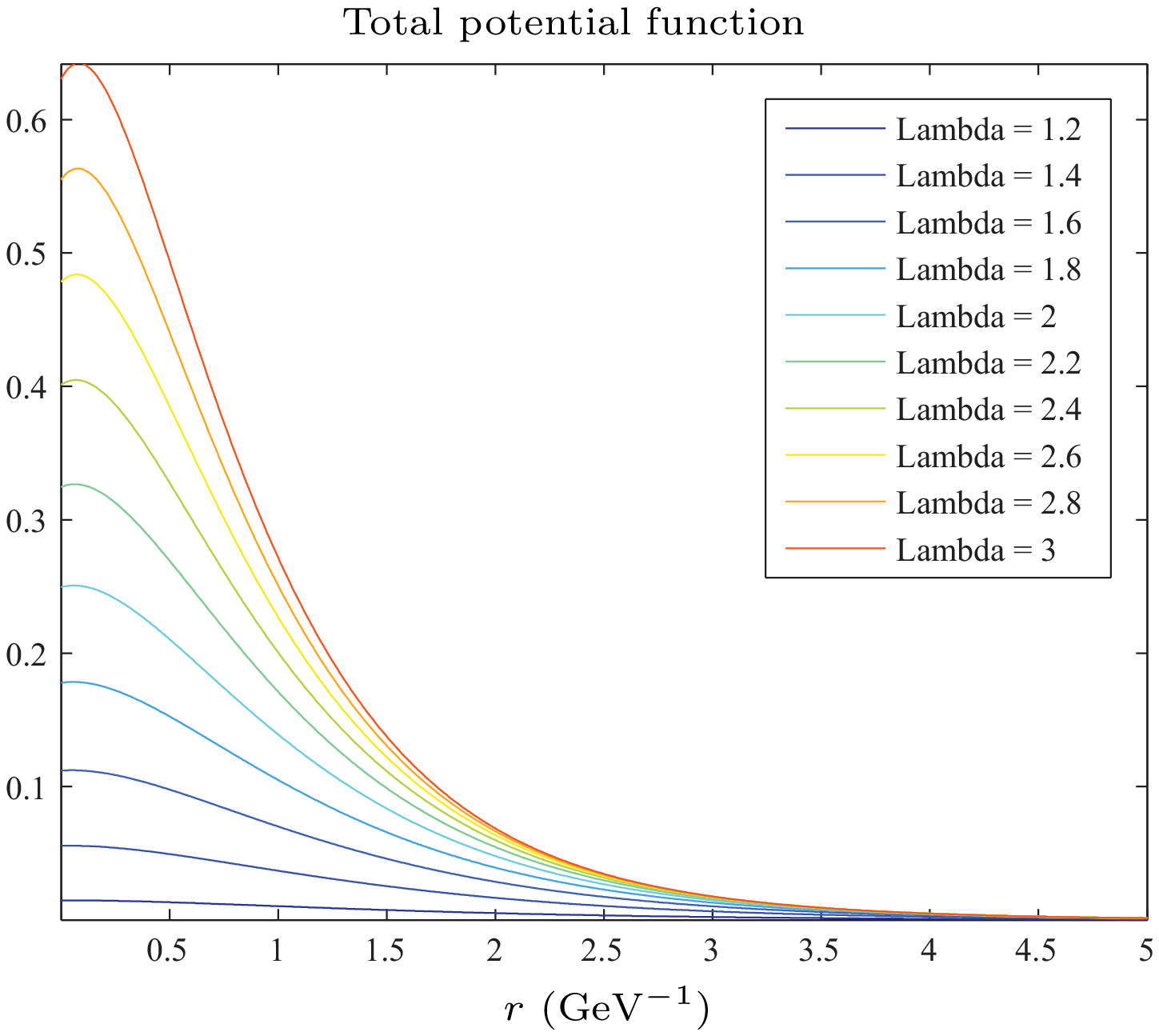}}&\scalebox{0.55}{\includegraphics{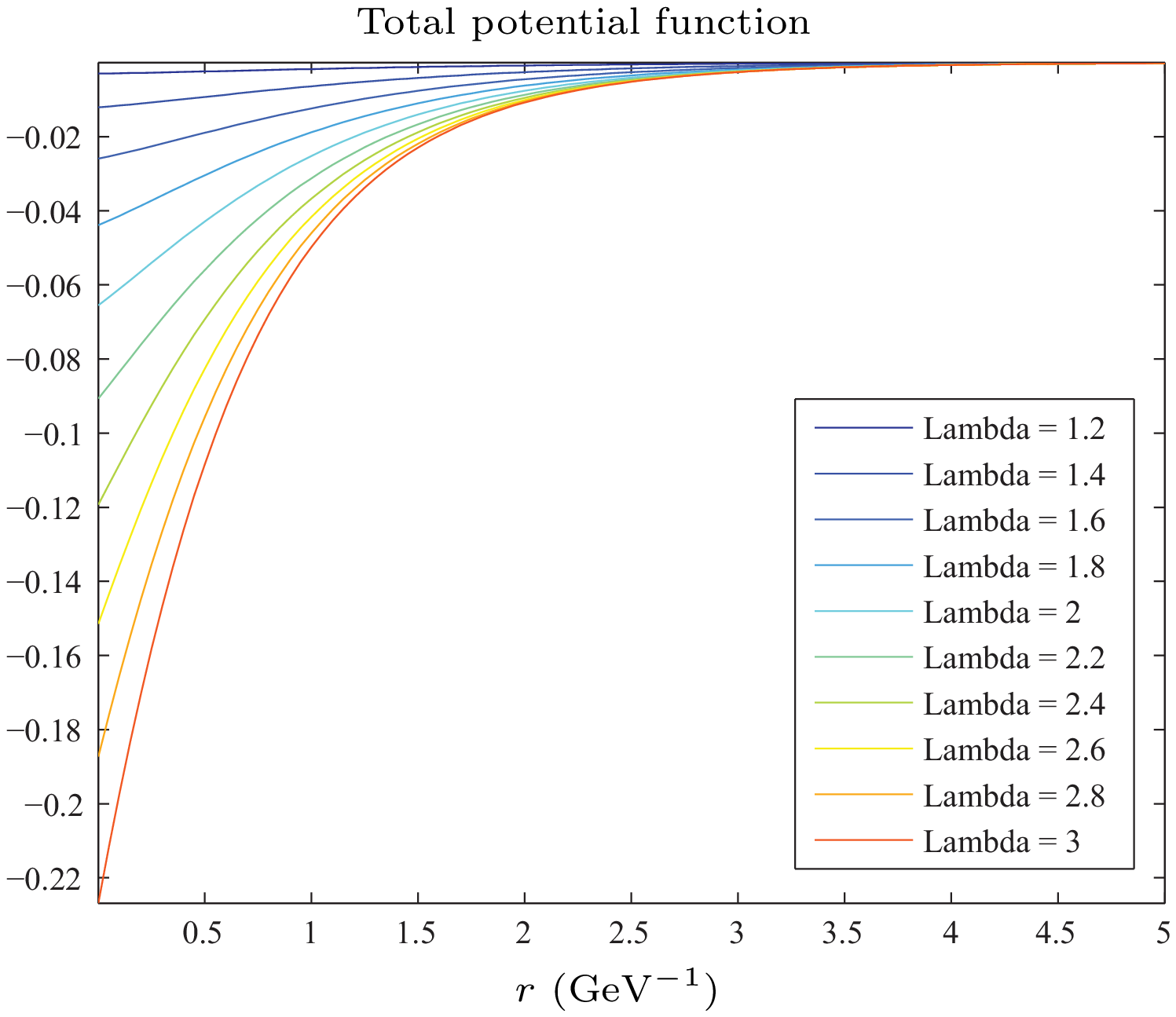}}\\
(c)&(d)
\end{tabular}
\caption{(a), (b), (c) and (d) are the potentials of the
$[D^*D_0^*]_{s1}^{0}$ with $(c=+1,\,\beta\beta^\prime>0)$,
$(c=+1,\,\beta\beta^\prime<0)$, $(c=-1,\,\beta\beta^\prime>0)$ and
$(c=-1,\,\beta\beta^\prime<0)$, respectively. Here, we take
$\zeta\varpi>0$ with $|\beta|=0.909$, $|\beta^\prime|=0.533$,
$|\zeta|=0.727$ and $|\varpi|=0.364$. \label{dsds0s3potential}}
\end{figure}
\end{center}

\begin{center}
\begin{figure}[htb]
\begin{tabular}{cccc}
\scalebox{0.55}{\includegraphics{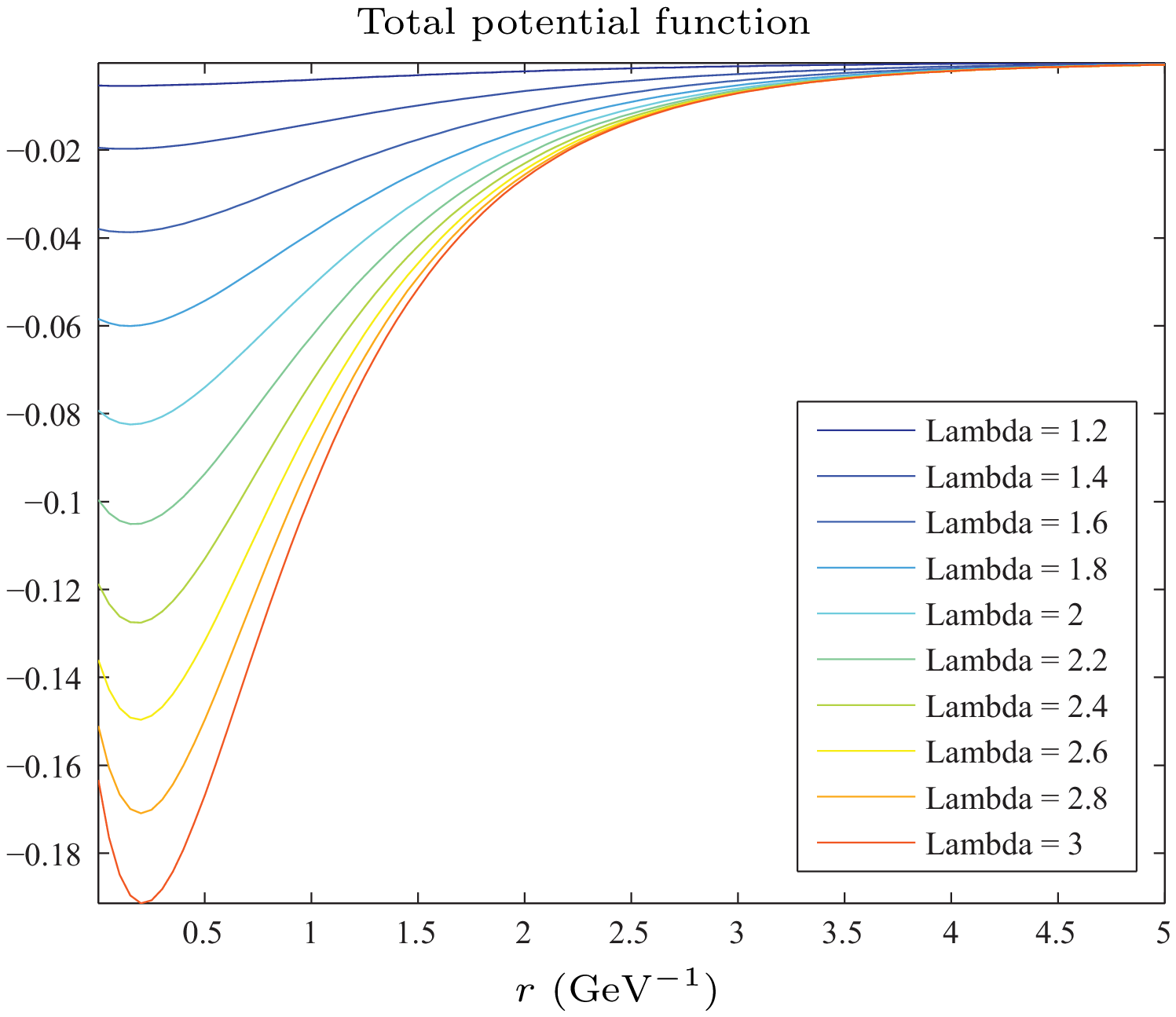}}&\scalebox{0.55}{\includegraphics{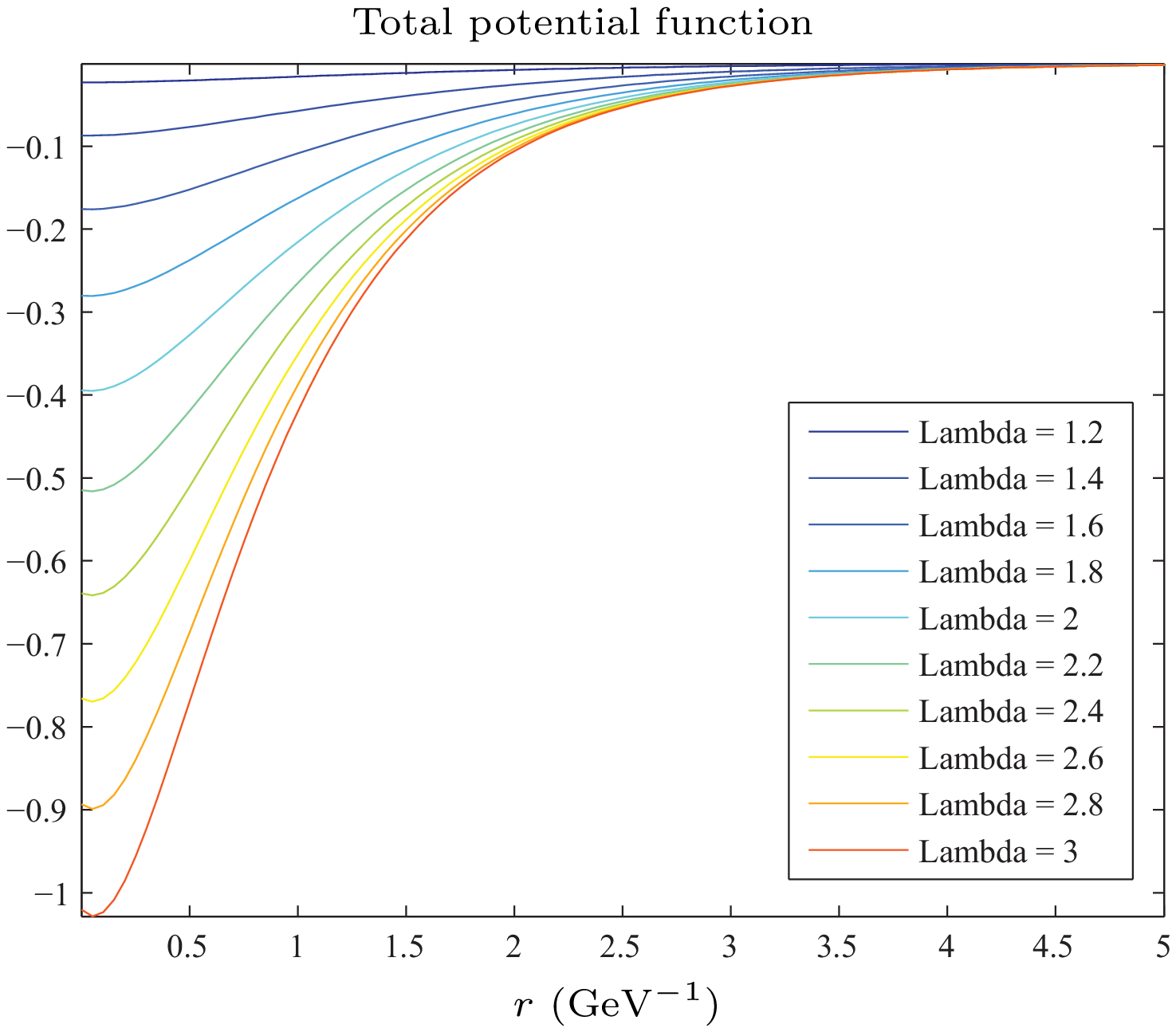}}\\
(a)&(b)\\
\scalebox{0.55}{\includegraphics{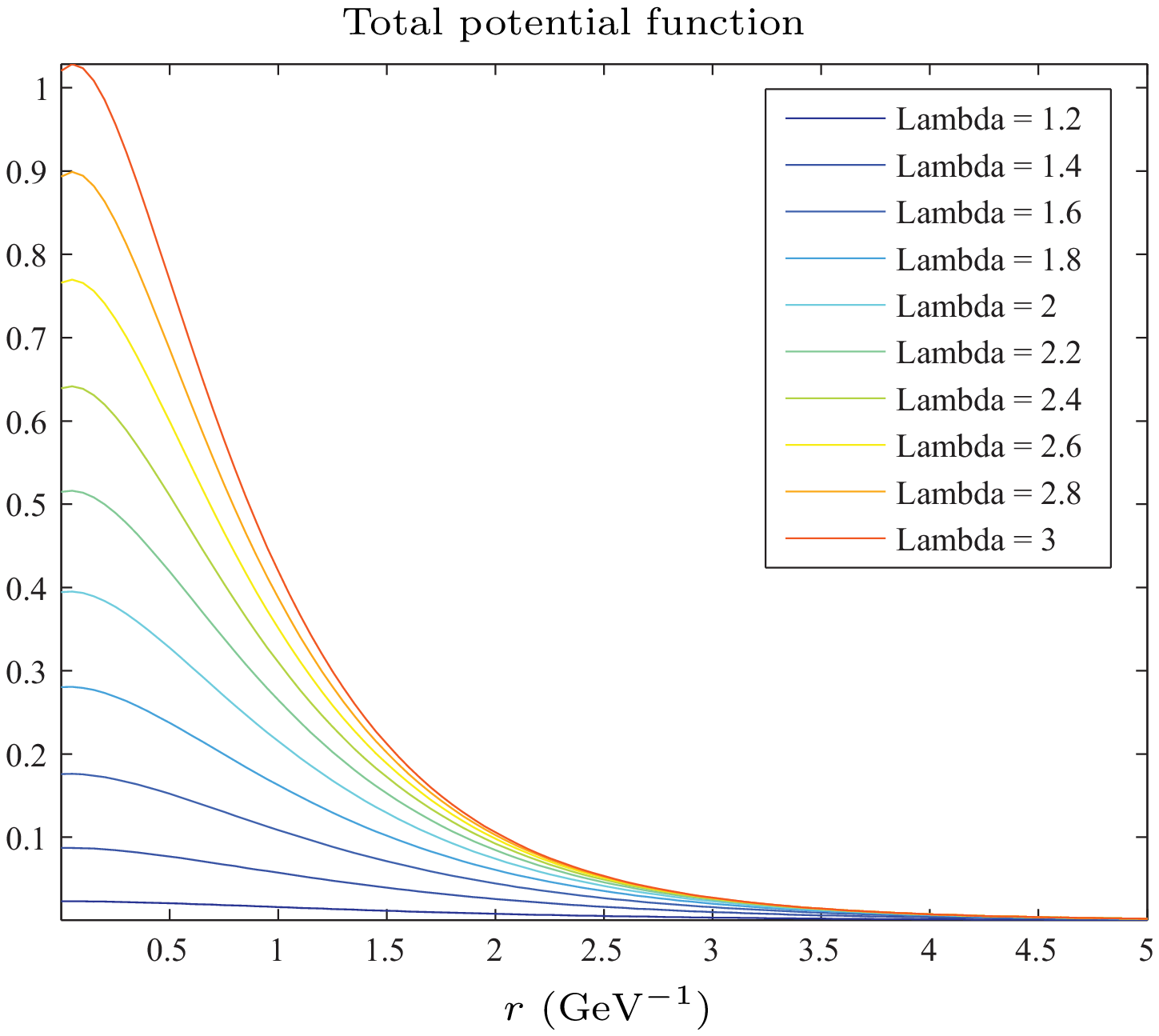}}&\scalebox{0.55}{\includegraphics{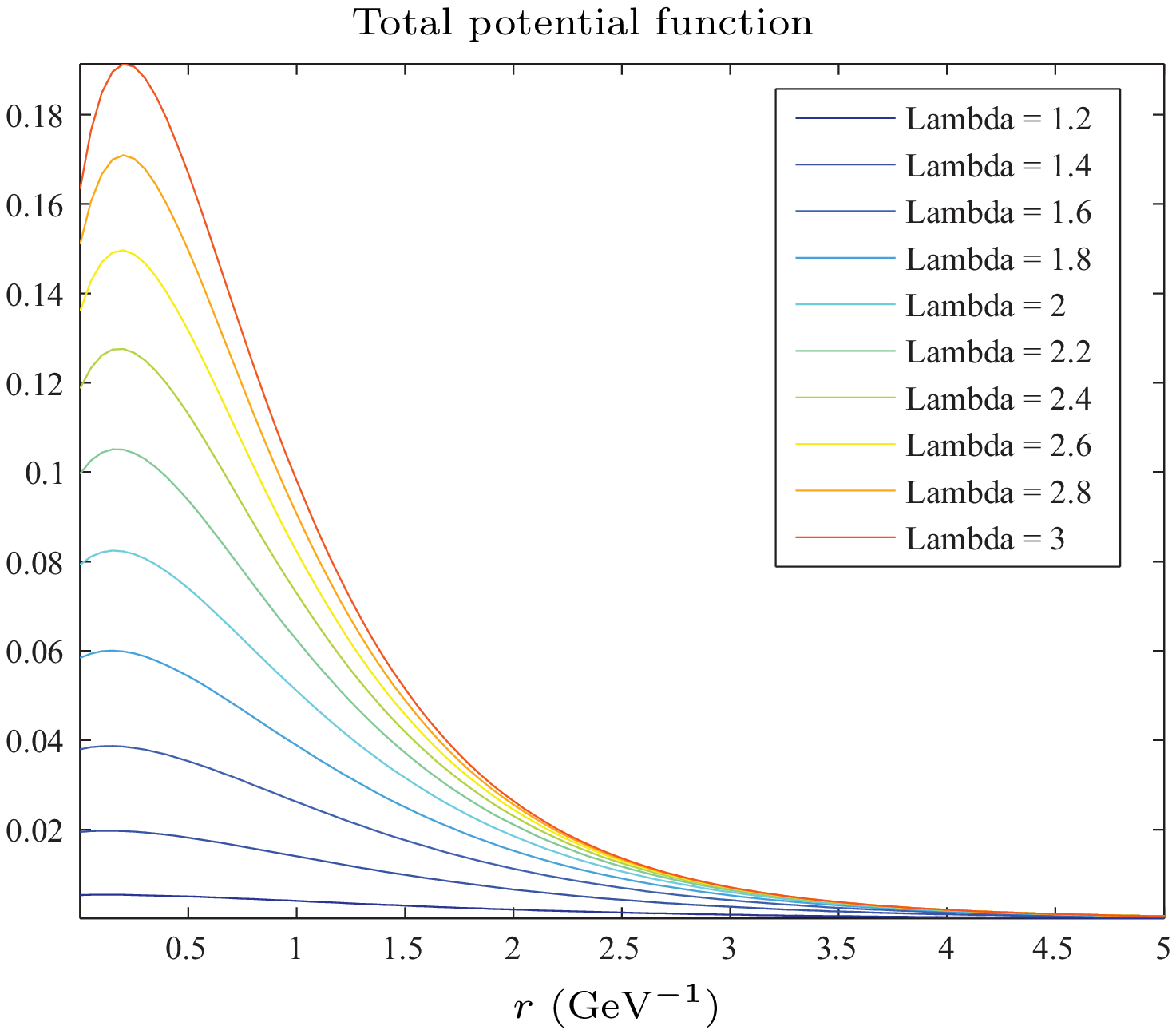}}\\
(c)&(d)
\end{tabular}
\caption{(a), (b), (c) and (d) are the potentials of the
$[D^*D_0^*]_{s1}^{0}$ system with $(c=+1,\,\beta\beta^\prime>0)$,
$(c=+1,\,\beta\beta^\prime<0)$, $(c=-1,\,\beta\beta^\prime>0)$ and
$(c=-1,\,\beta\beta^\prime<0)$, respectively. Here, we take
$\zeta\varpi<0$ with $|\beta|=0.909$, $|\beta^\prime|=0.533$,
$|\zeta|=0.727$ and $|\varpi|=0.364$. \label{dsds0s3potential-1}}
\end{figure}
\end{center}

Here, the effective potential of the $[P^*P_0^*]_{s1}^{0}$ system
results from the $\phi$ meson exchange. There exist eight
combinations of the parameters $c$, $\beta\beta^\prime$ and
$\zeta\varpi$. Under different parameter space, the dependence of
the effective potential of $[P^*P_0^*]_{s1}^{0}$ with $c=\pm 1$ on
$r$ is listed in Figs.
\ref{dsds0s3potential}-\ref{dsds0s3potential-1}.

\begin{center}
\begin{figure}[htb]
\begin{tabular}{cccc}
\scalebox{0.8}{\includegraphics{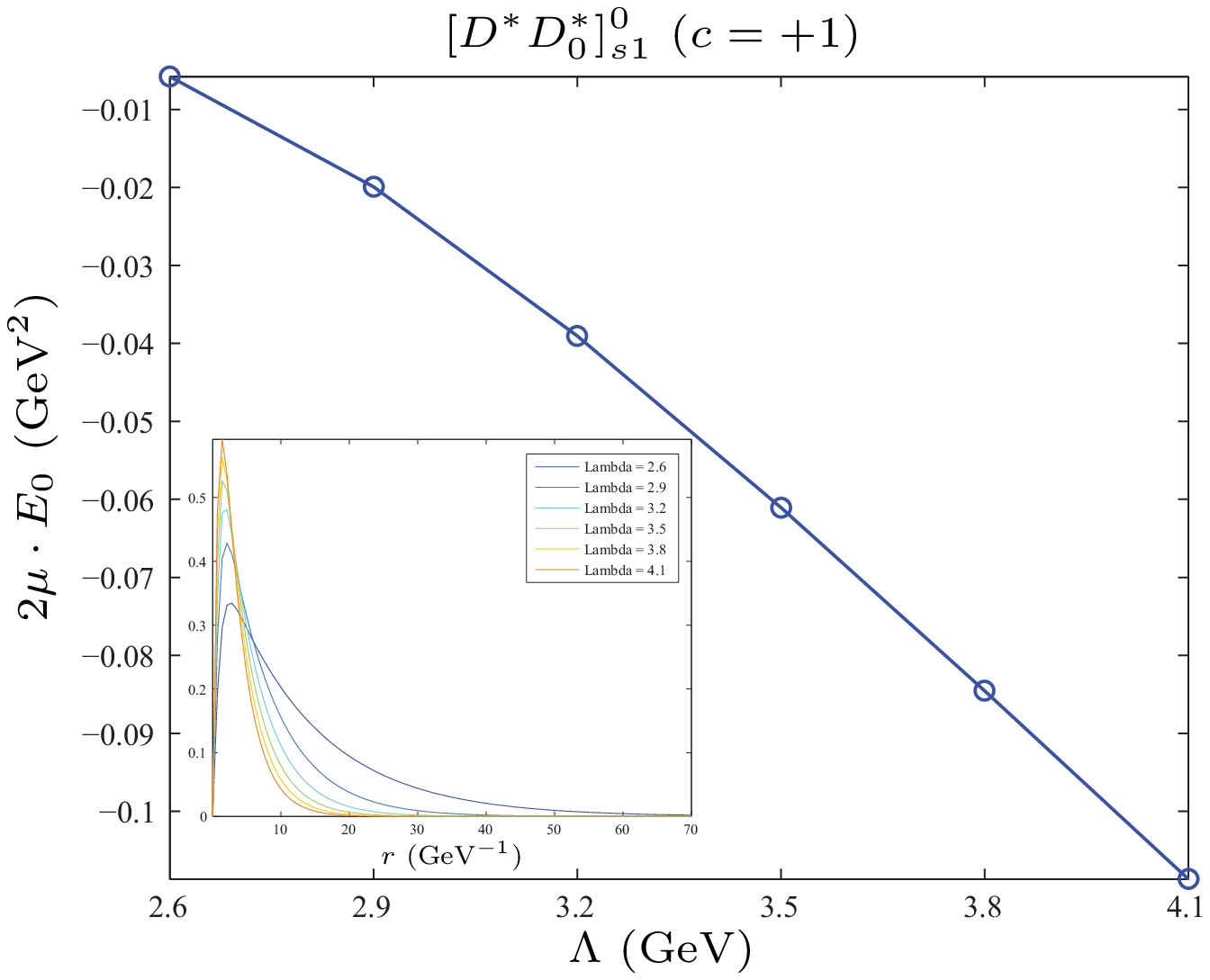}}
\end{tabular}
\caption{The dependence of the binding energy of
$[D^*D_0^*]_{s1}^{0}$ on $\Lambda$ with $c=+1$,
$\beta\beta^\prime<0$ and $\zeta\varpi<0$ with $|\beta|=0.909$,
$|\beta^\prime|=0.533$, $|\zeta|=0.727$ and $|\varpi|=0.364$.
\label{dsds0s3-E}}
\end{figure}
\end{center}

When solving the Schr\"{o}dinger equation, the cutoff $\Lambda$
should be larger than the mass of the $\phi$ meson. There exists a
bound state solution of the $[P^*P_0^*]_{s1}^{0}$ system only with
$c=+1$, $\beta\beta^\prime <0$ and $\zeta\varpi <0$, which is
shown in Fig. \ref{dsds0s3-E}.

\subsubsection{$[P^*P_0^*]_{8}^{0}$}

\begin{center}
\begin{figure}[htb]
\begin{tabular}{cccc}
\scalebox{0.55}{\includegraphics{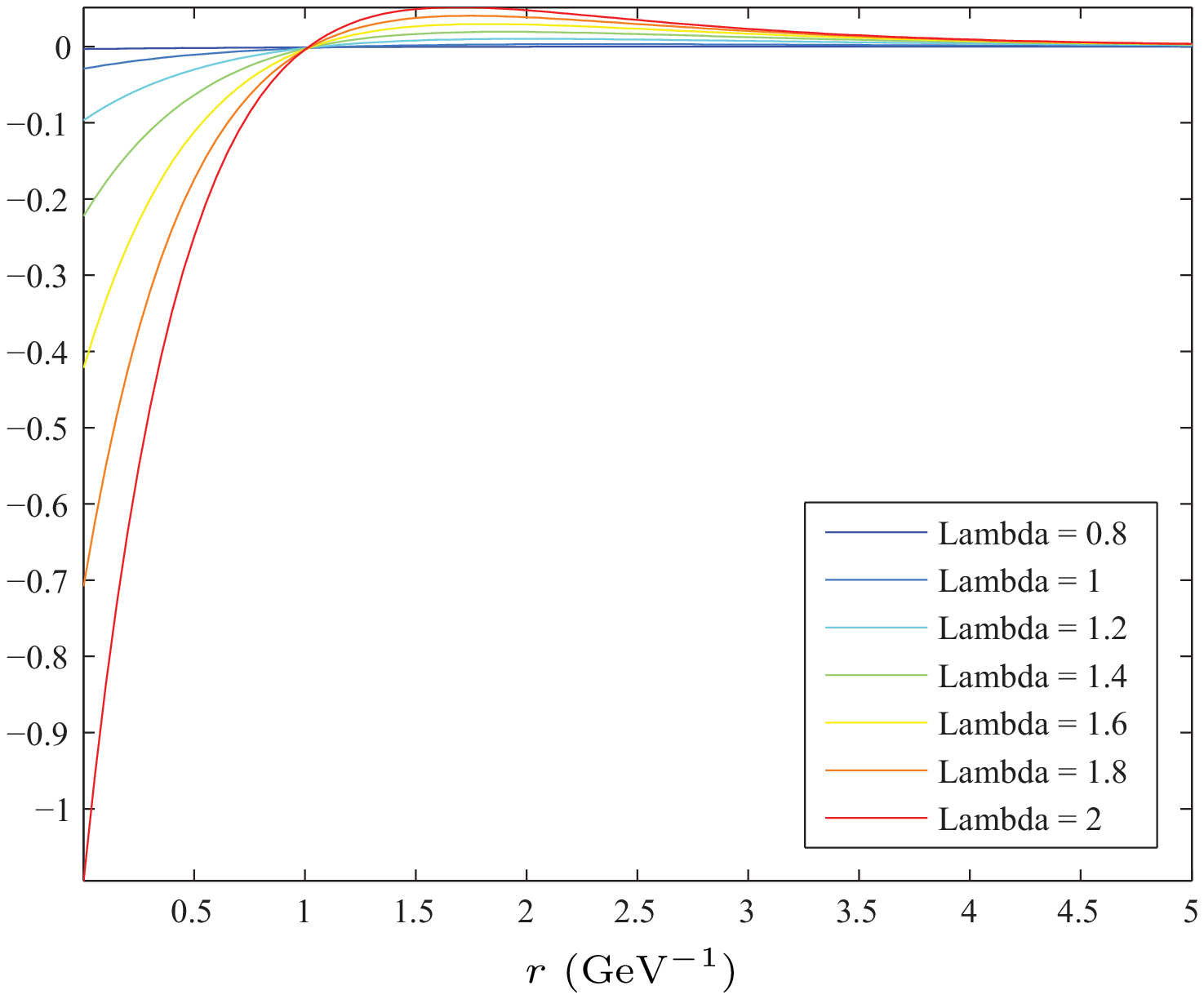}}&\scalebox{0.55}{\includegraphics{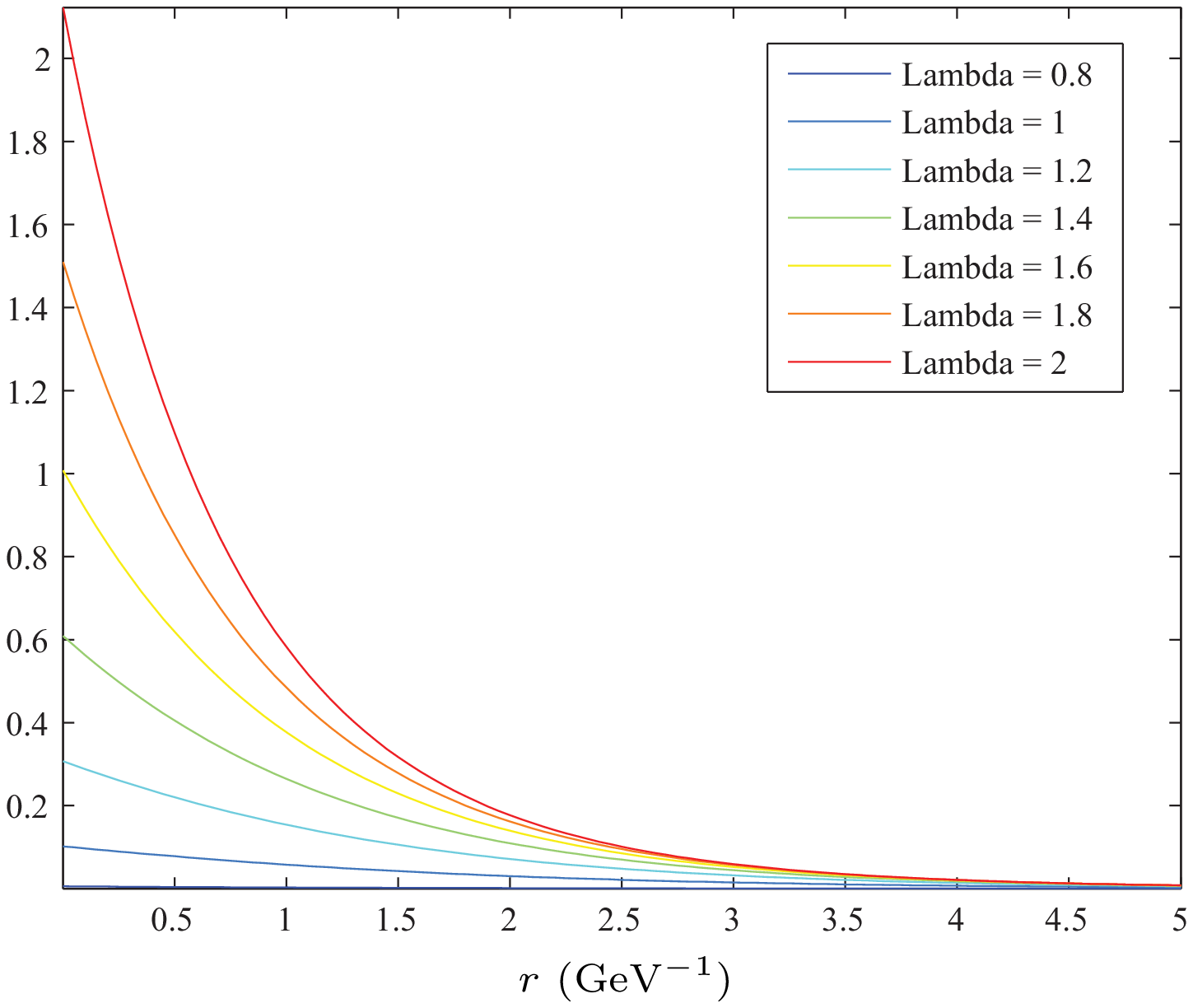}}\\
(a)&(b)\\
\scalebox{0.55}{\includegraphics{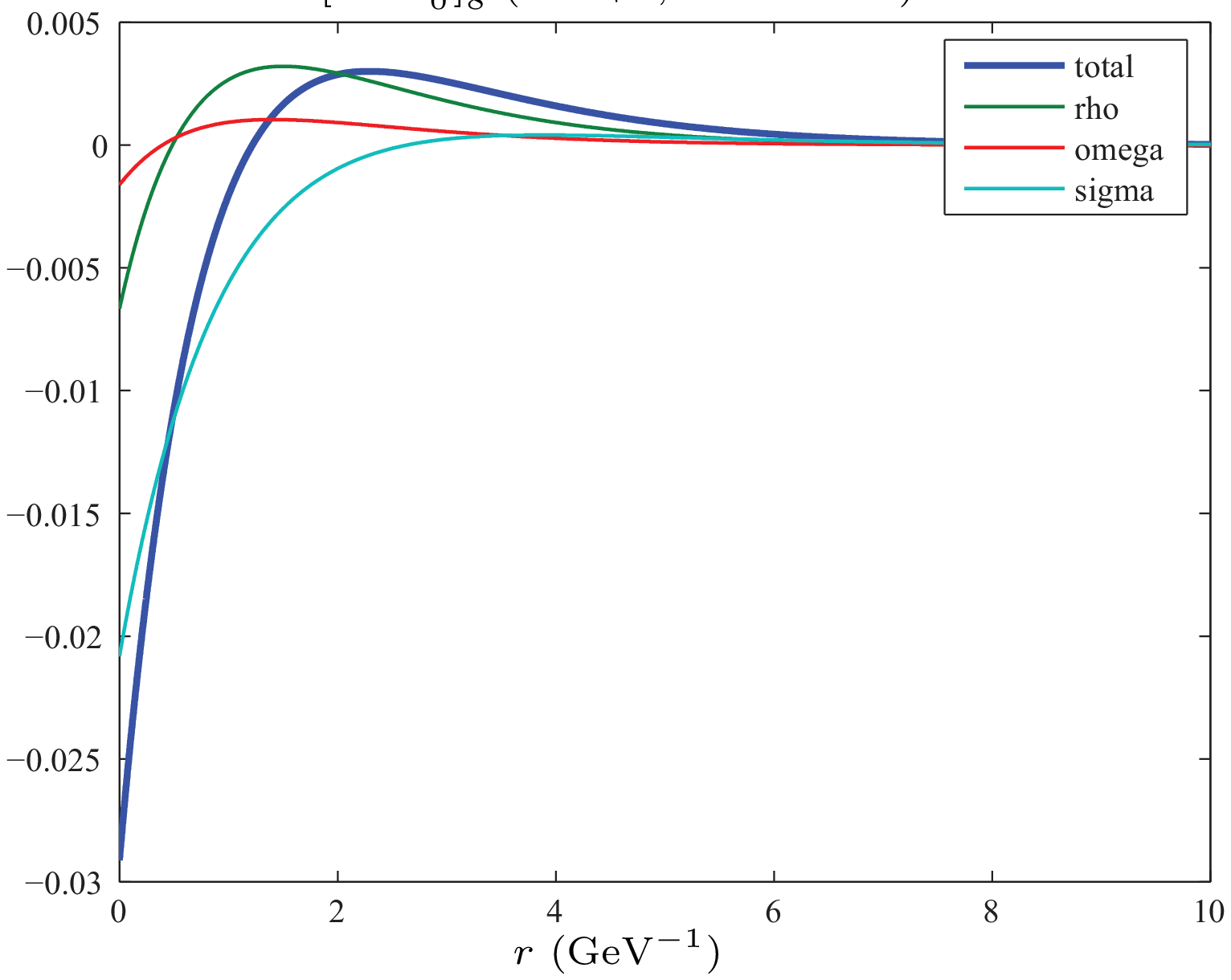}}&\scalebox{0.55}{\includegraphics{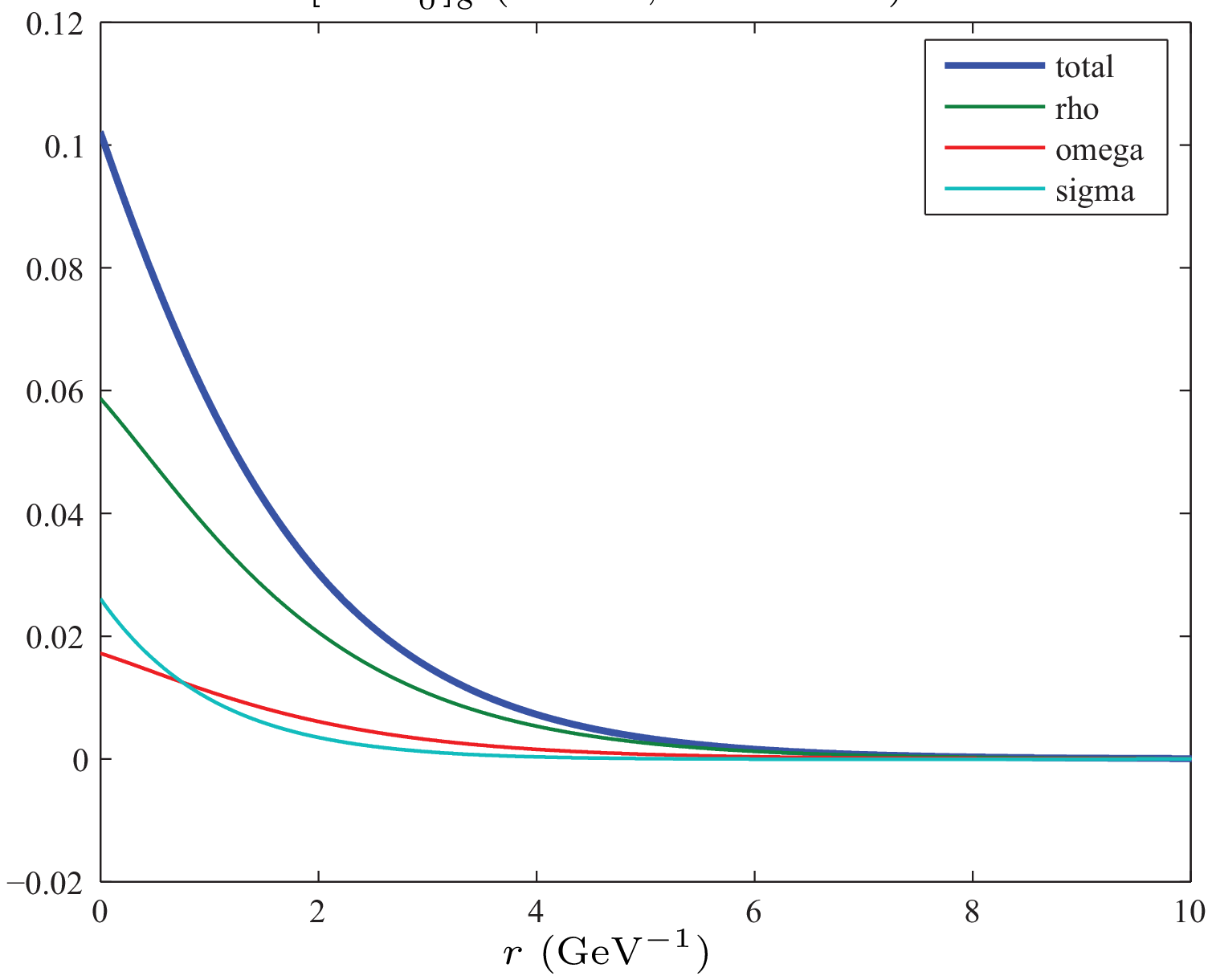}}\\
(c)&(d)
\end{tabular}
\caption{(a) and (b) are the total potentials of the
$[D^*D_0^*]_8^{0}$ system with $c=+1$ and $c=-1$ respectively with
$g_\sigma g_\sigma^\prime>0$, $\beta\beta^\prime>0$ and
$\zeta\mu>0$ with $|\beta|=0.909$, $|\beta^\prime|=0.533$,
$|\zeta|=0.727$ and $|\varpi|=0.364$. With $\Lambda=1$ GeV, we
show the comparison of the total potential and the
$\sigma,\,\rho,\,\omega$ exchange potentials, where (c) and (d)
illustrate the result of the $[D^*D_0^*]_8^{0}$ system with $c=+1$
and $c=-1$ respectively.\label{dsds0s4potential}}
\end{figure}
\end{center}

As indicated in Table \ref{potential}, the behavior of the $\rho$
meson exchange potential is similar to that of the $\omega$ meson
exchange potential. When exploring the $[P^*P_0^*]_{8}^{0}$
system, we need to consider the $\sigma$ exchange potential
together with both the $\rho$ and $\omega$ meson exchanges. The
relevant total and partial potentials are shown in Fig.
\ref{dsds0s4potential}.

\begin{center}
\begin{figure}[htb]
\begin{tabular}{cccc}
\scalebox{0.6}{\includegraphics{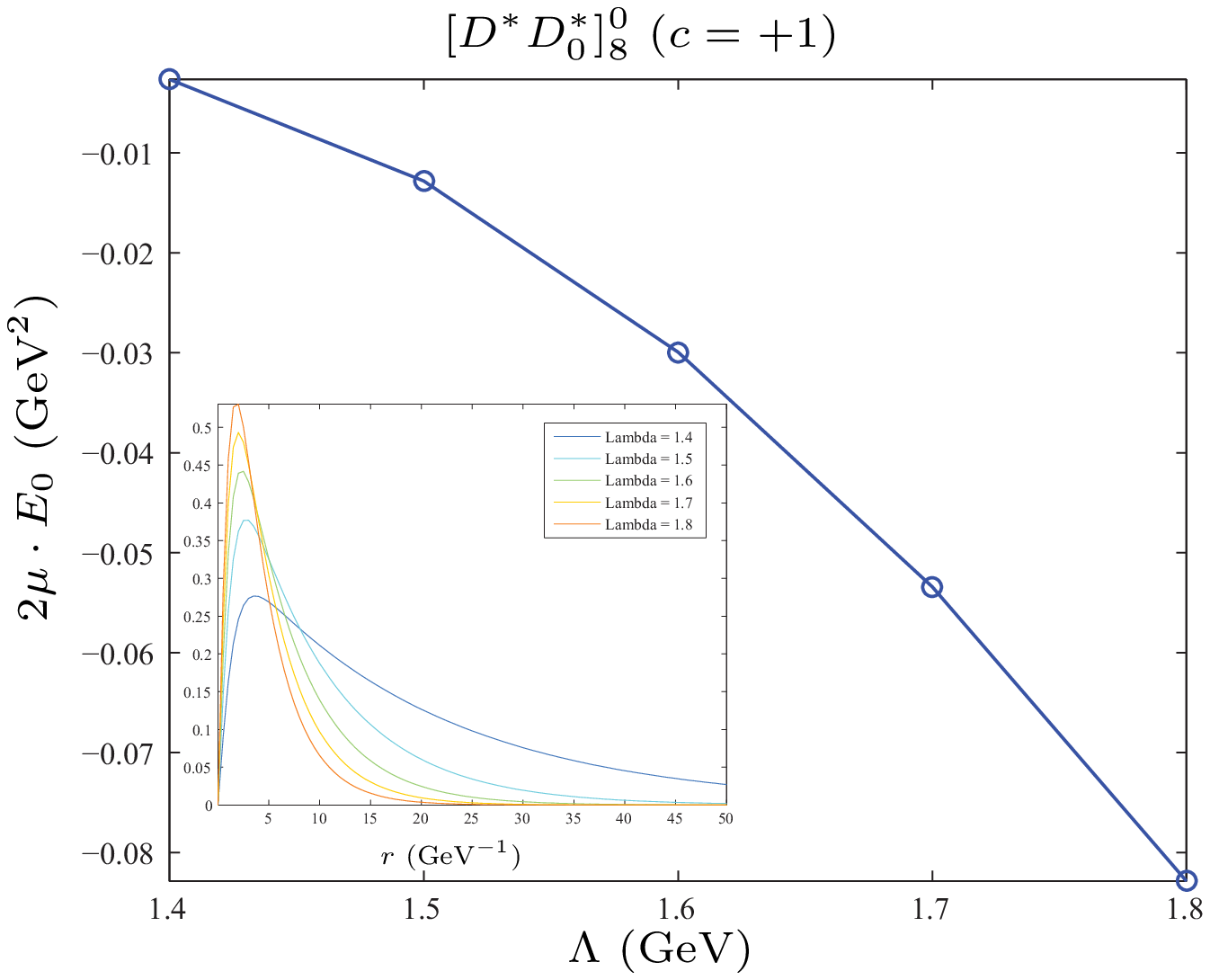}}&\scalebox{0.6}{\includegraphics{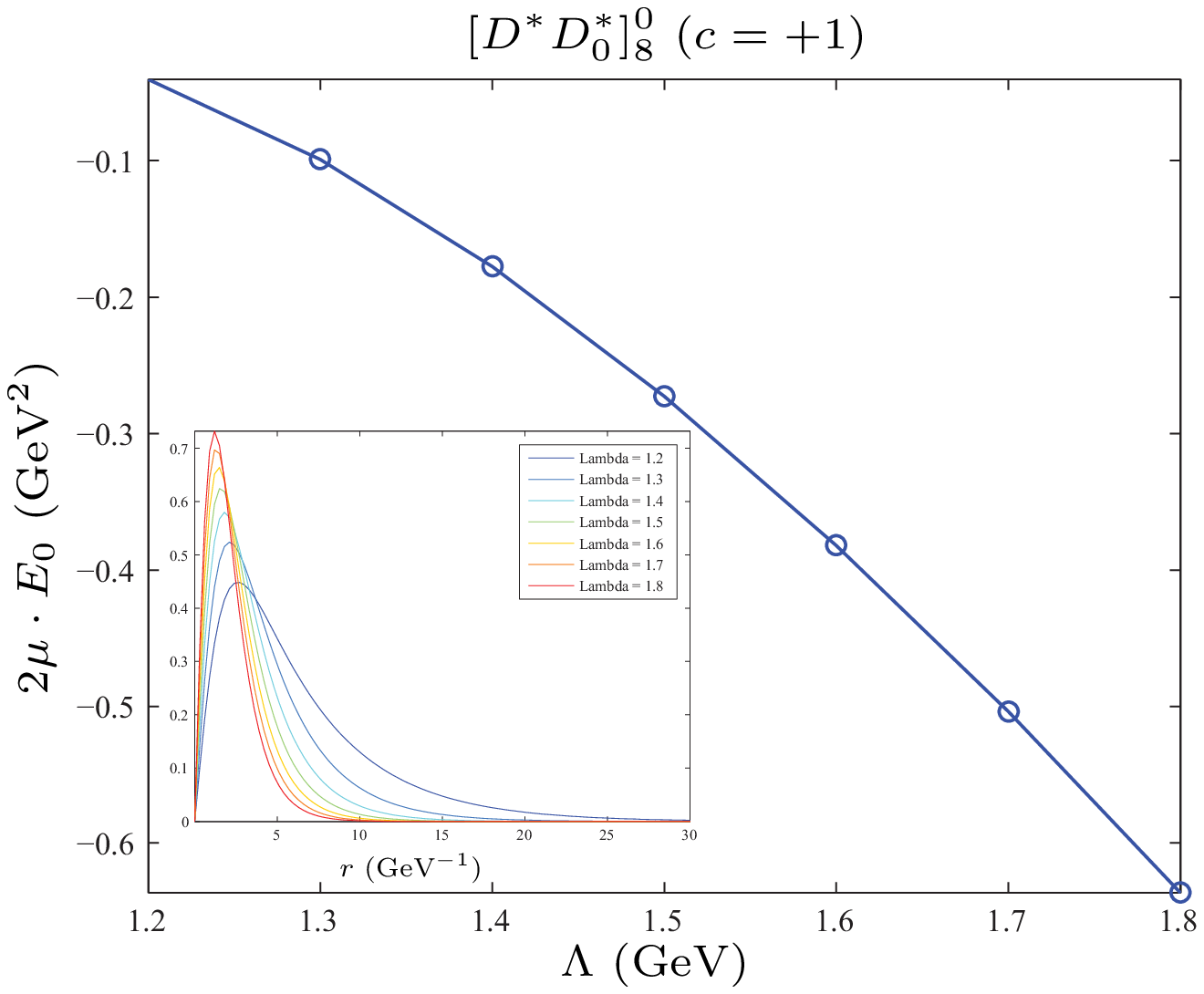}}\\
(a)&(b)\\
\scalebox{0.6}{\includegraphics{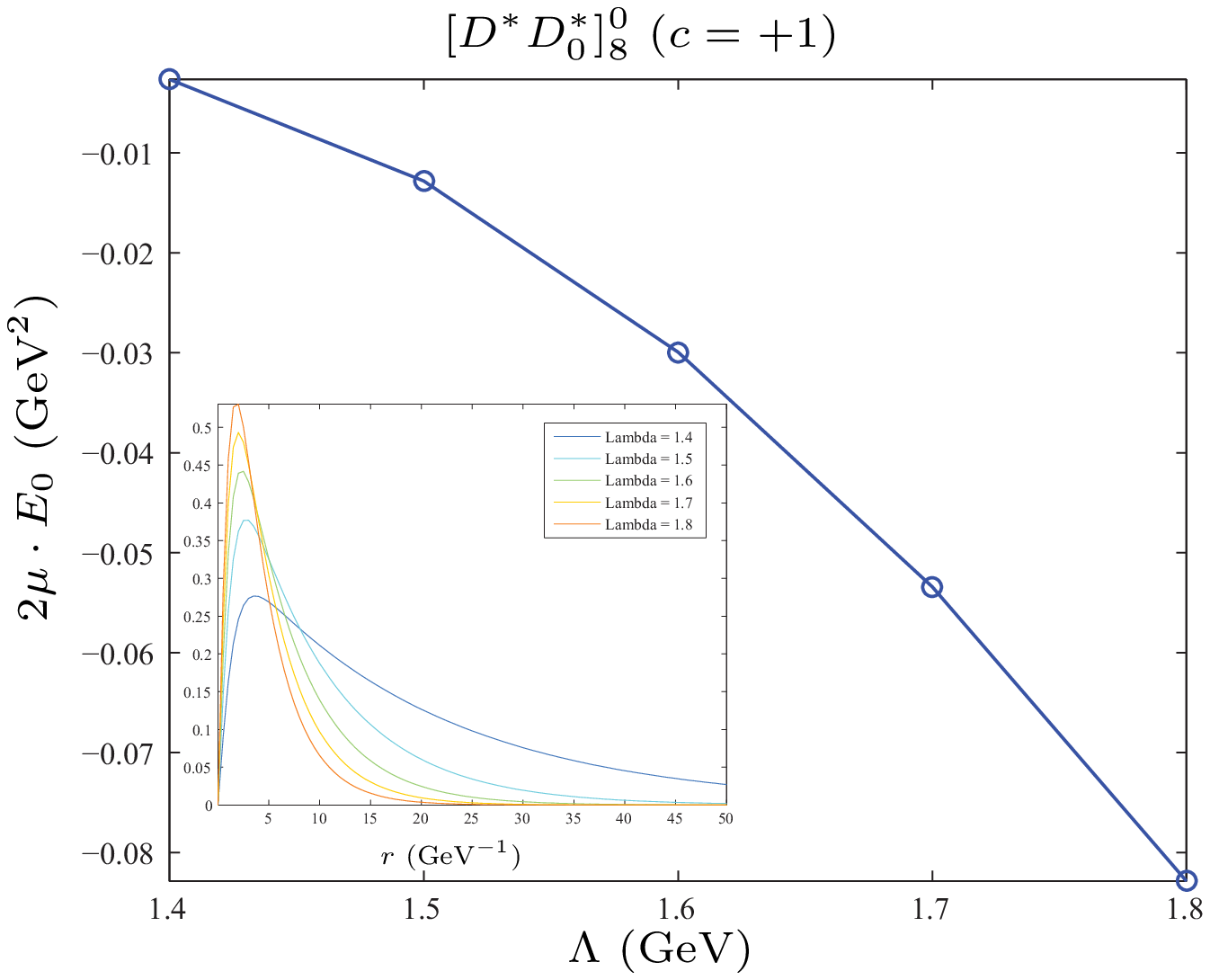}}&\scalebox{0.6}{\includegraphics{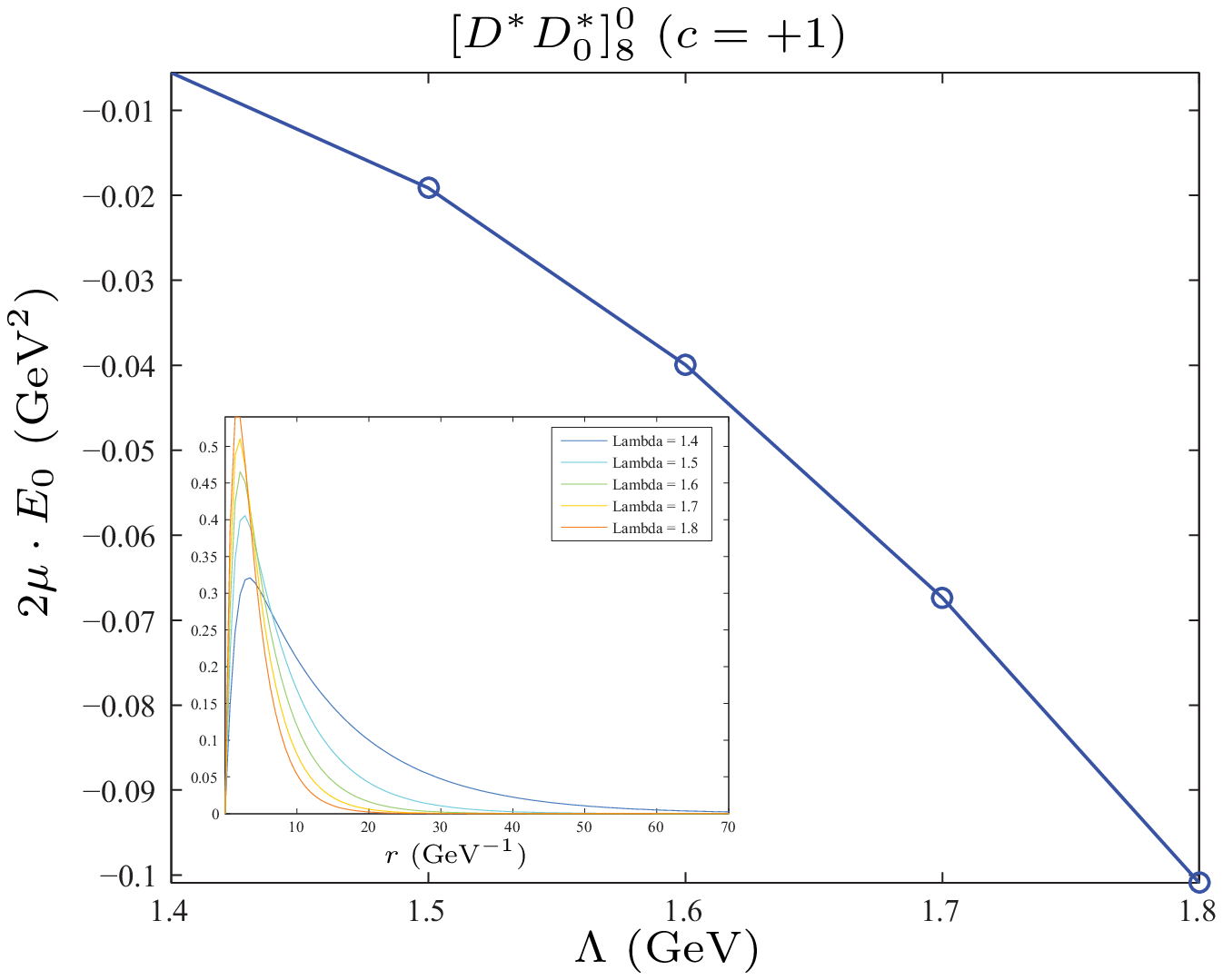}}\\
(c)&(d)\\
\scalebox{0.6}{\includegraphics{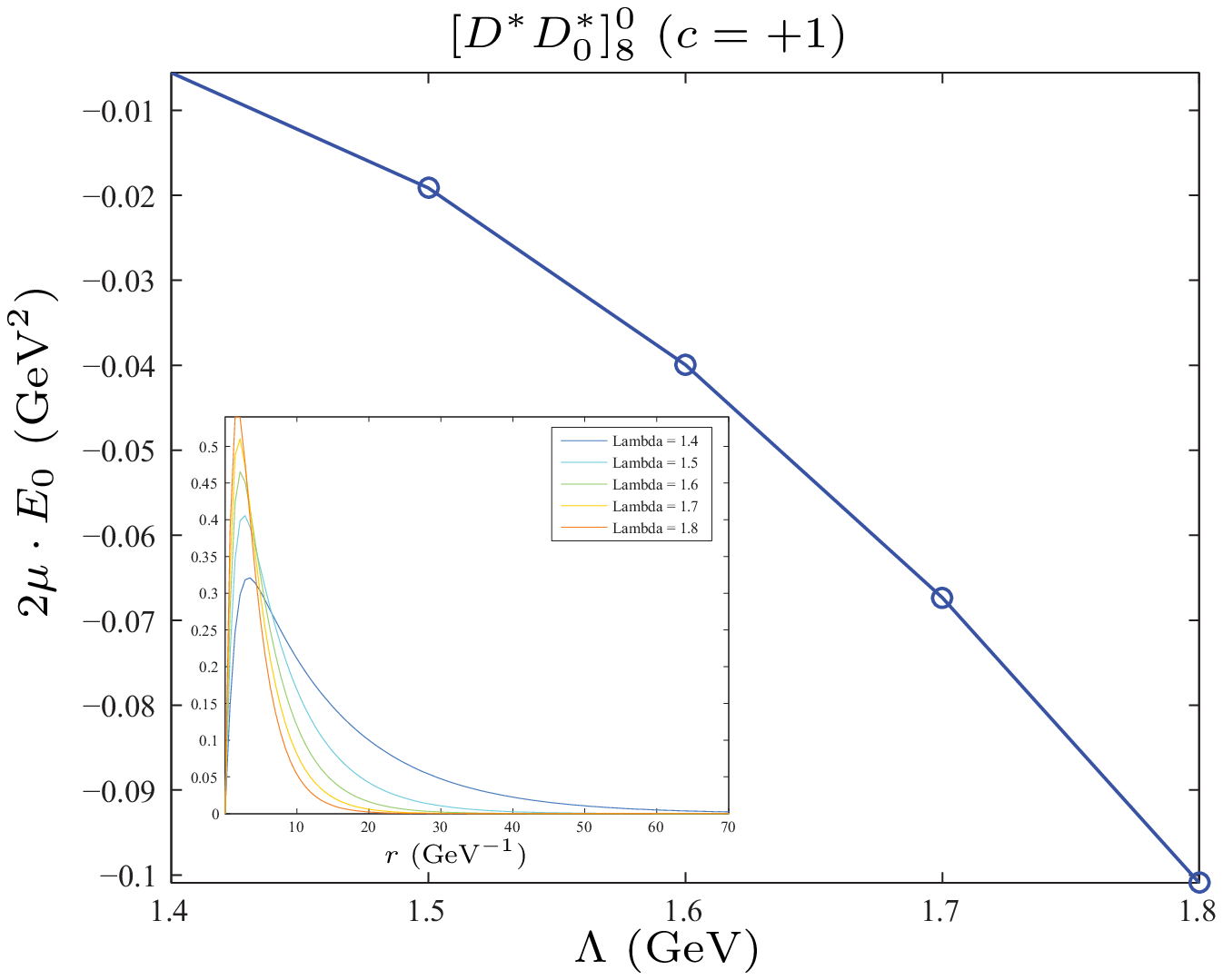}}&\scalebox{0.6}{\includegraphics{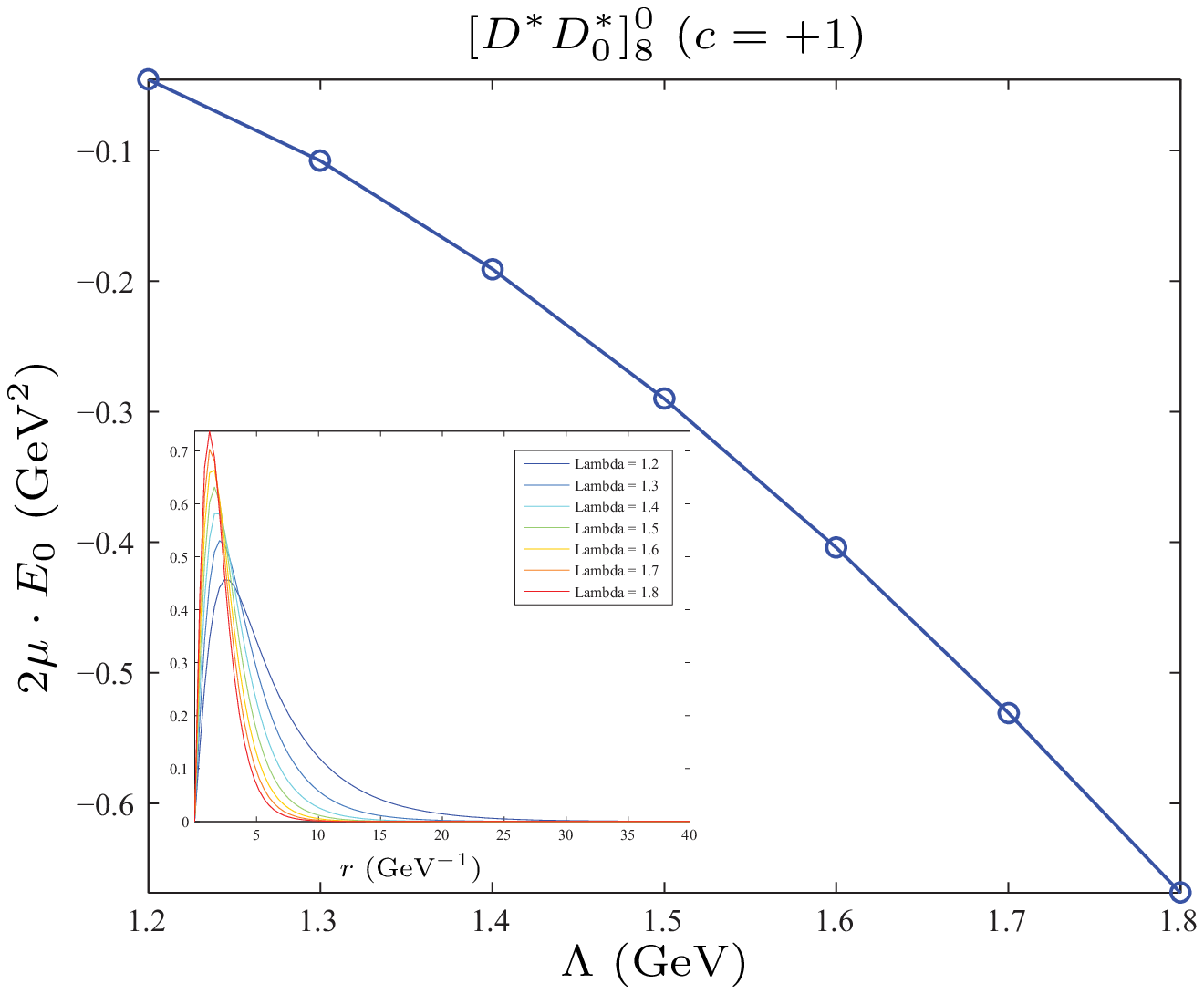}}\\
(e)&(f)\\
\end{tabular}
\caption{The bound state solution of the $[D^*D_0^*]_8^{0}$ with
$c=+1$. (a), (b), (c), (d), (e), and (f) correspond to the results
with $(g_\sigma g_\sigma^\prime>0,\, \beta\beta^\prime<0,\, \zeta
\varpi>0)$, $(g_\sigma g_\sigma^\prime>0,\, \beta\beta^\prime<0,\,
\zeta \varpi<0)$, $(g_\sigma g_\sigma^\prime>0,\,
\beta\beta^\prime>0,\, \zeta \varpi<0)$, $(g_\sigma
g_\sigma^\prime<0,\, \beta\beta^\prime<0,\, \zeta \varpi>0)$,
$(g_\sigma g_\sigma^\prime<0,\, \beta\beta^\prime>0,\, \zeta
\varpi<0)$ and $(g_\sigma g_\sigma^\prime<0,\,
\beta\beta^\prime<0,\, \zeta \varpi<0)$ respectively. When taking
$(g_\sigma g_\sigma^\prime>0,\, \beta\beta^\prime>0,\, \zeta
\varpi>0)$ or $(g_\sigma g_\sigma^\prime<0,\,
\beta\beta^\prime>0,\, \zeta \varpi>0)$, we can not find the bound
state solution. Here, $|g_\sigma|=0.76$, $|g_\sigma^\prime|=0.76$,
$|h_\sigma|=0.323$, $|\beta|=0.909$ and $|\beta^\prime|=0.533$,
$\zeta=0.727$ and $\varpi=0.364$. \label{dsds0s4-E}}
\end{figure}
\end{center}

Since the total potential of the $[P^*P_0^*]_{8}^{0}$ system with
$c=-1$ is repulsive, there does not exist a $[P^*P_0^*]_{8}^{0}$
bound state system with $c=-1$. In the following, we mainly focus
on the $[P^*P_0^*]_{8}^{0}$ system with $c=+1$, where the total
effective potential is attractive. As shown in Fig.
\ref{dsds0s4-E}, there exists the bound state solution of the
$[P^*P_0^*]_{8}^{0}$ system with $c=+1$.

\subsection{The $[P^*P_1^\prime]$ system}

In Refs. \cite{liu3,liu4,close1}, the $[P^*P_1^\prime]^{\pm,0}$
and $[P^*P_1^\prime]_{8}^{0}$ systems were studied. In this work
we will discuss the rest two cases of the $[P^*P_1^\prime]$
system, i.e.,
$[P^*P_1^\prime]_s^{\pm,0}/[P^*P_1^\prime]_{\hat{s}}^{0}$ and
$[P^*P_0^\prime]_{s1}^{0}$.

\subsubsection{$[P^*P_1^\prime]_s^{\pm,0},\,[P^*P_1^\prime]_{\hat{s}}^{0}$}

\begin{center}
\begin{figure}[htb]
\begin{tabular}{cccc}
\scalebox{0.5}{\includegraphics{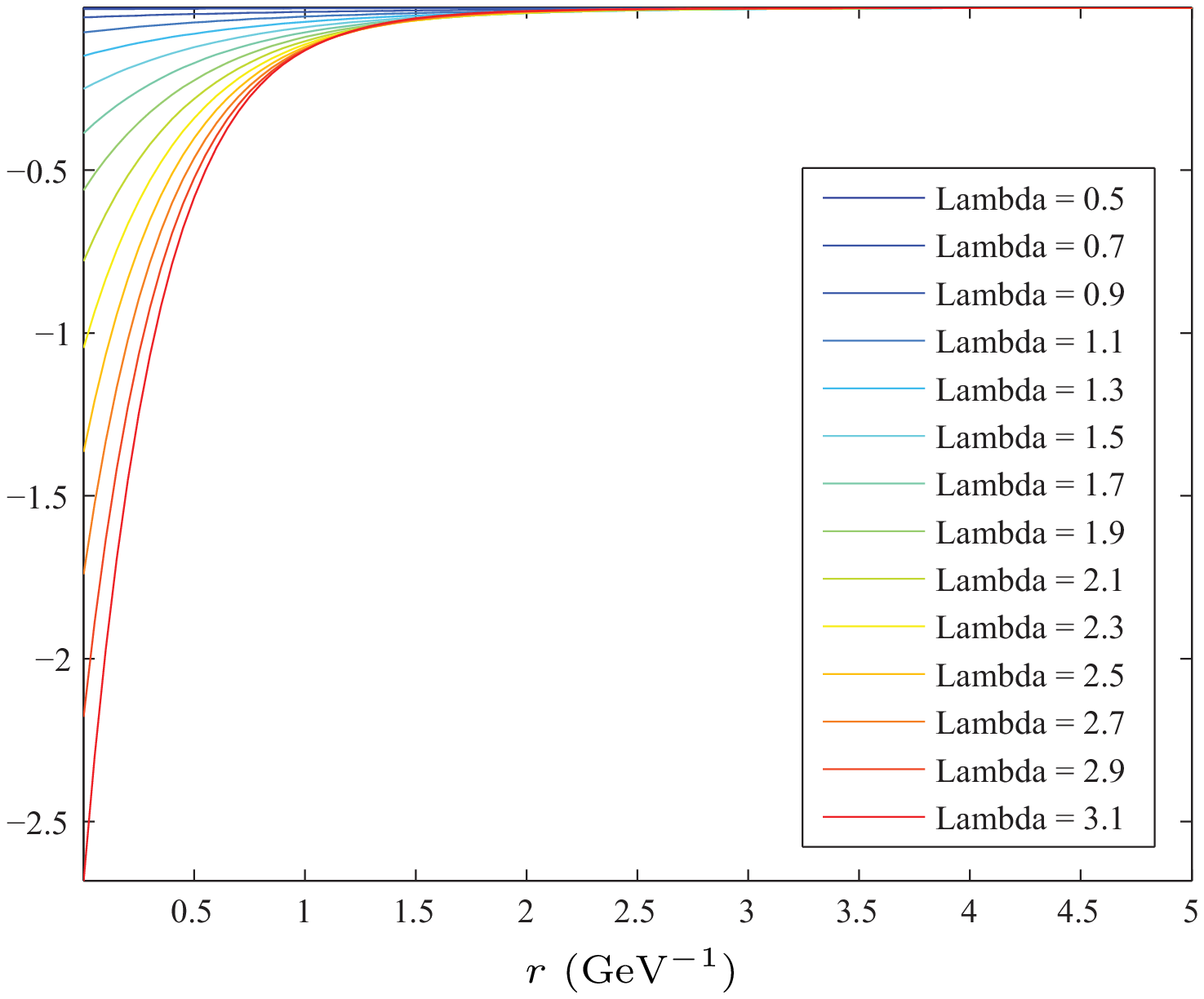}}&\scalebox{0.5}{\includegraphics{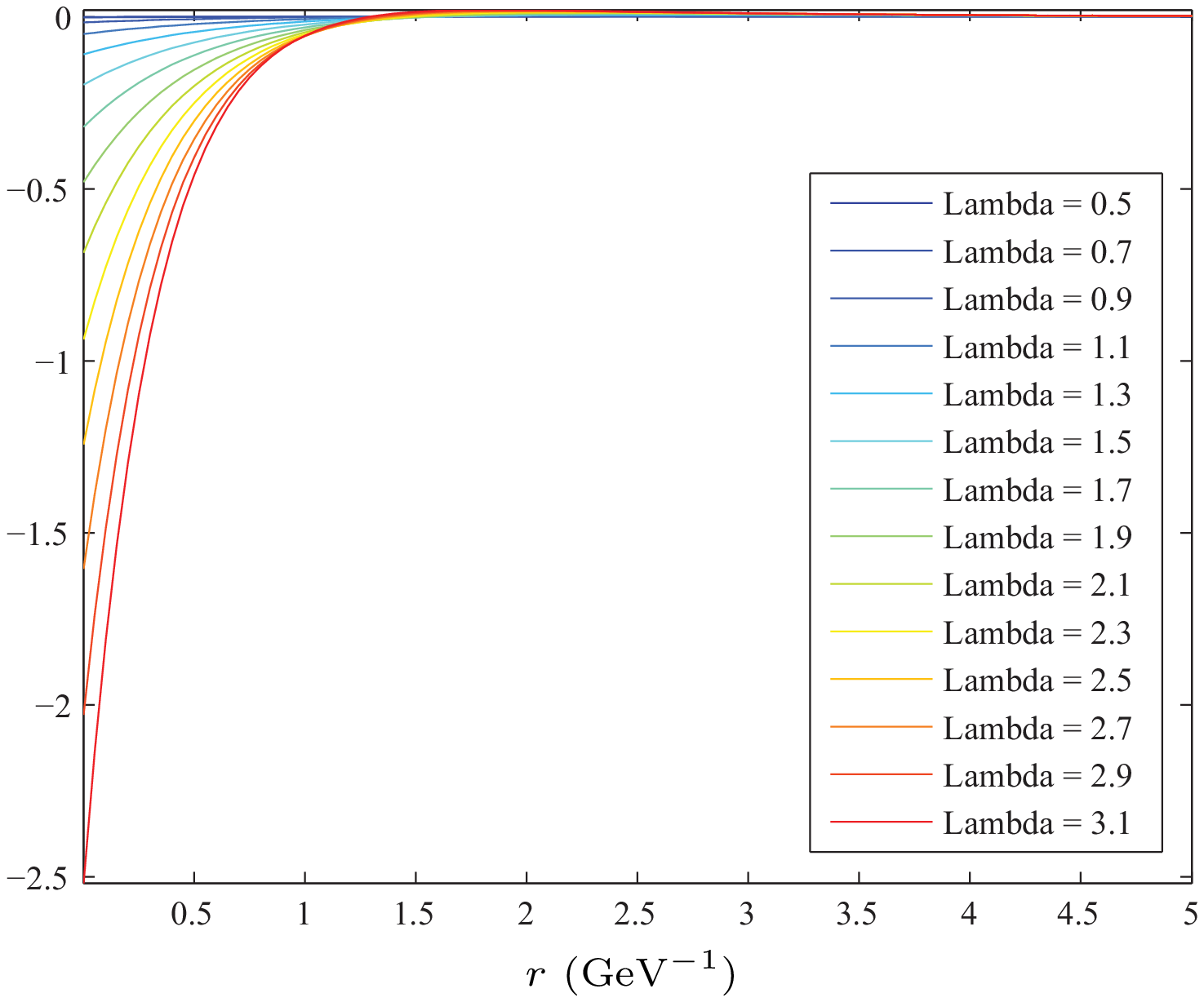}}\\
(a)&(b)\\
\scalebox{0.5}{\includegraphics{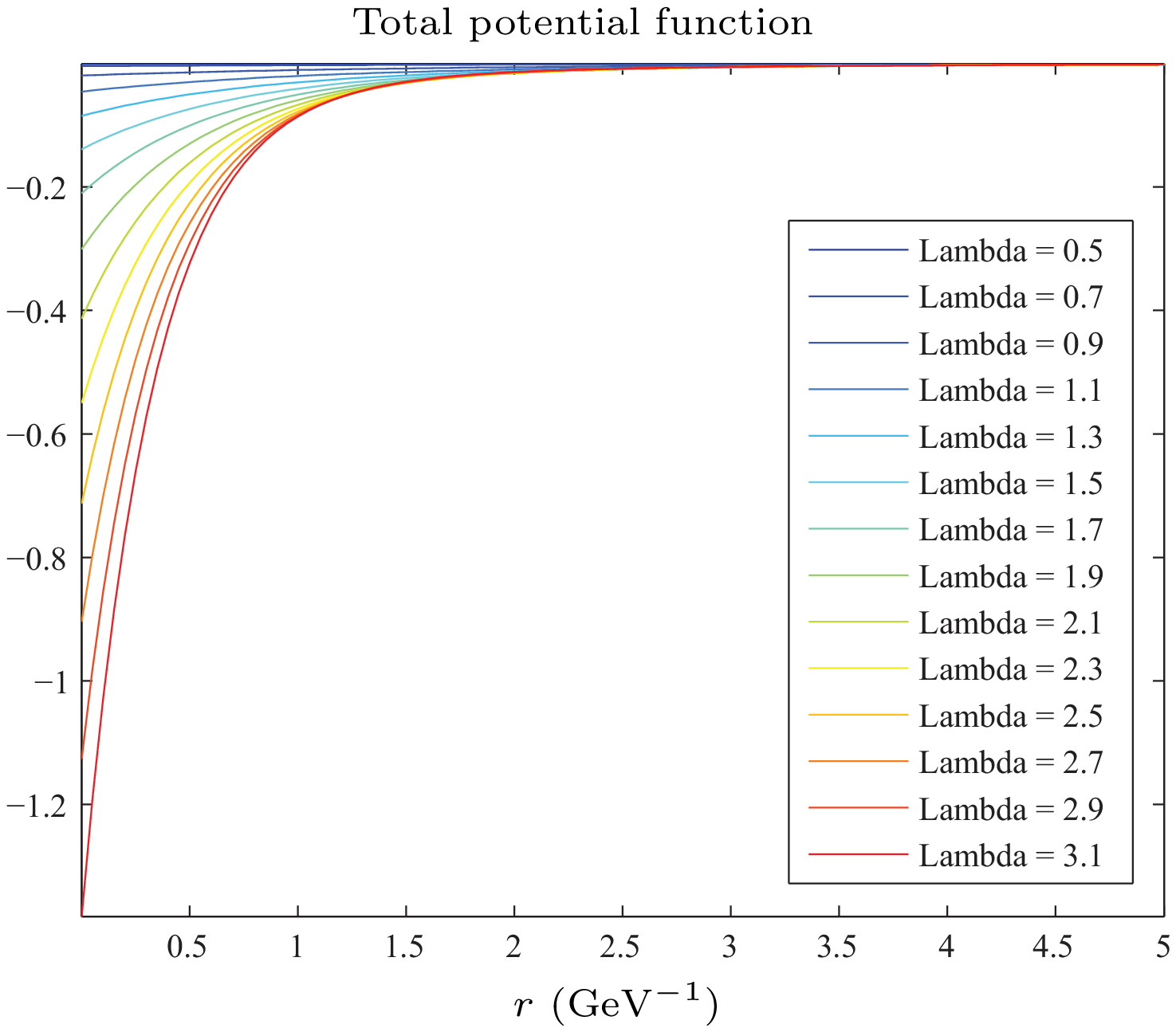}}&\scalebox{0.5}{\includegraphics{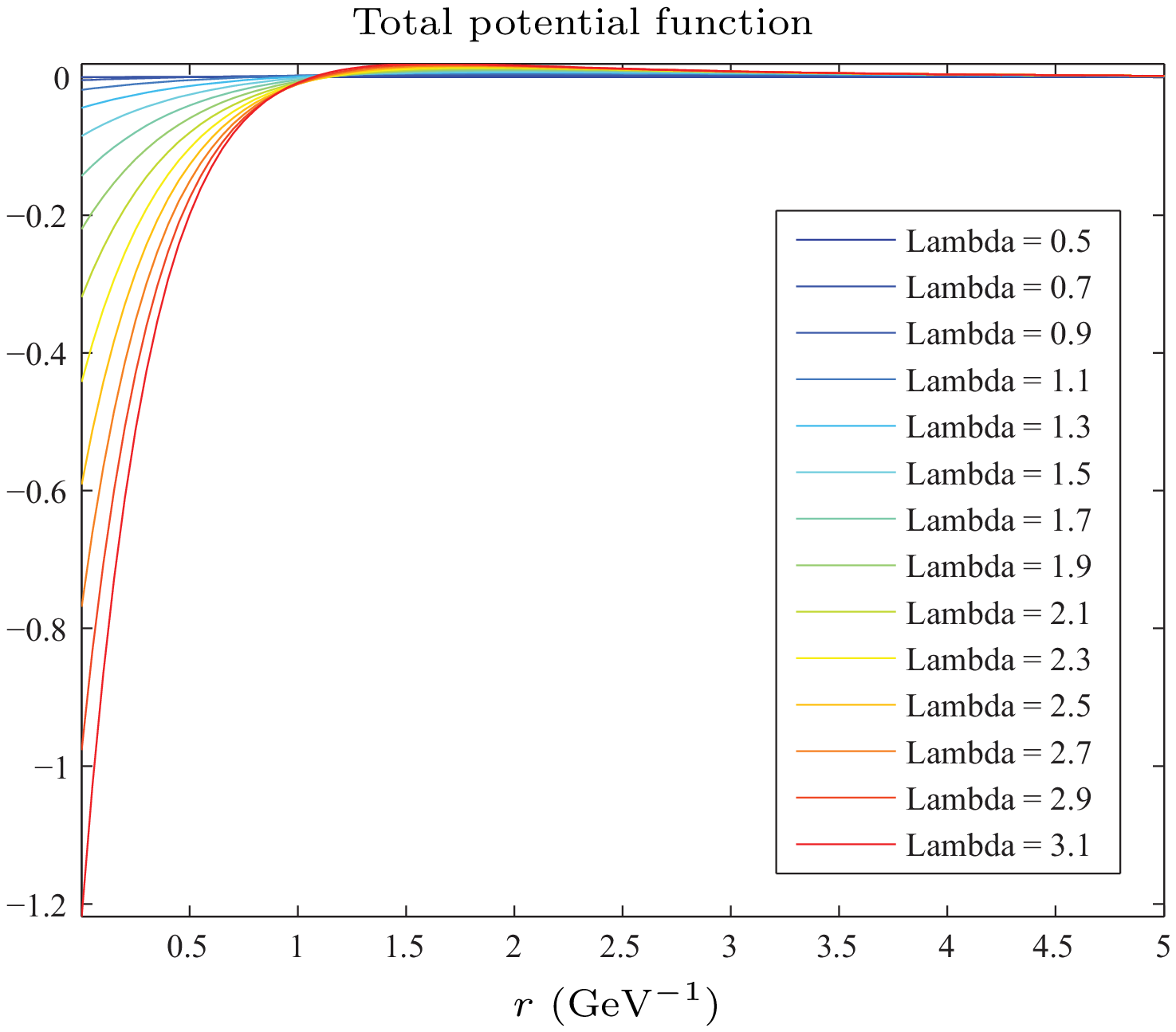}}\\
(c)&(d)\\
\scalebox{0.5}{\includegraphics{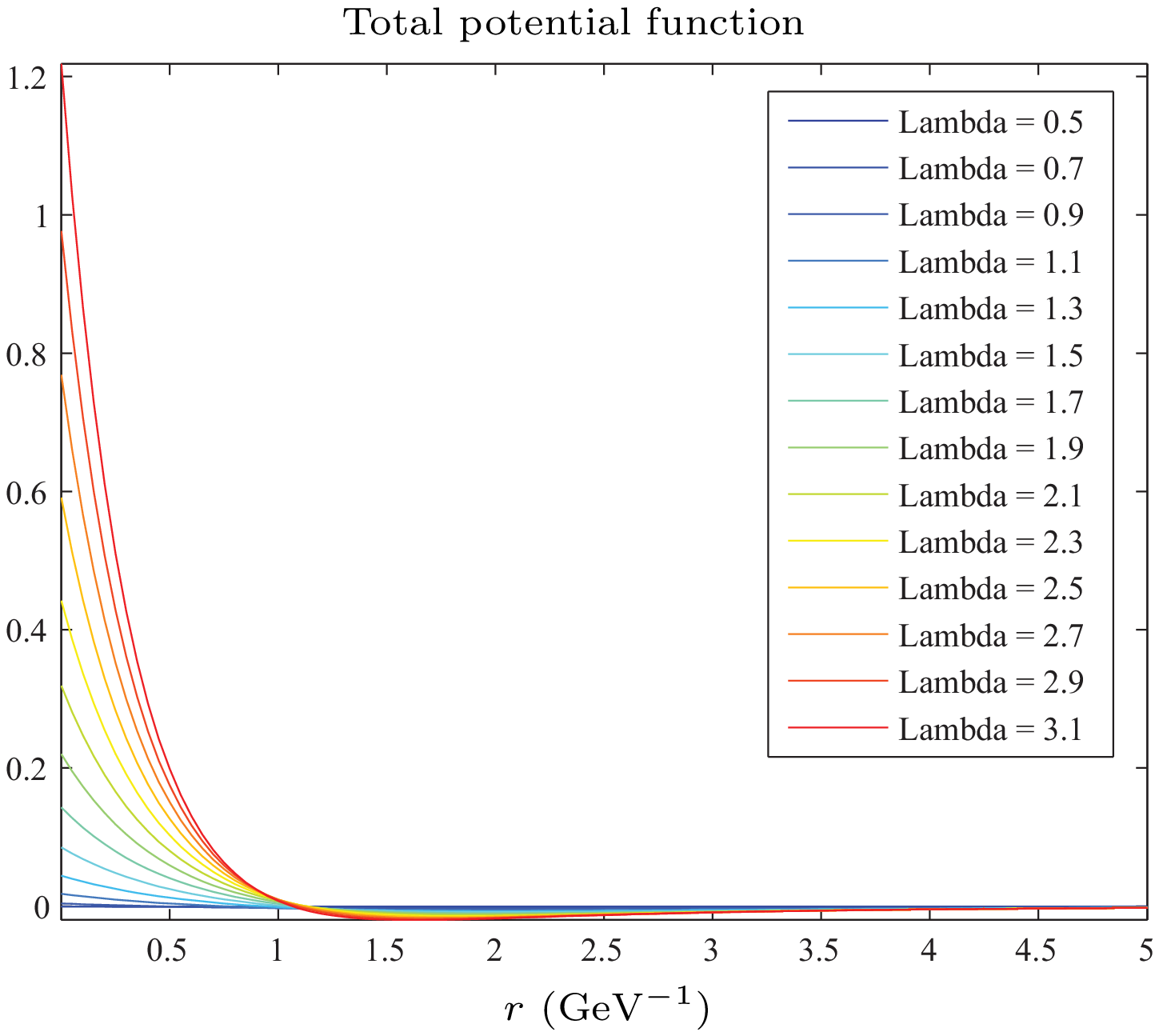}}&\scalebox{0.5}{\includegraphics{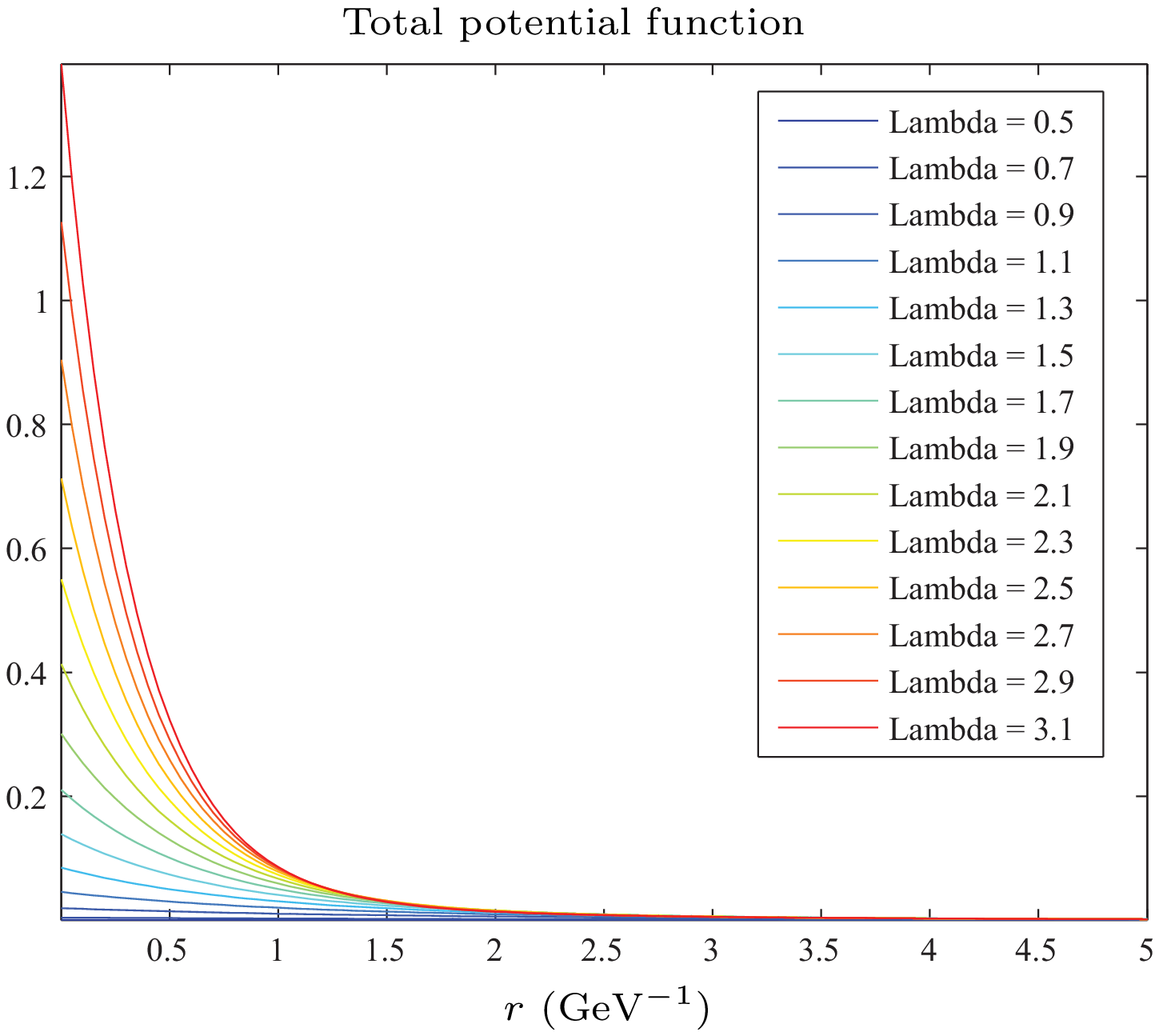}}\\
(e)&(f)\\
\end{tabular}
\caption{The variation of the potential of the
$[D^*D_1^\prime]_s^{\pm,0}$, $[D^*D_1^\prime]_{\hat s}^{0}$
system, where we only list several typical cases with parameters
$(c=+1,\,J=0,\,g g^\prime>0)$, $(c=-1,\,J=0,\,g g^\prime>0)$,
$(c=+1,\,J=1,\,g g^\prime>0)$, $(c=-1,\,J=1,\,g g^\prime>0)$,
$(c=+1,\,J=2,\,g g^\prime>0)$ and $(c=-1,\,J=0,\,g g^\prime>0)$,
which correspond to diagrams (a), (b), (c), (d), (e) and (f)
respectively. The above results are obtained with $1h$. Here,
$|h|=0.56$, $|g|=0.75$ and $|g^\prime|=0.25$.
\label{dsd1s2-potential}}
\end{figure}
\end{center}

The effective potential of the
$[P^*P_1^\prime]_s^{\pm,0},\,[P^*P_1^\prime]_{\hat{s}}^{0}$ system
arises from the $\eta$ meson exchange only, which is plotted in
Fig. \ref{dsd1s2-potential} under several typical combinations of
parameters.

\begin{center}
\begin{figure}[htb]
\begin{tabular}{cccc}
\scalebox{0.53}{\includegraphics{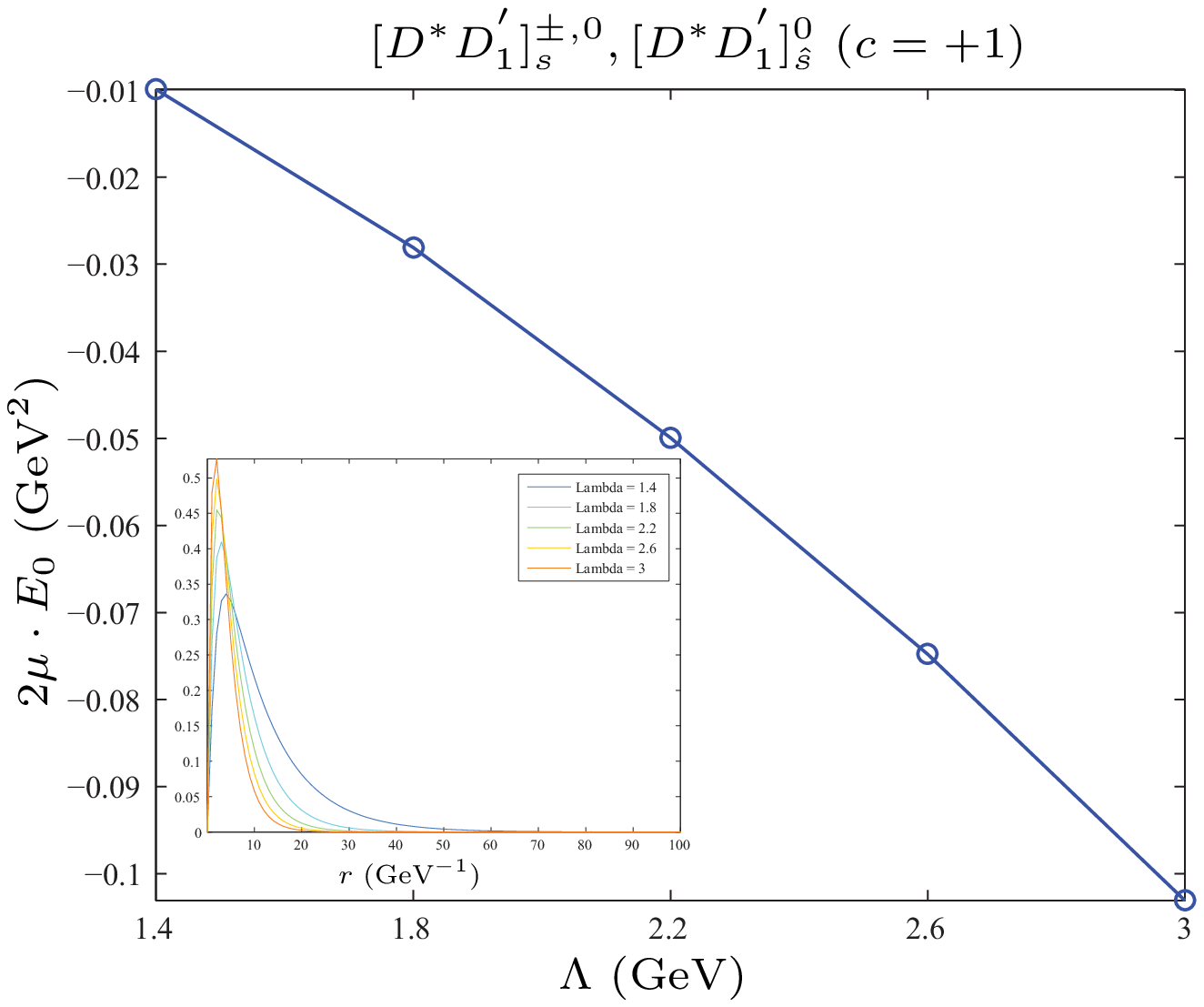}}&\scalebox{0.53}{\includegraphics{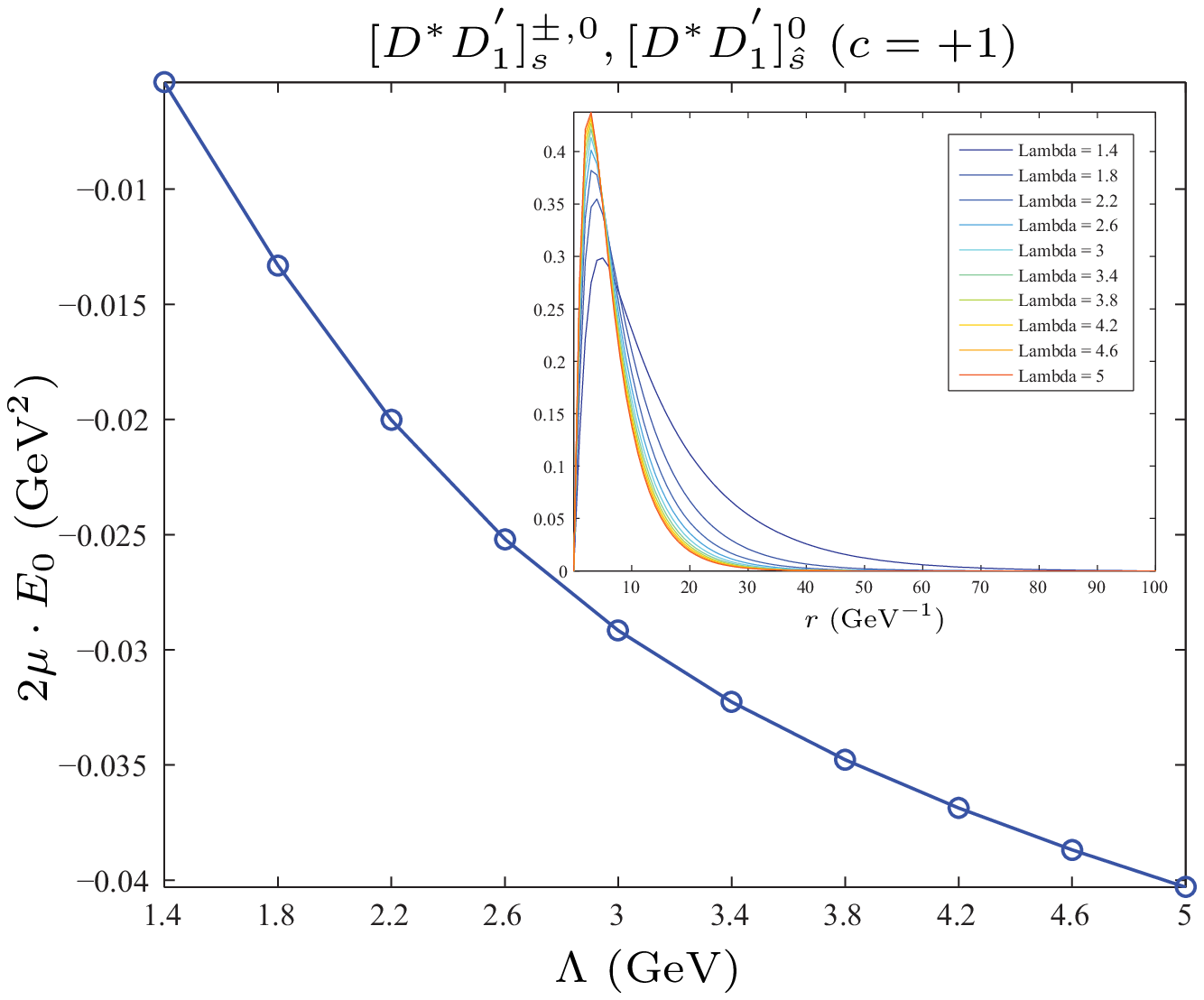}}\\
(a)&(b)\\
\scalebox{0.53}{\includegraphics{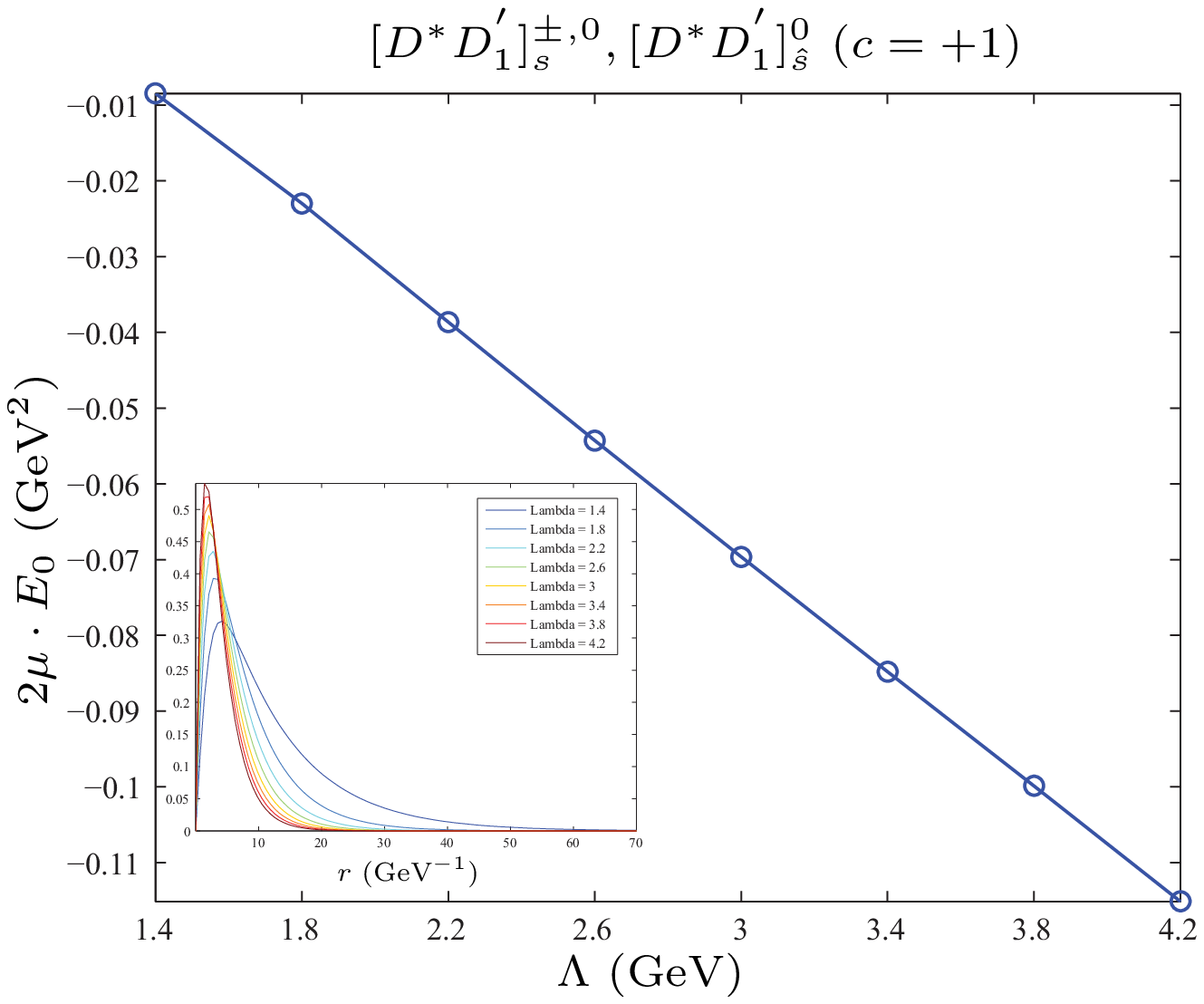}}&\scalebox{0.53}{\includegraphics{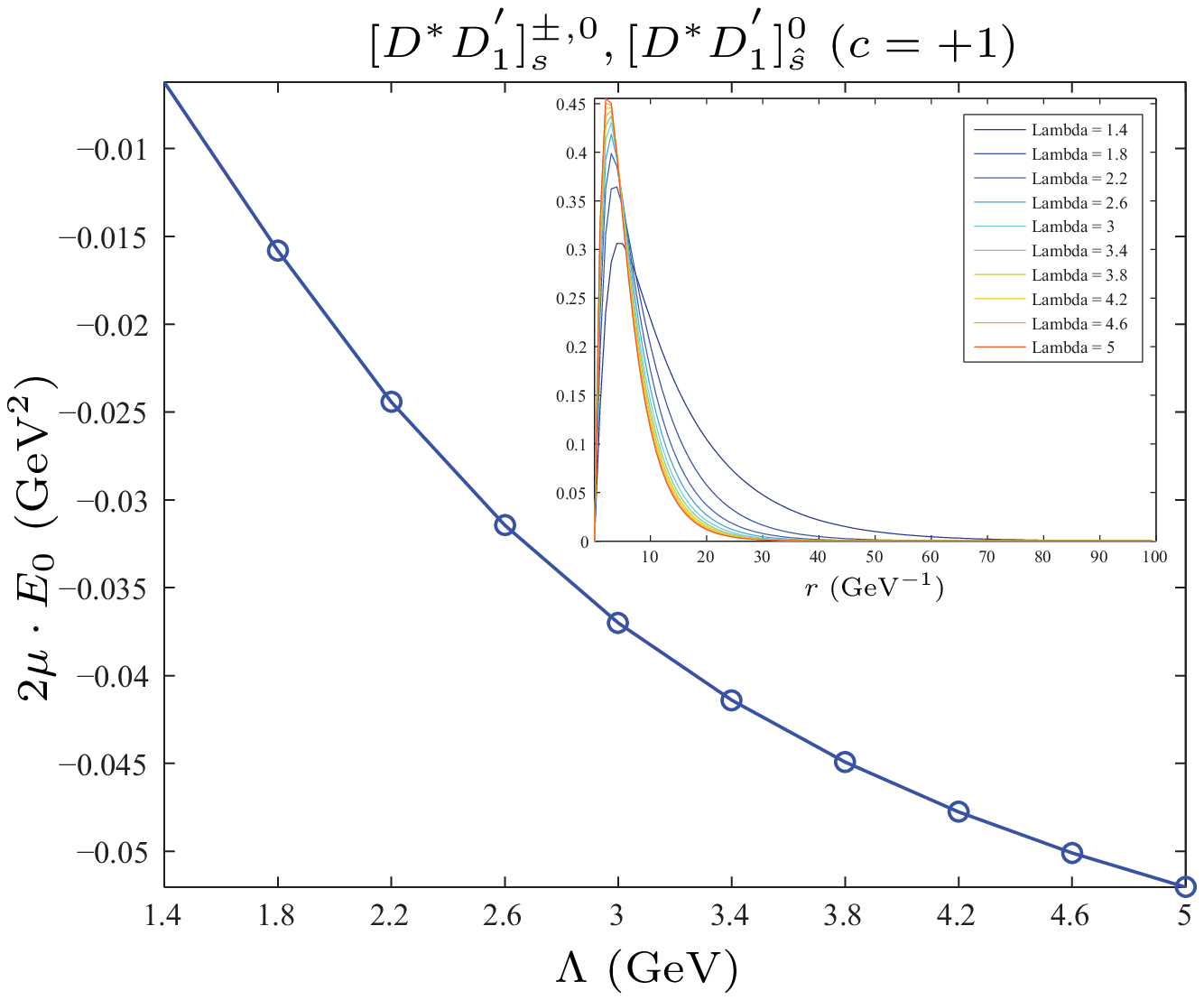}}\\
(c)&(d)\\
\scalebox{0.53}{\includegraphics{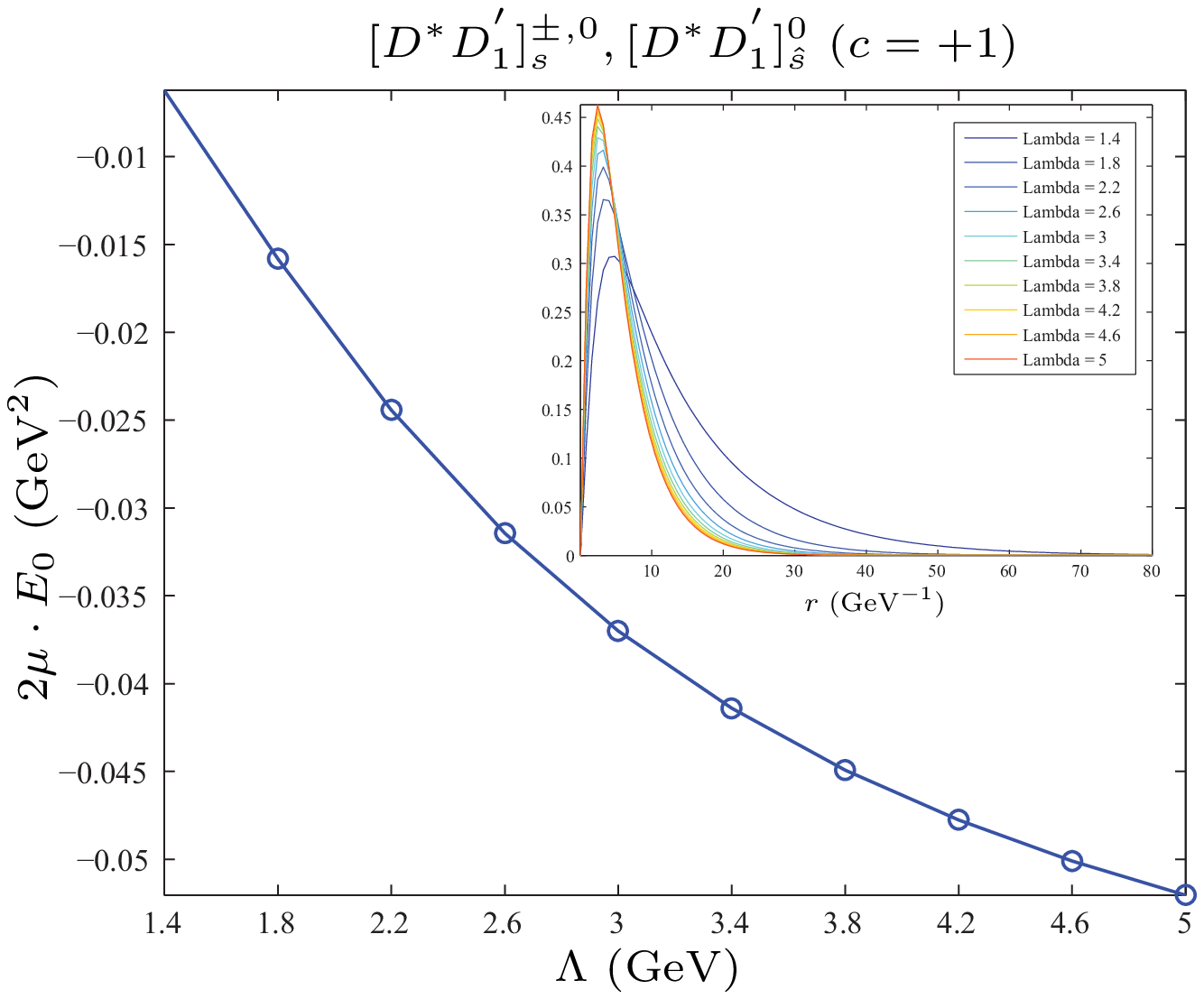}}&\scalebox{0.53}{\includegraphics{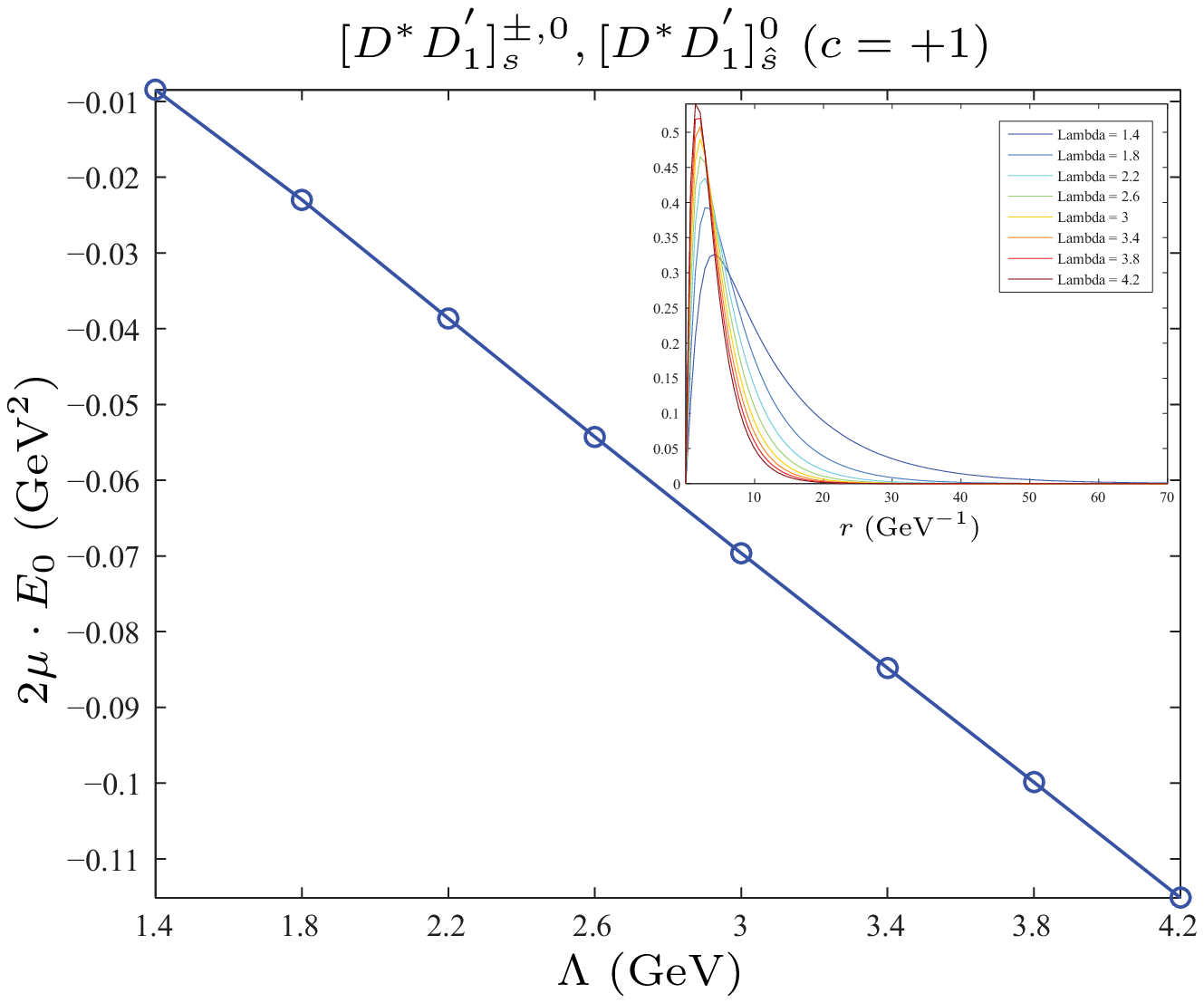}}\\
(e)&(f)\\
\end{tabular}
\caption{The bound state solutions of the
$[D^*D_1^\prime]_s^{\pm,0}$, $[D^*D_1^\prime]_{\hat s}^{0}$ system
with $(c=+1,\,J=0,\,g g^\prime>0)$, $(c=+1,\,J=0,\,g g^\prime<0)$,
$(c=+1,\,J=1,\,g g^\prime>0)$, $(c=+1,\,J=1,\,g g^\prime<0)$,
$(c=+1,\,J=2,\,g g^\prime>0)$ and $(c=+1,\,J=0,\,g g^\prime<0)$,
which correspond to diagrams (a), (b), (c), (d), (e) and (f)
respectively. The above numerical results are obtained with $3h$
while we can not find the bound state solution for $1h$ and $2h$.
Here, $|h|=0.56$, $|g|=0.75$ and $|g^\prime|=0.25$.
\label{dsd1s2-E}}
\end{figure}
\end{center}

When solving the the Schr\"{o}dinger equation, we obtain the bound
state solutions of the $[D^*D_1^\prime]_s^{\pm,0}$,
$[D^*D_1^\prime]_{\hat s}^{0}$ system, which are listed in Fig.
\ref{dsd1s2-E}. The bound state solution of the
$[D^*D_1^\prime]_s^{\pm,0}$, $[D^*D_1^\prime]_{\hat s}^{0}$ system
appears when $(c=+1,\,J=0,\,g g^\prime>0)$, $(c=+1,\,J=0,\,g g^\prime<0)$,
$(c=+1,\,J=1,\,g g^\prime>0)$, $(c=+1,\,J=1,\,g g^\prime<0)$,
$(c=+1,\,J=2,\,g g^\prime>0)$ and $(c=+1,\,J=0,\,g g^\prime<0)$. The above numerical results are obtained with $3h$
while we can not find the bound state solution for $1h$ and $2h$.

\subsubsection{$[P^*P_0^\prime]_{s1}^{0}$}

\begin{center}
\begin{figure}[htb]
\begin{tabular}{cccc}
\scalebox{0.5}{\includegraphics{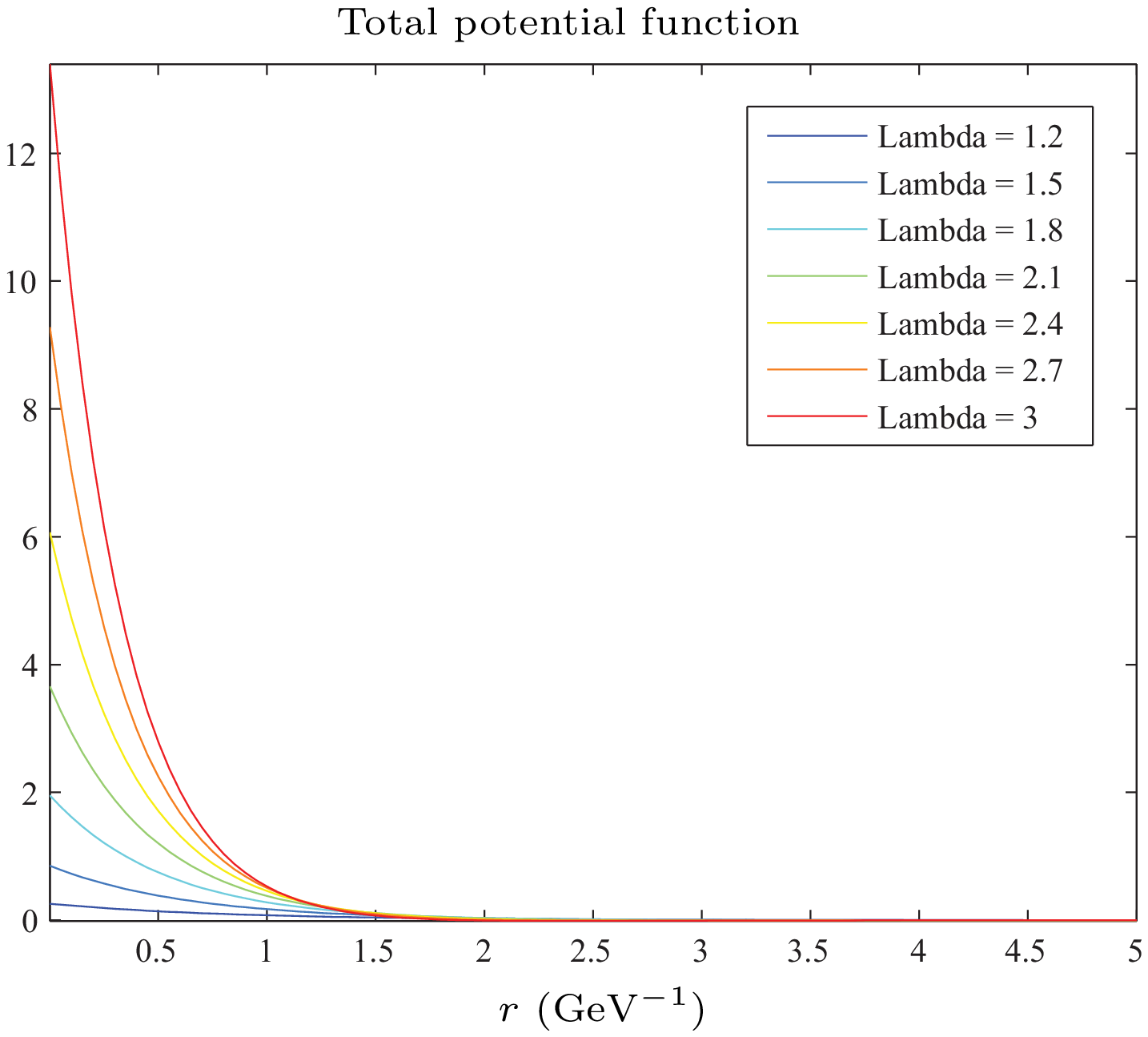}}&\scalebox{0.5}{\includegraphics{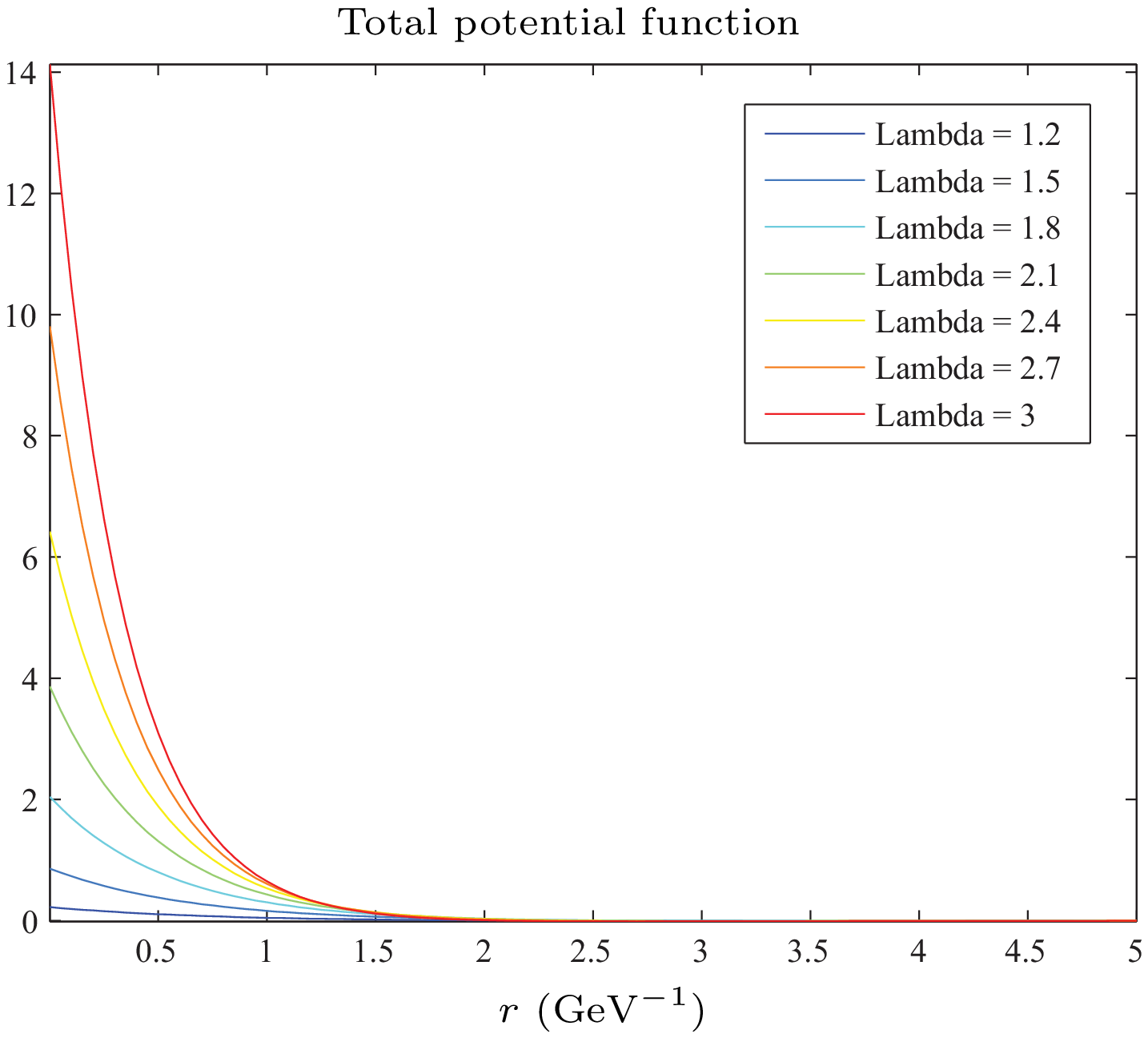}}\\
(a)&(b)\\
\scalebox{0.5}{\includegraphics{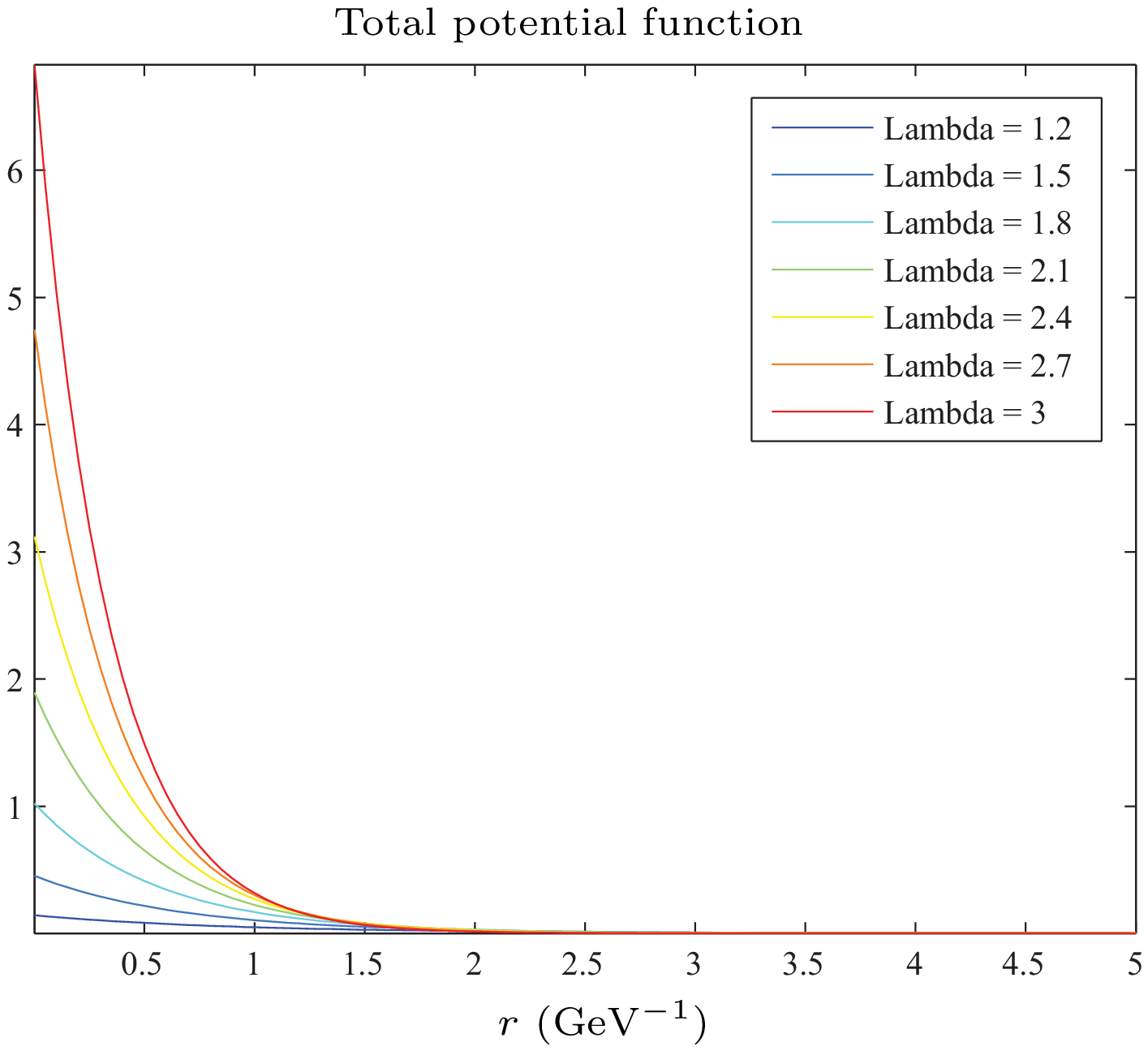}}&\scalebox{0.5}{\includegraphics{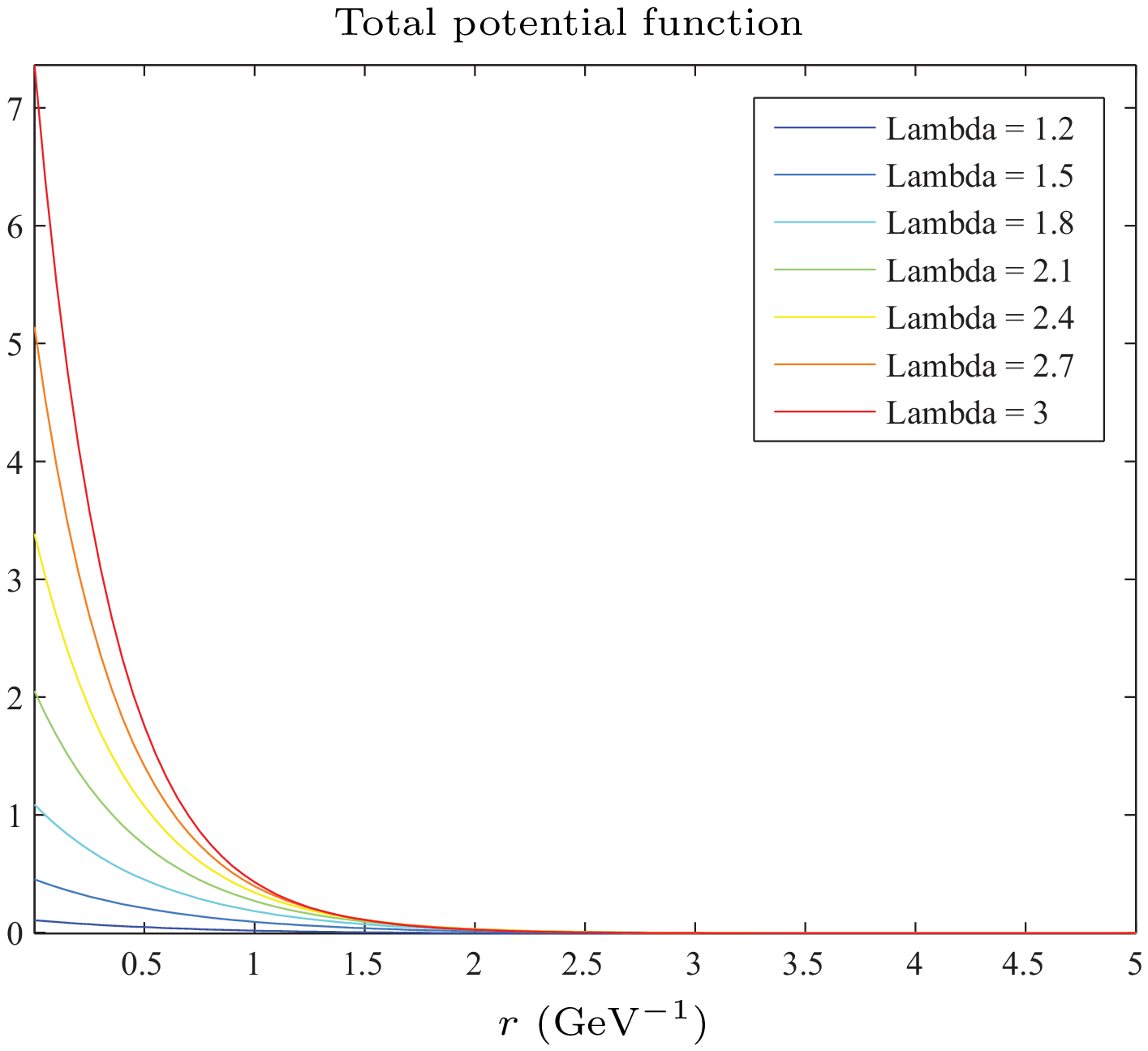}}\\
(c)&(d)\\
\scalebox{0.5}{\includegraphics{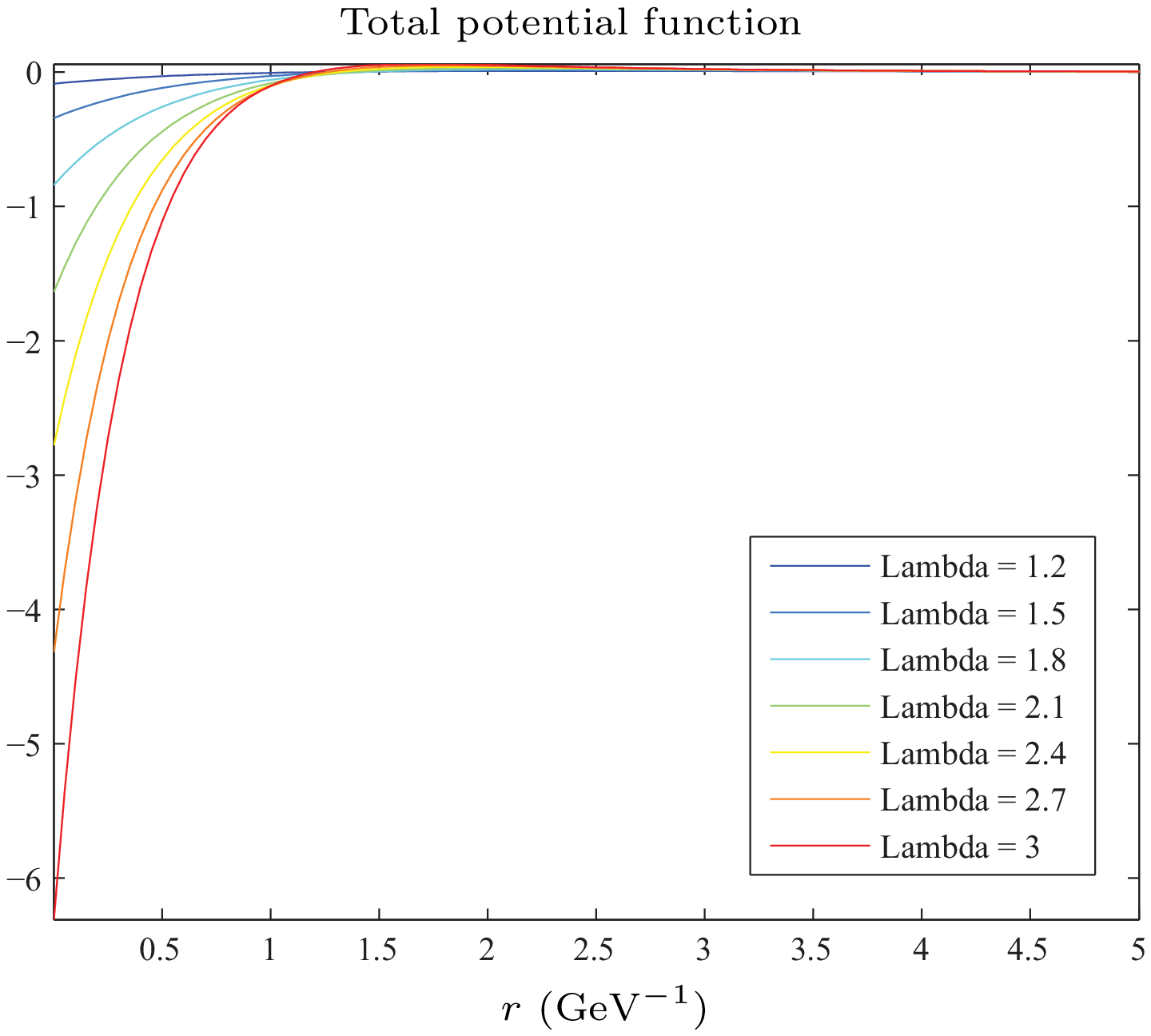}}&\scalebox{0.5}{\includegraphics{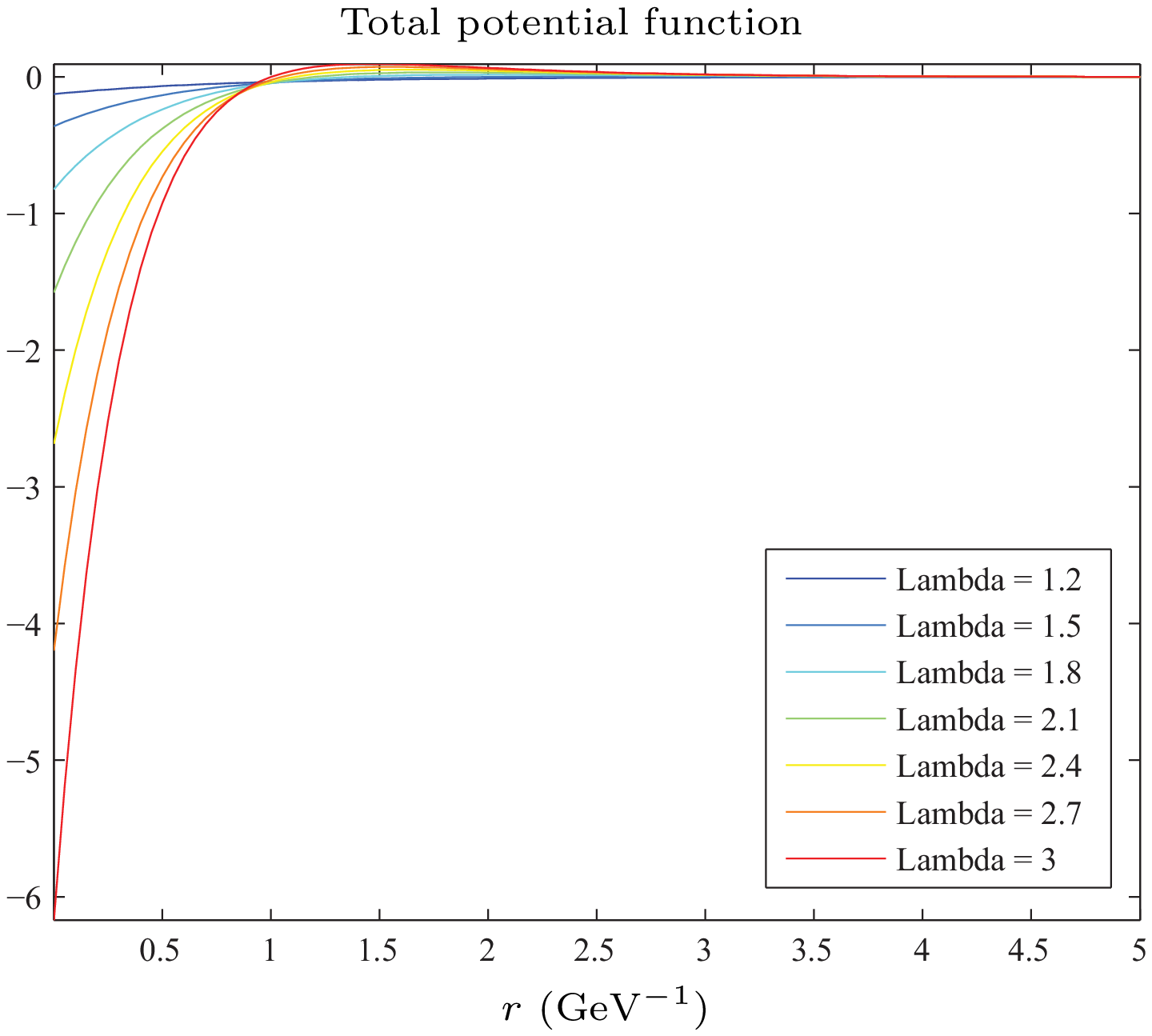}}\\
(e)&(f)\\
\end{tabular}
\caption{The variation of the potential of the
$[D^*D_1^\prime]_{s1}^{0}$ system with $r$. (a), (b), (c), (d),
(e) and (f) are for parameters
$(c=+1,\,J=0,\,\beta\beta^\prime>0,\,\lambda\lambda^\prime>0,\,\zeta\varpi>0)$,
$(c=-1,\,J=0,\,\beta\beta^\prime>0,\,\lambda\lambda^\prime>0,\,\zeta\varpi>0)$,
$(c=+1,\,J=1,\,\beta\beta^\prime>0,\,\lambda\lambda^\prime>0,\,\zeta\varpi>0)$,
$(c=-1,\,J=1,\,\beta\beta^\prime>0,\,\lambda\lambda^\prime>0,\,\zeta\varpi>0)$,
$(c=+1,\,J=2,\,\beta\beta^\prime>0,\,\lambda\lambda^\prime>0,\,\zeta\varpi>0)$
and
$(c=-1,\,J=2,\,\beta\beta^\prime>0,\,\lambda\lambda^\prime>0,\,\zeta\varpi>0)$,
respectively. Here, $|h|=0.56$, $|g|=0.75$, $|g^\prime|=0.25$,
$|\lambda|=0.4546$, $|\lambda^\prime|=0.2667$, $|\beta|=0.909$,
$|\beta^\prime|=0.533$, $|\zeta|=0.727$ and $|\varpi|=0.364$.
\label{dsd1s3-potential}}
\end{figure}
\end{center}

\begin{center}
\begin{figure}[htb]
\begin{tabular}{cccc}
\scalebox{0.5}{\includegraphics{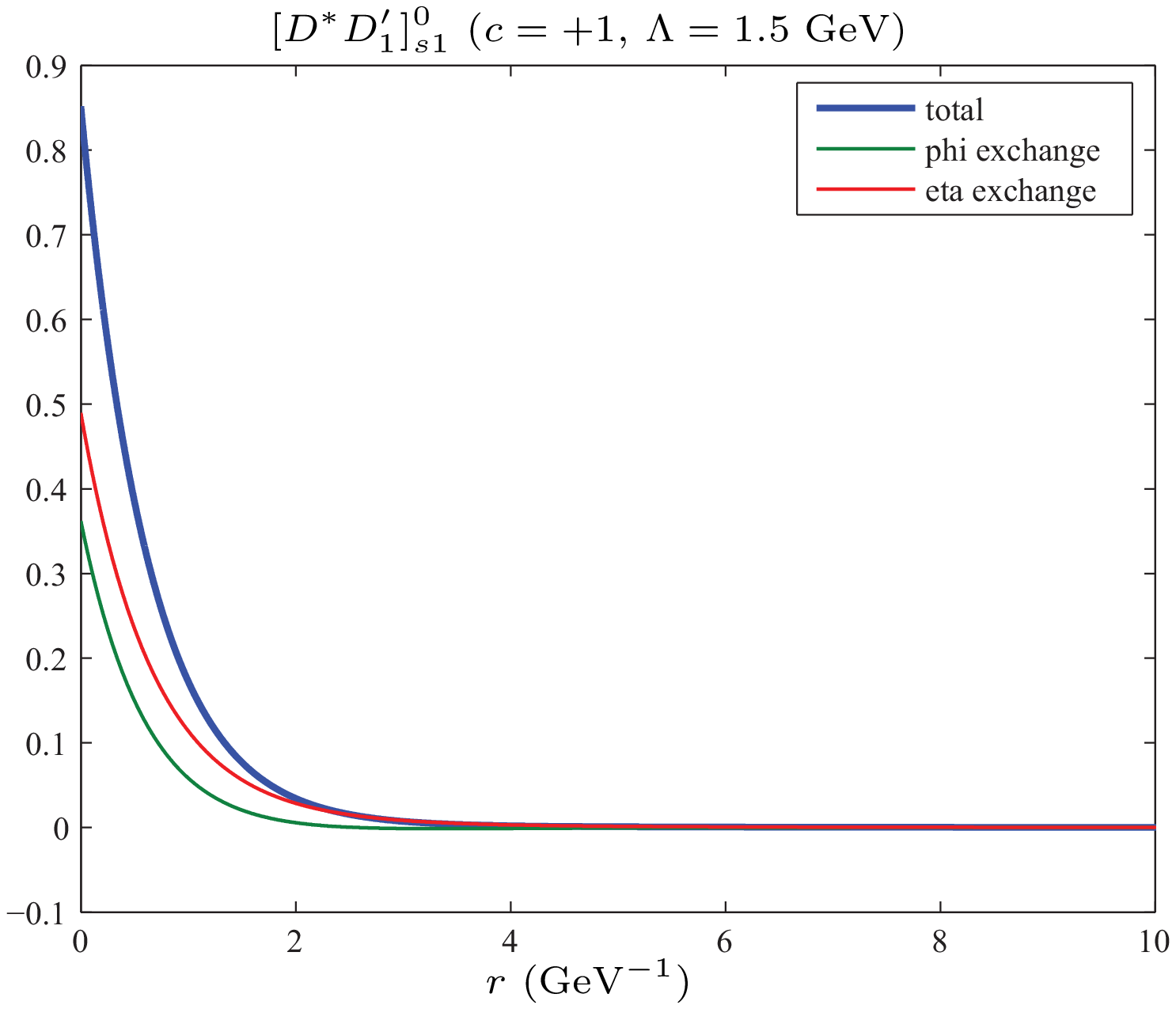}}&\scalebox{0.5}{\includegraphics{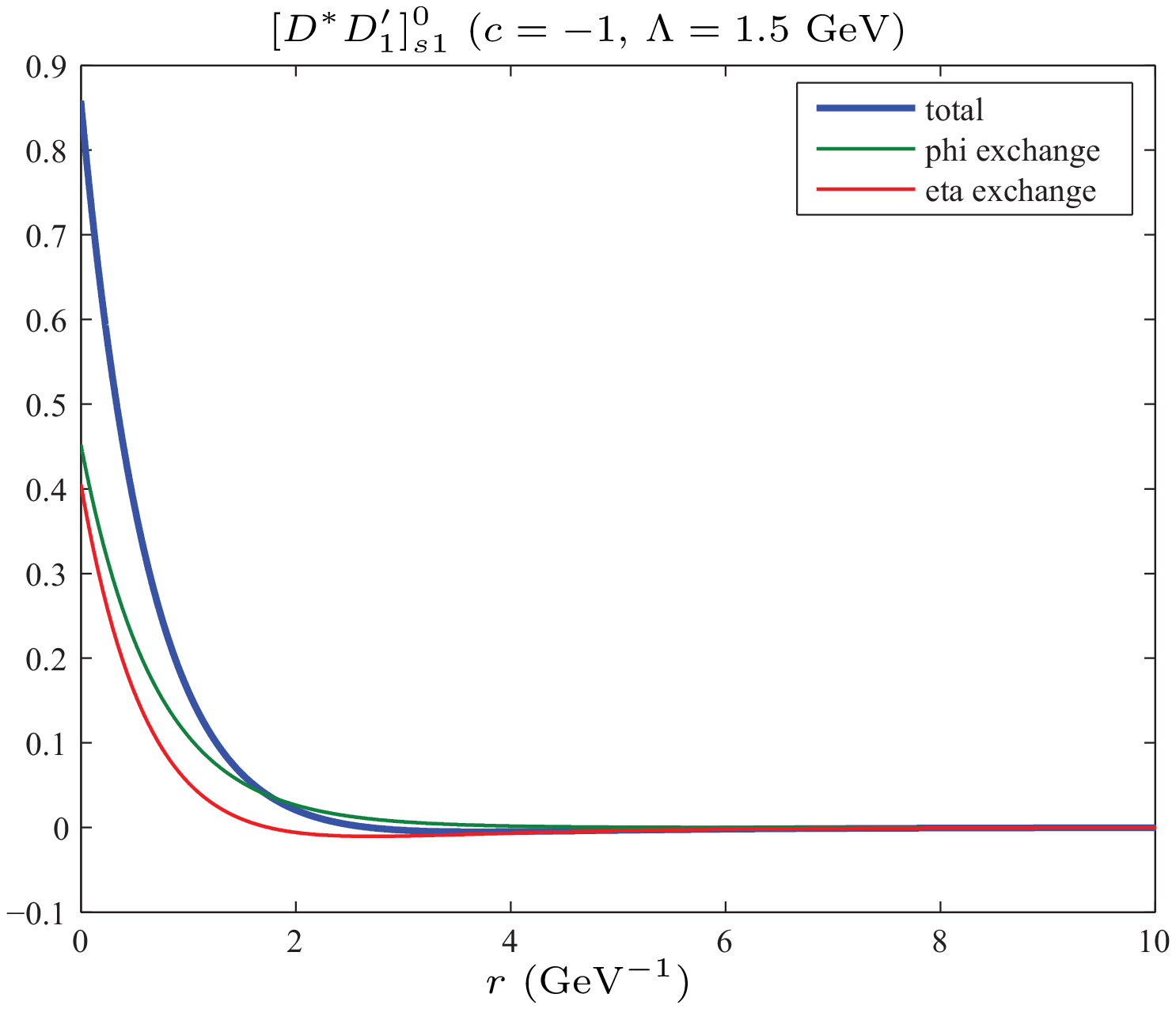}}\\
(a)&(b)\\
\scalebox{0.5}{\includegraphics{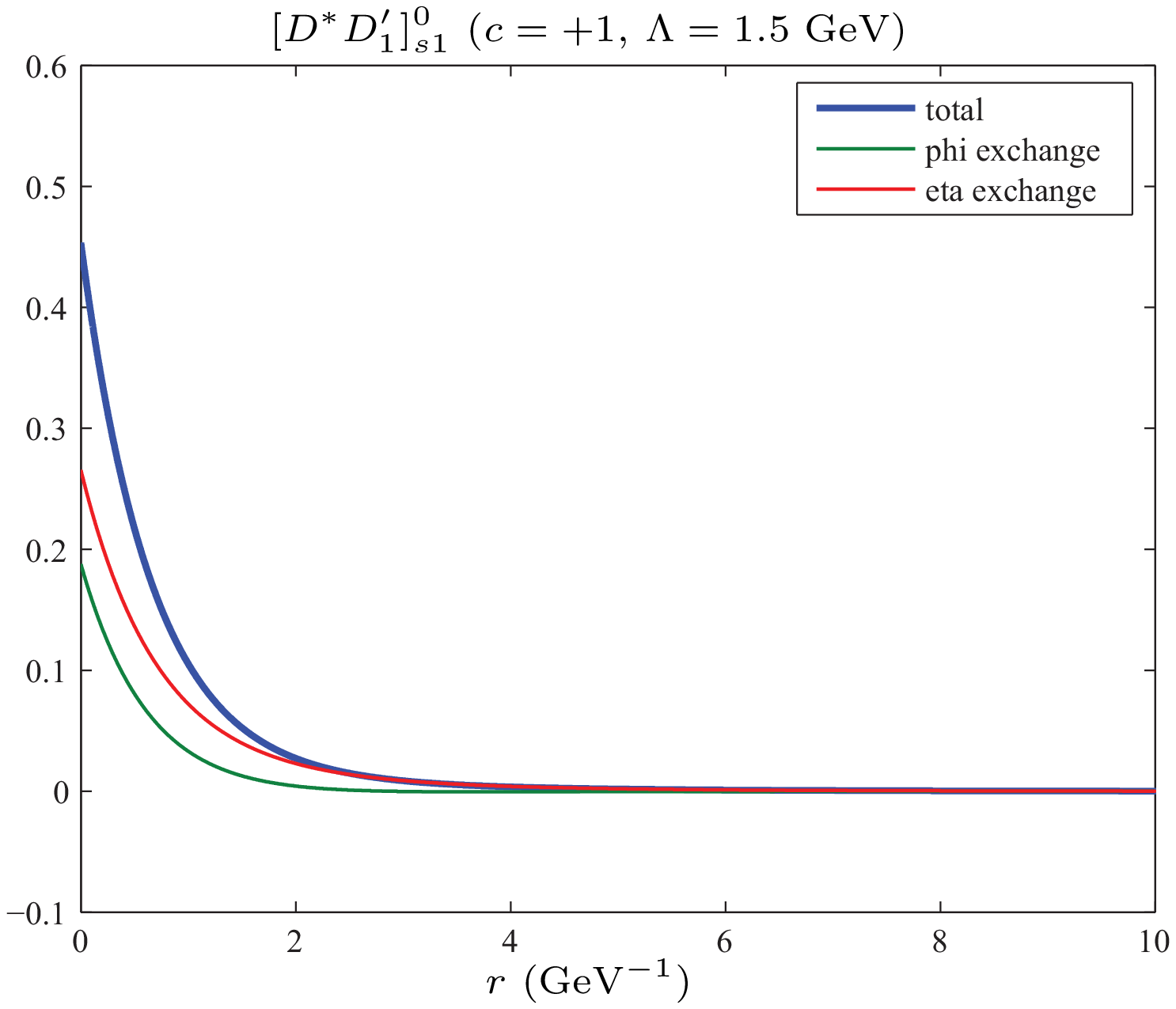}}&\scalebox{0.5}{\includegraphics{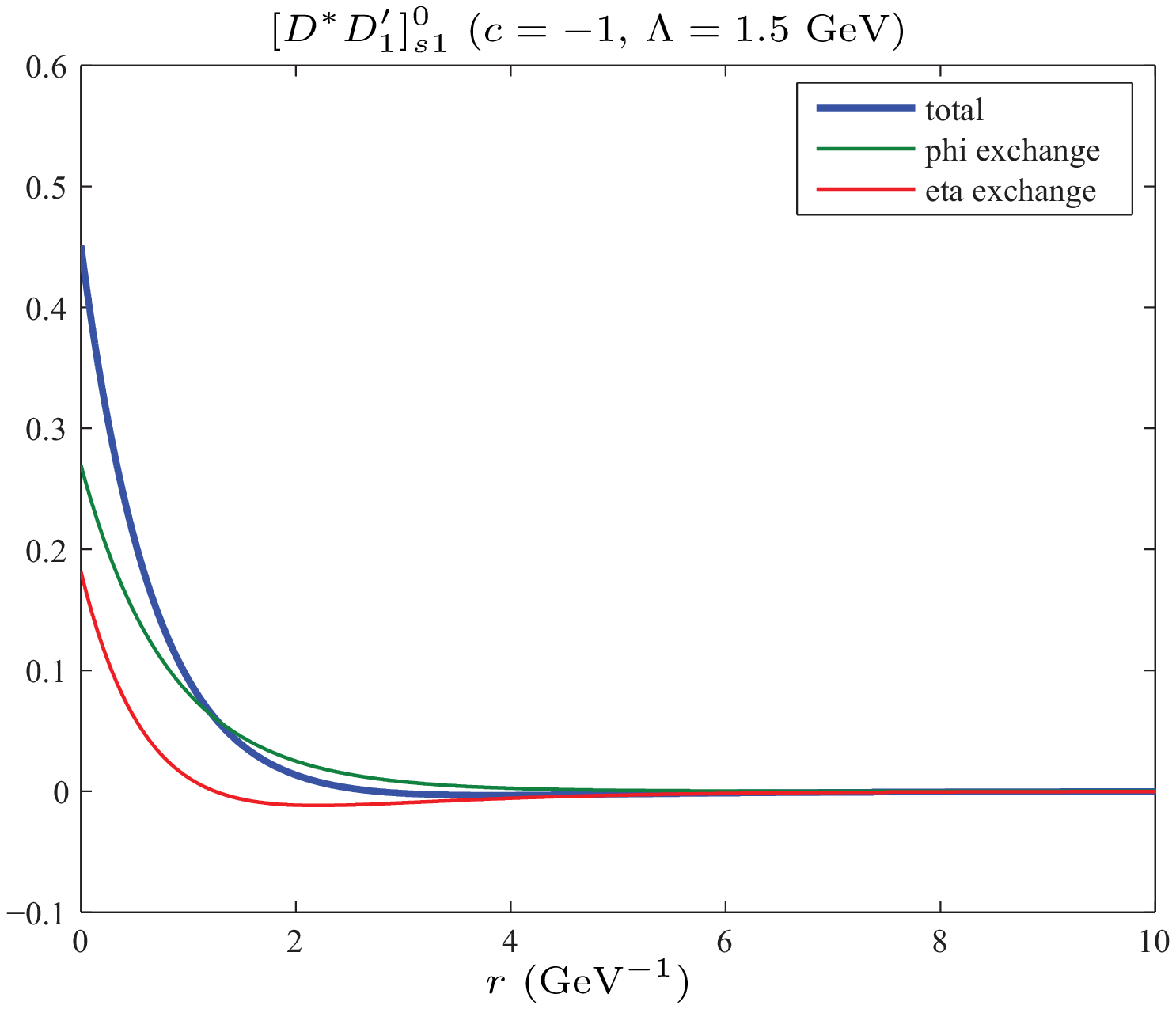}}\\
(c)&(d)\\
\scalebox{0.5}{\includegraphics{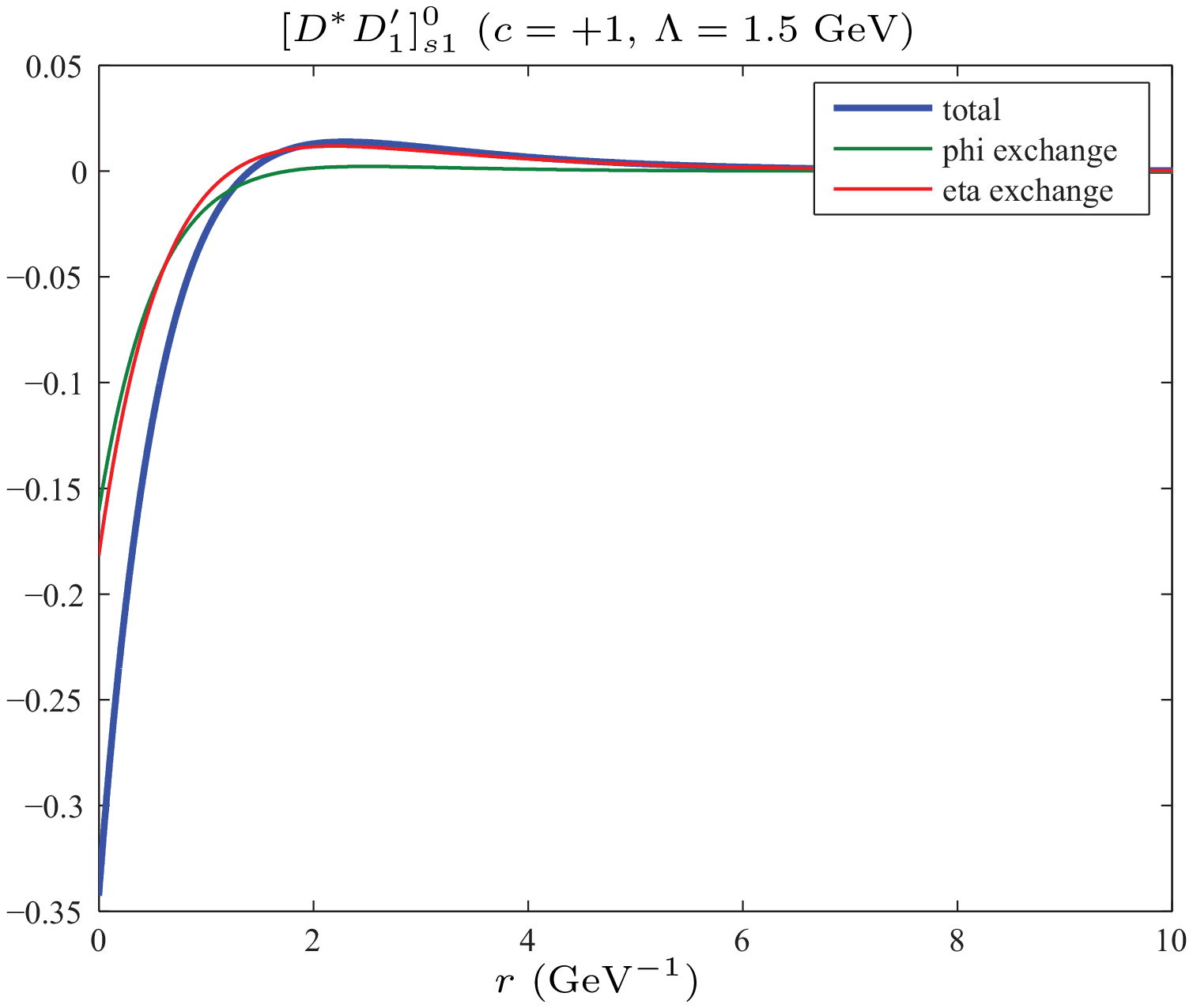}}&\scalebox{0.5}{\includegraphics{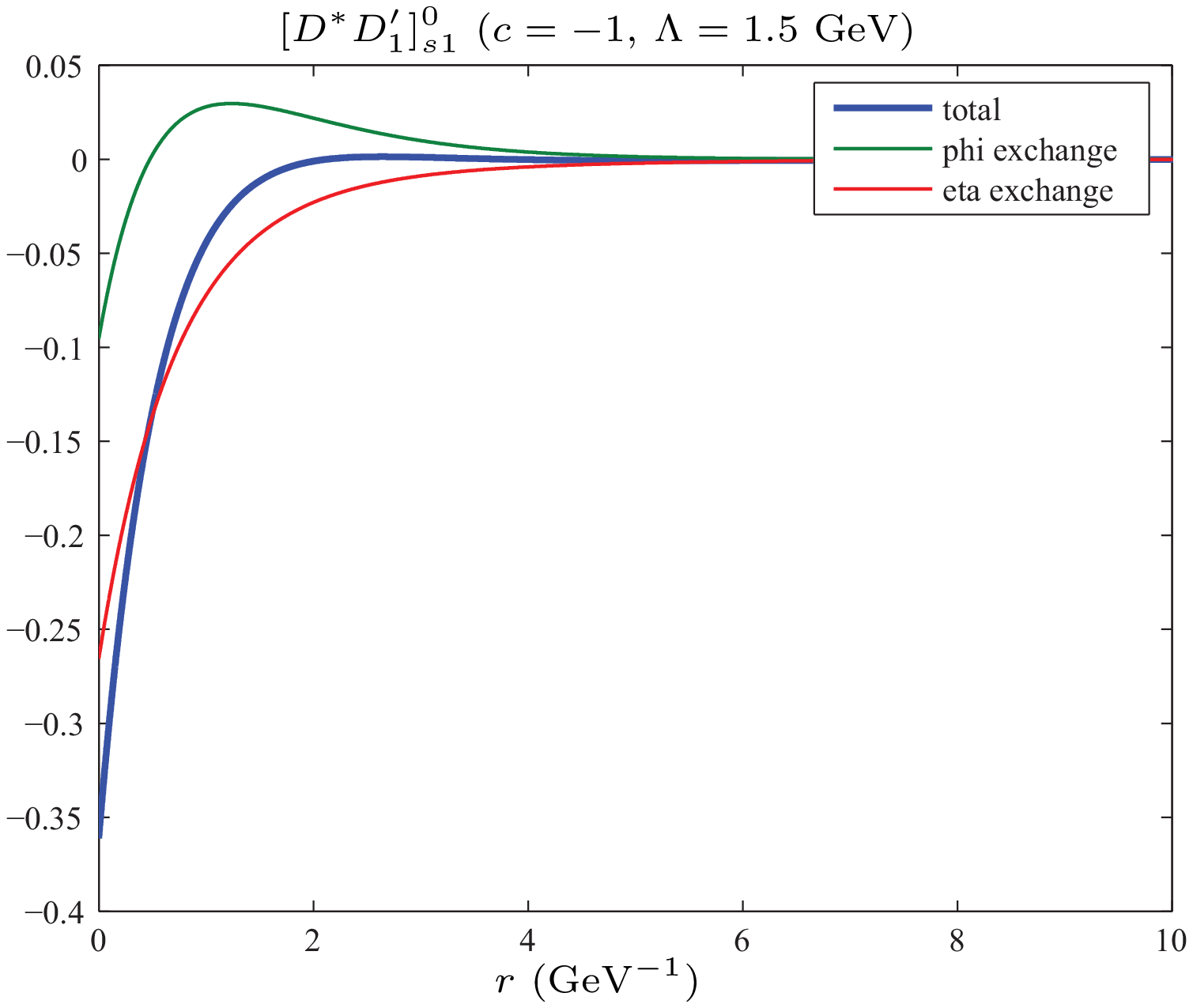}}\\
(e)&(f)\\
\end{tabular}
\caption{The comparison of the $\eta$ and $\phi$ exchange
potentials with the total effective potential for the
$[D^*D_1^\prime]_{s1}^{0}$ system with $\Lambda=1.5$ GeV.  (a),
(b), (c), (d), (e) and (f) are for parameters
$(c=+1,\,J=0,\,\beta\beta^\prime>0,\,\lambda\lambda^\prime>0,\,g
g^\prime>0,\,\zeta\varpi>0)$,
$(c=-1,\,J=0,\,\beta\beta^\prime>0,\,\lambda\lambda^\prime>0,\,g
g^\prime>0,\,\zeta\varpi>0)$,
$(c=+1,\,J=1,\,\beta\beta^\prime>0,\,\lambda\lambda^\prime>0,\,g
g^\prime>0,\,\zeta\varpi>0)$,
$(c=-1,\,J=1,\,\beta\beta^\prime>0,\,\lambda\lambda^\prime>0,\,g
g^\prime>0,\,\zeta\varpi>0)$,
$(c=+1,\,J=2,\,\beta\beta^\prime>0,\,\lambda\lambda^\prime>0,\,g
g^\prime>0,\,\zeta\varpi>0)$ and
$(c=-1,\,J=2,\,\beta\beta^\prime>0,\,\lambda\lambda^\prime>0,\,g
g^\prime>0,\,\zeta\varpi>0)$, respectively. Here, $|h|=0.56$,
$|g|=0.75$, $|g^\prime|=0.25$, $|\lambda|=0.4546$,
$|\lambda^\prime|=0.2667$, $|\beta|=0.909$,
$|\beta^\prime|=0.533$, $|\zeta|=0.727$ and $|\varpi|=0.364$.
\label{dsd1s3-potential-1}}
\end{figure}
\end{center}

The effective potential of the $[P^*P_1^\prime]_{s1}^{0}$ system
results from both the $\eta$ and $\phi$ meson exchanges as shown
in Figs. \ref{dsd1s3-potential} and \ref{dsd1s3-potential-1}.

\begin{center}
\begin{figure}[htb]
\begin{tabular}{cccccc}
\scalebox{0.3}{\includegraphics{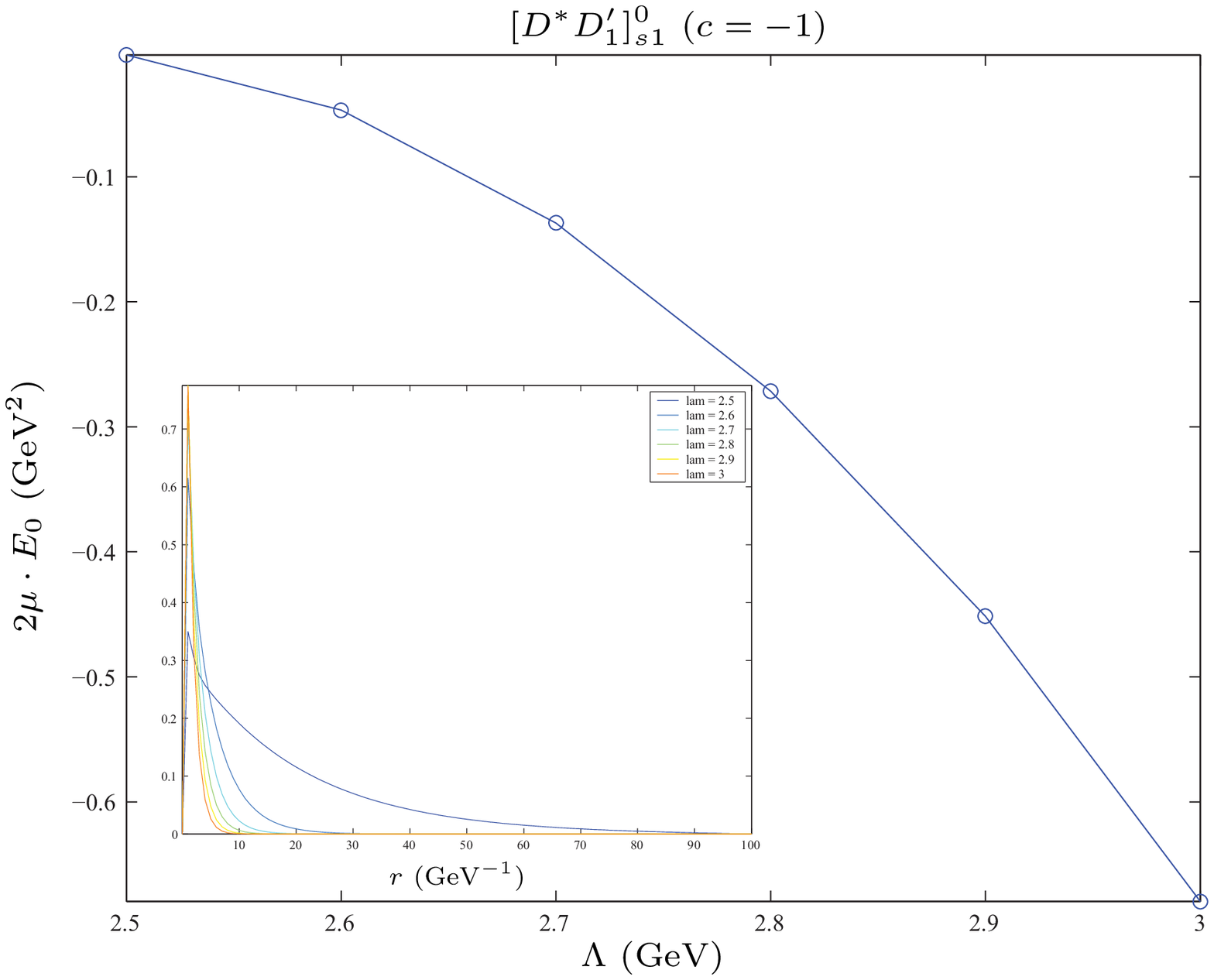}}&\scalebox{0.3}{\includegraphics{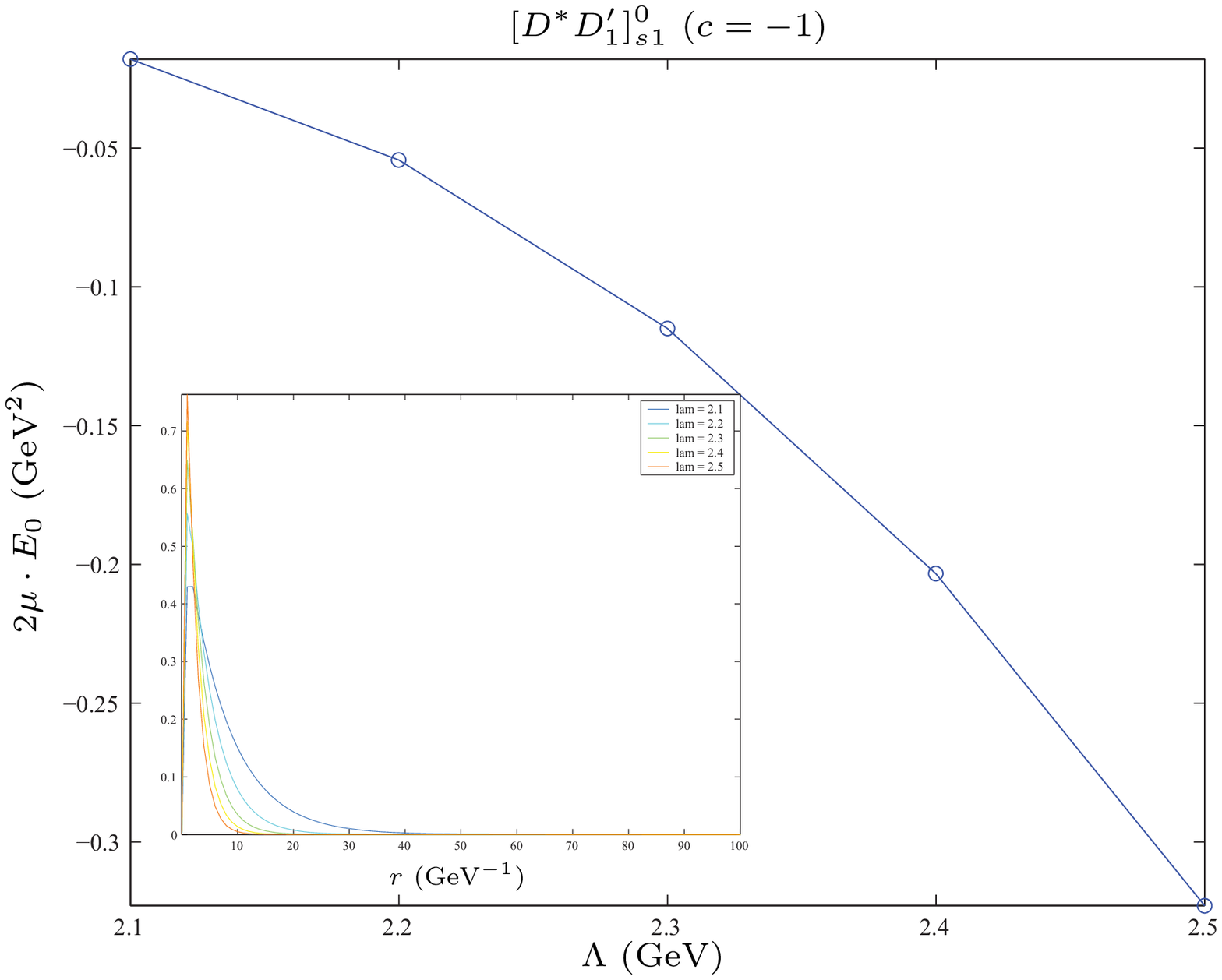}}&
\scalebox{0.3}{\includegraphics{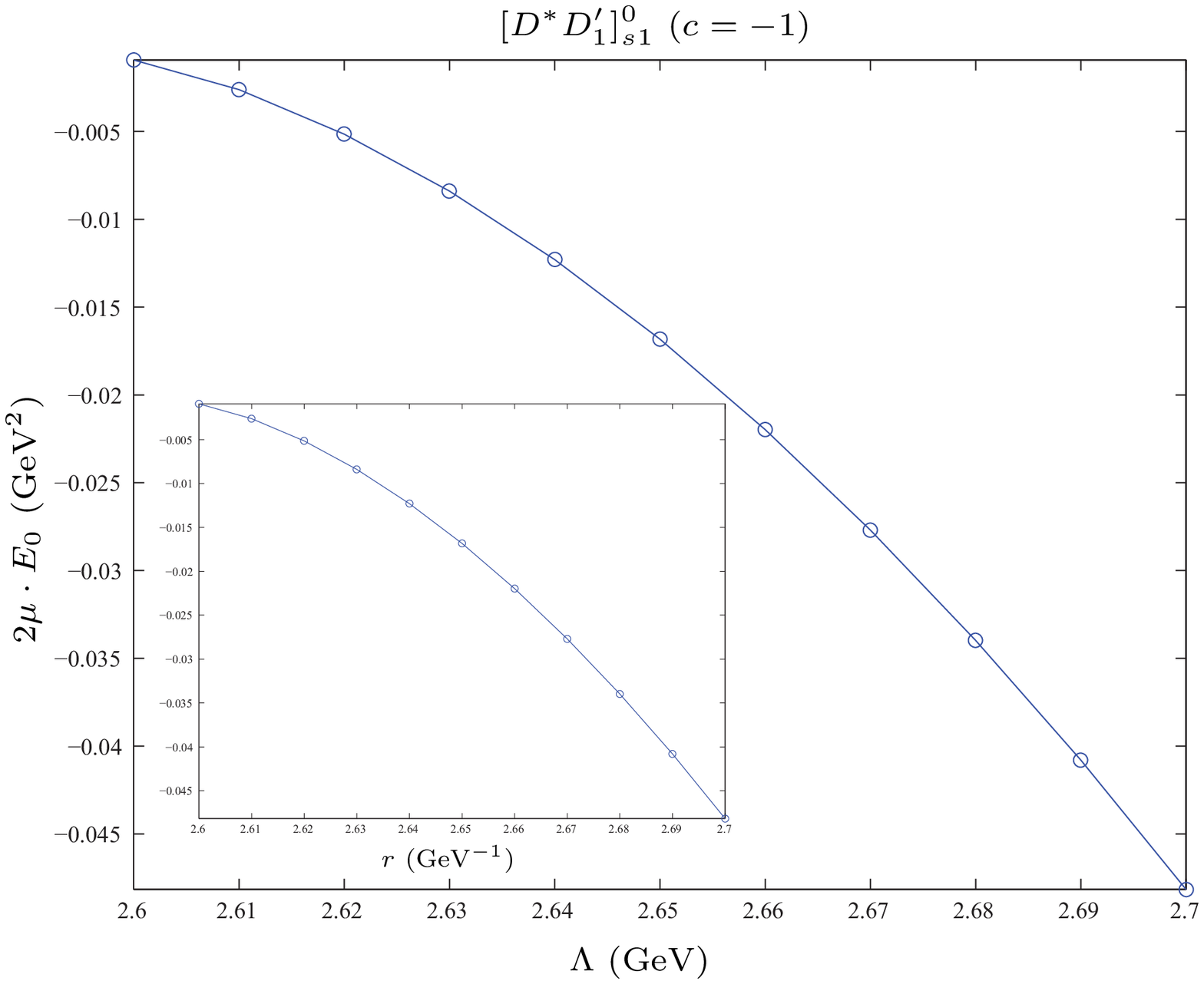}}\\
(a)&(b)&(c)\\
\scalebox{0.3}{\includegraphics{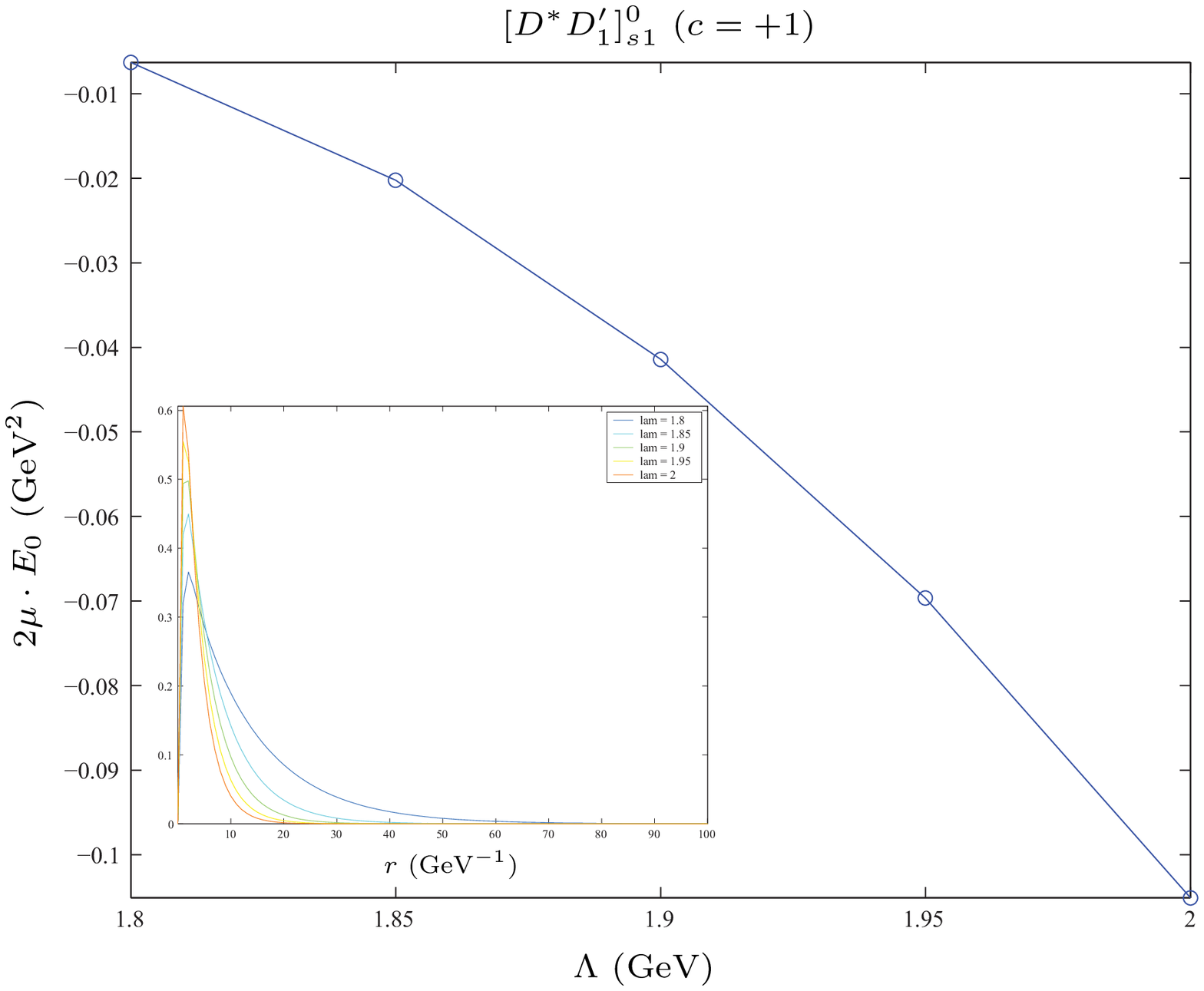}}&
\scalebox{0.3}{\includegraphics{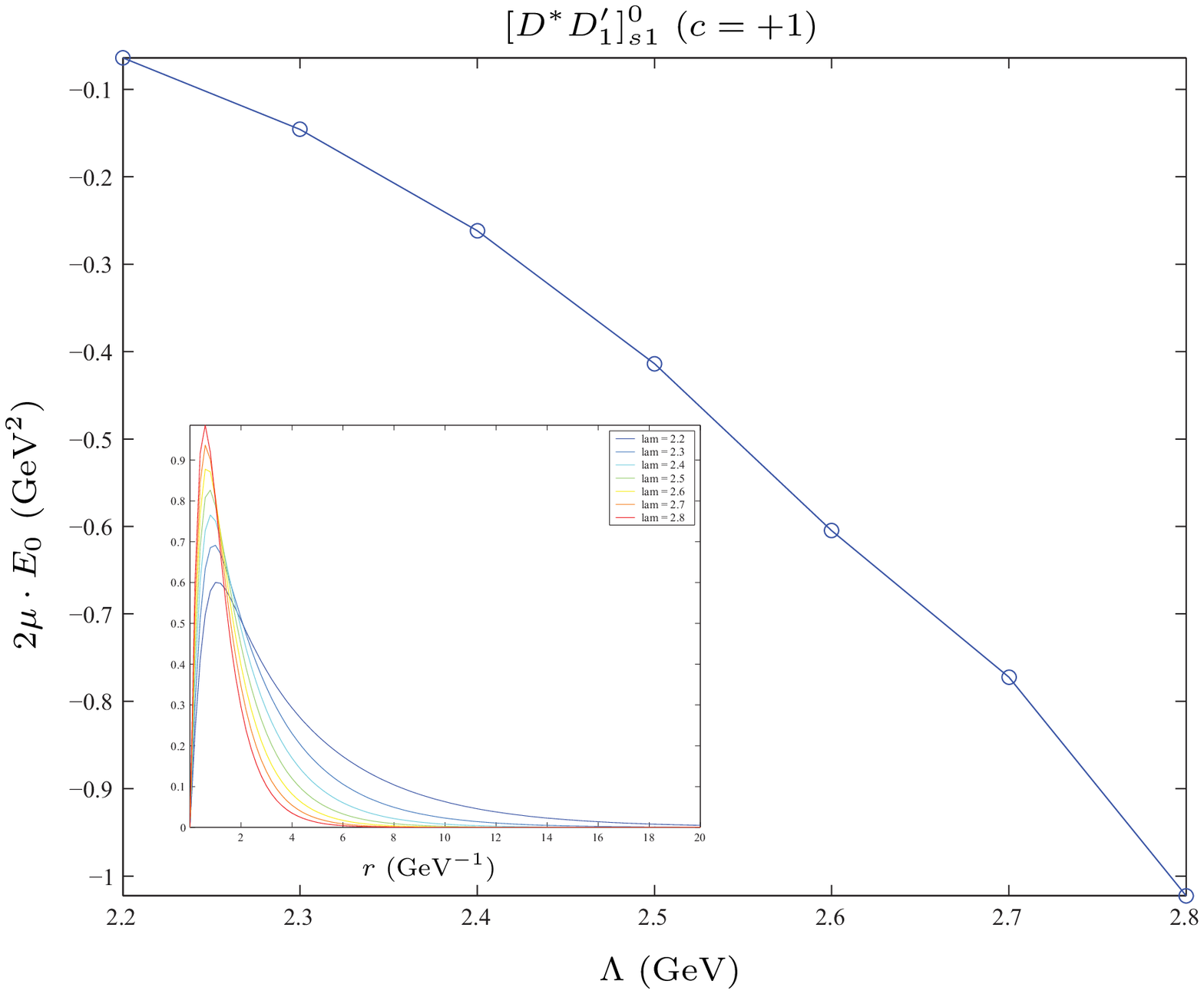}}&\scalebox{0.3}{\includegraphics{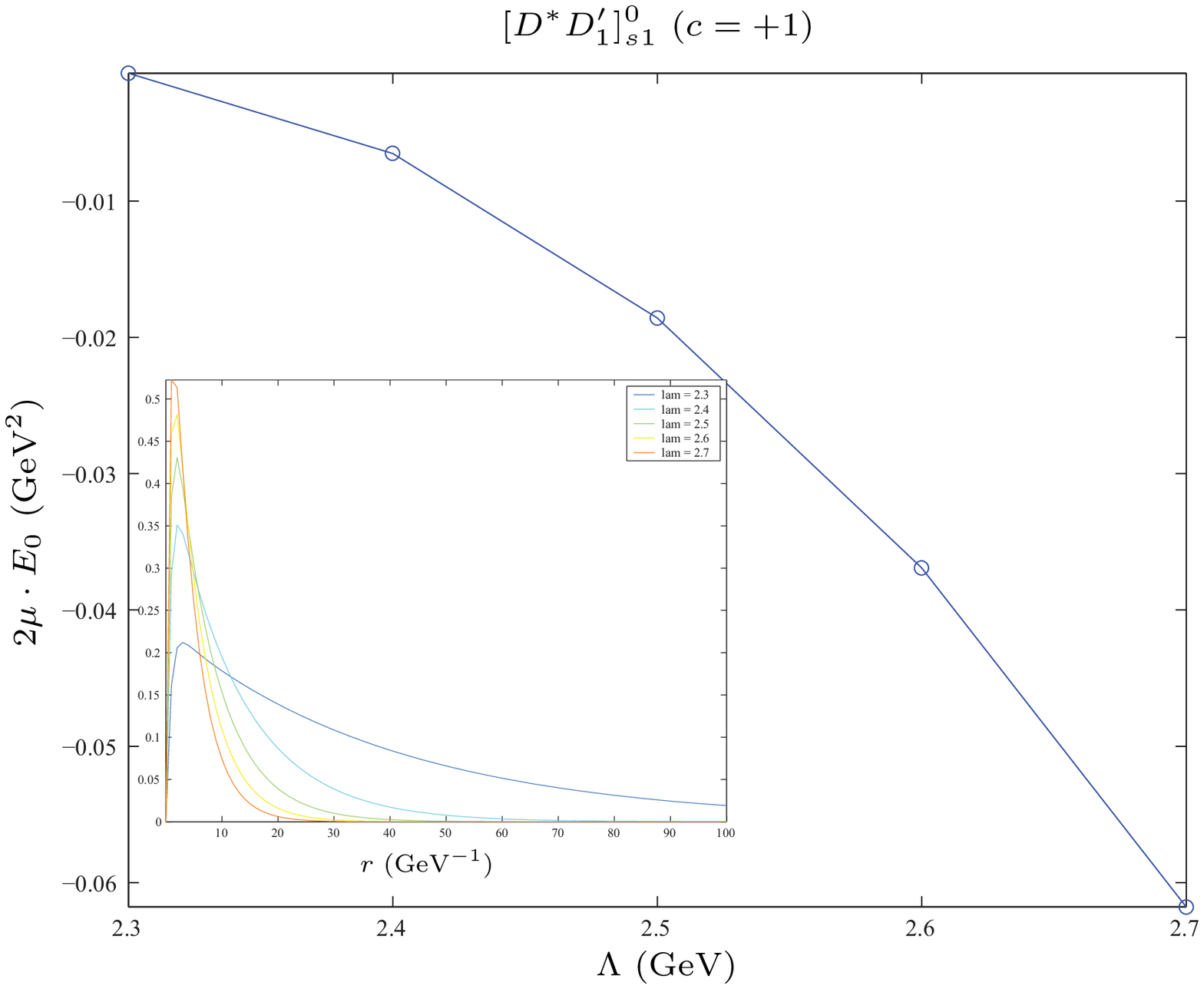}}\\
(d)&(e)&(f)\\
\scalebox{0.3}{\includegraphics{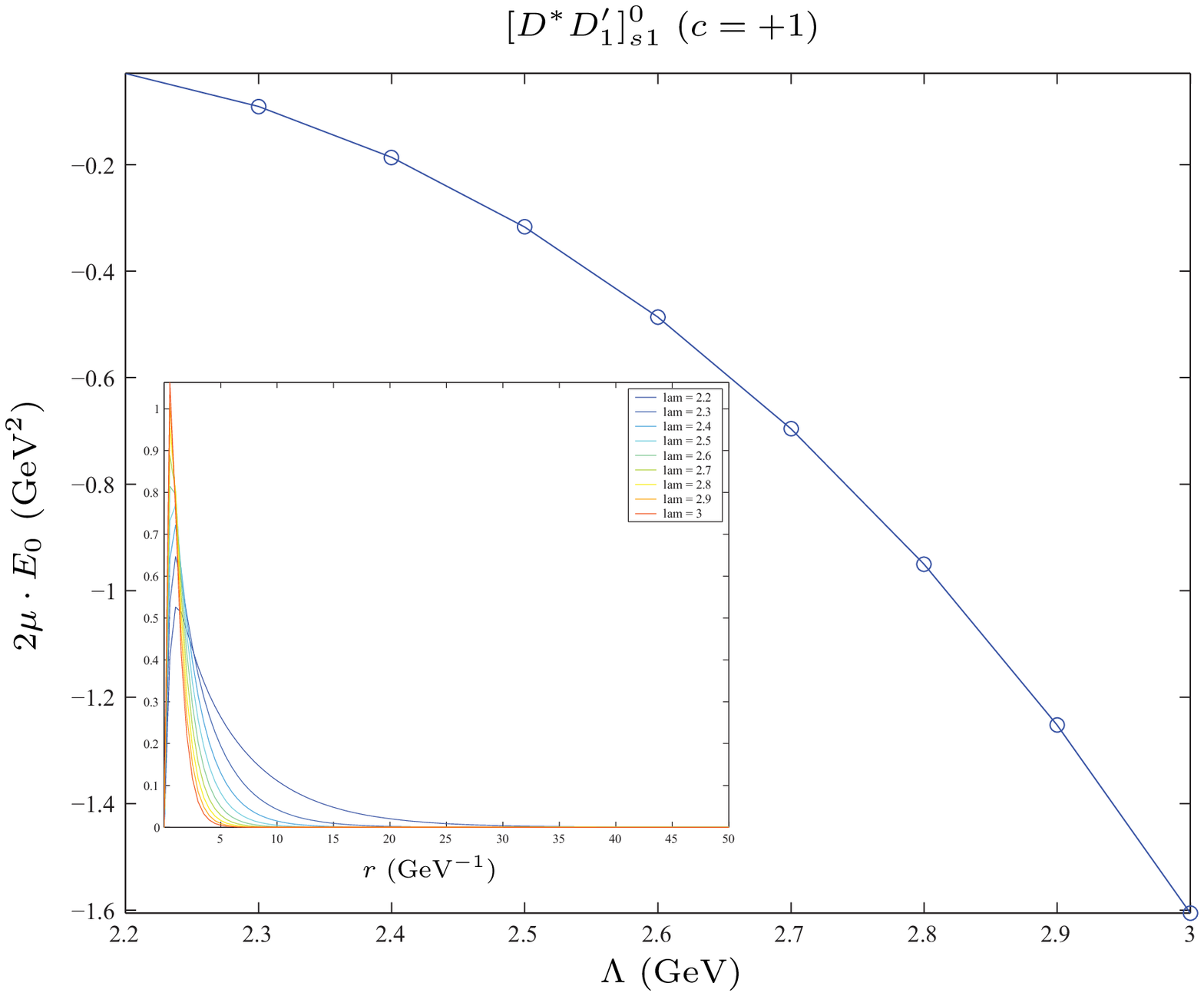}}&\scalebox{0.3}{\includegraphics{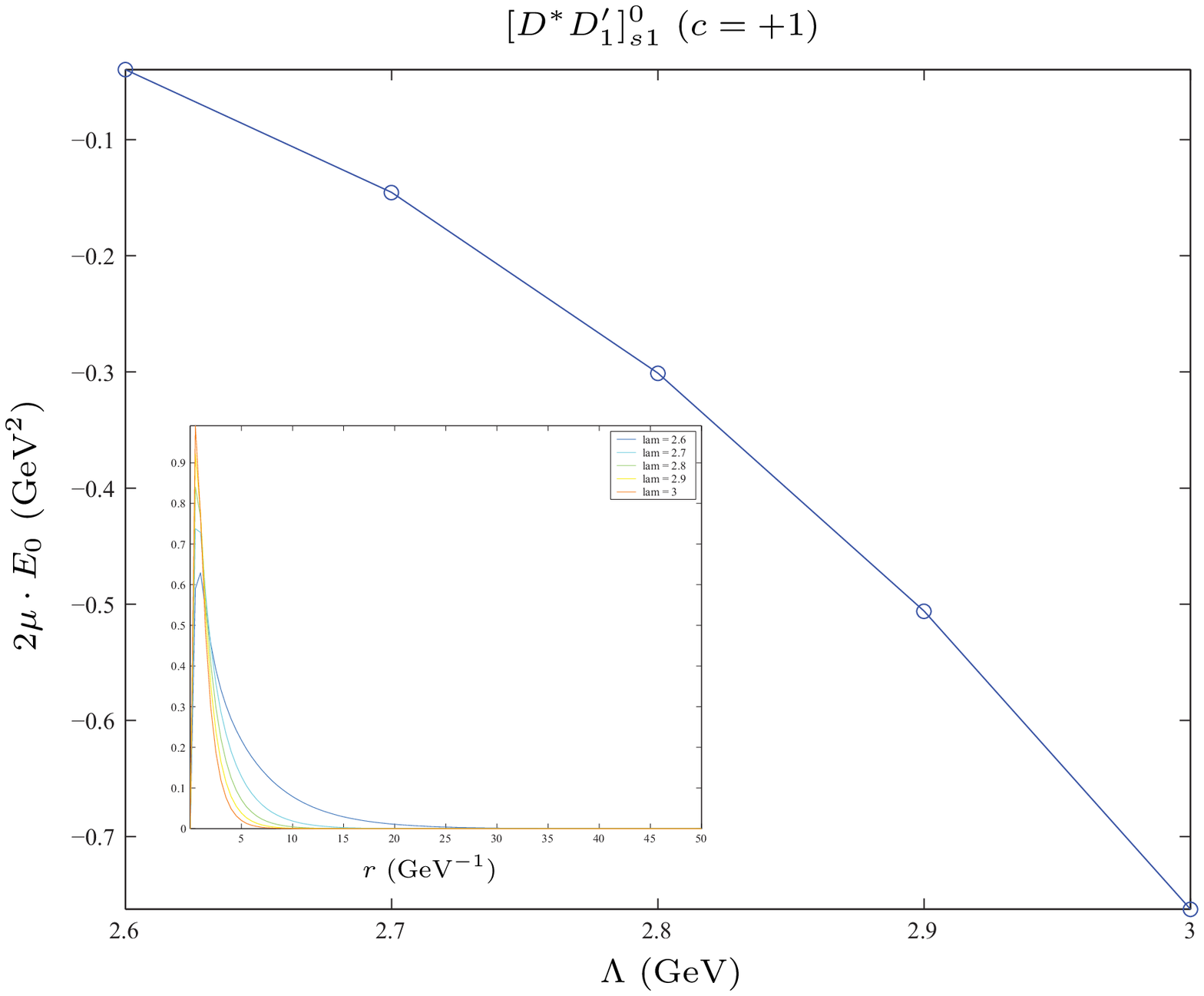}}&\\
(g)&(h)\\
\end{tabular}
\caption{The bound state solution of the
$[D^*D_1^\prime]_{s1}^{0}$ system with $(c=-1,\,J=0,\,\beta\beta^\prime<0,\,\lambda\lambda^\prime<0,\,g g^\prime<0,\,\zeta\varpi<0)$, $(c=-1,\,J=0,\,\beta\beta^\prime<0,\,\lambda\lambda^\prime<0,\,g g^\prime<0,\,\zeta\varpi>0)$, $(c=-1,\,J=0,\,\beta\beta^\prime>0,\,\lambda\lambda^\prime<0,\,g g^\prime<0,\,\zeta\varpi>0)$,
$(c=+1,\,J=0,\,\beta\beta^\prime<0,\,\lambda\lambda^\prime<0,\,g g^\prime<0,\,\zeta\varpi<0)$, $(c=+1,\,J=0,\,\beta\beta^\prime<0,\,\lambda\lambda^\prime<0,\,g g^\prime<0,\,\zeta\varpi>0)$,
 $(c=+1,\,J=0,\,\beta\beta^\prime<0,\,\lambda\lambda^\prime<0,\,g g^\prime>0,\,\zeta\varpi<0)$,
 $(c=+1,\,J=0,\,\beta\beta^\prime>0,\,\lambda\lambda^\prime<0,\,g g^\prime<0,\,\zeta\varpi<0)$ and
 $(c=+1,\,J=0,\,\beta\beta^\prime>0,\,\lambda\lambda^\prime<0,\,g g^\prime<0,\,\zeta\varpi>0)$,
 which correspond to diagrams (a), (b), (c), (d), (e), (f), (g) and (h) respectively.
 Here, we take $1h$.  $|h|=0.56$, $|g|=0.75$, $|g^\prime|=0.25$, $|\lambda|=0.4546$,
 $|\lambda^\prime|=0.2667$, $|\beta|=0.909$, $|\beta^\prime|=0.533$, $|\zeta|=0.727$
 and $|\varpi|=0.364$.  \label{dsd1s3-E}}
\end{figure}
\end{center}

\begin{center}
\begin{figure}[htb]
\begin{tabular}{cccc}
\scalebox{0.4}{\includegraphics{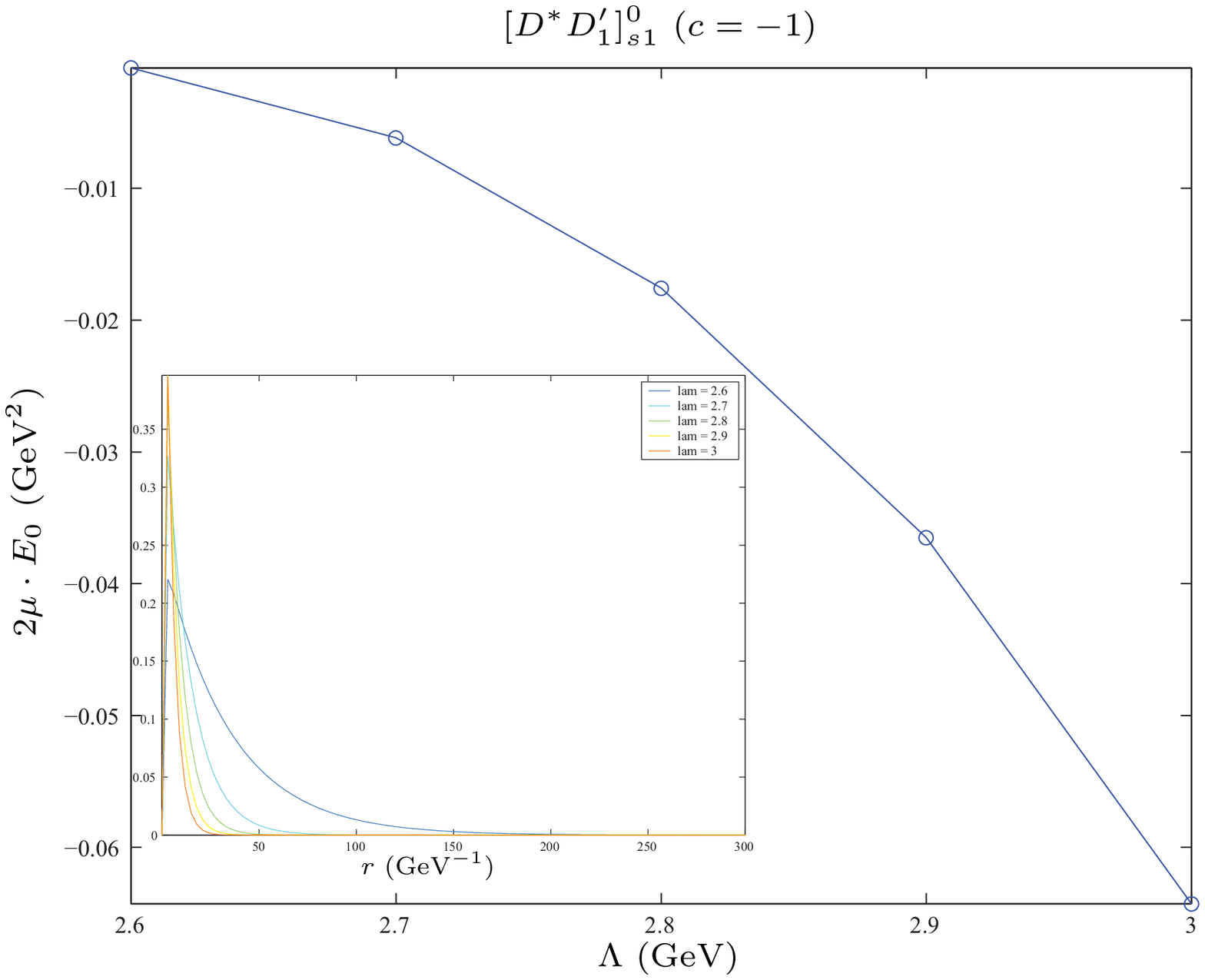}}&\scalebox{0.4}{\includegraphics{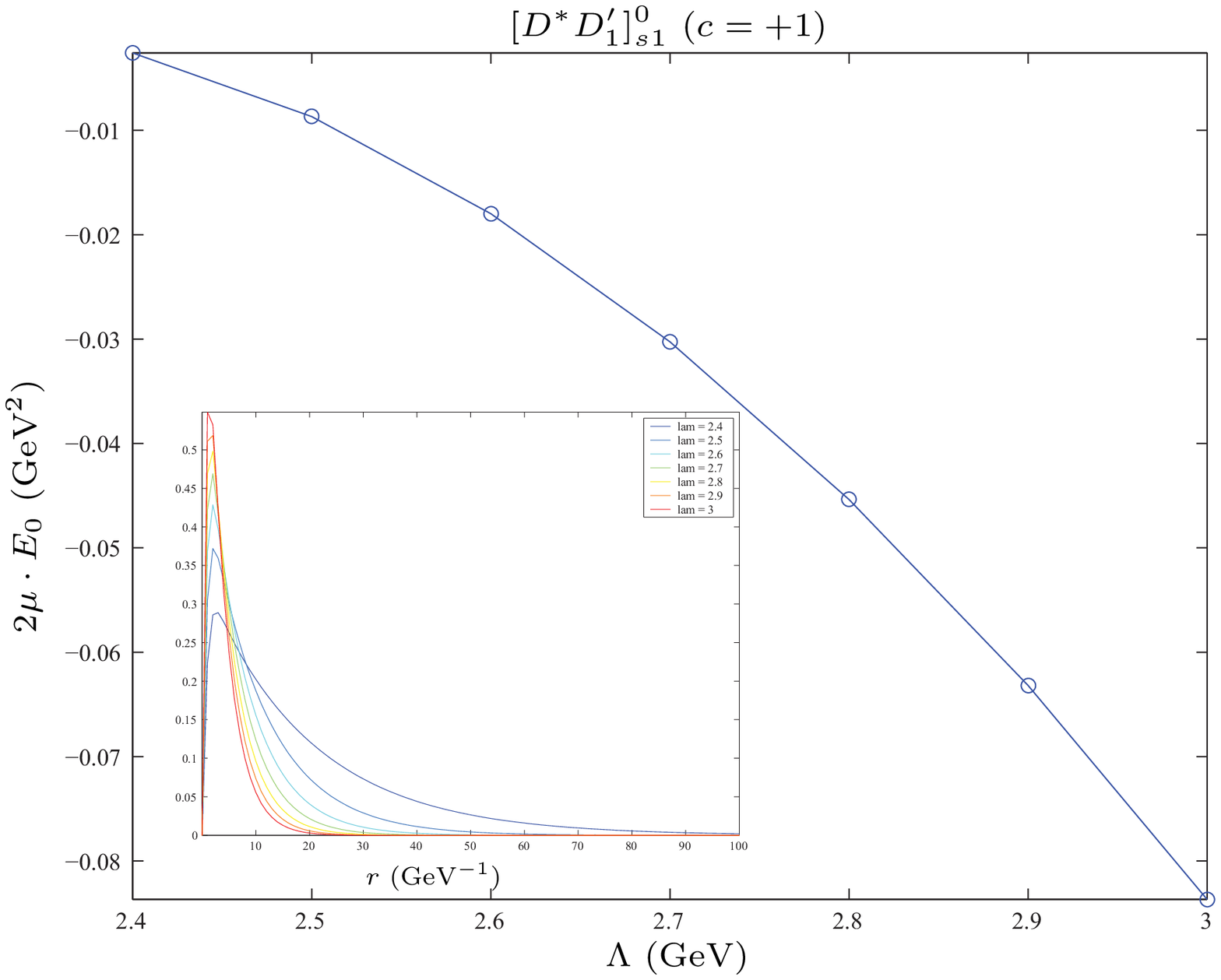}}&\\
(a)&(b)\\
\scalebox{0.4}{\includegraphics{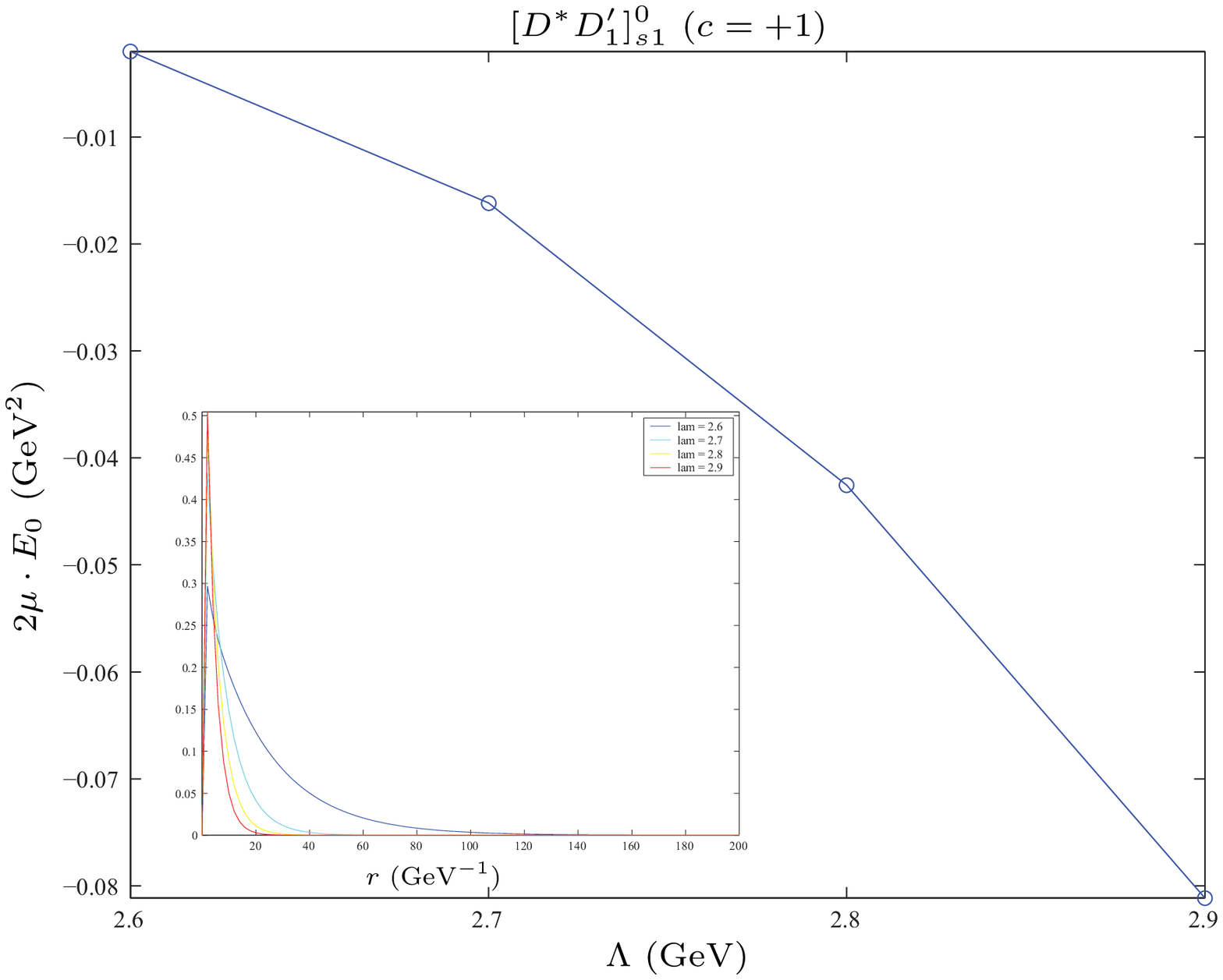}}&\scalebox{0.4}{\includegraphics{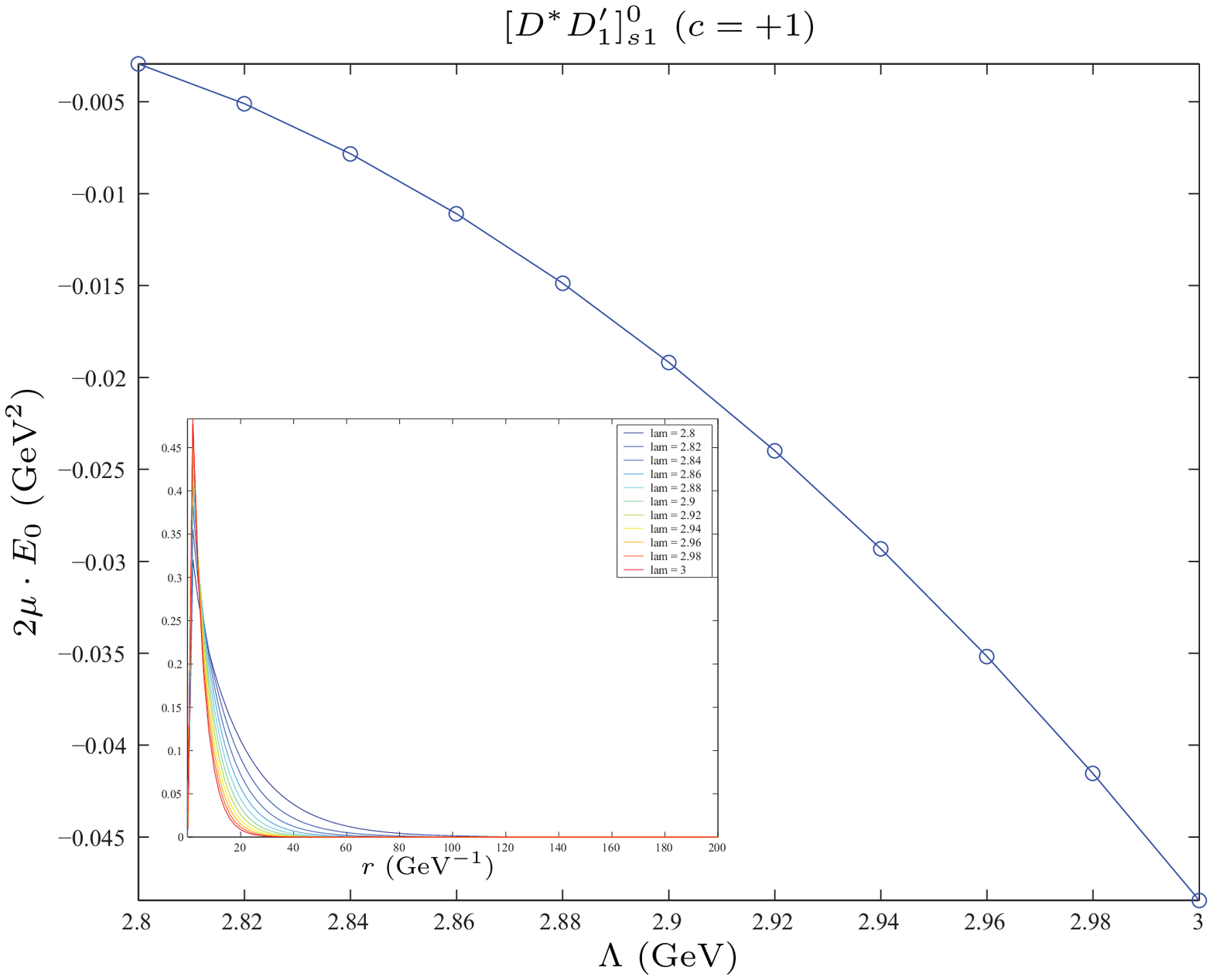}}\\
(c)&(d)\\
\scalebox{0.4}{\includegraphics{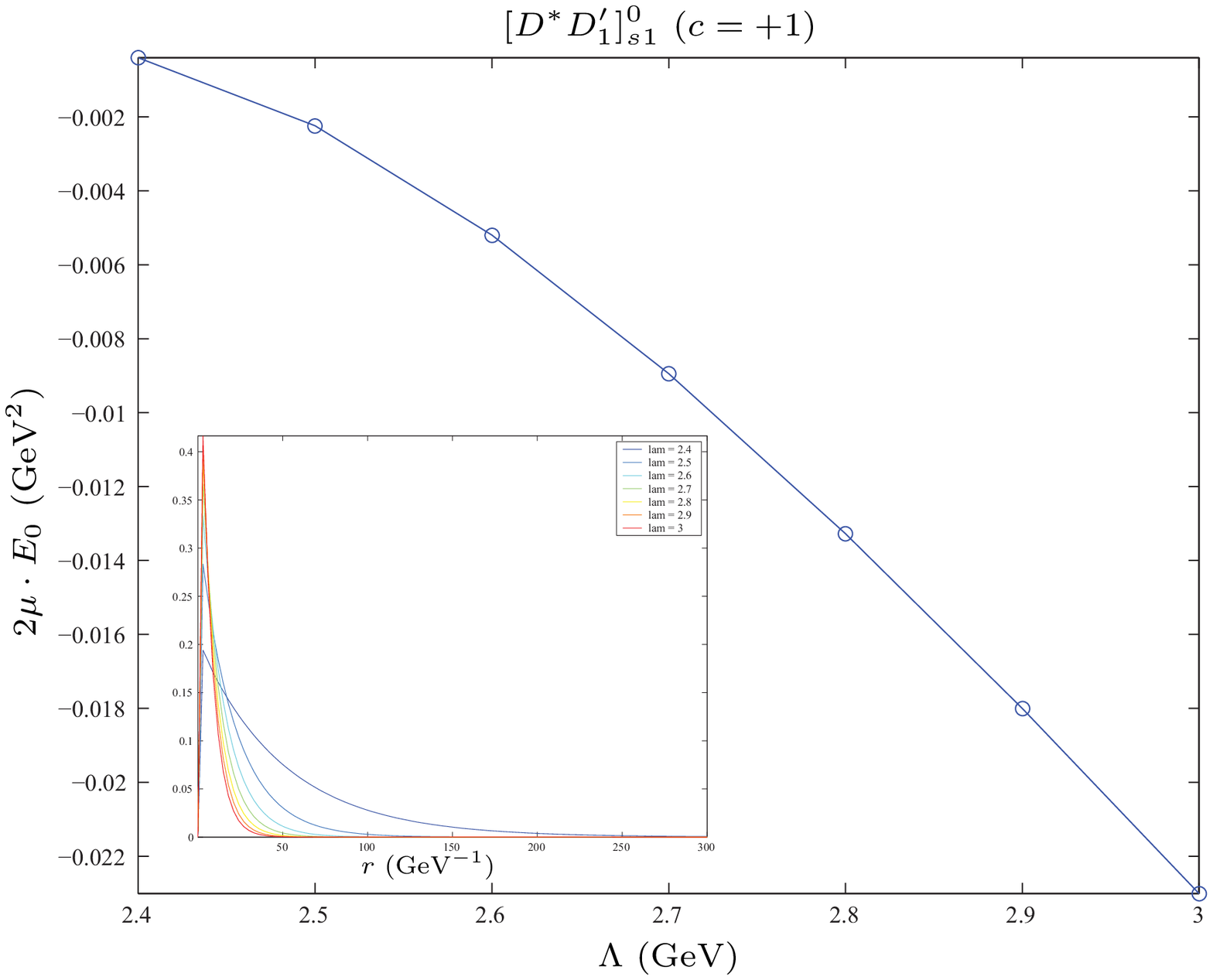}}&\scalebox{0.4}{\includegraphics{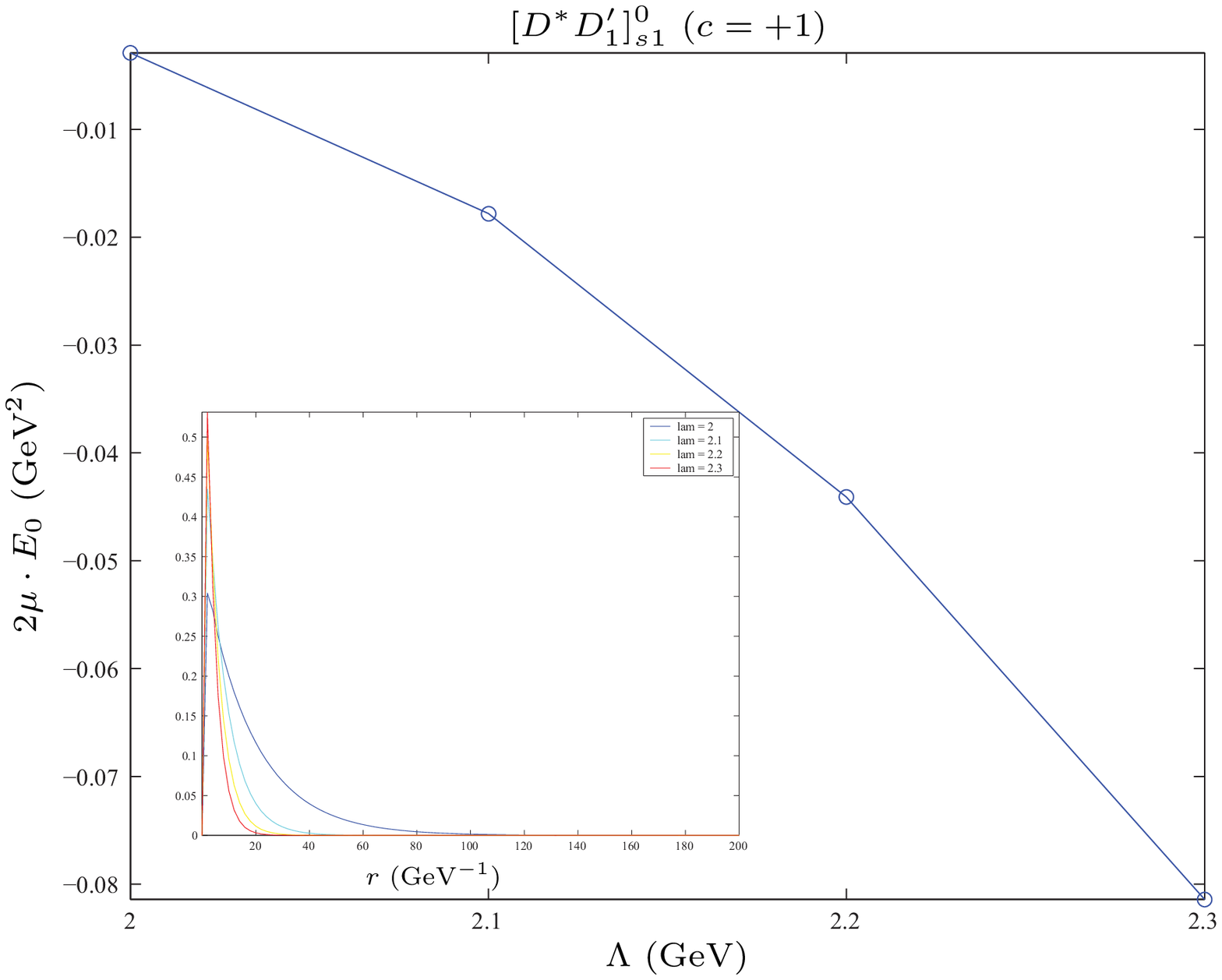}}&\\
(e)&(f)\\
\end{tabular}
\caption{The bound state solution of the
$[D^*D_1^\prime]_{s1}^{0}$ system with $(c=-1,\,J=1,\,\beta\beta^\prime<0,\,\lambda\lambda^\prime<0,\,g g^\prime<0,\,\zeta\varpi>0)$, $(c=+1,\,J=1,\,\beta\beta^\prime>0,\,\lambda\lambda^\prime<0,\,g g^\prime>0,\,\zeta\varpi<0)$, $(c=+1,\,J=1,\,\beta\beta^\prime>0,\,\lambda\lambda^\prime<0,\,g g^\prime<0,\,\zeta\varpi>0)$,
$(c=+1,\,J=1,\,\beta\beta^\prime>0,\,\lambda\lambda^\prime<0,\,g g^\prime<0,\,\zeta\varpi<0)$,  $(c=+1,\,J=1,\,\beta\beta^\prime<0,\,\lambda\lambda^\prime>0,\,g g^\prime<0,\,\zeta\varpi<0)$ and
 $(c=+1,\,J=1,\,\beta\beta^\prime<0,\,\lambda\lambda^\prime<0,\,g g^\prime<0,\,\zeta\varpi<0)$,
 which correspond to diagrams (a), (b), (c), (d), (e) and (f) respectively. Here,
 we use $1h$. $|h|=0.56$, $|g|=0.75$, $|g^\prime|=0.25$, $|\lambda|=0.4546$,
 $|\lambda^\prime|=0.2667$, $|\beta|=0.909$, $|\beta^\prime|=0.533$, $|\zeta|=0.727$
 and $|\varpi|=0.364$.  \label{dsd1s3-E-1}}
\end{figure}
\end{center}

\begin{center}
\begin{figure}[htb]
\begin{tabular}{cccc}
\scalebox{0.4}{\includegraphics{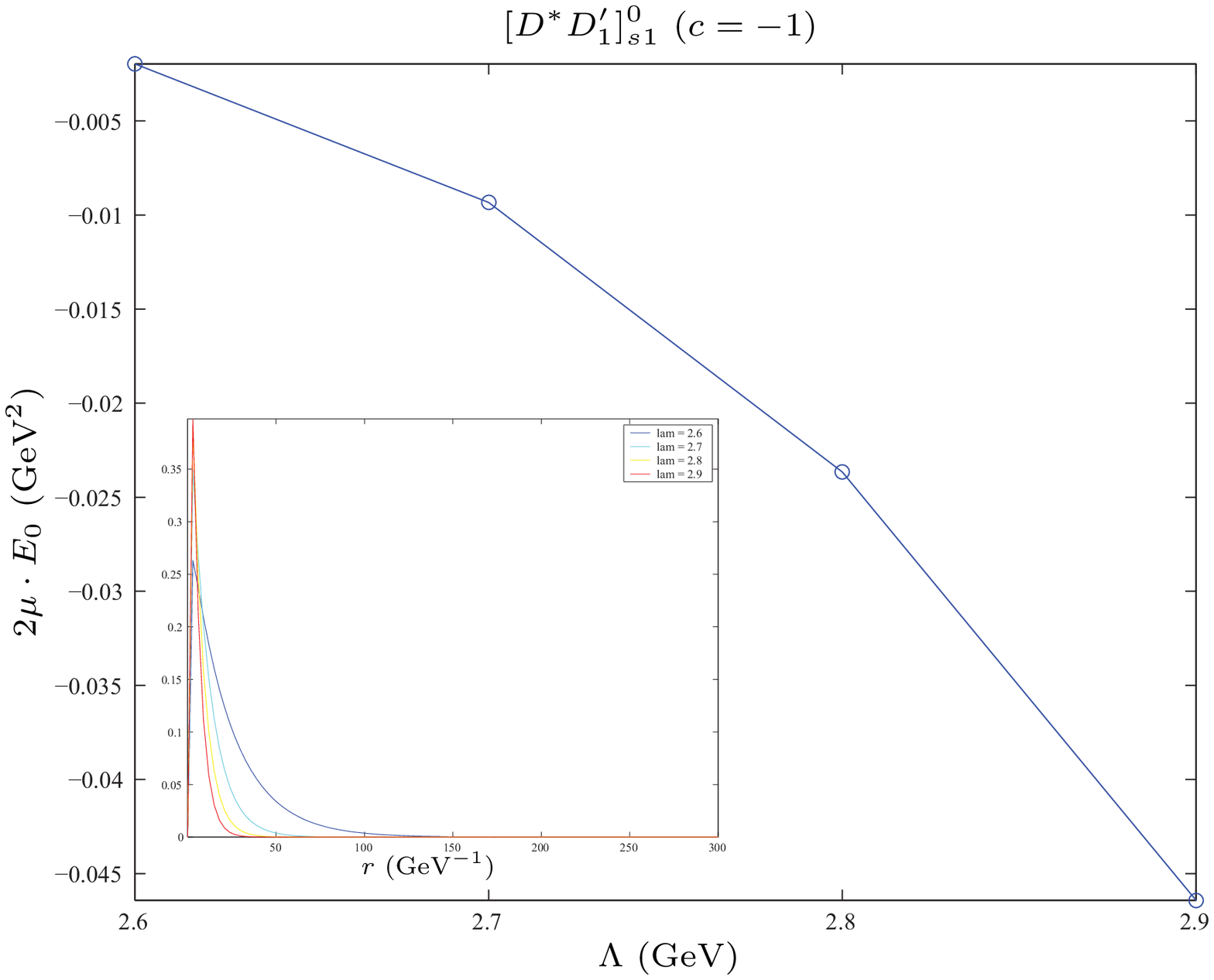}}&\scalebox{0.4}{\includegraphics{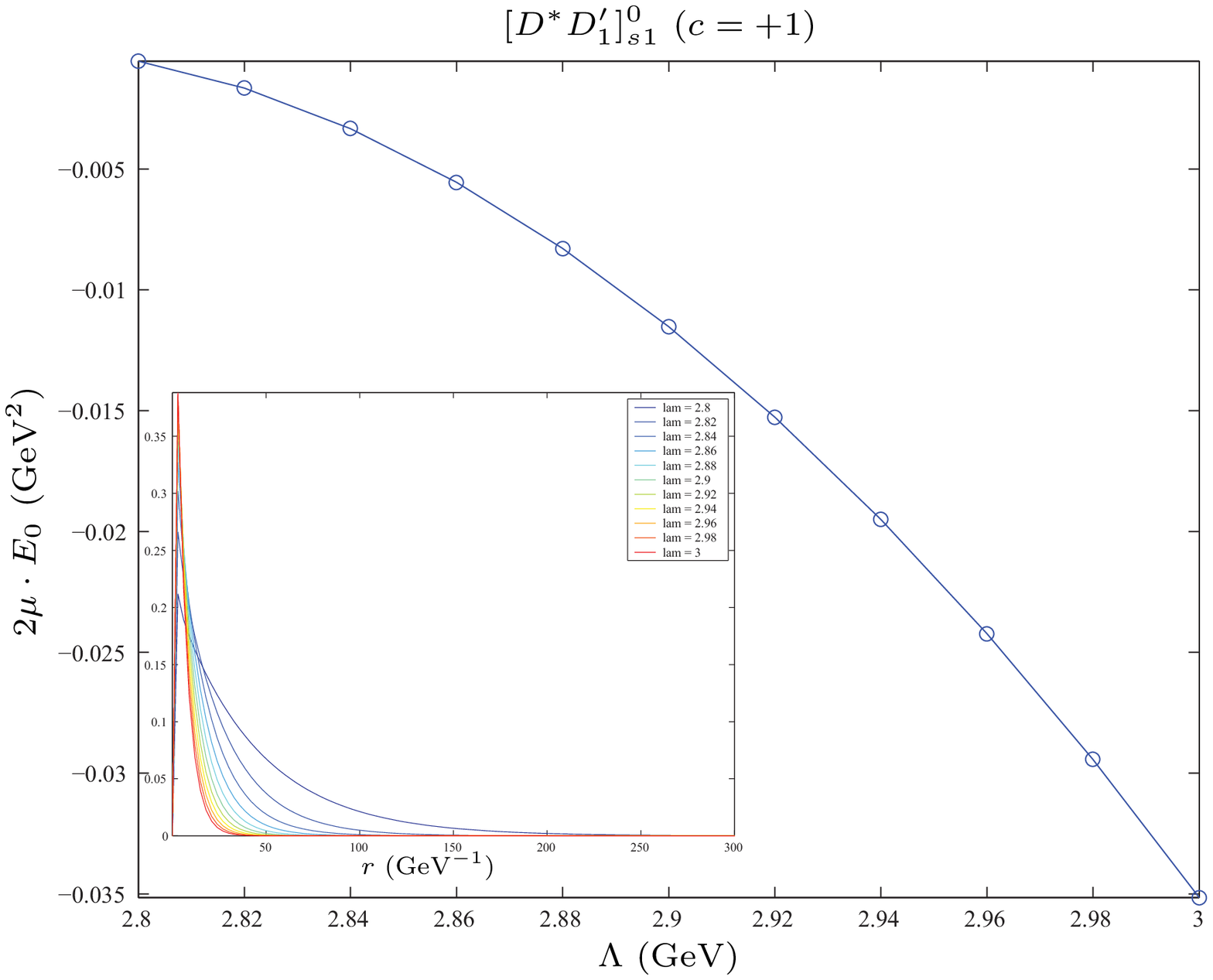}}&\\
(a)&(b)\\
\scalebox{0.4}{\includegraphics{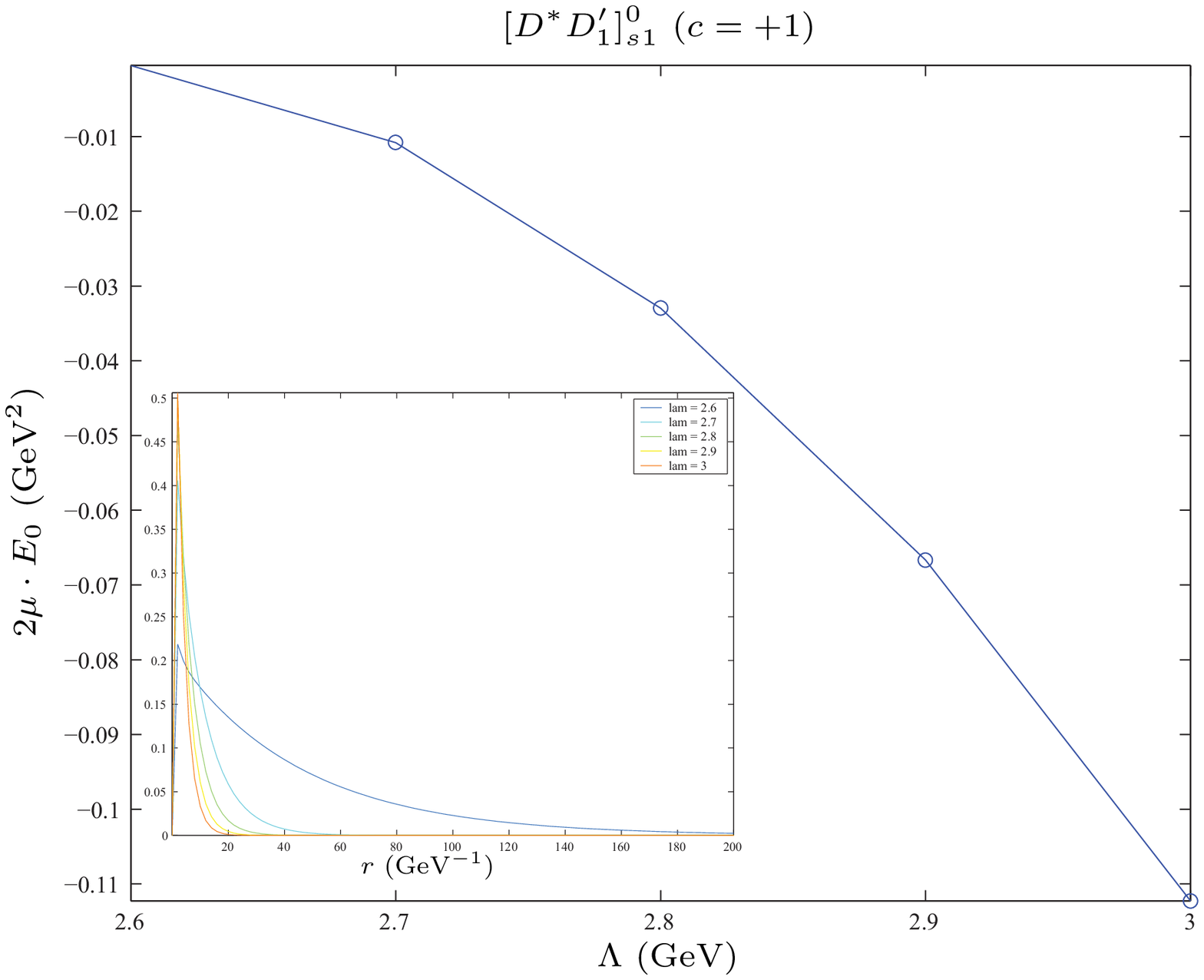}}&\scalebox{0.4}{\includegraphics{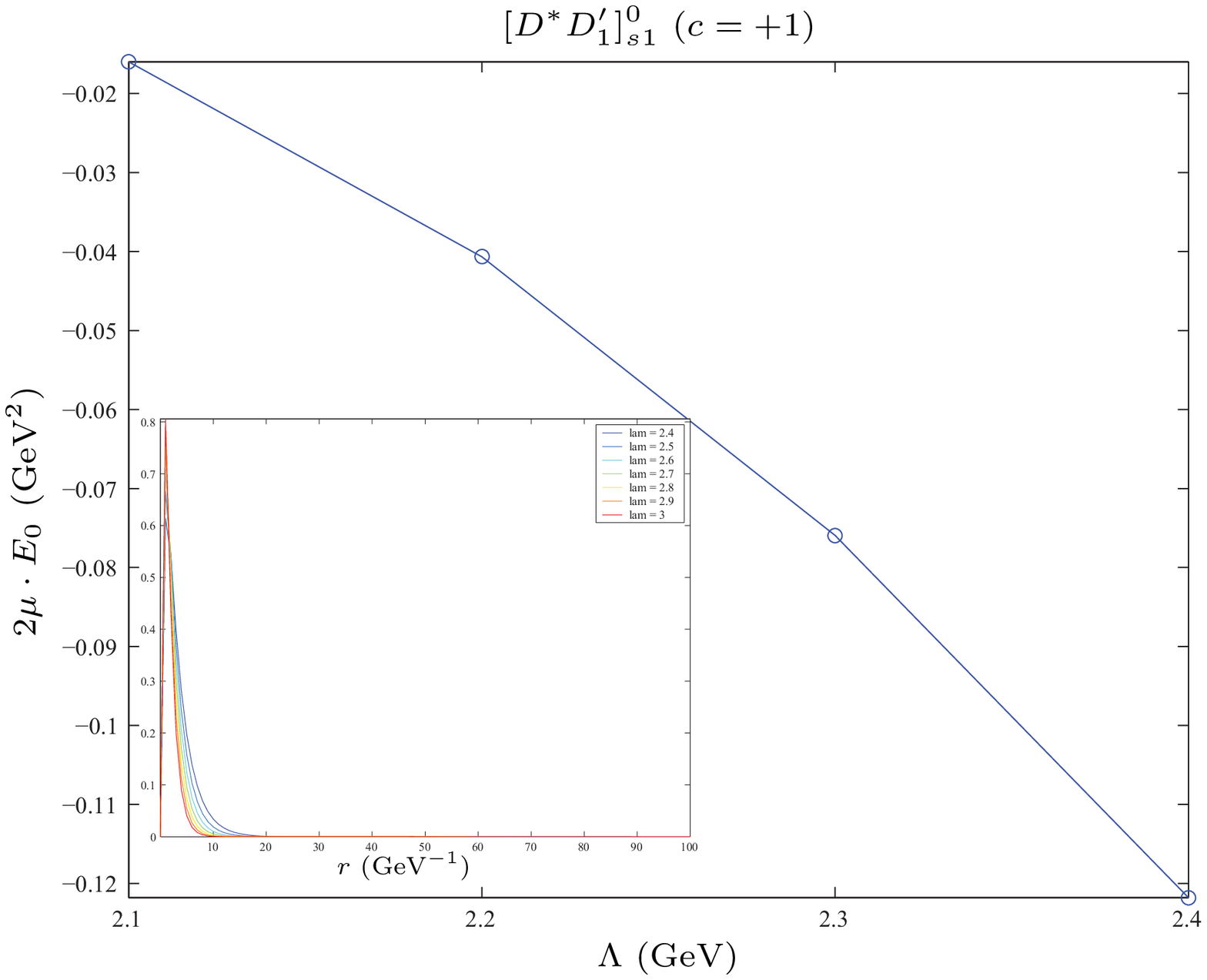}}\\
(c)&(d)\\
\scalebox{0.4}{\includegraphics{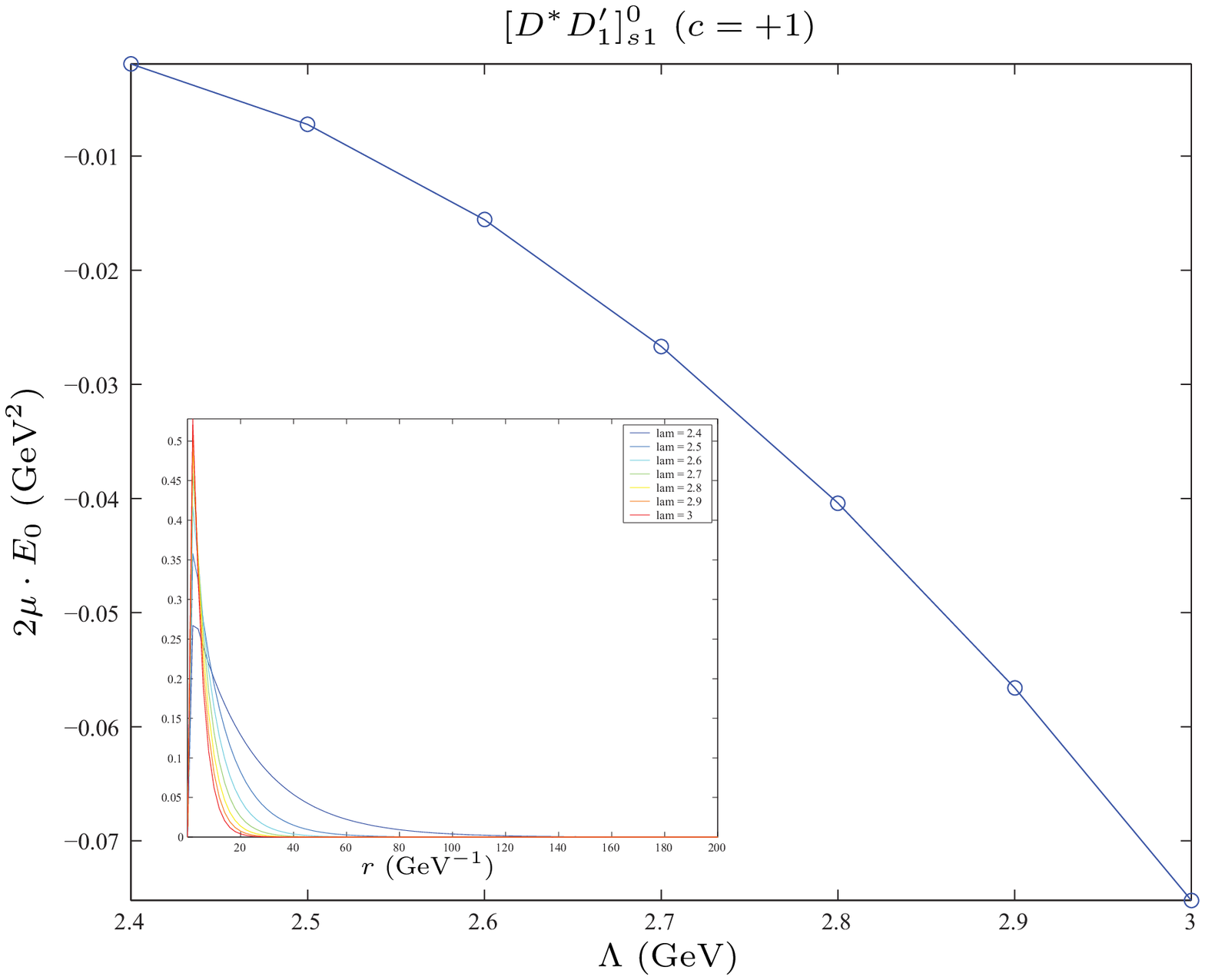}}&\scalebox{0.4}{\includegraphics{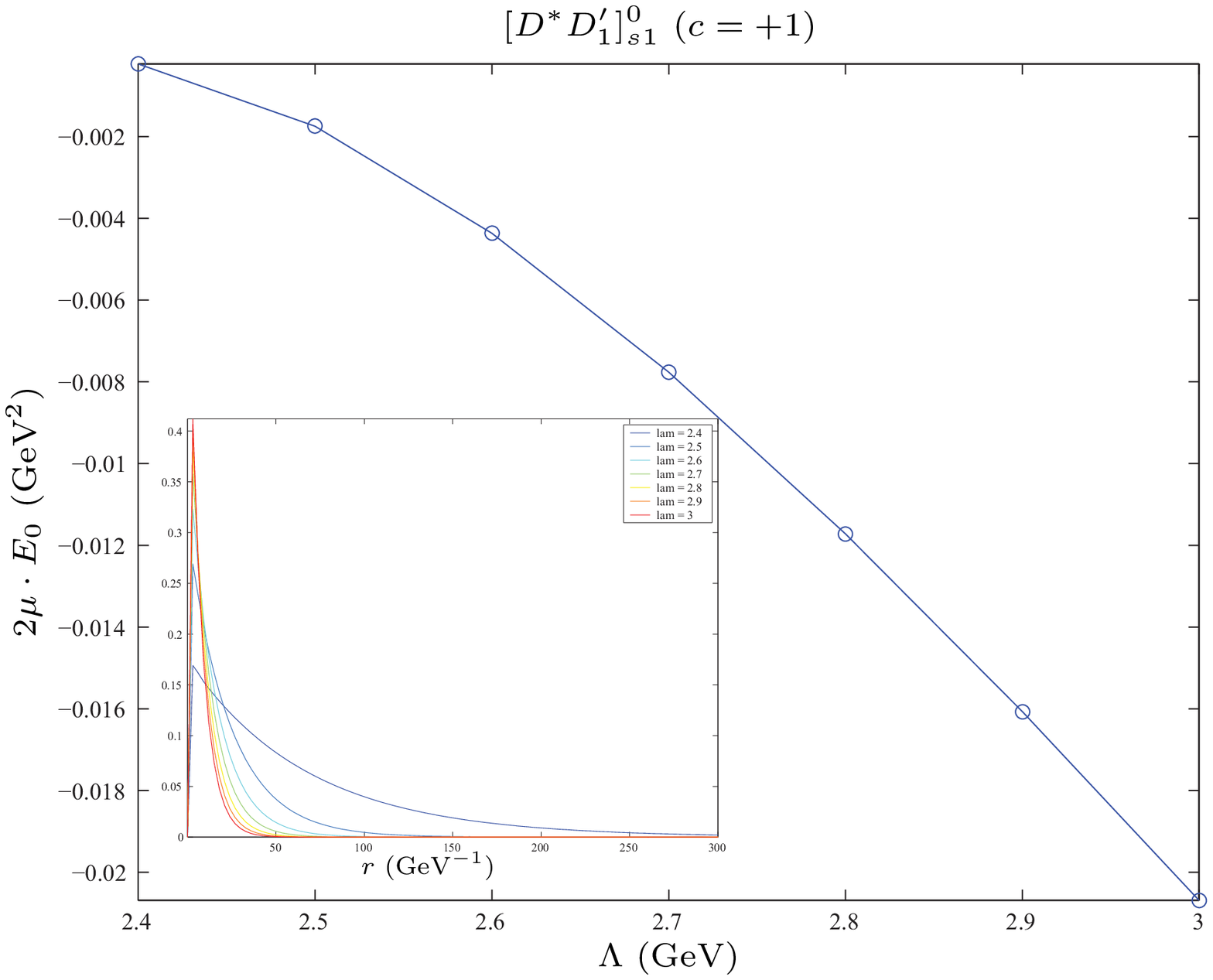}}&\\
(e)&(f)\\
\end{tabular}
\caption{The bound state solution of the
$[D^*D_1^\prime]_{s1}^{0}$ system with $(c=-1,\,J=2,\,\beta\beta^\prime<0,\,\lambda\lambda^\prime>0,\,g g^\prime>0,\,\zeta\varpi>0)$, $(c=+1,\,J=2,\,\beta\beta^\prime>0,\,\lambda\lambda^\prime>0,\,g g^\prime>0,\,\zeta\varpi<0)$, $(c=+1,\,J=2,\,\beta\beta^\prime<0,\,\lambda\lambda^\prime>0,\,g g^\prime>0,\,\zeta\varpi>0)$,
$(c=+1,\,J=2,\,\beta\beta^\prime<0,\,\lambda\lambda^\prime>0,\,g g^\prime>0,\,\zeta\varpi<0)$,  $(c=+1,\,J=2,\,\beta\beta^\prime<0,\,\lambda\lambda^\prime>0,\,g g^\prime<0,\,\zeta\varpi<0)$ and
$(c=+1,\,J=2,\,\beta\beta^\prime<0,\,\lambda\lambda^\prime<0,\,g g^\prime>0,\,\zeta\varpi<0)$,
 which correspond to diagrams (a), (b), (c), (d), (e) and (f) respectively. Here,
 we use $1h$. $|h|=0.56$, $|g|=0.75$, $|g^\prime|=0.25$, $|\lambda|=0.4546$,
 $|\lambda^\prime|=0.2667$, $|\beta|=0.909$, $|\beta^\prime|=0.533$, $|\zeta|=0.727$
 and $|\varpi|=0.364$.  \label{dsd1s3-E-2}}
\end{figure}
\end{center}

There exist 96 combinations of the parameters $c$,
$\beta\beta^\prime$, $gg^\prime$, $\lambda\lambda^\prime$ and
$\zeta\varpi$. When solving the Schr\"{o}dinger equation, the
cutoff $\Lambda$ should be larger than the mass of the $\phi$
meson. We only find bound state solutions of the
$[P^*P_1^\prime]_{s1}^{0}$ system under 20 combinations of $c$,
$\beta\beta^\prime$, $gg^\prime$, $\lambda\lambda^\prime$ and
$\zeta\varpi$, which are shown in Figs. \ref{dsd1s3-E}, \ref{dsd1s3-E-1}  and
\ref{dsd1s3-E-2} for $J=0,\, 1,\, 2$ cases respectively.

\section{Conclusion}\label{sec4}

We have investigated the possible molecular states composed of a
pair of heavy mesons in the $H$ and $S$ doublet. They are expected to
be loosely bound states mainly via the long range pseudoscalar
scalar meson exchange. We are interested in these shallow S-wave
states with $J^P=0^-, 1^-, 2^-$, especially those neutral ones
with either $C=+$ or $C=-$. Those states with $J^{PC}=0^{--},
1^{-+}$ are exotic. They may exist with very reasonable coupling
constants. For example, the $J^{PC}=0^{--}$ state is shown in Fig.
\ref{dds1e}.

The non-strange P-wave $(0^+, 1^+)$ heavy mesons are very broad
with a width around several hundred MeV \cite{pdg}. Instead of
forming a stable molecular state, the system containing one
non-strange P-wave heavy meson may decay rapidly. Experimental
identification of such a molecular state may be difficult. The
attractive interaction between the meson pair may lead to a
possible threshold enhancement in the production cross section.

In contrast, the P-wave charm-strange mesons $D_{s0}(2317)$ and
$D_{s1}(2460)$ lie below the $DK$ and $D^\ast K$ threshold
respectively. They are extremely narrow because their strong
decays violate the isospin symmetry. The future experimental
observation of the possible heavy molecular states composed of
$D_s/D_s^\ast$ and $D_{s0}(2317)/D_{s1}(2460)$ may be feasible if
they really exist.

There may exist (1) two $0^-$ states around 4.25 GeV with
different C-parity; (2) two $0^-$ states around 4.5 GeV with
different C-parity; (3) two $2^-$ states around 4.5 GeV with
different C-parity; (4) four $1^-$ states around 4.35 GeV with
different C-parity; (5) two $1^-$ states around 4.5 GeV with
different C-parity. The three neutral $J^{PC}=1^{--}$ states may
be searched for via the initial state radiation technique. We
notice that they are are rather close to those $1^{--}$ states
observed by Belle around this mass region. The other states might
be produced from $B$ or $B_s$ decays if kinematically allowed or
at other possible facilities such as RHIC, Tevatron and LHCb.

The dominant decay modes of the above states are the open-charm
modes $D_s^{(\ast)} {\bar D}_s^{(\ast)}$ if angular momentum
conservation, parity and C-parity symmetry allow. The other
characteristic decay modes are the hidden-charm modes containing
one $J/\psi $, or $\eta_c $ or $\chi_{cJ} $ etc. One may easily
exhaust the possible final states according to C/P parity and
angular momentum conservation and kinematical considerations. For
those non-exotic states, they may be significantly narrower than
the conventional charmonium around the same mass region because of
their molecular nature. The manifestly exotic states may be
searched for through their quantum number.

\section*{Acknowledgment}
The authors thank Wei-Zhen Deng for useful discussions. This
project is supported by the National Natural Science Foundation of
China under Grants No. 10625521, No. 10721063, No. 10705001, the
Ministry of Science and Technology of China (2009CB825200) and the
Ministry of Education of China (FANEDD under Grants No. 200924,
DPFIHE under Grants No. 20090211120029, NCET under Grants No.
NCET-10-0442).

\newpage

\section*{Appendix}



\end{document}